\documentclass[a4paper,11pt,fleqn]{article}

\usepackage{graphicx}
\usepackage{dcolumn}
\usepackage{bm}
\usepackage{amsmath}
\usepackage{amssymb}
\usepackage{amsfonts}
\usepackage{esvect}
\usepackage[dvipsnames]{xcolor}
\usepackage{fix-cm} 
\usepackage{cite}
\usepackage{soul}
\usepackage{mdframed}
\usepackage{lipsum}
\usepackage{multicol}
\usepackage{float}
\usepackage{longtable}
\usepackage{setspace}
\usepackage{tabularx}
\usepackage{ltablex}
\usepackage{newfloat}
\DeclareFloatingEnvironment[fileext=frm,placement={!ht},name=Box]{boxfloat}
\usepackage{capt-of}
\usepackage[htt]{hyphenat} 
\usepackage{hyperref}
\usepackage{url}
\usepackage{afterpage}
\usepackage[utf8]{inputenc}
\usepackage{longtable}
\usepackage[capitalise, english]{cleveref}
\usepackage{makecell}
\usepackage{verbatim}
\usepackage{graphicx}
\usepackage{subcaption}
\usepackage{tikz}
\usetikzlibrary{decorations.pathreplacing}
\usepackage{scalerel}

\usepackage{listings}
\usepackage{minted}
\newminted{shell-session}{
    frame=lines, breaklines=true}
\definecolor{mintbackground}{RGB}{246, 246, 246}

\newminted{C++}{framesep=2mm,
baselinestretch=1.2,
bgcolor=mintbackground,
fontsize=\small,breaklines=true}

\usepackage{empheq} 

\usepackage[symbol]{footmisc}

\newcommand{\CL}{$\mathcal{C}${\tt osmo}$\mathcal{L}${\tt attice}~}
\newcommand{\CLns}{${\mathcal C}${\tt osmo}${\mathcal L}${\tt attice}}

\newcommand{\dx}{\ensuremath{\delta x}}

\newcommand{\deta}{\delta\eta}
\newcommand{\bn}{{\bf n}}
\newcommand{\dd}{\text{d}}
\newcommand{\mpl}{m_p}

\newcommand{\be}{\begin{equation}}
\newcommand{\ee}{\end{equation}}
\newcommand{\bea}{\begin{eqnarray}}
\newcommand{\eea}{\end{eqnarray}}

\newcommand{\FFdual}{F_{\mu \nu}\,\tilde{F}^{\mu \nu}}

\definecolor{cborange}{HTML}{e69f00}
\definecolor{cbgreen}{HTML}{009e73}
\definecolor{cbyellow}{HTML}{f1dd42}
\definecolor{cblblue}{HTML}{56b4e9}
\definecolor{cbblue}{HTML}{0072b2}
\definecolor{defgrey}{HTML}{9f9f9f}
\definecolor{defgreen}{HTML}{8eba42}

\newcommand{\affil}[2]{%
  \parbox[c]{0.95\linewidth}{\centering
    \textit{$^{#1}$~}\ignorespaces#2%
  }%
}

\crefname{equation}{Eq.}{Eqs.}
\crefname{section}{Sect.}{Sects.}
\Crefname{equation}{Eq.}{Eqs.}
\Crefname{section}{Sec.}{Sects.}
\newcommand{\rcite}[1]{ref.~\cite{#1}}
\newcommand{\rrcite}[1]{refs.~\cite{#1}}

\newcommand{\cM}{\mathcal{M}}
\newcommand{\cK}{\mathcal{K}}
\newcommand{\cP}{\mathcal{P}}
\newcommand{\cO}{\mathcal{O}}

\newcommand{\fstar}{f_*}
\newcommand{\omegastar}{\omega_*}
\newcommand{\str}{\text{str}}

\newcommand{\addressDESY}{\textit{Deutsches Elektronen-Synchrotron DESY, Platanenallee 6, 15738 Zeuthen, Germany}}

\newcommand{\addressIFIC}{%
\textit{Instituto de Física Corpuscular (IFIC), Universitat de València (UV) and}\\
\textit{Consejo Superior de Investigaciones Científicas (CSIC), 46980, Valencia, Spain}%
}

\newcommand{\addressIFIConeLine}{\textit{Instituto de Física Corpuscular (IFIC), UV-CSIC, 46980, Valencia, Spain}}

\newcommand{\addressBielefeld}{\textit{Fakult\"at f\"ur Physik, Universit\"at Bielefeld, D-33615 Bielefeld, Germany}}

\newcommand{\addressEHU}{\textit{Department of Physics, University of the Basque Country UPV/EHU, 48080 Bilbao, Spain}}

\newcommand{\addressEHUQC}{%
\textit{EHU Quantum Center, University of the Basque Country UPV/EHU, Leioa, 48940 Biscay, Spain}%
}

\newcommand{\addressEHUoneLine}{\textit{Department of Physics and EHU Quantum Center, UPV/EHU, Leioa, 48940 Biscay, Spain}}

\newcommand{\addressFZU}{%
\textit{CEICO-FZU, Institute of Physics of the Czech Academy of Sciences,}\\
\textit{Na Slovance 1999/2, 182 00, Prague, Czechia}%
}

\newcommand{\addressSISSA}{\textit{SISSA -- International School for Advanced Studies}  
\textit{Via Bonomea 265, 34136, Trieste, Italy}}

\newcommand{\addressINFN}{\textit{INFN, Sezione di Trieste, Via Bonomea 265, 34136, Trieste, Italy}}

\newcommand{\addressUB}{%
\textit{Departament de F\'isica Qu\`antica i Astrof\'isica \& Institut de Ci\`encies del Cosmos}\\
\textit{(ICCUB), Universitat de Barcelona, Mart\'i i Franqu\`es 1, 08028 Barcelona, Spain}%
}

\newcommand{\addressUBoneLine}{\textit{Departament de F\'isica Qu\`antica i Astrof\'isica \& ICCUB, Universitat de Barcelona, Spain}}

\usepackage{etoolbox}
\makeatletter
\patchcmd{\@maketitle}
  {\vskip 2em}
  {\vskip-1cm{\hfill DESY-25-191}}
  {}
  {}
\makeatother

\begin{document}

\begin{figure}
\vspace*{-0.5cm}
\hspace{0.0cm}
\includegraphics[width = 8.6cm]{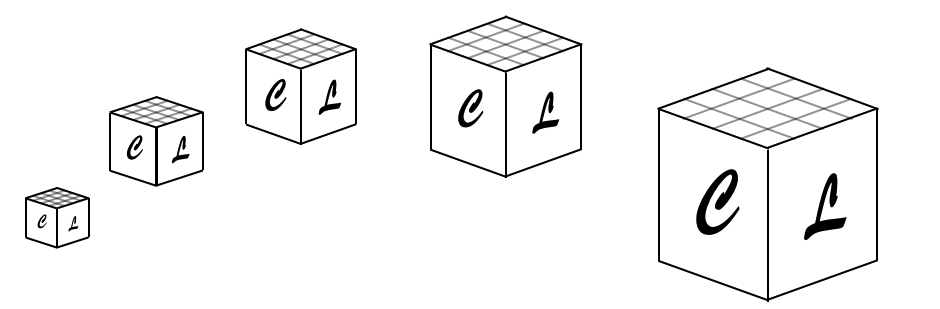}~
\includegraphics[width = 8.6cm]{CL_iconSequence.png}
\end{figure}

\begin{center}
\vspace*{-0.5cm}
{\fontsize{26.5}{0} \bf\textsf{The art of simulating the early Universe}}
\\[0.15cm]
$\hspace*{2mm}\left(~
\makecell{
\text{\LARGE \it A dissertation series about lattice techniques for simulating} \vspace{1mm}\\\text{\LARGE \it the dynamics of interacting fields in an expanding Universe}
} ~~\right)$
\vspace*{1.0cm}\\

{\fontsize{19.0}{0}\bf\textsf{Part II.\, Non-Canonical Cases 
\& Gravitational Waves}}
\\[1.25cm]

{\Large  Jorge Baeza-Ballesteros}\\
\addressDESY
\\[0.5cm]
{\Large \rm Daniel~G.~Figueroa}\\
\addressIFIConeLine
\\[0.5cm]
{\Large \rm Adrien Florio }\\
\addressBielefeld
\\[0.5cm]
{\Large \rm Joanes Lizarraga}\\
\addressEHUoneLine
\\[0.5cm]
{\Large \rm Nicol\'as Loayza}\\
\addressFZU 
\\[0.5cm]
{\Large \rm Kenneth Marschall}\\
\addressIFIConeLine
\\[0.5cm]
{\Large \rm Toby Opferkuch}\\
\addressSISSA \\
\addressINFN
\\[0.5cm]
{\Large \rm Ben A. Stefanek}\\
\addressIFIConeLine
\\[0.5cm]
{\Large \rm Francisco Torrent\'i}\\ 
\addressUBoneLine
\\[0.5cm]
{\Large \rm Ander Urio}\\
\addressEHUoneLine

\footnotetext{Corresponding author(s): jorge.baeza.ballesteros@desy.de, loayza@fzu.cz}

\vspace{.3cm}
\end{center}

\thispagestyle{empty}
\addtocounter{page}{-1}
\newpage
\,
\thispagestyle{empty}
\addtocounter{page}{-1}
\newpage

\setcounter{page}{1}

\title{\bf The art of simulating the early Universe. Part II}

\author{%
Jorge Baeza-Ballesteros\,$^{1}$, Daniel G. Figueroa\,$^{2}$, Adrien Florio\,$^{3}$, Joanes Lizarraga\,$^{4,5}$,\\
Nicol\'as Loayza\,$^{2,6}$, Kenneth Marschall\,$^{2}$, Toby Opferkuch\,$^{7,8}$, Ben A. Stefanek\,$^{2}$,\\
Francisco Torrent\'i\,$^{9}$, Ander Urio\,$^{4,5}$\vspace*{0.35cm}\\
\affil{1}{\small\addressDESY}\\
\affil{2}{\small\addressIFIC}\\
\affil{3}{\small\addressBielefeld}\\
\affil{4}{\small\addressEHU}\\
\affil{5}{\small\addressEHUQC}\\
\affil{6}{\small\addressFZU}\\
\affil{7}{\small\addressSISSA}\\
\affil{8}{\small\addressINFN}\\
\affil{9}{\small\addressUB}\\
}
\date{}
\maketitle

\begin{abstract}
We present 
a discussion on lattice techniques for the simulation of non-canonical field theory circumstances, complementing our previous monograph~\cite{Figueroa:2020rrl} on canonical cases. 
We begin by reviewing basic aspects of lattice field theory, including symplectic and non-symplectic evolution algorithms. We then introduce lattice implementations of non-canonical  
interactions, considering 
scalars with a non-minimal  coupling to gravity, $\phi^2R$, non-minimal scalar kinetic theories, $\mathcal{G}_{ab}(\lbrace\phi_c\rbrace)\partial_\mu\phi^a\partial^\mu\phi^b$, and axion-like particle (ALP) interactions with Abelian gauge fields, $\phi F_{\mu\nu}\tilde F^{\mu\nu}$. Next, we discuss methods to set up special field configurations, including the creation of cosmic defect networks 
towards scaling ({\it e.g.}~cosmic strings and domain walls), 
field configurations based on arbitrary power spectra 
or spatial profiles, 
and probabilistic methods as required {\it e.g.}~for thermal configurations. 
We further extend the notion of non-canonical theories, 
discussing the discretization of scalar field 
dynamics in $d + 1$ dimensions, with 
$d \neq 3$. Unrelated to non-canonical aspects, we also discuss implementation(s) of gravitational wave (GW) dynamics on the lattice. This document represents the theoretical basis for the 
non-canonical field theory aspects (interactions, initial conditions, dimensionality) and GW dynamics implemented in
{\color{blue}\ttfamily \href{http://www.cosmolattice.net}{${\mathcal C}$osmo${\mathcal L}$attice~v2.0}} to be released in 2026. 
\end{abstract}

\tableofcontents

\clearpage

\section*{Conventions and Notation}
\label{sec:Conventions}
\addcontentsline{toc}{section}{Conventions}

We consider the following conventions throughout the document:

\begin{itemize}

\item Natural units $c=\hbar=1$ and metric signature $(-1,+1,+1,+1)$.

\item The Newton constant $G$, the full Planck mass $M_p \simeq 1.22\cdot 10^{19}$ GeV, and the reduced Planck mass $m_p \simeq 2.44\cdot 10^{18}$ GeV, are used interchangeably, related through $M_p^2 = 8\pi m_p^2 = 1/G$.

\item Latin indices $i, j, k, ... = 1,2,3$ are reserved for spatial dimensions, and Greek indices $\alpha, \beta, \mu, \nu,... = 0,1,2,3$ for space-time dimensions. We use the {\it Einstein convention} of summing over repeated indices {\it only in the continuum}. {\bf In the lattice, unless stated otherwise, repeated indices do not represent summation}.

\item We consider a spatially-flat FLRW metric with line element $ds^2 = -a^{2\alpha}(\eta)d\eta^2 + a^2(\eta) \, \delta_{ij} \, dx^i dx^j$, with $\alpha \in \mathcal{R}$ a constant chosen conveniently in each scenario. For $\alpha = 0$, $\eta$ denotes the {\it cosmic time}, $t$, for $\alpha = 1$, $\eta$ denotes the {\it conformal time}, $\tau = \int {dt'\over a(t')}$, and for arbitrary $\alpha$, we refer to the time variable as the {\it $\alpha$-time}.

\item We reserve the notation $\dot{f}$ for derivatives of a function $f$ with respect to cosmic time with $\alpha = 0$, and $f'$ for derivatives with respect to $\alpha$-time with arbitrary $\alpha$, unless otherwise specified.

\item Physical momenta are represented by ${\bf p}$ and comoving momenta by ${\bf k}$. The $\alpha$-time Hubble rate is denoted as $\mathcal{H} = a'/a$, whereas the physical Hubble rate is  $H = \dot a / a \equiv \mathcal{H}|_{\alpha = 0}$.

\item Cosmological parameters, when used, are fixed to the CMB values given in \cite{Aghanim:2018eyx,Akrami:2018odb}.

\item Our Fourier transform convention in the continuum is given by
\begin{eqnarray}\label{eq:FTcont}
f({\bf x}) = \frac{1}{(2 \pi)^3} \int d^3 {\bf k} \, f({\bf k}) \, e^{+i {\bf k} {\bf x}}\, ~~~ \Longleftrightarrow ~~~  f({\bf k}) = \int d^3 {\bf x} \, f ( {\bf x}) \, e^{-i {\bf k} {\bf x}}\,.\nonumber
\end{eqnarray}

\item Our discrete Fourier transform (DFT) is defined as
\begin{eqnarray}\label{eq:FTdiscreteAux}
f({\bf n}) \equiv {1\over N^3}\sum_{\tilde n} e^{+i{2\pi\over N} {\bf \tilde n n}} f({\bf \tilde n}) ~~~~ \Longleftrightarrow ~~~~  f({\bf \tilde n}) \equiv \sum_{n} e^{-i{2\pi\over N} {\bf n \tilde n} }f({\bf n})\,.\nonumber
\end{eqnarray}

\item A scalar field living in a generic lattice site $n = (n_0,\bn) = (n_0,n_1,n_2,n_3)$, i.e.~$\phi_n = \phi(n)$, will be simply denoted as $\phi$. If the point is displaced in the $\mu$-direction by one unit lattice spacing/time step, $n + \hat\mu$, with $\hat{\mu}$ a unit vector in the direction $\mu$, we will then use the notation $n+\mu$ or simply by $+\mu$ to indicate this, so that the field amplitude in the new point is expressed as $\phi_{+\mu} = \phi_{n+\mu}  \equiv \phi(n+\hat\mu)$.

\item When representing explicitly gauge fields on a lattice, we will automatically understand that they live in the middle of lattice points, i.e.~$A_{\mu} \equiv A_{\mu}(n+{1\over2}\hat\mu)$. It follows then that e.g.~$A_{\mu,+\nu} \equiv A_{\mu}\big(n + {1\over2}\hat\mu +  \hat\nu\big)$, $A_{\mu,-\mu} \equiv A_{\mu}\big(n - {1\over2}\hat\mu\big)$, etc. In the case of links, we will use the notation $U_\mu \equiv U_{\mu,n} \equiv U_\mu(n+{1\over2}\hat\mu)$, and hence $U_{\mu,\pm\nu} = U_{\mu,n\pm\nu} \equiv U_\mu(n + {1\over2}\hat\mu \pm \hat\nu)$.

\item Even though the {\it time step} $\delta \eta$ is typically chosen to be smaller than the the {\it lattice spacing} $\dx$, they are often of the same order, unless otherwise specified. We may therefore speak loosely of corrections of 
$\mathcal{O}(\dx^\mu)$, independently of whether we are referring to the lattice spacing or the time step. Only if strictly necessary we will make a distinction. 

\end{itemize}

\newpage
\,
\thispagestyle{empty}
\addtocounter{page}{-1}
\newpage

\section{Introduction and purpose of this monograph} 
~~~~~{\it Inflation}~\cite{Guth:1980zm, Linde:1981mu, Albrecht:1982wi,Brout:1977ix,Starobinsky:1980te,Kazanas:1980tx,Sato:1980yn}, an early period of accelerated expansion in the Universe, is a cornerstone of modern 
cosmology that overcomes the limitations of the hot Big Bang framework. Remarkably, inflation also provides a natural explanation for the origin of the primordial perturbations~\cite{Mukhanov:1981xt, Guth:1982ec,Starobinsky:1982ee, Hawking:1982cz,Bardeen:1983qw}, which are 
required to explain the large scale structure in the Universe and have been 
accurately measured 
in the form of anisotropies in the cosmic microwave background (CMB)~\cite{Planck:2018nkj,Aghanim:2018eyx,Akrami:2018odb}. Standard scenarios propose an inflationary epoch driven by a scalar field---the \textit{inflaton}---, with a potential appropriately chosen to support accelerated expansion for approximately 50-60 e-folds. Following inflation, a \textit{reheating} phase is required to transfer the energy available into other particle species, which subsequently form a relativistic thermal ensemble that dominates the energy budget of the Universe. This marks the onset of the standard {\it hot Big Bang} framework for the expansion of the Universe. Comprehensive reviews of inflation and reheating can be found in \cite{Lyth:1998xn, Riotto:2002yw, Bassett:2005xm, Linde:2007fr, Baumann:2009ds} and~\cite{Allahverdi:2010xz, Amin:2014eta, Lozanov:2019jxc, Allahverdi:2020bys}, respectively. 

    The phenomenology of the early Universe, both during and after inflation, is rich and often involves complex non-linear processes. These include preheating and other particle production mechanisms~\cite{Traschen:1990sw, Kofman:1994rk, Shtanov:1994ce, Kaiser:1995fb, Kofman:1997yn, Greene:1997fu, Kaiser:1997mp, Kaiser:1997hg, Greene:1998nh, Greene:2000ew, Peloso:2000hy, Berges:2010zv, Enqvist:2012tc,Figueroa:2015rqa}, the generation of scalar metric perturbations~\cite{Bassett:1998wg, Bassett:1999mt, Bassett:1999ta, Finelli:2000ya, Chambers:2007se, Bond:2009xx,Linde:2012bt,Imrith:2019njf, Musoke:2019ima, Martin:2020fgl, Adshead:2023mvt}, and the possibility to form primordial black holes~\cite{Cotner:2019ykd, Martin:2019nuw, GarciaBellido:1996qt, Green:2000he, Hidalgo:2011fj, Torres-Lomas:2014bua, Suyama:2004mz, Suyama:2006sr, Cotner:2018vug}. Additionally, phenomena like phase transitions~\cite{Guth:1982pn,Witten:1984rs,Kosowsky:1991ua,Kosowsky:1992rz,Kamionkowski:1993fg,Rajantie:2000fd,Hindmarsh:2001vp,Copeland:2002ku,GarciaBellido:2002aj,Niemi:2018asa,Mazumdar:2018dfl, Hindmarsh:2020hop,Brandenburg:2017neh, Brandenburg:2017rnt}, or the creation and evolution of cosmic defects~\cite{Hindmarsh:1994re, Felder:2000hj, Hindmarsh:2000kd, Rajantie:2001ps, Rajantie:2002dw, Donaire:2004gp, Copeland:2009ga, Hiramatsu:2012sc, Kawasaki:2014sqa, Fleury:2016xrz, Moore:2017ond} and of soliton-like objects like oscillons~\cite{Gleiser:1993pt,Copeland:1995fq,Amin:2010dc,Amin:2011hj,Gleiser:2011xj,Antusch:2015ziz,Lozanov:2017hjm,Hasegawa:2017iay,Amin:2018xfe,Kitajima:2018zco,Antusch:2019qrr,Ibe:2019lzv,Sang:2019ndv,Kou:2019bbc,Nazari:2020fmk,Sang:2020kpd,Aurrekoetxea:2023jwd,Mahbub:2023faw,Piani:2023aof,Shafi:2024jig,Drees:2025iue,Piani:2025dpy}, are also part of this non-linear landscape. These phenomena can have significant observational consequences, including the production of gravitational wave backgrounds~\cite{Caprini:2018mtu,Khlebnikov:1997di,Easther:2006gt,Easther:2006vd,Garcia-Bellido:2007nns,GarciaBellido:2007af,Dufaux:2007pt,Dufaux:2008dn,Dufaux:2010cf,Figueroa:2012kw, Hiramatsu:2013qaa,Hindmarsh:2013xza,Zhou:2013tsa, Bethke:2013aba, Bethke:2013vca,Hindmarsh:2015qta,Figueroa:2016ojl,Antusch:2016con, Hindmarsh:2017gnf, Antusch:2017flz,Antusch:2017vga, Figueroa:2017vfa,Cutting:2018tjt, Liu:2018rrt,Lozanov:2019ylm, Adshead:2019lbr,Adshead:2019igv, Cutting:2019zws,Pol:2019yex, Figueroa:2020lvo,Cutting:2020nla, Figueroa:2022iho,Cosme:2022htl, Klose:2022knn,Cui:2023fbg, Baeza-Ballesteros:2023say,Baeza-Ballesteros:2024otj,Servant:2023tua}, the generation of dark matter relics~\cite{Garcia:2018wtq, Garcia:2021iag, Garcia:2022vwm, Lebedev:2022vwf, Zhang:2023xcd}, the realization of cosmological magnetogenesis~\cite{DiazGil:2005qp, DiazGil:2007qx, DiazGil:2007dy, DiazGil:2008tf, Fujita:2016qab, Adshead:2016iae, Vilchinskii:2017qul}, the creation of the baryon asymmetry in the Universe~\cite{Kolb:1996jt, Kolb:1998he, GarciaBellido:1999sv, Allahverdi:2000zd, Rajantie:2000nj, Cornwall:2001hq, Copeland:2001qw, Smit:2002yg, GarciaBellido:2003wd, Tranberg:2003gi, Tranberg:2009de, Kamada:2010yz, Lozanov:2014zfa}, and the determination of the equation of state after inflation, with implications for CMB observations~\cite{Podolsky:2005bw,Dufaux:2006ee, Lozanov:2016hid, Figueroa:2016wxr, Krajewski:2018moi, Maity:2018qhi, Antusch:2020iyq, Saha:2020bis, Antusch:2021aiw, Mansfield:2023sqp,Garcia:2023eol,Garcia:2023dyf,Antusch:2025ewc}. 

In summary, the high-energy physics phenomenology of the early universe is vast, intricate, and often involves non-linear dynamics. As the latter are frequently too complex to be effectively captured by analytical methods, developing numerical techniques becomes essential to gain a comprehensive understanding of the non-linearities arising in any given scenario. Consequently, it is vital to develop numerical methods that are both efficient and reliable for simulating nonlinear field theory phenomena, and which adhere to physical principles, such as {\it e.g.}~energy conservation or gauge invariance, while minimizing numerical integration errors. Additionally, developing a variety of different techniques for tackling a given problem is also relevant for the validation and cross-checking of simulation results. Only through a robust development of these techniques, we will achieve a solid foundation for the prediction of `observables' from the early Universe. 

\begin{mdframed}
{\bf Note -.} Work focused on developing and applying numerical methods to address non-linear field theory phenomena in the early Universe is rapidly evolving, leading to the emergence of a research field of its own. We like to refer to this field as {\bf Lattice Cosmology}, and to the methods used as {\bf Lattice Cosmology Techniques (LCT)}. From now on, we will adopt this terminology, which will be used constantly throughout this document.
\end{mdframed}

Lattice Cosmology has gained significant attention in the recent times, as reflected by the number of specialized LCT packages created over the last years~\cite{Felder:2000hq,Felder:2007nz,Frolov:2008hy,Sainio:2009hm,Easther:2010qz,Huang:2011gf,Sainio:2012mw,Daverio:2015ryl,Lozanov:2019jff,Giblin:2019nuv,Andrade:2021rbd,Caravano:2025klk}. We expect Lattice Cosmology to become an increasingly influential approach in determining observational strategies to probe the early Universe. It is in this context that our package \CL (\href{http://www.cosmolattice.net}{\color{blue} http://www.cosmolattice.net}) was originally developed by some of us~\cite{Figueroa:2021yhd}, created purposely to explore the phenomenology and observational implications of non-linearities in field theory early Universe scenarios. Contrary to standard codes focused in a given set of equations and observables, \CL is rather a {\it platform} for the implementation of any field theory system~\cite{Figueroa:2023xmq}, characterized by partial differential equations suitable for
discretization on a lattice. Written in C++, \CL uses a modular structure to separate technical details from the physics, establishing a unique symbolic language wherein field variables and their associated operations are defined in close resemblance to the continuum. 

The release of \CL in 2021~\cite{Figueroa:2021yhd} was preceded by the publication of a monographic review on LCT: {\it The Art of Simulating the early Universe - Part I}~\cite{Figueroa:2020rrl}. Such monograph, which we like to refer colloquially 
as {\tt The Art\,-\,I}, constituted the theoretical basis for the physics cases initially released in the {\tt v1.0} of the code: namely, {\it canonical} scalar-singlet and $SU(2)$$\times$$U(1)$ scalar-gauge theories in a spatially-flat expanding background. In particular, in {\tt The Art\,-\,I} we constructed symplectic explicit-in-time evolution algorithms for theories with canonically normalized kinetic terms, with accuracies ranging from $\mathcal{O}(dt^{2})$ to $\mathcal{O}(dt^{10})$, and with self-consistent expansion of the Universe sourced by all dynamical fields present. The algorithms were designed to simulate singlet scalar and/or Abelian $U(1)$ and non-Abelian $SU(2)$ scalar-gauge interactions, preserving the Gauss constraint to machine precision [for either $U(1)$ or $SU(2)$]. 

The present document, {\it The Art of Simulating the early Universe - Part II}, or for short, {\tt The Art\,-\,II}, constitutes a new monographic review that serves for a twofold purpose: it extends our previous dissertation on LCT, introducing a variety of  
physics cases and circumstances not considered in {\tt The Art\,-\,I}, and it constitutes the theoretical basis for new modules added in \CL{\tt v2.0}, which will be publicly released soon after the publication of this monograph. In contrast to {\tt The Art\,-\,I}, which focused on canonical field theories, in {\tt The Art\,-\,II} we present lattice methods for non-canonical aspects in field theory. By these we refer {\it e.g.}~to theories with non-minimal gravitational couplings, or non-minimal kinetic terms with non-trivial field metrics, as considered for instance in~\cite{DeCross:2015uza,DeCross:2016fdz,DeCross:2016cbs,Nguyen:2019kbm,vandeVis:2020qcp,Figueroa:2021iwm,Figueroa:2024asq}. Non-canonical scenarios may also include interactions between field variables and their conjugate momenta, 
as naturally arising in derivative couplings between an axion-like particle and gauge fields, see {\it e.g.}~\cite{Adshead:2015pva,Figueroa:2017qmv,Cuissa:2018oiw,Adshead:2018doq,Caravano:2022epk,Figueroa:2023oxc,Figueroa:2024rkr,Lizarraga:2025aiw,Caravano:2024xsb,Sharma:2024nfu}. Non-canonical interactions can be numerically complicated to deal with, and usually require involved integration methods (often non-symplectic), {\it e.g.}~with high memory requirements or non-explicit in time. It is precisely because of these circumstances that we naturally separated the methods for canonical 
theories presented in {\tt The Art\,-\,I}, from the methods 
required for non-canonical circumstances, which we present here in {\tt The Art\,-\,II}. 

In addition to 
interactions, we also extend the  non-canonical notion to the initial configuration, considering cases with initial conditions different than standard. In canonical set-up's like {\it e.g.}~in preheating scenarios~\cite{Antusch:2020iyq,Antusch:2021aiw,Antusch:2022mqv}, it is customary to initialize simulations by setting a homogeneous scalar field configuration first, and adding on top small random fluctuations, obtained from sampling a Gaussian distribution with variance equal to the spectrum of quantum vacuum modes (see Section 7 of {\tt The Art\,-\,I} for details). In other systems, however, a different procedure is needed to initialize fields in coordinate-space (as opposed to Fourier-space), before one can run their ordinary dynamics on a lattice. This is the case, for example, of cosmic defect networks (cosmic strings, domain walls, etc), see~\cite{Vincent:1997cx,Bevis:2006mj,Hindmarsh:2014rka,Daverio:2015nva,Lizarraga:2016onn,Lopez-Eiguren:2017dmc,Hindmarsh:2018wkp,Eggemeier:2019khm,Hindmarsh:2019csc,Gorghetto:2018myk,Baeza-Ballesteros:2023say,Baeza-Ballesteros:2024otj,Dankovsky:2024zvs,Dankovsky:2024ipq,Correia:2024cpk,Correia:2025nns}. In this monograph we present algorithms to set up properly the initial spatial configuration of cosmic defect networks in a lattice, so that their evolution can be run afterwards with the canonical integrators already presented in {\tt The Art\,-\,I}. We extend further the notion of non-canonical initial conditions, presenting also procedures for setting up arbitrary initial configurations on the lattice, either based on a spatial field profile (as {\it e.g.}~in~\cite{Cutting:2018tjt,Cutting:2020nla,Matsunami:2019fss,Saurabh:2020pqe,Baeza-Ballesteros:2023say,Baeza-Ballesteros:2024otj}), or set by an arbitrary spectrum that represents modes excited above vacuum (as {\it e.g.}~in~\cite{Figueroa:2021iwm,Figueroa:2023oxc,Figueroa:2024asq,Figueroa:2024yja,Figueroa:2024rkr}). We also discuss algorithmic procedures, used for example to create field configurations from the sampling of a probability distribution, as {\it e.g.}~in thermal initial conditions. 

Furthermore, a remarkable feature of our code 
is that 
it can actually work in an arbitrary number of spatial dimensions $d \neq 3$. 
While this feature has not been exploited in previous versions, \CL {\tt v2.0} will include a new module to run simulations of scalar fields in $d = 1$ and $d = 2$ spatial dimensions, including appropriate observables adapted to the reduced dimensionality. Thus, in this review we also present the theoretical basis for such lower-dimensional lattice simulations, extending in this way, the non-canonical notion to the dimensionality of space. 

Finally, we also note that since the initial release of \CL {\tt v1.0}, new features have been added to the code, 
for example the capability to simulate gravitational waves (GWs) sourced by scalar singlets (\CL {\tt v1.1}), 
or by a $U(1)$ gauge sector (\CL {\tt v1.2}). 
While 
admittedly this is unrelated to non-canonical aspects on which this review is centered, we simply take the opportunity to present in this monograph the theoretical basis of our implementation(s) of GW dynamics on a lattice~\cite{GWinCL}, including 
a new algorithm that improves the memory requirements 
of previous methods commonly used in the literature.  

To conclude this introduction, we note that 
there are a number of LCT of utmost interest, which are neither considered here in {\tt The Art-\,II}, nor in {\tt The Art\,-\,I}. Some of us are currently working in two more monographs, which will represent the theoretical foundation for lattice implementations of {\it scalar-gauge-fluid} (SGF) dynamics and {\it gravitational dynamics}, respectively. We postpone the release of these monographs and their code implementations for the near future. 

\begin{mdframed}
{\bf Note -.} All the physics and technical aspects that we present here in {\tt The Art\,-\,II}, are explicitly implemented in \CL{\tt v2.0}, which will be publicly released soon after this monograph is published. However, the presentation of the techniques themselves in the monograph, should be understood as a theoretical review on non-canonical LCT, complementing our previous review on canonical LCT, {\tt The Art\,-\,I}. The techniques presented here are therefore detached from \CLns, as they represent general LCT, which could be implemented in other codes.
\end{mdframed}

\section{Lattice Field Theory}
\label{sec:LatticeTechniques}

~~~~Ideally, the reader of this document is already familiarized with canonical LCT, like those presented in {\tt The Art\,-\,I}~\cite{Figueroa:2020rrl}. 
As we prefer not to make such assumption, we summarize in this section some of the key concepts discussed in {\tt The Art\,I}, which are necessary to set up basic terminology and concepts that will be used in the remainder of this monograph: field dynamics in the continuum (Sect.~\ref{subsec:eomCont}), definition of a lattice (Sect.~\ref{subsec:Lattice}), definition of power spectrum and initial conditions (Sect.~\ref{subsec:PS}), and evolution algorithms (Sect.~\ref{subsec:Algorithms}). 

\subsection{Canonical field theory in the continuum} \label{subsec:eomCont}

Throughout this monograph, we consider 
a homogeneous and isotropic universe described by the Friedmann-Lemaitre-Robertson-Walker (FLRW) line element, which we write as
\be
\dd s^2 \equiv g_{\mu\nu}\dd x^\mu\dd x^\nu = - \dd t^2 + a^2(t) \delta_{ij} \dd x^i \dd x^j \ , \label{eq:FLRWmetric}
\ee
with $t$ the cosmic time and $a(t)$ the scale factor. It is useful to define a new time coordinate $\eta$ through the relation
\be \dd \eta \equiv a^{-\alpha} (t) dt \ ,\ee
where $\alpha$ is a real number conveniently chosen for a given problem at hand. We denote it as $\alpha$\textit{-time}, and it represents cosmic and conformal time for $\alpha=0,1$, respectively. We will often let $\alpha$ be an unspecified real number and write all the relevant equations in terms of $\alpha$-time. 

The evolution of the scale factor is dictated by the stress-energy tensor of the matter fields, which in order to be compatible with homogeneity and isotropy, must take the following \textit{perfect fluid} form,
\be {\bar T}_{\mu \nu} \equiv (\bar\rho + \bar p )u_{\mu} u_{\nu} + \bar p g_{\mu \nu}  \ , \hspace{0.4cm} g_{\mu \nu} u^{\mu} u^{\nu} = -1 \hspace{0.4cm} \Longrightarrow \hspace{0.4cm}   \begin{cases}
    \bar\rho \equiv a^{-2 \alpha}\,{\bar T}_{00} \ , \vspace{0.2cm}\\
    \bar p \equiv {1\over 3a^2} \sum_j {\bar T}_{jj} \ ,
\end{cases} \label{eq:stresstensor} \ee
where $\bar{p}$ and $\bar{\rho}$ are the background values of the total pressure and energy densities of the matter sector, and $u_{\mu} = (a^{\alpha},0,0,0)$
is the four-velocity of a fluid at rest. The evolution of the scale factor is determined by the first and second Friedmann equations,
\begin{eqnarray}
    \mathcal{H}^2 \equiv \left({a'\over a}\right)^2 = a^{2 \alpha} \frac{\bar {\rho}}{3 m_p^2} 
    \,,~~~~~~~
    {a''\over a} = \frac{a^{2 \alpha}}{6 m_p^2}\Big[ (2 \alpha - 1) \bar{\rho} - 3 \bar{ p} \Big]\,, \label{eq:Friedmann-full}
\end{eqnarray}
where we remind the reader that in our notation ${f}' \equiv \text{d} f/ \text{d} \eta$ and $m_p$ is the reduced Planck mass. 

In this work we present the lattice formulation of different theories involving scalar and/or gauge fields. In order to introduce notation, we review here the equations of motion of a gauge-invariant theory formed by three kinds of canonically-normalized scalar fields: a real scalar singlet $\phi$, a $U(1)$-charged complex scalar field $\varphi$, and a $[SU(N) \times U(1)]$-charged scalar field $\Phi$. In the last two cases, there are also present, respectively, Abelian and non-Abelian gauge fields, $A_{\mu}$ and $C_{\mu} \equiv C_{\mu}^a T_a$, with $T_a$ the $N^2-1$ group generators of $SU(N)$. The scalar fields can be explicitly written in terms of real components as follows,
\begin{eqnarray} \label{eq:ChargedScalars}
    \begin{array}{ccccc}
        \phi \in \mathcal{R}e & , &  \varphi \equiv {1\over\sqrt{2}}(\varphi_1 +i\varphi_2) & , & \Phi = \left(
        \begin{array}{c}
            \varphi^{(1)} \\ \varphi^{(2)} \\ \vdots \\ \varphi^{(N)}
        \end{array}
        \right) =
        {1\over\sqrt{2}}
        \left(
        \begin{array}{c}
            \varphi_1 +i\varphi_2 \vspace*{0.1cm}\\ \varphi_3 +i\varphi_3 \\ \vdots \\ \varphi_{2N -1} +i\varphi_{2N}
        \end{array}
        \right) \,.
    \end{array}
\end{eqnarray}

\noindent More specifically, for canonical scalar-gauge theories we consider the action $S = \int d^4x  \sqrt{-g}\, \mathcal{L}$, with $g \equiv |{\rm det} (g_{\mu \nu})|$, and the Lagrangian
\begin{align} 
-\mathcal{L} = \frac{1}{2}\partial_{\mu} \phi \partial ^{\mu}\phi + (D_{\mu}^A \varphi)^{*}(D_A^{\mu} \varphi) +  (D_{\mu}\Phi )^{\dagger} (D^{\mu} \Phi) + \frac{1}{4} F_{\mu \nu} F^{\mu \nu} + \frac{1}{2}{\rm Tr}\{G_{\mu \nu}G^{\mu \nu}\} + V \ ,
\label{eq:lagrangian} 
\end{align}
with $V \equiv V(\phi,|\varphi|, |\Phi|)$ the potential describing the interactions between the scalar fields, and where the {\it covariant derivatives} and {\it field strength tensors} associated to the gauge fields, are defined as
\begin{eqnarray}
D_{\mu}^{\rm A}  &\equiv &  \partial _{\mu} - i  g_A Q_AA_\mu \ , \hspace{4cm} F_{\mu \nu}\equiv  \partial_{\mu}  A_{\nu} - \partial_{\nu} A_{\mu} \ , \label{eq:AbCovDerivCont} \\
D_{\mu} & \equiv  &
\mathcal{I}D^{\rm A}_\mu
- i g_C Q_C C_{\mu}^a \,T_a
\ , \hspace{3.13cm}  G_{\mu \nu} \equiv \partial_{\mu} C_{\nu} - \partial_{\nu} C_{\mu} - i[C_\mu,C_\nu]\,,  \label{eq:CovDerivCont}  
\end{eqnarray}
with $g_{A}$ and $g_C$ the Abelian and non-Abelian gauge couplings, $Q_{A}$ and $Q_C$ the Abelian and non-Abelian charges of the scalar fields, $\mathcal{I}$ the $N\times N$ identity matrix and $[\cdot,\cdot]$ the usual matrix commutator. The gauge-invariant electric and magnetic fields associated to the Abelian and non-Abelian fields can be written, respectively, as follows,
\be\label{eq:ElectricMagneticDefs}
E_i \equiv F_{0i} , \,\,\,\,\,\,\,\,  B_i \equiv \frac{1}{2} \epsilon_{i j k} F^{j k} , \,\,\,\,\,\,\,\,   E_i^a \equiv G_{0i}^a , \,\,\,\,\,\,\,\,  B_i^a \equiv \frac{1}{2} \epsilon_{i j k} G^{j k}_a \ , \ee
where $\epsilon_{ijk}$ is the Levi-Civita symbol in three dimensions with normalization $\epsilon_{123}=+1$, and $G_{\mu \nu}^a \equiv {\rm Tr}(2G_{\mu \nu} T_a)$.

The equations of motion for the matter fields and the scale factor have been derived in more detail in {\tt The Art\,I}. Here we simply quote their resulting form, which reads
\begingroup
\allowdisplaybreaks
\begin{eqnarray}
    \phi'' - a^{-2(1 - \alpha)} {\vv\nabla}^{\,2} \hspace{-1mm}\phi + (3 - \alpha)\mathcal{H} {\phi'} &=& - a^{2 \alpha} V_{,\phi} \ , \label{eq:singlet-eom} \\
    \varphi'' - a^{-2(1 - \alpha)} {\vv D}_{\hspace{-0.5mm}A}^{\,2}\varphi + (3 - \alpha) \mathcal{H}  {\varphi'} &=& - \frac{a^{2 \alpha}V_{,|\varphi|} }{2} \frac{\varphi}{|\varphi |} \ , \label{eq:higgsU1-eom}\\
    \Phi'' - a^{-2(1 - \alpha)} {\vv D}^{\,2}\Phi + (3 - \alpha) \mathcal{H}  {\Phi'} &=& - \frac{a^{2 \alpha} V_{,|\Phi|}}{2} \frac{\Phi}{|\Phi |} \ , \label{eq:higgsSU2-eom}
    \\
    \partial_0 F_{0i} - a^{-2(1 - \alpha )}\partial_j F_{ji} + (1 - \alpha) \mathcal{H} F_{0i} &=&
    a^{2 \alpha}J^A_i \ , \label{eq:U1eom}
    \\
    (\mathcal{D}_0 )_{a b} (G_{0i})^b - a^{-2(1 - \alpha )} ( \mathcal{D}_j )_{a b} (G_{ji} )^b + (1 - \alpha) \mathcal{H} (G_{0i} )^b &=& a^{2 \alpha}(J_i)_a \ , \label{eq:SU2eom}
    \\
    \partial_i F_{0i} &=& a^2J^A_0 \ , \label{eq:GaussU1-eom}\\
    (\mathcal{D}_i )_{a b} (G_{0i})^b &=& a^2(J_0)_a \ , \label{eq:GaussSU2-eom}
\end{eqnarray}
\endgroup
where we have introduced the derivative operators $(\mathcal{D}_{\nu}O)_a = (\mathcal{D}_{\nu})_{a b}O_b \equiv ( \delta_{a b}  \partial_{\nu} - f_{abc} C_{\nu}^c ) O_b$, and defined the matter currents
\begin{eqnarray}
    \label{eq:AbelianCurrent}
    J_A^\mu & \equiv & 2g_A Q_A^{(\varphi)} \mathcal{I}m [ \varphi^{*} ( D_A^{\mu} \varphi )] + 2g_A Q_A^{(\Phi)} \mathcal{I}m [ \Phi^\dag (D^{\mu} \Phi  )]\,,\\
    \label{eq:NonAbelianCurrent}
    J_a^\mu & \equiv & 2g_C Q_C\mathcal{I}m [ \Phi^{\dag} T_a( D^{\mu} \Phi )]\,.
\end{eqnarray}
Note that Eqs.~\eqref{eq:GaussU1-eom} and \eqref{eq:GaussSU2-eom} are the Gauss constraint of the Abelian and non-Abelian sectors, respectively, which must be preserved at all times during the evolution.

\noindent The energy-momentum tensor of a system characterized by a lagrangian $\mathcal{L}$, is given by 
\begin{equation}
T_{\mu \nu} \equiv -\frac{2}{\sqrt{g}}\frac{\delta(\sqrt{g} \mathcal{L})}{\delta g^{\mu \nu}}\,.
\end{equation}
This definition leads, using Eqs.~(\ref{eq:stresstensor}) and~(\ref{eq:lagrangian}), to {\it local} expressions for the field's energy and pressure densities, 
\begin{eqnarray}
    \hspace*{-0.5cm}\rho &=& {K}_{\phi} + {K}_{\varphi} + {K}_{\Phi} + {G}_{\phi} + {G}_{\varphi} + {G}_{\Phi} + {K}_{U(1)} + {G}_{U(1)} + {K}_{SU(2)} + {G}_{SU(2)} + {V},  \label{eq:rhoLocal}\\
    \hspace*{-0.5cm}p &=& {K}_{\phi} + {K}_{\varphi} + {K}_{\Phi} -{1\over3}({G}_{\phi} + {G}_{\varphi} + {G}_{\Phi}) + {1\over3}({K}_{U(1)} + {G}_{U(1)}) + {1\over3}({K}_{SU(2)} + {G}_{SU(2)}) - {V},  \label{eq:pLocal}
\end{eqnarray}
with the different energy density contributions given by
\bea
\begin{array}{l} \label{eq:energy-contrib}
    {K}_{\phi} = \frac{1}{2 a^{2\alpha} } {\phi'}^2 \vspace{0.1cm}\\
    {K}_{\varphi} = \frac{1}{a^{2\alpha} } (D_0^A \varphi)^*(D_0^A \varphi)
    \vspace{0.1cm}\\
    {K}_{\Phi} = \frac{1}{a^{2\alpha} } (D_0 \Phi )^\dag(D_0 \Phi)
    \vspace{0.1cm}\\
\end{array}
\hspace{0.1cm};\hspace{0.4cm}
\begin{array}{l}
    {G}_{\phi} = \frac{1}{2 a^2} \sum_i (\partial_i \phi)^2
    \vspace{0.1cm}\\
    {G}_{\varphi} = \frac{1}{a^2} \sum_i (D_i^A \varphi)^*(D_i^A \varphi)
    \vspace{0.1cm}\\
    {G}_{\Phi} = \frac{1}{a^2} \sum_i (D_i \Phi)^\dag(D_i \Phi)
    \vspace{0.1cm}\\
\end{array}
\hspace{0.1cm};\hspace{0.4cm}
\begin{array}{l}
    {K}_{U(1)} = \frac{1}{2 a^{2 + 2 \alpha}}  \sum_{i} F_{0i}^2
    \vspace{0.1cm}\\
    {K}_{SU(2)} = \frac{1}{2 a^{2 + 2 \alpha}}  \sum_{a,i} (G_{0i}^a)^2
    \vspace{0.1cm}\\
    {G}_{U(1)} = \frac{1}{2 a^4}  \sum_{i,j<i} F_{ij}^2
    \vspace{0.1cm}\\
    {G}_{SU(2)} = \frac{1}{2 a^4}  \sum_{a,i,j<i}  (G_{ij}^a)^2  \, . \vspace*{0.2cm}\\
\end{array}
\nonumber\\
{\rm(Kinetic-Scalar)} \hspace*{2.15cm} {\rm(Gradient-Scalar)} \hspace*{2.00cm} {\rm (Electric ~\&~ Magnetic)} \hspace{1.0cm} \nonumber\\
\eea
If the fields dominate the energy budget of the Universe, the expansion rate can be determined through the Friedmann equations~\eqref{eq:Friedmann-full}, which in our case can be written as
\begin{eqnarray}\label{eq:FriedmannHubble}
    &&\hspace*{-1.0cm}\left({a'\over a}\right)^2 =  \frac{a^{2 \alpha}}{3 m_p^2}\left\langle {K}_{\phi} + {K}_{\varphi} + {K}_{\Phi} + {G}_{\phi} + {G}_{\varphi} + {G}_{\Phi} + {K}_{U(1)} + {G}_{U(1)} + {K}_{SU(2)} + {G}_{SU(2)} + {V}\right\rangle\hspace*{-0.5mm},
    \\
    \label{eq:FriedmannDDa}
    &&\hspace*{-0.33cm}{a''\over a} = \frac{a^{2 \alpha}}{3 m_p^2}\big\langle (\alpha-2)({K}_{\phi} + {K}_{\varphi} + {K}_{\Phi}) + \alpha({G}_{\phi} + {G}_{\varphi} + {G}_{\Phi}) + (\alpha + 1)V \\
    &&\hspace*{5.75cm} \left. +~ (\alpha-1)({K}_{U(1)} + {G}_{U(1)} + {K}_{SU(2)} + {G}_{SU(2)}) \right\rangle,\nonumber
\end{eqnarray}
with $\langle \dots \rangle$ denoting an average over a sufficiently large volume to encompass all relevant wavelengths of the fields. Typically, we use Eq.~(\ref{eq:FriedmannDDa}) to solve for the scale factor, while monitoring that the constraint equation (\ref{eq:FriedmannHubble}) is verified throughout the evolution to some desired accuracy, see Sect.~\ref{subsec:Algorithms}. 

\subsection{Characterization of a lattice} \label{subsec:Lattice}

~~~~A regular cubic lattice in $d=3$ spatial dimensions is fully characterized by two parameters: the number of points per dimension $N$, and the length of each side $L$. The total number of lattice sites 
is therefore $N^3$. 
The ratio between $L$ and $N$ represents the \textit{lattice spacing},
\begin{eqnarray}
    \dx \equiv {L\over N}\,,
\end{eqnarray}
which is the minimum distance resolved in a lattice. The different lattice sites can be tagged by a vector ${\bf n}$ as follows,
\begin{eqnarray}
    {\bf n} = (n_1,n_2,n_3),~~~~ {\rm with}~~ n_i = 0,1,...,N-1 \,,~~~i = 1,2,3\,.
\end{eqnarray}

A function ${\tt f}(\bf x)$ defined in the continuum can be represented on a lattice by a discrete function $f({\bf n})$, which takes the same value as the continuum function 
{\it i.e.}~$f({\bf n}) \equiv {\tt f}({\bf x} = {\bf n} \, \dx)$. Unless explicitly stated otherwise, we assume {\it periodic boundary conditions} through all three spatial directions, meaning that the function satisfies $f({\bf n} + \hat{\imath} N) = f({\bf n})$ for $i = 1,2,3$, with $ \hat{\imath}$ the unitary vector in the $i$-spatial direction. 
In general, spatially dependent functions ${f}({\bf n})$ on a lattice represent field amplitudes at a given time, so their value will change as the simulation progresses. Therefore, functions on a lattice will depend not only on spatial coordinates $\bf n$ (or reciprocal coordinates $\tilde{\bf n}$, as introduced in the next subsection), but also on a discrete time variable $n_0 = 0, 1, 2, \dots$, counting the number of evolution time-steps. A given time is thus indicated as $\eta = \eta_* + n_0 \delta \eta$, where $\delta \eta$ is the temporal step and $\eta_*$ an initial time. We will therefore treat fields as four-dimensional functions and write them as $f(n) = f(n_0,{\bf n})$ or $f(\tilde n) = f(n_0,\tilde{\bf n})$. Additionally, we denote a one-step time advance using $\hat{0}$, so that, for example, $f(n+\hat 0) = f(n_0+1,{\bf n})$. We will write $f(n+\hat{\mu})$ representing either $f(n_0+1,{\bf n})$ or $f(n_0,{\bf n}+\hat{\imath})$, depending on whether $\mu = 0$ or $\mu = i$. 

\subsubsection{Reciprocal lattice}
\label{sub:reciprocal}

For any lattice we can define its \textit{reciprocal lattice} in momentum space, which is also regular and whose points can be tagged by a vector $\tilde{\bf n}$ with the following components,
\begin{eqnarray}
    \tilde{\bf n} = (\tilde n_1, \tilde n_2, \tilde n_3), ~~~~{\rm with}~~
    \tilde n_i = -\frac{N}{2}+1, -\frac{N}{2}+2, ... ,-1,0,1, ... , \frac{N}{2} - 1, \frac{N}{2}  \,,~~~ i  = 1,2,3\,.
\end{eqnarray}
We define the discrete Fourier transform (DFT) as follows,
\begin{eqnarray}\label{eq:FTdiscrete}
    f({\bf n}) \equiv {1\over N^3}\sum_{\tilde {\bf n}} e^{-i{2\pi\over N} {\bf \tilde n n}} f({\bf \tilde n}) ~~~~ \Longleftrightarrow ~~~~  f({\bf \tilde n}) \equiv \sum_{\bf n} e^{+i{2\pi\over N} {\bf n \tilde n} }f({\bf n})\,,
\end{eqnarray}
where the weight in the first expression is a consequence of the identity $ \sum_{\bf n} e^{i{2\pi\over N} {\bf n} \tilde {\bf n} } = N^3\delta_{{\bf 0}, \tilde {\bf n}} $. As expected, the Fourier-transformed functions exhibit periodicity in the reciprocal lattice, {\it i.e.}~satisfy periodic boundary condition $ f({\bf\tilde{n}} + {\hat \imath} N) =  f({\bf\tilde{n}}) $.

The \textit{infrared} and \textit{ultraviolet} cutoffs in a lattice, {\it i.e.}~the minimum and maximum momentum resolved by the reciprocal lattice in each dimension, are
\be \label{eq:IRandUVmodes}
k_{\rm IR} \equiv \frac{2\pi}{L} = \frac{2\pi}{N\dx}\,, \hspace{0.7cm}  k_{\rm UV} \equiv {N\over2}k_{\rm IR} = {\pi\over \dx} \ . 
\ee
We note that $k_{\rm UV}$ is also known as the {\it Nyquist} frequency.

The reciprocal lattice hence captures a range of discrete momenta, 
\begin{eqnarray}
{\bf k} = k_{\rm IR} (\tilde{n}_1,\tilde{n}_2,\tilde{n}_3)\,,
\end{eqnarray}
with the maximum modulus corresponding to the diagonal of the reciprocal lattice $k_{\rm max}  = \sqrt{3}{N \over 2}k_{\rm IR} = \sqrt{3}\pi / \dx$. The modulus of momentum will be indicated as $k = k(\tilde n) 
\equiv k_{\rm IR}|\tilde {\bf n}|$, where $\tilde n = |\tilde {\bf n}| \equiv (\tilde{n}_1^2+\tilde{n}_2^2+\tilde{n}_3^2)^{1/2}$. We note that while the
the number of modes with approximately the same modulus 
grows roughly as $4\pi |\tilde{\bf n}|^2$ for sub-Nyquist modes ($k < k_{\rm UV}$), 
this number  starts decaying abruptly for supra-Nyquist modes ($k > k_{\rm UV}$) as we approach $k_{\rm max}$.

\subsubsection{Program variables}
\label{subsubsec:ProgramVariables}

When simulating the evolution of interactive fields on a lattice, it is convenient to work with \textit{program variables}, which are a set of dimensionless field and spacetime variables defined as follows,
\begin{align}
    \hspace*{-0.4cm}\delta\tilde\eta \equiv a^{- \alpha} \omega_* \delta t\ , \hspace{0.4cm}
    \delta\tilde x^i \equiv \omega_* \delta x\ ,
    \hspace{0.4cm}
    \tilde\phi = \frac{\phi}{f_*} \ , \hspace{0.4cm}
    \tilde\varphi = \frac{\varphi}{f_*} \ , \hspace{0.4cm} \widetilde{\Phi} = \frac{\Phi}{f_*} \ , \hspace{0.4cm}  \widetilde{A}_\mu=\frac{A_\mu }{\omega_*} \ , \hspace{0.4cm} \widetilde C_{\mu}^a = \frac{C_{\mu}^a}{\omega_*} \ , \label{eq:GaugeProgramVar}
\end{align}
where $\delta t$ and $\delta x$ are the time-step and lattice spacing used for solving the field dynamics, respectively, and $f_*$ and $\omega_*$ are constants of dimension mass +1. For each problem, one can choose $f_*$ and $\omega_*$ appropriately so that the program variables take numerical values of order unity during the fields' evolution. It is also convenient to define the \textit{program potential} as
\be\label{eq:ProgramPotMultiScalar}
\widetilde{V} (\tilde{\phi}, |\tilde{\varphi}|, |\widetilde{\Phi}|) \equiv \frac{1}{f_*^2 \omega_*^2} V(f_* \tilde \phi, f_* |\tilde \varphi|, f_* |\widetilde \Phi|  )\,, \  \ee
as well as program variables for the field strengths and covariant derivatives as follows,
\bea
\widetilde{F}_{\mu \nu} \equiv \frac{F_{\mu \nu}}{ \omega_*^2}\ , \hspace{0.4cm} \widetilde{G}_{\mu \nu}^a \equiv \frac{G_{\mu \nu}^a}{ \omega_*^2}\ , \hspace{0.4cm} \widetilde{D}_{\mu}^A \equiv \frac{D_{\mu}^A}{ \omega_*}\ , \hspace{0.4cm} \widetilde{D}_{\mu} \equiv \frac{D_{\mu}}{ \omega_*}\,.
\eea
Finally, we note that as the definition of linear momentum in Sect.~\ref{sub:reciprocal} scales as $k\propto {1/\delta x}$, we naturally normalize the linear momentum on the lattice using the inverse re-scaling for $\delta x$, {\it i.e.}
\begin{eqnarray}
    \kappa \equiv \frac{k}{\omega_*}\,.
\end{eqnarray}

\subsubsection{Gradients and lattice momentum}\label{subsec:LatticeMomentum}

When writing the fields' equation of motion (EOM) on the lattice, we need to substitute continuum derivatives by discretized operations that must reproduce the continuum expressions up to some order of accuracy in the lattice spacing/time step. For example, the derivative of a continuous function {\tt f}  
can be approximated by the following \textit{neutral} or \textit{centered} difference,
\begin{equation}
    \label{eq:neutrald}
    [\nabla^{(0)}_\mu {\tt f}] = \frac{{\tt f}({n}+\hat\mu) - {\tt f}({n}-\hat\mu)}{2\dx ^\mu} ~~\longrightarrow ~~ \partial_{\mu} {\tt f}({x})\big|_{{x}\,\equiv\, {\bf n}\dx+n_0\deta} + \mathcal{O}(\dx_\mu^2)\,,
\end{equation}
where $\delta x^{\mu}$ represents either the time step $\delta \eta$ (for $\mu = 0$) or the lattice spacing $\delta x$ (for $\mu = i$). The expression is symmetric around the lattice point $n$, and recovers the continuum expression up to $\mathcal{O}(\delta x_\mu^2)$. We could also approximate the continuous derivative by the following \textit{charged} difference,
\begin{eqnarray}
    \label{eq:forwardbackwardd}
    [\nabla^\pm_\mu {\tt f}] = \frac{\pm {\tt f}({n}\pm \hat\mu) \mp {\tt f}({n})}{\dx^\mu} ~~\longrightarrow ~~ \left\lbrace\begin{array}{l}
        \partial_{\mu} {\tt f}({x})\big|_{{x}\,\equiv\, {\bf n}\dx+n_0\deta} + \mathcal{O}(\dx_\mu)\,,  \vspace*{0.2cm}\\
        \partial_{\mu} {\tt f}({x})\big|_{{x}\,\equiv\, ({n} \pm \hat\mu/2)\dx^\mu} + \mathcal{O}(\dx_\mu^2)\,,
    \end{array}\right.
\end{eqnarray}
where $\nabla^+_\mu {\tt f}$ and $\nabla^-_\mu {\tt f}$ are called the \textit{forward} and \textit{backward} derivatives, respectively. Compared to the neutral derivative, they have the advantage of being sensitive to the minimum space interval captured by a lattice, {\it i.e.}~to the lattice spacing. These expressions, if expanded around an actual lattice site ${\bf n}$,  only recover the continuum derivative up to $\mathcal{O}({\delta x}_\mu)$. However, if expanded in between the two lattice sites involved, they approximate the continuum expression to $\mathcal{O}({\delta x}_\mu^2)$. One can also implement discrete derivatives of higher order at either grid or half-grid points, involving field values of at more lattice points \cite{DiscreteDerivatives}. 

Associated to each spatial lattice derivative, we can define a {\it lattice momentum} ${\bf k_\text{L}}$ through the following relation in Fourier space,
  \be  \label{eqn:latticemomentum}  [\nabla_i f]({\tilde{\bf n}}) = -i{\bf k}_{\text{L}}({\tilde{\bf n}}) f({\tilde{\bf n}}) \:. \ee
For example, the cartesian components of the lattice momentum for the derivative defined in (\ref{eq:neutrald}) is
\begin{equation}
    k^0_{\text{L}, i} = \dfrac{\sin (2\pi \Tilde{n}_i/N)}{\delta x} \: .\label{eqn:k0}\\
\end{equation}
while for the charged derivative \eqref{eq:forwardbackwardd} we have instead
\begin{align}
    k^{\pm}_{\text{L}, i} &= 2 e^{\mp i\pi \tilde{n}_i/N} \dfrac{\sin(\pi \tilde{n}_i/N)}{\delta x} = \dfrac{\sin(2 \pi \tilde{n}_i/N)}{\delta x} \mp i \dfrac{1-\cos(2 \pi \tilde{n}_i/N)}{\delta x} \: \ , \label{eqn:kpm} \\ 
    k_{{\rm L},i}^{\pm} &= 2\frac{\sin (\pi \tilde{n}_i / N)}{ \delta x} \: \ , \label{eq:latticeMomentum0}
\end{align}
where the first expression applies when $\nabla^\pm_i {\tt f}$ is defined on integer lattice sites, and the second when it is defined on half-integer lattice sites.

Finally, we mention that when one wishes to simulate scalar-gauge theories on a lattice, 
it is important to preserve gauge invariance. For such purpose, one needs to discretize the theory more carefully, in particular using {\it links} and {\it plaquettes}, which are quantities purposely defined to build gauge-invariant versions of discretized gauge theories. In this review, only the particular case of local strings 
in Sect.~\ref{sec:DefectsV}, involve scalar-gauge interactions that require to use these techniques, so we do not dwell on them here. We refer the reader to Sect.~3.2 of {\tt The Art\,I} for a discussion on lattice gauge-invariant techniques. 

\subsection{Power spectrum and initial conditions}
\label{subsec:PS}

Given a continuous function ${\tt f} ({\bf x})$, we are often interested on its power spectrum $\Delta_{\tt f}(k)$, which is defined in terms of its ensemble average $\langle {\tt f}^2 \rangle$ as follows,
\be
\langle {\tt f}^2 \rangle = \int d\log k~\Delta_{\tt f}(k)~~, ~~~\Delta_{\tt f}(k) \equiv {k^3\over 2\pi^2}\mathcal{P}_{\tt f}(k)~~,~~~ \langle {\tt f}_{\bf k} {\tt f}_{{\bf k}^{\prime}}^* \rangle = (2\pi)^3 \mathcal{P}_{\tt f}(k) \delta (\mathbf{k}-\mathbf{k^{\prime}})~. \label{eq:continuumPS}
\ee

\noindent On the lattice we substitute the previous ensemble average by an average over the lattice volume $V$, 
\be
\langle {f}^2 \rangle_V = \frac{\dx^3}{V}\sum_{\bf n} {f}^2({\bf n}) = \frac{1}{N^3}\sum_{\bf n} {f}^2({\bf n})~\,, \label{eq:Averagef2}
\ee
and using the definition of the discrete Fourier transform \eqref{eq:FTdiscrete}, re-write this expression as
\be
\langle {f}^2 \rangle_V = \frac{1}{2\pi}\sum_{|\tilde{\bf n}|}\Delta\log k(\tilde{\bf n}) ~k(\tilde{\bf n})\frac{\delta x}{N^5} \#_{R(\tilde{\bf n})} \big\langle \big|{f}(\tilde{\bf n})\big|^2\big\rangle_{R(\tilde{\bf n})}~\,,
\label{eq:discretePSaux}
\ee
where we have defined $\Delta \log k(\tilde{\bf n}) \equiv k_{\rm IR}/k(\tilde{\bf n})$, and introduced $\langle ( ... ) \rangle_{R(\tilde{\bf n})} \equiv \frac{1}{\#_{R(\tilde{\bf n})}}\sum_{\tilde{\bf n}^{\prime}\in R(\tilde{\bf n})}( ... )$ representing an angular average over a spherical shell, $R(\tilde{\bf n})$, that contains all sites with radius $|\tilde{\bf n}^{\prime}| \in \big[|\tilde{\bf n}|,|\tilde{\bf n}|+ \Delta\tilde{n}\big)$, with $\Delta\tilde{n}$ a given radial binning, and $\#_{R(\tilde{\bf n})} $ the {\it multiplicity}, {\it i.e.}~the number of sites contained within the spherical shell. Comparing Eq.~(\ref{eq:discretePSaux}) with Eq.~(\ref{eq:continuumPS}), we can define the lattice power spectrum as follows 
\begin{eqnarray}\label{eq:discretePST1}
\Delta_{f}(k(|{\bf \tilde{n}}|)) \equiv {k(\tilde{\bf n})\over 2\pi}\frac{\delta x}{N^5} \#_{R(\tilde{\bf n})} \big\langle \big|{f}(\tilde{\bf n})\big|^2\big\rangle_{R(\tilde{\bf n})} = \frac{k^3(\tilde {\bf n})}{2\pi^2}\;{\Upsilon_{|\tilde{\bf n}|}}\;\left(\frac{\delta x}{N}\right)^3 \big\langle \big|{f}(\tilde{\bf n})\big|^2\big\rangle_{R(\tilde{\bf n})}\,,
\end{eqnarray}
where 
\begin{eqnarray}\label{eq:Upsilon}
    \Upsilon_{|\tilde{\bf n}|} \equiv \frac{\#_{R(\tilde{\bf n})}}{4\pi|\tilde{\bf n}|^2}\;.
\end{eqnarray}
While the most precise evaluation of Eq.~(\ref{eq:discretePST1}) 
requires to compute $\Upsilon_{|\tilde{\bf n}|}$ exactly (for each bin) according to Eq.~(\ref{eq:Upsilon}), many works in the past commonly used the multiplicity approximation $\#_{R(\tilde{\bf n})} \simeq 4\pi |\tilde{\bf n}|^2$, so that $\Upsilon_{|\tilde{\bf n}|} \simeq 1$, hence dropping $\Upsilon_{|\tilde{\bf n}|}$ from Eq.~(\ref{eq:discretePST1}). While this is only an approximation, for historical reasons we define two types of power spectra, depending on the multiplicity assumption, 
\begin{eqnarray}\label{eq:TypeIandIIPS}
\left\lbrace
    \begin{array}{ccll}
         \Upsilon_{|\tilde{\bf n}|} \equiv  \frac{\#_{R(\tilde{\bf n})}}{4\pi|\tilde{\bf n}|^2} \neq 1 & \Rightarrow & \Delta_{f}(k(|{\bf \tilde{n}}|)) \equiv {k(\tilde {\bf n})\delta x\over 2\pi N^5}\#_{|\tilde{\bf n}|}\big\langle\left|f (\tilde{\bf n})\right|^2\big\rangle_{R(\tilde{\bf n})} &  \text{\tt [Type-I]} \vspace*{4mm}\\
          \Upsilon_{|\tilde{\bf n}|} = 1 \;, & \Rightarrow & \Delta_{f}(k(|{\bf \tilde{n}}|)) \simeq \frac{k^3(\tilde {\bf n})}{2\pi^2}\left(\frac{\delta x}{N}\right)^3 \big\langle \big|{f}(\tilde{\bf n})\big|^2\big\rangle_{R(\tilde{\bf n})} &  \text{\tt [Type-II]}
    \end{array}\right.\,.
\end{eqnarray}
The definition of {\tt Type-I} spectrum naturally incorporates the exact multiplicity, and hence the actual lack of statistical sampling of supra-Nyquist frequencies $k > {N\over2}k_{\rm IR}$ on a lattice\footnote{\CL uses by default {\tt Type-I} spectra for its output, but allows the user to switch to {\tt Type-II} if desired.}.
{\tt Type-II} can be actually seen as a good approximation of {\tt Type-I} spectra for many points in the (reciprocal) lattice, namely in the bins where $\#_{R(\tilde{\bf n})} \approx 4\pi |\tilde{\bf n}|^2$ holds. While such approximation is quite good at intermediate scales on a lattice, it fails moderately for the most infrared modes, and most noticeably it fails significantly for the ultraviolet modes above the {\it Nyquist} frequency, {\it i.e.}~$k > {N\over2}k_{\rm IR}$. For further details on these aspects, see~\cite{TechnicalNoteI}. 

The notion of power spectrum is particularly useful to initialize fundamental fields on a lattice. In the case of a scalar field, it is common to consider quantum vacuum fluctuations, characterized by a vacuum expectation value (continuum variance) as
\begin{eqnarray}
\mathcal{P}_{\tt f} (k) = \frac{1}{a^3} \frac{1}{2\omega_k}\,,~~~~~~ \mathcal{P}_{\tt f'} (k) = \frac{1}{a^{3-2\alpha}} {\omega_k\over 2}\,, ~~~~{\rm with}~~  \omega_k \equiv \left[(k/a)^2 + m_{\tt f}^2\right]^{1/2}\,,
\end{eqnarray}
where $m_{\tt f}^2 \equiv \frac{\partial^2 V({\tt f})}{\partial {\tt f}^2} > 0$. One initializes the field amplitudes ${f}(\tilde{\bf n})$ and time derivatives ${f}'(\tilde{\bf n})$ on the lattice by sampling from a Gaussian distributions $\mathcal{N}[\mu_k,\sigma_k]$ with vanishing mean ($\mu_k = 0$) and variance given by the power spectrum ($\sigma_k^2 \equiv \mathcal{P}_{\tt f}(k)$ or $\sigma_k^2 \equiv \mathcal{P}_{\tt f'}(k)$, respectively), so that
\begin{eqnarray}\label{eq:varICs}
|\tilde {f}(\tilde{\bf n})|^2  \equiv  {1\over \Upsilon_{|\tilde{\bf n}|}} \left(\frac{{N}}{\delta \tilde{x}}\right)^3 \mathcal{N}\big[0,\mathcal{P}_{\tt f}^{1/2}(k(|{\bf \tilde{n}}|))\big]^2\,,~~~~ |\tilde {f}'(\tilde{\bf n})|^2 \equiv {1\over \Upsilon_{|\tilde{\bf n}|}} \left(\frac{{N}}{\delta \tilde{x}}\right)^3 \mathcal{N}\big[0,\mathcal{P}_{\tt f'}^{1/2}(k(|{\bf \tilde{n}}|))\big]^2\;.
\end{eqnarray}
For a broader discussion on initial conditions from a generic power spectrum, see Sect.~\ref{subsec:ArbitrarySpectrum}.

\subsection{Evolution algorithms} \label{subsec:Algorithms}

The equations of motion (EOM) of canonical relativistic fields, Eqs.~(\ref{eq:singlet-eom})–(\ref{eq:SU2eom}), form a system of coupled second-order hyperbolic partial differential equations (PDE). As the fields propagate in an expanding background, we need to simultaneously consider  the EOM for the scale factor, Eq.~(\ref{eq:FriedmannDDa}). To solve all these equations on a lattice, we need to construct discretized versions of the EOM, and choose suitable integration schemes that satisfy the Hubble constraint in Eq.~(\ref{eq:FriedmannHubble}), and in the case of gauge theories, the Gauss constraints in Eqs.~(\ref{eq:GaussU1-eom})-(\ref{eq:GaussSU2-eom}). 
During the evolution 
we need to track 
the field amplitudes $\{ f_i \}$ and their corresponding conjugate momenta $\{ \pi_i \}$ ($\pi_i \propto \dot f_i$), both of which are evaluated at each lattice site. We also need to track 
the scale factor amplitude $a(\eta)$ and its conjugate momentum $\pi_a \equiv a'(\eta)$, which contrary to the fields, are homogeneous functions. 

The number of field amplitudes $\{ f_i \}$ defines the number of degrees of freedom ({\it dof}) in the system. 
The EOM of the fields then take the general form
\begin{eqnarray}\label{eq:SchemeContVirgin1}
\pi_a(\eta) &=& a'(\eta)\,,\\
\label{eq:SchemeContVirgin2}
\pi_a'(\eta) &=& \mathcal{K}_a[a(\eta), E_V(\eta), E_K(\eta), E_G(\eta)]\,,\\
\label{eq:SchemeContVirgin3}
\pi_i({\bf x},\eta) &=& \mathcal{D}_i[f_i'({\bf x},\eta),a(\eta),\pi_a(\eta);\lbrace f_{j}({\bf x},\eta) \rbrace, \lbrace f'_{j\neq i}({\bf x},\eta) \rbrace]\,,\\
\label{eq:SchemeContVirgin4}
\pi_i'({\bf x},\eta) &=& \mathcal{K}_i[f_i({\bf x},\eta),\pi_i({\bf x},\eta),a(\eta),\pi_a(\eta);\lbrace f_{j\neq i}({\bf x},\eta) \rbrace, \lbrace \pi_{j\neq i}({\bf x},\eta) \rbrace]\,,
\end{eqnarray}
where primes denote differentiation with respect to $\alpha$-time. Here $\mathcal{D}_i[...]$ is a functional---the {\it drift}---that defines the conjugate momentum of the $i$th $dof$, and $\mathcal{K}_i[...]$ is another functional---the {\it kernel} or {\it kick}---, that determines the interactions of the $i$th $dof$ with the rest of $dof's$ (possibly including itself). The kernel of the scale factor,  $\mathcal{K}_a[...]$, is given by the $rhs$ of Eq.~(\ref{eq:FriedmannDDa}), based on the volume average $\langle ... \rangle$ 
of the potential, kinetic and gradient energy densities of the $dof$ involved in the problem, namely $E_V \equiv \langle V \rangle$, $E_{K} \equiv \langle  \sum_j K_{j}\rangle $ and $E_{G} \equiv \langle \sum_j  G_{j}\rangle$.

This section reviews time-integration algorithms suitable for both canonical and non-canonical systems. While the algorithms we discuss can be adapted for any system of interactive fields, for clarity we illustrate each method adapting the algorithm to the the case of $N_s$ 
canonically normalized interacting scalar fields $\{\phi_i\}$.  
This are characterized by an action $S = - \int d^4x\, \sqrt{-g}\left(\frac{1}{2}\partial_{\mu} \phi_i \partial^{\mu} \phi_i + V(\lbrace \phi_j \rbrace) \right)$,
which, when specialized into a flat FLRW background given in Eq.~(\ref{eq:FLRWmetric}) and 
re-casted in terms of the program variables defined in Eqs.~(\ref{eq:GaugeProgramVar})-(\ref{eq:ProgramPotMultiScalar}), can be re-written as
\bea
\tilde S  = \left( \frac{\omega_*}{f_*}\right)^2 S = \int d^3\tilde x d \tilde\eta \left\{ \frac{1}{2} a^{3 - \alpha}\sum_i\left({\tilde\phi}_{i}\right)'^{\,2} - \frac{1}{2} a^{1 + \alpha} \sum_{i,k} (\tilde\partial_k \tilde\phi_{i})^2 - a^{3 + \alpha} \widetilde V(\lbrace \tilde\phi_{j} \rbrace) \right\} \, .
 \label{eq:ActionScalar}
\eea

While there is no unique way to obtain the discrete version of the EOM (see {\tt The Art\,I} for discussion on this), in this review we adopt a {\it hybrid} scheme, where at the level of the action only spatial derivatives are discretized\footnote{We demand  to recover the continuum limit at the level of the action at least to order $\mathcal{O}(dx^2)$}, while the temporal coordinate is treated as a continuous variable. Thus, the action for our reference example, using for example forward derivatives, reads
\be
\widetilde S^{\rm L} = \int d\tilde\eta\sum_{\bn} \delta \tilde x^{\,3} \left\{\frac{ 1}{2} a^{3 - \alpha}\sum_i( \tilde\phi_{i}')^2 - \frac{1}{2} a^{1 + \alpha}   \sum_{i,k} (\widetilde\nabla_k^+ \tilde\phi_{i})^2 - a^{3 + \alpha} \widetilde V (\lbrace \tilde\phi_j \rbrace ) \right\} \, , \label{eq:ActionScDiscHybrib}
\ee
and the scalar fields EOM 
\bea
\label{eq:EOMScalar-Discr_Hybrid}
\left(a^{3 - \alpha} \tilde\phi_{i}' \right)' & = & a^{1 + \alpha} \sum_k \widetilde\nabla_k^- \widetilde\nabla_k^+ \tilde\phi_{i}  -  a^{3 + \alpha} \widetilde V_{,\tilde\phi_{i}}\,,~~~~ i = 1, 2, ..., N_s\, , \\
\label{eq:EOMScaleFactor-Discr_Hybrid}
a'' & = &  \frac{1}{3} \left( \frac{ f_*}{m_p} \right)^2 a^{1+2\alpha}\Big[ (\alpha - 2){\widetilde E}_{K}  + \alpha {{\widetilde E}_{G}} + (\alpha + 1 ) {{\widetilde E}_V} \Big] \,,
\eea
with 
\bea\label{eq:EK_EG_EV_Discrete}
{\widetilde E}_K \equiv \frac{1}{2 a^{2\alpha}}\sum_{i}\left\langle (\tilde \phi_i')^2 \right\rangle\,,~~~ {\widetilde E}_G \equiv \frac{1}{2 a^2 }\sum_{i,k} \left\langle (\widetilde \nabla_k^+ \tilde \phi_{i})^2 \right\rangle\,, ~~~{\widetilde E}_V \equiv \left\langle \tilde{V}(\lbrace \tilde\phi_i\rbrace) \right\rangle\,.
\eea
From here, one can choose a suitable evolution algorithm to solve Eqs.~(\ref{eq:EOMScalar-Discr_Hybrid})-(\ref{eq:EOMScaleFactor-Discr_Hybrid}). 
As we will see, the hybrid prescription is specially suitable for the examples considered in this review, as it allows for a flexible choice of the time-integrator. 

In the following we present a collection of algorithms, 
divided into {\it symplectic} and {\it non-symplectic} integrators. Symplectic integrators include the {\it Leapfrog} and {\it Position-} and {\it Velocity-Verlet} methods, which are very stable numerical algorithms for canonical field theories, 
allowing for large-time evolution. They can also be extended 
to higher-order accuracy evolvers, know as {\it Yoshida} integrators, through recursive compositions of sub-steps. Non-symplectic integrators, on the other hand, are suitable for more general applications, including systems with non-canonical kinetic terms, dissipative dynamics, or interactions containing canonical momenta in the kernels. These algorithms include explicit {\it Runge–Kutta} schemes of various orders and multi-stage algorithms, which naturally allow for adaptive time-stepping and the use of auxiliary fields to handle intermediate sub-steps. Many of the complex models discussed in this review are non-canonical, and hence non-symplectic integrators are necessary to maintain accuracy and stability of the numerical solutions.

\subsubsection{Symplectic integrators}
\label{subsubsec:SymplecticInt}

Time integrators that are symplectic represent a class of algorithms tailored for the integration of Hamiltonian systems. The core principle of symplectic methods stems from {\it Liouville’s theorem}, which states that the phase-space volume must remain conserved throughout the system’s evolution. As a result, the field amplitudes and the corresponding conjugate momenta remain bounded, and they accurately preserve key constraints of the system, such as energy conservation. This property makes them particularly suitable for problems where a long-term dynamical behaviour is of primary importance.

One subtlety of these integrators lies in the importance of a wise choice of the conjugate momenta associated to the {\it dof}. 
An `improper' choice may lead to the loss of 
`symplecticity', resulting in a degradation of the 
desired 
numerical stability. This situation arises, for instance, when the choice of conjugate momentum $\pi_i$ associated to the $dof$ $f_i$, leads to a kernel that contains such momentum. 
In those cases, the application of a symplectic algorithm will not lead to an accurate (or even stable) solution. 

Following with the example of interacting scalar fields introduced in the previous subsection, a convenient choice for the conjugate momenta is the canonical choice (see {\tt The Art\,I} for a detailed discussion) 
\be
\tilde\pi_{i} \equiv a^{3-\alpha}\tilde\phi_i'\, ,
\ee
whereas for the scale factor we use
\be
 b = \pi_a \equiv a' \, .
\ee
The evolution kernels for our canonically normalised fields read therefore
\bea
\label{eq:EOMScalar-LatKernel}
\mathcal{K}^{\rm L}_{i}[a,\lbrace \tilde\phi_j \rbrace] & = & a^{1 + \alpha} \sum_k \widetilde\nabla_k^- \widetilde\nabla_k^+ \tilde\phi_{i}  -  a^{3 + \alpha} \widetilde V_{,\tilde\phi_{i}}\,,~~~~ i = 1, 2, ..., N_s\,, \\
\label{eq:EOMScaleFactor-LatKernel}
\mathcal{K}^{\rm L}_{a}[a,{\widetilde E}_K,{\widetilde E}_G,{\widetilde E}_V] & = & \frac{1}{3} \left( \frac{ f_*}{m_p} \right)^2 a^{1+2\alpha}\Big[ (\alpha - 2){\widetilde E}_{K}  + \alpha {{\widetilde E}_{G}} + (\alpha + 1 ) {{\widetilde E}_V} \Big] \,,
\eea

We review now representative cases of symplectic integrators (for en extensive discussion on these, see {\tt The Art-I}). To this end, we present their concrete implementation for the reference case of scalar interactive singlets.\\

\textbf{I) (Staggered) Leapfrog}. The {\it leapfrog} algorithm is one of the simplest methods for solving second order differential equations that ensures order $\mathcal{O}(\delta \eta^2)$. It requires that the field amplitudes and their conjugate momenta are displaced between each other by a half-time step $\delta\eta/2$. The same applies to the scale-factor and its derivative. In our case of reference, a convenient choice of the conjugate momenta is
\be
\tilde\pi_{i,+0/2}=a_{+0/2}^{3 - \alpha} \widetilde\nabla_0^+ \tilde\phi_{i}\, ,
\ee
with the subindex $_{+0/2}$ indicating that the evaluation must be done half time-step ahead. The algorithm consists of a `kick-drift' scheme with discretized EOM as
\bea
\widetilde\nabla_0^- [\tilde\pi_{i,+0/2} ] & = & \mathcal{K}^{\rm L}_{i}[a,\lbrace \tilde\phi_j \rbrace]\, ,~~~~ i = 1, 2, ..., N_s\,,\\
\label{eq:EOMScalar-Discr}
b' & = &  \mathcal{K}^{\rm L}_{a}[a,\overline{{\widetilde E}_{K}},{\widetilde E}_G,{\widetilde E}_V] \,, ~~~~{\rm with}~~ \overline{{\widetilde E}_{K}} \equiv \left({\widetilde E}_{K} + {\widetilde E}_{K,-0/2} \right)/2\,,
\eea
where
\bea\label{eq:EK_EG_EV_Discrete_2}
{\widetilde E}_K \equiv \frac{1}{2 a^{2\alpha}_{+0/2} }\sum_{i}\left\langle (\widetilde\nabla_0^+\tilde \phi_i)^2 \right\rangle\,,~~~ {\widetilde E}_G \equiv \frac{1}{2 a^2 }\sum_{i,k} \left\langle (\widetilde \nabla_k^+ \tilde \phi_{i})^2 \right\rangle\,, ~~~{\widetilde E}_V \equiv \left\langle \tilde{V}(\lbrace \tilde\phi_i\rbrace) \right\rangle\, .
\eea
An iterative scheme is then written as
\begin{eqnarray}
&& \hspace*{2mm}IC  :  \lbrace \tilde\phi_i,a \rbrace {\rm ~at~} \tilde\eta_0, ~~~\lbrace \tilde\pi_{i,-{0}/2},b_{-{0}/2}\rbrace {\rm ~at~} \tilde\eta_0-0.5\delta\tilde\eta\, , \nonumber\\[1mm]
&& \left\lbrace
\begin{array}{rcl}
\tilde\pi_{i,+0/2} & = & \tilde\pi_{i,-0/2} + \delta\tilde\eta\mathcal{K}^{\rm L}_{i}[a,\lbrace \tilde\phi_j \rbrace]\, , \vspace*{0.15cm}\\
b_{+0/2} &=& b_{-0/2} + \delta\tilde\eta \mathcal{K}^{\rm L}_{a}[a,\overline{{\widetilde E}}_{K},{\widetilde E}_G,{\widetilde E}_V]\, , \vspace*{0.15cm}\\
a_{+0} &=&  a + \delta\tilde\eta\, b_{+0/2},\ ~~~~ \longrightarrow ~~~~ a_{+0/2} \equiv (a_{+0} + a)/2\,,\\
\tilde\phi_{i,+0} &=& \tilde\phi_a + \delta\tilde\eta\,\tilde\pi_{i,+0/2}a_{+0/2}^{-(3-\alpha)}\,,\vspace*{0.15cm}\\
\end{array}
\right. \\[1mm]
&& \hspace*{2mm}HC : b_{+0/2}^2 = \frac{1}{3} \left( \frac{ f_*}{m_p} \right)^2a_{+0/2}^{2(\alpha+1)} \Big({{\widetilde E}_{K}} + \overline{{\widetilde E}}_{G} + \overline{{\widetilde E}}_{V} \,\Big)\,,\nonumber
\label{eq:HCschemeIII}
\end{eqnarray}
where $\overline{{\widetilde E}}_{K} \equiv \left({\widetilde E}_{K, -0/2} + {\widetilde E}_{K,+0/2} \right)/2$, $\overline{{\widetilde E}}_{G} \equiv \left({\widetilde E}_{G} + {\widetilde E}_{G,+0} \right)/2$ and $\overline{{\widetilde E}}_{V} \equiv \left({\widetilde E}_{V} + {\widetilde E}_{V,+0} \right)/2$. Above $IC$ represents the {\it initial conditions}, whereas $HC$ stands for {\it Hubble Constraint}.\\

\textbf{II) Velocity- and Position-Verlet}. Verlet methods eliminate the half–time-step offset in the leapfrog method between field amplitudes and conjugate momenta, by either applying the velocity part of the leapfrog algorithm at two successive half–time steps but with a single position update in between, or by applying the coordinate part of the leapfrog algorithm at two successive half–time steps with one velocity update in between. The former prescription is known as known as the {\it Velocity-Verlet} (VV) or “kick–drift–kick”  scheme, whereas the latter is known as the {\it Position-Verlet} (PV) or “drift-kick–drift” scheme. Through the intermediate steps both position and velocity can be obtained after the three steps at integer times, with an accuracy up to order $\mathcal{O}(\delta \eta^2)$. For our reference example of singlet fields, the Verlet iterative schemes read\\

\newpage
\noindent
\begin{minipage}[t]{0.5\textwidth}
\raggedright
\hspace{0.15cm}\textbf{II-1)} Velocity Verlet
\end{minipage}
\hfill
\begin{minipage}[t]{0.47\textwidth}
\raggedright
\textbf{II-2)} Position Verlet
\end{minipage}

\vspace{0.4cm}

\begin{minipage}[t]{0.5\textwidth}
\raggedright
\hspace{0.45cm}$IC: \{\tilde{\phi}_i,\tilde{\pi}_i,a,b\}\ \text{at}\ \tilde{\eta}_0\, ,$
\end{minipage}
\hfill
\begin{minipage}[t]{0.47\textwidth}
\raggedright
\hspace{0.25cm}$IC:  \{\tilde{\phi}_i,\tilde{\pi}_i,a,b\}\ \text{at}\ \tilde{\eta}_0\, ,$
\end{minipage}

\vspace{-0.5cm}
\noindent
\begin{minipage}[t]{0.5\textwidth}
\raggedright
\begin{equation}
\vcenter{\hbox{%
$\displaystyle
\left\lbrace
\begin{array}{@{}l}
b_{+0/2} = b +{\dfrac{\delta\tilde\eta}{2}} 
\mathcal{K}^{\rm L}_{a}[a,{\widetilde E}_{K},{\widetilde E}_G,{\widetilde E}_V]\,,\\\vspace{0.15cm}
\tilde\pi_{i,+0/2} = \tilde\pi^{(b)} + 
{\dfrac{\delta\tilde\eta}{2}} \mathcal{K}^{\rm L}_{i}[a,\{\tilde\phi_j\}]\,,~\\\vspace{0.15cm}
a_{+0} =  a +  {\delta\tilde\eta}b_{+0/2}\,,~a_{+0/2} = \dfrac{a_{+0}+a}{2}\,,\\\hspace{0.15cm}
\tilde\phi_{i,+0} = \tilde\phi_i + 
\delta\tilde\eta\,\tilde\pi_{i,+0/2}a_{+0/2}^{-(3-\alpha)}\,,\\\vspace{0.15cm}
\tilde\pi_{i,+0} = \tilde\pi_{i,+0/2} +
{\dfrac{\delta\tilde\eta}{2}} \mathcal{K}^{\rm L}_{i}[a,\{\tilde\phi_j\}]\big|_{+0}\,,\\\vspace{0.15cm}
b_{+0} = b_{+0/2} + {\dfrac{\delta\tilde\eta}{2}}  
\mathcal{K}^{\rm L}_{a}[a,{\widetilde E}_{K},{\widetilde E}_G,{\widetilde E}_V]\big|_{+0}\,,
\end{array}
\right.
$}}\nonumber
\end{equation}
\end{minipage}
\hfill
\begin{minipage}[t]{0.47\textwidth}
\raggedright
\begin{equation}
\vcenter{\hbox{%
$\displaystyle
\left\lbrace
\begin{array}{@{}l}
a_{+0/2} =  a + {\dfrac{\delta\tilde\eta}{2}} b\,,\\\vspace{0.15cm}
\tilde\phi_{i,+0/2} = \tilde\phi_i + 
{\dfrac{\delta\tilde\eta}{2}}\,\tilde\pi_i a^{-(3-\alpha)}\,,\\\vspace{0.15cm}
\tilde\pi_{i,+0} = \tilde\pi_i +
{\delta\tilde\eta}\,\mathcal{K}^{\rm L}_{i}[a,\{\tilde\phi_j\}]\big|_{+0/2}\,,\\\vspace{0.15cm}
b_{+0} = b +{\delta\tilde\eta}\,
\mathcal{K}^{\rm L}_{a}[a,\overline{\widetilde{E}}_{K},{\widetilde E}_G,{\widetilde E}_V]\big|_{+0/2}\,,\\\hspace{0.15cm}
a_{+0} =  a_{+0/2} + {\dfrac{\delta\tilde\eta}{2}} b_{+0}\,,\\\vspace{0.15cm}
\tilde\phi_{i,+0} = \tilde\phi_{i,+0/2} + 
{\dfrac{\delta\tilde\eta}{2}}\,\tilde\pi_{i,+0}a_{+0}^{-(3-\alpha)}\,,
\end{array}
\right.
$}}
\end{equation}
\end{minipage}

\vspace{0.15cm}
\begin{minipage}[t]{0.5\textwidth}
\raggedright
\hspace{0.45cm}$HC: b^2 = \dfrac{1}{3}\!\left(\dfrac{ f_*}{m_p}\right)^2
a^{2(\alpha+1)} \big({\widetilde E}_{K} + {\widetilde E}_{G} + {\widetilde E}_{V}\big)\, ,$
\end{minipage}
\hfill
\begin{minipage}[t]{0.47\textwidth}
\raggedright
\hspace{0.25cm}$HC: b^2 = \dfrac{1}{3}\!\left(\dfrac{ f_*}{m_p}\right)^2
a^{2(\alpha+1)} \big({\widetilde E}_{K} + {\widetilde E}_{G} + {\widetilde E}_{V}\big)\, ,$\\
\end{minipage}
\vspace{0.2cm}
with $\overline{{\widetilde E}}_{K} \equiv \left({\widetilde E}_{K} + {\widetilde E}_{K,+0} \right)/2$.

\textbf{III) Yoshida: Verlet Integration of $\mathcal{O}(\delta\eta^n)$}. The Verlet integration methods can be used recursively to construct higher–order (even) integrators with accuracy $\mathcal{O}(\delta \eta^n)$, with $n = 4, 6, 8, ...$. A single time step $\delta \eta$ is decomposed into $s$ sub-steps, $\delta \eta_p = w_p \delta \eta$, with $\sum_{p=1}^s w_p = 1$, and the corresponding Verlet algorithm is applied sequentially in each sub-step. For instance, the Velocity-Verlet version of this scheme can be written as
\be
IC  :  \lbrace \tilde \phi_i^{(0)},\tilde\pi_i^{(0)},a^{(0)},b^{(0)}\rbrace {\rm ~at~} \tilde\eta_0\,,\\\nonumber
\ee
\vspace*{-0.75cm}
\begin{equation}
\left\lbrace
\begin{array}{rcl}
b^{(p)}_{1/2} &=& b^{(p-1)} + \omega_p{\delta\tilde\eta\over 2}\mathcal{K}_{ a}^{{\rm L},(p-1)}\, ,\vspace*{0.15cm}\\
\tilde\pi^{(p)}_{i,1/2} &=& \tilde\pi_i^{(p-1)} + \omega_p{\delta\tilde\eta\over 2}\mathcal{K}_{i}^{{\rm L},(p-1)}\, ,\vspace*{0.15cm}\\
a_{1/2}^{(p)} &=&  a^{(p-1)} + b_{1/2}^{(p)}\omega_p{\delta\tilde\eta\over2}\, ,\vspace*{0.15cm}\\
\tilde\phi^{(p)}_{i} &=& \tilde\phi^{(p-1)}_i + \omega_p\delta\tilde\eta\,\tilde\pi_{i,1/2}^{(p)}(a_{1/2}^{(p)})^{-(3-\alpha)}\, ,\vspace*{0.15cm}\\
a^{(p)} &=& a^{(p)}_{1/2} +  b^{(p)}_{1/2}\omega_p{\delta\tilde\eta\over2}\,,\vspace*{0.15cm}\\
\tilde\pi_{i}^{(p)} & = & \tilde\pi^{(p)}_{i,1/2} + \omega_p{\delta\tilde\eta\over 2}\mathcal{K}_{i}^{{\rm L}, (p)}\, ,
\vspace*{0.15cm}\\
b^{(p)} &=& b^{(p)}_{1/2} + \omega_p{\delta\tilde\eta\over 2}\mathcal{K}_{a}^{{\rm L}, (p)}\, ,
\end{array}
\right\rbrace_{p\,=\,1,\, ...,\, s} \hspace*{-1cm}
\Longrightarrow
\left\lbrace
\begin{array}{rcl}
\tilde\phi_{i,+0} &=&  \tilde\phi_i^{(s)}\, , \vspace*{0.15cm}\\
a_{+0} &=& a^{(s)}\, , \vspace*{0.15cm}\\
\tilde\pi_{i,+0} &=& \tilde\pi_i^{(s)}\, , \vspace*{0.15cm}\\
b_{+0} &=& b^{(s)}\, ,\vspace*{0.15cm} 
\end{array}
\right.
\end{equation}
\vspace*{-0.5cm}
\be
HC : b^2 = \frac{1}{3} \left( \frac{ f_*}{m_p} \right)^2a^{2(\alpha+1)} \Big({{\widetilde E}_{K}} + {{\widetilde E}_{G}} + {{\widetilde E}_{V}} \Big)\, ,\nonumber
\ee
where we have compacted the notation introducing  $\mathcal{K}^{{\rm L}, (l)}_{a}=\mathcal{K}^{\rm L}_{a}[a^{(l)},{\widetilde E}_K^{(l)},{\widetilde E}_G^{(l)},{\widetilde E}_V^{(l)}]$ and $\mathcal{K}^{{\rm L}, (l)}_{i}= \mathcal{K}^{\rm L}_{i}[a^{(l)},\lbrace\tilde\phi_{j}^{(l)}\rbrace]$. The sub-index $_{1/2}$ represents intermediate updates of the variables at each iteration step and should not be confused with a half–time-step displacement. Using the appropriate coefficients $w_p$, see Table~\ref{tab:VVnCoeffs} of the Appendix, one achieves a cancellation of truncation errors up to order $\mathcal{O}(\delta \eta^{n})$, with $n = 4, 6, 8,$ and $10$, corresponding to $s = 3, 7, 15,$ and $31$ sub-steps, respectively.

\subsubsection{Non-symplectic integrators}
\label{subsubsec:NonSymplecticInt}

Non-symplectic integrators form a versatile set 
of methods with broad applicability, performing well across a variety of systems for which symplectic methods are less suitable, including non-Hamiltonian, dissipative, or stiff ones with canonical momenta appearing in the kernels. Non-symplectic methods can naturally accommodate adaptive time-stepping. 

In these  
schemes, both 
the field amplitudes and their 
conjugate momenta are defined 
at the same time step. 
On the other hand, because these methods involve the execution of intermediate sub-steps, {\it auxiliary fields} are required to store the information at each stage. In the case of interacting scalar fields, a simple choice for the conjugate momenta,  
\be
\tilde{\pi}_i \equiv \tilde{\phi}_i' \, ,
\ee
suffices. The evolution kernel takes then the form
\be\label{eq:scalar_singlet_eom}
\tilde{\pi}'_i = \mathcal{K}^{\rm L}_{i}[a,\lbrace\tilde\phi_{j}\rbrace, b, \tilde{\pi}_{i}] \equiv -(3 - \alpha)\frac{a'}{a}\tilde{\pi}_{i} + a^{-2 (1  - \alpha )} \sum_i \tilde{\nabla}_i^-\tilde{\nabla}_i^+ \tilde{\phi}_i - \widetilde V_{,\tilde\phi_{i}} \; , ~~~~ i = 1, 2, ..., N_s\,.\\
\ee
which we note it depends explicitly on $\tilde{\pi}_{i}$.\\ 

\textbf{I) Runge-Kutta 2nd order (RK2)}. These algorithms provide an evolution scheme accurate to $\mathcal{O}(\delta \eta^2)$ by introducing one intermediate step, whose information is stored in auxiliary fields, one per field \textit{dof}. While there exist several implementations, here we review one of the most common ones, known as the \textit{modified Euler} method,
\be
IC  :  \lbrace \tilde \phi_i,\tilde\pi_i,a,b\rbrace {\rm ~at~} \tilde\eta_0\,,\\ \nonumber
\ee
\vspace{-0.75cm}
\begin{equation}
\vspace{-0.75cm}
\left\lbrace
\begin{array}{rcl}
\tilde{\phi_i}^{(1)} = \tilde{\phi_i}\,, & \tilde{\phi_i}^{(2)} = \tilde{\phi_i}^{(1)} + \delta\tilde{\eta}\tilde{\pi}^{(1)}_{i}\,,\vspace*{0.15cm}\\
\tilde{\pi}^{(1)}_{i} = \tilde{\pi}_{i} & \tilde{\pi}^{(2)}_{i} = \tilde{\pi}^{(1)}_{i} + {\delta\tilde{\eta}}\mathcal{K}_{i}^{{\rm L},(1)} \,,
\vspace*{0.15cm}\\
a^{(1)} = a\,, & a^{(2)} = a^{(1)} + {\delta\tilde{\eta}}\tilde{\pi}^{(1)}_a\,,\vspace*{0.15cm}\\
b^{(1)} = b\,, & b^{(2)} = b^{(1)} + {\delta\tilde{\eta}}\mathcal{K}_{a}^{{\rm L}, (1)}\,,
\end{array}
\right\rbrace\Longrightarrow
\left\lbrace
\begin{array}{rcl}
\tilde{\phi}_{i,+{0}} &=&
\tilde{\phi}^{(1)}_i
+ \frac{1}{2}\delta\tilde{\eta}
\left[\tilde{\pi}_i^{(1)}+\tilde{\pi}_i^{(2)}\right]\,, \\[2mm]
a_{+{0}} &=&
a^{(1)} + \frac{1}{2}\delta\tilde{\eta}
\left[b^{(1)}+b^{(2)}\right]\,, \\[2mm]
\tilde{\pi}_{i,+{0}} &=&
\tilde{\pi}^{(1)}_{i}
+ \frac{1}{2}\delta\tilde{\eta}
\left[\mathcal{K}_{i}^{{\rm L},(1)}+\mathcal{K}_{i}^{{\rm L},(2)}\right]\,, \\[2mm]
b_{+{0}} &=&
b^{(1)} + \frac{1}{2}\delta\tilde{\eta}
\left[\mathcal{K}_{a}^{{\rm L},(1)}+\mathcal{K}_{a}^{{\rm L},(2)}\right]\,,
\end{array}\label{eq:RK2algorithm_1}
\right.
\end{equation}
\vspace{-0.5cm}
\bea
HC : b^2 = \frac{1}{3} \left( \frac{ f_*}{m_p} \right)^2a^{2(\alpha+1)} \Big({{\widetilde E}_{K}} + {{\widetilde E}_{G}} + {{\widetilde E}_{V}} \Big)\,,\nonumber
\eea
where again we use $\mathcal{K}^{{\rm L},(l)}_{ i} = \mathcal{K}_{i}^{\rm L}[a^{(l)},\lbrace\tilde\phi_{j}^{(l)}\rbrace,b^{(l)},\tilde{\pi}^{(l)}_{i}]$ and $\mathcal{K}_{a}^{{\rm L}, (l)}=\mathcal{K}_{a}^{\rm L}[a^{(l)},{\widetilde E}_K^{(l)},{\widetilde E}_G^{(l)},{\widetilde E}_V^{(l)}]$.\\

\textbf{II) Runge-Kutta 4th order (RK4)}. The accuracy can be increased to $\mathcal{O}(\delta\eta^4)$ by adding a weighted average of four derivative stages in the previous Runge-Kutta algorithm of 2nd order. This leads to the renowned RK4 algorithm as
\be
IC :  \lbrace \tilde \phi_i,\tilde\pi_i,a,b\rbrace {\rm ~at~} \tilde\eta_0\,,\\ \nonumber
\ee
\vspace*{-0.7cm}
\begin{equation}
\left.
\hspace{7.5mm}
\begin{array}{c}
\left\lbrace
\begin{array}{llll}
\tilde{\phi}^{(1)}_i = \tilde{\phi}_i\,, & \tilde{\phi}^{(2)}_i = \tilde{\phi}^{(1)}_i + {{\delta\tilde{\eta}}\over2}\tilde{\pi}^{(1)}_{i}\,, & \tilde{\phi}^{(3)}_i = \tilde{\phi}^{(1)}_i + {{\delta\tilde{\eta}}\over2}\tilde{\pi}^{(2)}_{i}\,, & \tilde{\phi}^{(4)}_i = \tilde{\phi}^{(1)} + {\delta\tilde{\eta}}\tilde{\pi}^{(3)}_{i}\,,\vspace*{0.15cm}\\
\tilde{\pi}^{(1)}_{i} = \tilde{\pi}_{i} & \tilde{\pi}^{(2)}_{i} = \tilde{\pi}^{(1)}_{i} + {{\delta\tilde{\eta}}\over2}\mathcal{K}^{{\rm L}, (1)}_{i}\,, & \tilde{\pi}^{(3)}_{i} = \tilde{\pi}^{(1)}_{i} + {{\delta\tilde{\eta}}\over2}\mathcal{K}^{{\rm L}, (2)}_{i}\,, & \tilde{\pi}^{(4)}_{i} = \tilde{\pi}^{(1)}_{i} + {\delta\tilde{\eta}}\mathcal{K}^{{\rm L}, (3)}_{i}\,,\vspace*{0.15cm}\\
a^{(1)} = a\,, & a^{(2)} = a^{(1)} + {{\delta\tilde{\eta}}\over2}b^{(1)}\,, & a^{(3)} = a^{(1)} + {{\delta\tilde{\eta}}\over2}b^{(2)}\,, & a^{(4)} = a^{(1)} + {\delta\tilde{\eta}}b^{(3)}\,,\vspace*{0.15cm}\\
b^{(1)} = b\,, & b^{(2)} = b^{(1)} + {{\delta\tilde{\eta}}\over2}\mathcal{K}_{a}^{{\rm L}, (1)}\,, & b^{(3)} = b^{(1)} + {{\delta\tilde{\eta}}\over2}\mathcal{K}_{a}^{{\rm L},(2)}\,, & b^{(4)} = b^{(1)} + {\delta\tilde{\eta}}\mathcal{K}_{a}^{{\rm L}, (3)}\,, 
\end{array}
\right\rbrace\Longrightarrow\nonumber
\end{array}
\right.
\end{equation}
\vspace*{-0.5cm}\\
\begin{equation}
\Longrightarrow
\left\lbrace
\begin{array}{rcl}
\tilde{\phi}_{i,+0} &=& \tilde{\phi}^{(1)}_i + {1\over6}\delta\tilde{\eta}\left[\tilde{\pi}^{(1)}_{i}+2\tilde{\pi}^{(2)}_{i}+2\tilde{\pi}^{(3)}_{i}+\tilde{\pi}^{(4)}_{i}\right]\,,\vspace*{0.15cm}\\
a_{+0} &=& a^{(1)} + {1\over6}\delta\tilde{\eta}\left[b^{(1)}+2b^{(2)}+2b^{(3)}+b^{(4)}\right]\,,\vspace*{0.2cm}\\
\tilde{\pi}_{i,+0}&=&\tilde{\pi}^{(1)}_{i}+{1\over6}\delta\tilde{\eta}\left[\mathcal{K}^{{\rm L},(1)}_{i}+2\mathcal{K}^{{\rm L}, (2)}_{i}+2\mathcal{K}^{{\rm L}, (3)}_{i}+\mathcal{K}^{{\rm L}, (4)}_{i}\right]\,,\vspace*{0.15cm}\\
b_{+0}&=&b^{(1)}+{1\over6}\delta\tilde{\eta}\left[\mathcal{K}_{a}^{{\rm L}, (1)}+2\mathcal{K}_{a}^{{\rm L}, (2)}+2\mathcal{K}_{a}^{{\rm L},(3)}+\mathcal{K}_{a}^{{\rm L}, (4)}\right]\,,\vspace*{0.15cm}\\
\end{array}\right. \label{eq:RK4algorithm_1}
\end{equation}
\vspace*{-0.75cm}\\
\bea
HC : b^2 = \frac{1}{3} \left( \frac{ f_*}{m_p} \right)^2a^{2(\alpha+1)} \Big({{\widetilde E}_{K}} + {{\widetilde E}_{G}} + {{\widetilde E}_{V}} \Big)\,.\nonumber
\eea

\textbf{III) Low-storage Runge-Kutta}. These methods represent a refined version of the previous schemes, in which the number of auxiliary fields is reduced while maintaining the integration accuracy of $\mathcal{O}(\delta \eta^n)$ \cite{Carpenter1994Thirdorder2R,Carpenter1994Fourthorder2R,Bazavov:2021pik,Bazavov:2025dzo,Bazavov:2025exj}. This is achieved by introducing $s$ intermediate sub-stages, each with its corresponding weight coefficient.
\be
IC  :  \lbrace \tilde \phi_i^{(0)},\tilde\pi_i^{(0)},a^{(0)},b^{(0)}\rbrace {\rm ~at~} \tilde\eta_0\,,\\ \nonumber
\ee
\vspace*{-1cm}\\
\begin{equation}
\left\lbrace
\begin{array}{rcl}
\Delta \tilde\phi^{(p)}_i
&=& A_p \Delta \tilde\phi^{(p-1)}_i
+ \delta \tilde \eta \tilde\pi_{\phi i}^{(p-1)} \, , \\[1mm]
\Delta\tilde\pi_{i}^{(p)}
&=& A_p\Delta\tilde\pi_{i}^{(p-1)}
+ \delta \tilde \eta \mathcal{K}^{{\rm L}, (p-1)}_{i} \, , \\[1mm]
\Delta a^{(p)}
&=& A_p \Delta a^{(p-1)}
+ \delta \tilde \eta b^{(p-1)} \, , \\[1mm]
\Delta b^{(p)}
&=& A_p\Delta b^{(p-1)}
+ \delta \tilde \eta \mathcal{K}^{{\rm L}, (p-1)}_{a} \, , 
 \end{array}
 \Longrightarrow
 \begin{array}{rcl}
       \tilde\phi^{(p)}_i &=& \tilde\phi^{(p-1)}_i + B_p  \Delta\tilde\phi^{(p)}_i\, ,  \vspace*{0.15cm}\\
        \tilde\pi_{i}^{(p)} &=&\tilde\pi_{i}^{(p-1)}+  B_p \Delta\tilde\pi_{i}^{(p)}\, ,   \vspace*{0.15cm}\\
        a^{(p)} &=&a^{(p-1)} +B_p  \Delta a^{(p)}\, ,  \vspace*{0.15cm}\\
        b^{(p)} &=&b^{(p-1)} +B_p  \Delta b^{(p)}\, , \vspace*{0.15cm}
\end{array}
\right\rbrace_{p\,=\,1,\, ...,\, s} \Longrightarrow
\hspace*{-1cm}\nonumber
\end{equation}
\vspace{-0.5cm}
\begin{equation}
\Longrightarrow
\left\lbrace
\begin{array}{rcl}
\tilde\phi_{i,+0} &=&  \tilde \phi_i^{(s)}\, , \vspace*{0.15cm}\\
a_{+0} &=& a^{(s)}\,, \vspace*{0.15cm}\\
\tilde\pi_{i,+0} &=& \tilde\pi_i^{(s)}\, , \vspace*{0.15cm}\\
b_{+0} &=& b^{(s)}\, ,\\
\end{array}\label{eq:RKLSalgorithm_1}
\right.
\end{equation}
\vspace*{-0.75cm}\\
\bea
HC : b^2 = \frac{1}{3} \left( \frac{ f_*}{m_p} \right)^2a^{2(\alpha+1)} \Big({{\widetilde E}_{K}} + {{\widetilde E}_{G}} + {{\widetilde E}_{V}} \Big)\,.\nonumber
\eea

The auxiliary fields $\{\Delta\tilde{\phi}^{(p)}, \Delta\tilde{\pi}^{(p)}_{i}, \Delta a^{(p)}, \Delta b^{(p)}\}$ are updated with information at each stage $p$ up to a total of $s$ stages, without the need to define additional kernels.

For instance, using 2 intermediate stages and  $\{(A_p, B_p)\} = \{ (0, 1), (-1, 1/2) \}$, we recover the explicit 2nd order RK. Moreover, the accuracy can be increased to third order with 4 intermediate stages  using
$\{(A_p, B_p)\}~ = ~ \{ \,(\,0.0, 0.06688758201974097\,)\,,\,(\,-0.7825460361923583\,, \,2.876554598956719\,),\, ...$
\\$ ~(\,-2.042914325731225\,,\,0.5534657361343982\,),(\,-1.799337253940777\,,\, 0.3912730180961791\,)\,\}\,$, for example. We refer the reader to the Table \ref{tab:RKlsCoefficients} from the Appendix to find the necessary coefficients for other orders/stages.

\section{Non-minimally coupled scalar fields}
\label{sec:NMCoupled_scalars}
~~~~Here we discuss scalar field dynamics with non-canonical interactions. We consider scalar fields non-minimally coupled (NMC) to gravity through a term of the form $\propto \phi^{2}R$, with $R$ the Ricci scalar, and scalar fields with non-minimal kinetic (NMK) terms of the form $f(\phi)X$, with $X \equiv \partial^{\mu}\phi\partial_{\mu}\phi$. These two possibilities have been extensively studied in early Universe scenarios, both analytically and numerically, see, {\it e.g.}~\cite{Bassett:1997az,Tsujikawa:1999jh,Tsujikawa:1999iv,Watanabe:2006ku,Garcia-Bellido:2008ycs,Figueroa:2016dsc,DeCross:2015uza,DeCross:2016fdz,DeCross:2016cbs,Ema:2016dny,Figueroa:2017slm,Sfakianakis:2018lzf,Nguyen:2019kbm,Fu:2019qqe,Opferkuch:2019zbd,Dimopoulos:2018wfg,Bezrukov:2020txg,vandeVis:2020qcp,Bettoni:2021zhq,Figueroa:2021iwm,Laverda:2023uqv,Laverda:2024qjt,Figueroa:2024asq,Figueroa:2024yja} for the former, and~\cite{Lachapelle:2008sy,Child:2013ria,Rahmati:2014cwa,Li:2019ncw,Adshead:2023nhk,Adshead:2024ykw,Huang:2024amu} for the latter. 

\subsection{Non-minimal coupling to gravity} \label{sec:NMC}

Quantized scalar fields $\phi$ on a curved spacetime background require an interaction term with the Ricci scalar $\xi R \phi^2$ in order to renormalize the theory. The (dimensionless) coupling $\xi$ is therefore a running parameter that cannot be set to zero at every energy scale. In many cosmological applications, such as Higgs inflation or (p)reheating, the inclusion of this term is crucial. If one considers all operators respecting the symmetries of the Standard Model (SM) and gravity, the operator $\xi|\Phi|^2 R$, with $\Phi$ the SM Higgs, is the only renormalizable one not already present in the SM, so one can consider $\xi$ to be the last unknown parameter of the SM. It is traditionally dealt with via a transformation to the Einstein frame using a conformal rescaling of the metric such that the gravitational action takes the standard Einstein--Hilbert form; however, here we work directly in the original Jordan frame, preserving the non-minimal coupling (NMC) explicitly. We first cover the continuum theory and then present our lattice discretization.
We keep the presentation compact; for details on the derivations, one should refer to~\cite{Figueroa:2021iwm}.

We now describe the continuum (Jordan-frame) dynamics of a scalar $\phi$ non–minimally
coupled to gravity. 
Our starting point is an action with a curvature interaction $\propto \xi R\phi^2$,
and a generic potential $V(\phi,\{\varphi_{\rm m}\})$ for $\phi$ and the remaining 
matter sector fields, $\{\varphi_{\rm m}\}$, represented by $\mathcal{L}_{\rm m}$,
\begin{align}
S &=
\int d^{4}x \sqrt{-g} \left( \frac{1}{2}m_p^2R - \frac{1}{2}\xi R \phi^{2} -\frac{1}{2} g^{\mu\nu}\partial_{\mu}\phi\partial_{\nu}\phi - V(\phi,{\{\varphi_{\rm m}\}})
+ \mathcal{L}_{\rm m}\right) \,.
\label{eq:action}
\end{align}
Here $R$ is the Ricci scalar and the parameter $\xi$ is the non-minimal coupling. Neglecting gravitational perturbations and specializing the metric to a spatially-flat FLRW background, the equation of motion for $\phi$ reads 
\begin{align}
\phi'' +(3-\alpha) \frac{a'}{a}\phi' - a^{-2(1-\alpha)}\nabla^2\phi  + a^{2\alpha} \left(\xi \bar R \phi + \frac{\partial V}{\partial \phi}\right) =0 \,,
\label{eq:eom}
\end{align}
where the (background) Ricci scalar takes the form
\begin{equation}
\bar R = \frac{6}{a^{2\alpha}} \left[\frac{a''}{a} +(1-\alpha)\left( \frac{a'}{a}\right)^{2}\right] \,.
 \label{eq:cosmic_R}
\end{equation}

Since we assume homogeneity and isotropy on large scales, the energy-momentum tensor $T^{\mu}_{\,\,\,\,\nu} = {\rm diag}\left\{-\bar\rho(\eta),\bar p(\eta),\bar p(\eta),\bar p(\eta) \right\}$ takes the perfect-fluid form. Here $\bar p, \bar \rho$ are the background pressure and energy densities, which we decompose into the NMC scalar field plus matter contributions, $\bar p = \bar p_\phi + \bar p_{\rm m}$, $\bar\rho = \bar\rho_\phi + \bar \rho_{\rm m}$. Although all fields may develop local spatial inhomogeneities, the homogeneous and isotropic pressure and energy density are defined as volume averages ($\langle \cdots \rangle$) of the inhomogeneous field quantities. The Einstein equations then simplify to the Friedmann equations in $\alpha$-time, {\it c.f.}~Eqs.~(\ref{eq:Friedmann-full}), 
\begin{align}
\mathcal{H}^{2} &\equiv \left(\frac{a'}{a}\right)^2 =   \frac{a^{2\alpha}}{3m_p^2} (\bar\rho_\phi + \bar \rho_{\rm m})  \,,
\label{eq:Hu}
\\
\frac{a''}{a} &= -\frac{a^{2\alpha}}{6m_{p}^2} \Big[(1-2\alpha)(\bar\rho_\phi + \bar \rho_{\rm m}) + 3(\bar p_\phi + \bar p_{\rm m}) \Big]\,,
 \label{eq:2FE}
\end{align}
where the energy density and pressure of the NMC scalar field are~\cite{Figueroa:2021iwm}
\begin{align}
\bar \rho_{\phi}(\eta) &= \frac{1}{2a^{2\alpha}} \langle \phi'^2\rangle + \frac{1}{2a^2} \langle( \nabla\phi)^2\rangle + \langle V(\phi)\rangle +  \frac{3\xi}{a^{2\alpha}}\mathcal{H}^{2}\langle \phi^2\rangle  +\frac{6\xi}{a^{2\alpha}} \mathcal{H} \langle \phi \phi'\rangle - \frac{\xi}{a^{2}}\langle \nabla^2\phi^2\rangle \,,\label{eq:nmcrho}
\\[10pt]
\bar p_{\phi}(\eta) &= \frac{(1-4\xi)}{2a^{2\alpha}} \langle \phi'^2\rangle - \frac{(1-12\xi)}{6a^2} \langle( \nabla\phi)^2\rangle - \langle V(\phi)\rangle +~\frac{2\xi}{a^{2\alpha}} \mathcal{H} \langle \phi \phi'\rangle -  \frac{\xi}{3a^{2}}\langle \nabla^2\phi^2\rangle   \nonumber\\
&+ 2\xi\langle\phi V_{,\phi} \rangle +\frac{\xi}{a^{2\alpha}}\left[\mathcal{H}^2 + 12\left(\xi-{\frac{1}{6}}\right)\left( {\frac{a''}{a}}+ (1-\alpha)\mathcal{H}^{2}\right)\right]\langle \phi^2\rangle \,,\label{eq:nmcp}
\end{align}
where $V_{,\phi} \equiv \partial V/\partial\phi$. The evolution of the scale factor can be done, in principle, using either \cref{eq:Hu} or \cref{eq:2FE}. Alternatively, we can use another approach  
based on the trace of the energy-momentum tensor of the NMC field,
\begin{align}
T_{\phi} = \left(6\xi -1\right) \left(\partial^{\mu}\phi\partial_{\mu}\phi  + \xi R\phi^2 \right) + 6\xi\phi \frac{\partial V}{\partial \phi} - 4V \,.
\label{eq:4dtrT}
\end{align}
As the traced Einstein equations read
\begin{align}
  R &= -\frac{1}{m_p^2}g^{\mu\nu}\left( T^\phi_{\mu\nu} +  T^{\rm m}_{\mu\nu}\right) = -\frac{1}{m_p^2}\left( T_\phi +  T_{\rm m}\right) \,,
  \label{eq:EFEtr}
\end{align}
we can write the background curvature as
\begin{align}
m_p^2\bar R &= -\left\langle T_\phi +  T_{\rm m}\right\rangle \equiv (1-6\xi)\left(\langle\partial^{\mu}\phi\partial_{\mu}\phi\rangle +\xi \bar R \langle \phi^2\rangle \right) - 6\xi\langle \phi V_{,\phi}\rangle + 4 \langle V\rangle - \langle T_{\rm m} \rangle\,. 
  \label{eq:EFEtrBack}
\end{align}
Solving for $\bar R$ leads to
\begin{align}
\bar R = \frac{F(\phi)}{m_p^{2}}\Big[\left(1-6\xi \right) \langle\partial^{\mu}\phi\partial_{\mu}\phi\rangle  + 4 \langle V\rangle- 6\xi\langle \phi V_{,\phi}\rangle-\langle T_{\rm m} \rangle \Big]\,,\label{eq:eomR}
\end{align}
with
\begin{align}
    F(\phi) \equiv \frac{1}{1 + \left(6\xi -1\right)\xi \langle\phi^2\rangle /m_p^2 } \,. 
\end{align}
From here, using  \cref{eq:cosmic_R}, we obtain a differential equation for the scale factor as
\begin{align}
   \frac{a''}{a} +(1-\alpha)\left( \frac{a'}{a}\right)^{2} =
   \frac{a^{2\alpha}F(\phi)}{6m_p^{2}}\Big[\left(1-6\xi \right) \langle\partial^{\mu}\phi\partial_{\mu}\phi\rangle  + 4 \langle V\rangle- 6\xi\langle \phi V_{,\phi}\rangle-\langle T_{\rm m} \rangle \Big]\,. \label{eq:piadot}
\end{align}
This equation can then be solved simultaneously with the EOM of the NMC and matter fields. 

\subsubsection{Lattice implementation}

We first write the continuum equations in terms of dimensionless program variables $\tilde\phi =\phi/{f_*},~\dd \tilde \eta = \omega_* \dd \eta\,,~\dd \tilde x_i = \omega_* \dd x_i\text{ and }\tilde{R}=\omega_*^{-2}R$. In terms of the conjugate momentum for $\tilde{\phi}$,
\begin{align}\label{eq:piTilde}
  \tilde\pi_{\phi} = a^{3-\alpha} \tilde\phi' \,,
\end{align}
the evolution of $\tilde\phi$ follows the first-order differential equations
\begin{align}
\left\lbrace
\begin{array}{l}
\tilde\phi' = a^{\alpha-3}\tilde\pi_{\phi}\,, \vspace*{0.2cm}\\
\tilde\pi'_{\phi} =  \mathcal{K}_\phi[a,\tilde\phi,\{\tilde\varphi_{\rm m}\},\tilde{\bar R}]\,,~~~~{\rm with}~~~~ \mathcal{K}_\phi[a,\tilde\phi,\{\tilde\varphi_{\rm m}\},\tilde{\bar R}] \equiv a^{1+\alpha}\, \tilde{\nabla}^2 \tilde\phi -a^{3+\alpha} \left(\xi \tilde{\bar R} \tilde\phi + \frac{\partial\widetilde V}{\partial \tilde\phi}  \right)\,,
\end{array}\right.
\label{eq:EOMpi}
\end{align}
while the remaining matter fields follow the canonical EOM, see {\tt the Art I} and \cref{sec:LatticeTechniques}. 
The scale factor is evolved using \cref{eq:piadot}, with the conjugate momentum defined as
\begin{eqnarray}
b=a^{1-\alpha} a'\,,
\end{eqnarray}
yielding the first-order differential equations
\begin{align}
\left\lbrace
\begin{array}{l}
a' = a^{\alpha-1}b\,, \vspace*{0.2cm}\\
b' = \mathcal{K}_a[a,\tilde{\bar R}]\,,~~~~~~~{\rm with}~~~~~~~\mathcal{K}_a[a,\tilde{\bar R}] \equiv \frac{a^{{2+\alpha}}}{6}  \tilde{\bar R} \ .
\end{array}
\right.
\label{eq:pia}
\end{align}
All we need now is an expression for $\tilde{\bar R}$ to be used in the kernels $\mathcal{K}_{\phi},~\mathcal{K}_a$. Using \cref{eq:eomR}, we have
\begin{align}
   \tilde{\bar R} &=\frac{f_*^2}{m_p^2} \left[\frac{2\left(1-6\xi \right) \big\langle\tilde  G^{\phi} -\tilde K^{\phi}\big\rangle  + 4\langle \tilde V\rangle- 6\xi\langle \tilde\phi \,\tilde V_{,\tilde\phi}\rangle+({\tilde{\bar\rho}}_{\rm  m}-3{\tilde{\bar p}}_{\rm  m})}{1 + \left(6\xi -1\right)\xi \langle\tilde\phi^2\rangle (f_*^2/m_p^2)}\right]\label{eq:Rnew} \,,
\end{align}
with
\begin{eqnarray}
    \tilde K^{\phi} \equiv \frac{1}{2a^{2\alpha}} \tilde\phi'^2\, , \quad \tilde G^{\phi} \equiv \frac{1}{2a^{2}} (\tilde\nabla
     \tilde\phi)^2 \, , 
\end{eqnarray}
where we have defined $\langle T_{\rm m} \rangle =3\bar p_{\rm  m} - \bar \rho_{\rm  m}$, see \cref{eq:rhoLocal,eq:pLocal} for a definition of the energy density and pressure of the matter fields, respectively. 

We have arrived at a system of equations that is suitable for discretization. To evolve equations \cref{eq:EOMpi,eq:pia} on the lattice, using Eq.~(\ref{eq:Rnew}), we need a spatial discretization prescription. We choose to discretize gradients using forward derivatives, {\it c.f.}~Eq.~(\ref{eq:forwardbackwardd}), while for the Laplacian we employ a symmetric implementation
\begin{align}
\partial_i\phi\partial_i\phi  &~~\longrightarrow~~ \sum_i\nabla_i^+\phi\nabla_i^+\phi \,, \\
\vec\nabla^2 \phi &~~\longrightarrow~~ \sum_i\nabla^-_i \nabla_i^+\phi\, .
\end{align}
We write discrete versions of the kernels as
\begin{align}
  \mathcal{K}^{\rm L}_{\phi}\left[a,\tilde \phi,\{\tilde\varphi_{\rm m}\}, \tilde{\bar R}\right] &= a^{1+\alpha} \sum_i\tilde\nabla^-_i \tilde\nabla_i^+\tilde\phi -a^{3+\alpha} \left(\xi \tilde{\bar R} \tilde\phi + \frac{\partial\widetilde V}{\partial \tilde\phi}  \right)\,, \\
  \mathcal{K}^{\rm L}_{a}\left[a,\tilde{\bar R}\right] &= \frac{a^{{2+\alpha}}}{6}  \tilde{\bar R} \, .
\end{align}

Because the background Ricci scalar $\tilde{\bar R} = \tilde {\bar R}[\tilde\phi, \tilde\pi_{\phi},\{\tilde\varphi_{\rm m}\},\{\tilde\pi_{\varphi_{\rm m}}\}]$ depends on (volume averages of) the fields and their conjugate momenta, the kernel for $\tilde\phi$ also depends on its own conjugate momentum. This means that symplectic algorithms should not be used, so we employ instead explicit {\it Runge-Kutta} (RK) algorithms, as those described in Sec.~\ref{subsubsec:NonSymplecticInt}. For example, the RK2 scheme leads to 
\begin{eqnarray}\label{eq:RKLSalgorithmNMC_2}
\hspace*{-0.2cm}\left\lbrace\hspace*{-0.2cm}
\begin{array}{ll}
\tilde{\phi}^{(1)}_a = \tilde{\phi}_a\,, & \hspace*{-0.2cm} \tilde{\phi}^{(2)}_a \hspace*{-0.1cm}= \tilde{\phi}^{(1)}_a + \delta\tilde{\eta}\;\left(a^{(1)}\right)^{\alpha-3}\hspace*{-0.1cm}\tilde{\pi}^{(1)}_{b}\,,\vspace*{0.2cm}\\
\tilde{\pi}^{(1)}_{a} = \tilde{\pi}_{a} & \hspace*{-0.2cm} \tilde{\pi}^{(2)}_{a} \hspace*{-0.1cm}= \tilde{\pi}^{(1)}_{a} + {\delta\tilde{\eta}}\;\mathcal{K}^{{\rm L}, (1)}_{\phi_a}\,,\vspace*{0.2cm}\\
a^{(1)} = a\,, & \hspace*{-0.2cm} a^{(2)}\hspace*{-0.1cm} = a^{(1)} + {\delta\tilde{\eta}}\;\left(a^{(1)}\right)^{\alpha-1}\hspace*{-0.1cm}b^{(1)}\,,\vspace*{0.2cm}\\
b^{(1)} = b\,, & \hspace*{-0.2cm} b^{(2)} \hspace*{-0.1cm} = b^{(1)} + {\delta\tilde{\eta}}\;\mathcal{K}^{{\rm L}, (1)}_a\,,\vspace*{0.2cm}
\end{array}\right. \hspace*{-0.7cm} ~~\Longrightarrow~~ 
\hspace*{-0.3cm}\left\lbrace\hspace*{-0.2cm}
\begin{array}{rcl}
\tilde{\phi}_{a,+0}\hspace*{-0.2cm} &=&\hspace*{-0.2cm} \tilde{\phi}_a^{(1)} + \hspace*{-0.1cm}{1\over2}\delta\tilde{\eta}\left[\left(a^{(1)}\right)^{\alpha-3}\hspace*{-0.1cm}\tilde{\pi}^{(1)}_{b}+\left(a^{(2)}\right)^{\alpha-3}\hspace*{-0.1cm}\tilde{\pi}^{(2)}_{b}\right]\,,\vspace*{0.2cm}\\
a_{+0} \hspace*{-0.2cm}&=&\hspace*{-0.2cm} a^{(1)} + \hspace*{-0.1cm}{1\over2}\delta\tilde{\eta}\left[\left(a^{(1)}\right)^{\alpha-1}\hspace*{-0.1cm}b^{(1)}+\left(a^{(2)}\right)^{\alpha-1}\hspace*{-0.1cm}b^{(2)}\right]\,,\vspace*{0.2cm}\\
\tilde{\pi}_{a,+0}\hspace*{-0.2cm}&=&\hspace*{-0.2cm}\tilde{\pi}^{(1)}_{a}\hspace*{-0.1cm}+\hspace*{-0.1cm}{1\over2}\delta\tilde{\eta}\left[\mathcal{K}^{{\rm L}, (1)}_{\phi_a}+\mathcal{K}^{{\rm L}, (2)}_{\phi_a}\right]\,,\vspace*{0.2cm}\\
b_{+0}\hspace*{-0.2cm}&=&\hspace*{-0.2cm}b^{(1)}\hspace*{-0.1cm}+\hspace*{-0.1cm}{1\over2}\delta\tilde{\eta}\left[\mathcal{K}^{{\rm L}, (1)}_{a}+\mathcal{K}^{{\rm L}, (2)}_{a}\right]\,.\vspace*{0.2cm}
\end{array}\right.\nonumber\\
\end{eqnarray}
The evolution scheme for the matter sector $\{\tilde\varphi_{\rm m}\}$ follows from Eqs.~\eqref{eq:scalar_singlet_eom} and~\eqref{eq:RK2algorithm_1}. Similarly, the approximation order can be improved to 4th or $n$th order by using the equivalent adapted versions of the algorithms of Eqs.~(\ref{eq:RK4algorithm_1}) and (\ref{eq:RKLSalgorithm_1}), respectively.

\subsection{Non-minimal kinetic theories}\label{sec:NMKfields}

In this subsection we consider an action for interacting real scalar fields, $\{\phi_a\}$, where we allow for an internal metric in field space ${\mathcal G}_{ab}$, which in principle can be a function of both field amplitudes and conjugate momenta, ${\mathcal G}_{ab}\equiv {\mathcal G}_{ab}[\lbrace \phi_a\rbrace, \lbrace \pi_{\phi_a}\rbrace]$. We write the action of this model as
\begin{eqnarray} \label{eq:ScalarActionNonCanonicalCont}
S_{\rm NMK} = - \int d^4 x  \sqrt{-g} \left\{\frac{1}{2}g^{\mu \nu}{\mathcal G}_{ab}\partial_{\mu} \phi_a \partial_{\nu}\phi_b + V(\lbrace \phi\rbrace)\right\}\,,
\end{eqnarray}
where ${\mathcal G}_{ab}$ is assumed to be symmetric in field space ${\mathcal G}_{ab}={\mathcal G}_{ba}$, and summation over field indices is assumed throughout this section. If ${\mathcal G}_{ab} = \delta_{ab}$, we recover the case of canonically-normalized scalar fields. If $\mathcal{G}_{ab} = \beta \delta_{ab}$ with $ \beta \neq 1$ and $\beta > 0$ a positive constant, we still recover canonically normalized scalar fields 
after a simple field redefinition $\phi_a \longrightarrow \sqrt{\beta}\phi_a$, which brings back the kinetic term into its canonical form. Only if ${\mathcal G}_{ab}$ exhibits an explicit dependence on field amplitudes and/or derivatives, do we say the scalar fields are {\it non-canonically normalized}. In that case we assume the metric determinant to be non-vanishing ${\rm det}(\mathcal{G}_{ab}) \neq 0$, so that an inverse metric $\mathcal{G}
^{-1}_{ab}$ exists, with $\mathcal{G}_{ac}\mathcal{G}
^{-1}_{cb} = \delta_{ab}$.

Varying the action~(\ref{eq:ScalarActionNonCanonicalCont}) leads to the EOM of the fields as
\begin{eqnarray}\label{eq:EOMflatScalarFlds}
\frac{1}{\sqrt{-g}}\partial_{\mu}(\sqrt{-g} \; g^{\mu\nu} \;{\mathcal G}_{ab} \partial_{\nu} \phi_b) -  \frac{1}{2}\,g^{\mu\nu}\,{\mathcal G}_{bc,a}\partial_{\mu}\phi_b\partial_{\nu} \phi_c   - \frac{\partial{V}}{\partial \phi_a}  =  0  \; ,
\end{eqnarray}
where $\mathcal{G}_{bc,a} \equiv \partial \mathcal{G}_{bc}/\partial\phi_a$. In a FLRW background, the EOM 
are conveniently rewritten in the form
\begin{eqnarray}\label{eqn:FLRWeqnforNMK}
\mathcal{G}_{ab}\phi''_b + (3-\alpha) \frac{a'}{a}\mathcal{G}_{ab}\phi'_{b} - a^{-2(1-\alpha)}\mathcal{G}_{ab}({\nabla}^{\,2}\phi_b) + \gamma_{abc}(\phi'_b\phi'_c- a^{-2(1-\alpha)}\,\vec{\nabla}\phi_b \cdot \vec{\nabla}\phi_c) + a^{2\alpha}\frac{\partial{V}}{\partial \phi_a} = 0\,,\nonumber\\
\end{eqnarray}
where we have defined 
\begin{eqnarray}    
\gamma_{abc} = \left(\mathcal{G}_{ab,c}-\frac{1}{2}\mathcal{G}_{bc,a}\right)\;.
\end{eqnarray}
Introducing the notation for the conjugate momenta $\pi_a\equiv \pi_{\phi_a}$, Eq.~\eqref{eqn:FLRWeqnforNMK} can be recast as
\begin{eqnarray}\label{eqn:NMKcontinuum}
\begin{cases}
{\phi}'_a  \equiv  \mathcal{G}_{ab}^{-1}\pi_b\,,\vspace*{2mm}\\
\pi'_a + (3-\alpha)\frac{a'}{a}\pi_a - (\mathcal{F}_a)_{bc}\pi_b\pi_c  = a^{-2(1-\alpha)}[\mathcal{G}_{ab}{\nabla}^{\,2}\phi_b + \gamma_{abc}\vec{\nabla}\phi_b \cdot \vec{\nabla}\phi_c] - a^{2\alpha}\frac{\partial{V}}{\partial \phi_a}\,,
\end{cases}
\end{eqnarray}
where 
\begin{eqnarray}
(\mathcal{F}_a)_{bc} \equiv \mathcal{G}_{ae}\;\mathcal{G}_{ec,d}^{-1}\;\mathcal{G}^{-1}_{db} + \left(\mathcal{G}_{ae,d}-\frac{1}{2}\mathcal{G}_{ed,a}\right)\mathcal{G}_{ec}^{-1}\mathcal{G}_{db}^{-1}\;, 
\end{eqnarray}
Finally, in order to obtain a complete numerical scheme, it is necessary to write down the EOM for the scale factor.  
The evolution of the scale factor is given by Eq.~\eqref{eq:FriedmannDDa}.  
In the case of scalar fields with non-canonical kinetic term, both the kinetic and gradient energy density contributions [compared to Eq.~\eqref{eq:energy-contrib}] are modified in the following manner,
\begin{eqnarray}\label{eq:energy-contrib-NMK}
    {K}_{{\rm NMK}} = \frac{1}{2 a^{2\alpha} } \mathcal{G}_{ab}{\phi'}_a{\phi'}_b \; ,~~~~
    {G}_{\rm NMK} = \frac{1}{2a^{2} } \mathcal{G}_{ab}\vec{\nabla}{\phi}_a \cdot \vec{\nabla}{\phi}_b \;,
\end{eqnarray}
so that the Friedmann equations read 
\begin{eqnarray}\label{eq:FriedmannHubble-NMK}
    \left({a'\over a}\right)^2 &=&  \frac{a^{2 \alpha}}{3 m_p^2}\left\langle {K}_{{\rm NMK}} + {G}_{{\rm NMK}} + {V}\right\rangle \,,
    \\
    \label{eq:FriedmannDDa-NMK}
    {a''\over a} &=& \frac{a^{2 \alpha}}{3 m_p^2}\left\langle (\alpha-2)\;{K}_{{\rm NMK}} + \alpha\;{G}_{{\rm NMK}} + (\alpha + 1)V \right\rangle
\end{eqnarray}
where $\langle \dots \rangle$ denotes, as usual, volume average over regions sufficiently large to encompass all fields' relevant wavelengths. 

\subsubsection{Lattice implementation}

Spatial discretization of a theory with scalar fields with non-canonical kinetic terms is actually straightforward. We can simply replace spatial derivatives by finite differences, for example by $ \partial_i \rightarrow \nabla_i^\pm$, {\it c.f.}~Eq.~(\ref{eq:forwardbackwardd}), and the spatial Laplacian by the lattice operator $\nabla^2 \rightarrow \sum_i \nabla^-_i \nabla^+_i$. On the other hand, the non-canonical kinetic term leads to evolution kernels that depend on the conjugate momenta, so the use of non-symplectic integrators becomes necessary. 

We present next an evolution scheme for a scalar sector with non-canonical kinetic terms, suitable for non-symplectic integrators. We write the lattice version of the EOM in a hybrid formulation, where the spatial derivatives are replaced by their lattice counterpart, but the time derivatives are kept continuous. Additionally, we make a transformation of coordinates to program variables using $\tilde\phi = {\phi}/{f_*}\;, d\tilde x_i = \omega_* dx_i\text{ and }d\tilde \eta = \omega_* d\eta$. The dynamics of the NMK scalar fields are controlled by 
\begin{eqnarray}\label{eqn:NMKL}
\begin{cases}
\tilde{\phi}'_a  =  \mathcal{G}_{ab}^{-1}\tilde\pi_b\,,\vspace*{0.25cm}\\
\tilde\pi'_a = \mathcal{K}^{\rm L}_{\phi_a}[a,\tilde\phi_a,b,\tilde\pi_a] \equiv -(3-\alpha)\frac{a'}{a}\tilde\pi_a + (\mathcal{F}_a)_{bc}\tilde\pi_b\tilde\pi_c   \\ \vspace*{0.25cm}
\hspace{4.0cm} +~ a^{-2(1-\alpha)}[\mathcal{G}_{ab}\sum_i \nabla^-_i \nabla^+_i\tilde\phi_b + \gamma_{abc} \sum_i\nabla^+_i\tilde\phi_b \nabla^+_i\tilde\phi_c] - a^{2\alpha}\frac{\partial{\tilde V}}{\partial \tilde\phi_a}\,,
\end{cases}
\end{eqnarray}
while the evolution of the scale factor is then governed by
\begin{eqnarray}\label{eqn:SF_NMK_L}
\begin{cases}
a'  =  b\,,\\
b' = \mathcal{K}^{\rm L}_a[a,b,\tilde{K}_{\rm NMK},\tilde{G}_{\rm NMK},\tilde{V}] \equiv \left(\frac{f_*}{m_p}\right)^2 \frac{a^{2 \alpha}}{3} \left\langle (\alpha-2)\;\tilde{K}_{{\rm NMK}} + \alpha\;\tilde{G}_{{\rm NMK}} + (\alpha + 1)\tilde V \right\rangle\,,
\end{cases}
\end{eqnarray}
with
\begin{eqnarray}\label{eqn:energy-contrib-NMK-L}
    \tilde{K}_{{\rm NMK}} = \frac{1}{2 a^{2\alpha} } \mathcal{G}_{ab}{\tilde\phi'}_a{\tilde\phi'}_b \; ,~~~~
    \tilde{G}_{\rm NMK} = \frac{1}{2a^{2} } \mathcal{G}_{ab} \sum_i\nabla^+_i\tilde\phi_a \nabla^+_i\tilde\phi_b\;.
\end{eqnarray}

The system of coupled differential equations~(\ref{eqn:SF_NMK_L}) and (\ref{eqn:NMKL}) can be solved simultaneously using any of the non-symplectic integrators described in Sec.~\ref{subsubsec:NonSymplecticInt}. For instance, the RK2 algorithm for this example can be written as
\begin{eqnarray}
\hspace*{-0.2cm}\left\lbrace
\begin{array}{ll}
\hspace*{-0.2cm}\tilde{\phi}^{(1)}_a = \tilde{\phi}_a\,, & \hspace*{-0.2cm} \tilde{\phi}^{(2)}_a = \tilde{\phi}^{(1)}_a + \delta\tilde{\eta}\;\left(\mathcal{G}_{ab}^{(1)}\right)^{-1}\tilde{\pi}^{(1)}_{b}\,,\vspace*{0.2cm}\\
\hspace*{-0.2cm}\tilde{\pi}^{(1)}_{a} = \tilde{\pi}_{a} & \hspace*{-0.2cm} \tilde{\pi}^{(2)}_{a} = \tilde{\pi}^{(1)}_{a} + {\delta\tilde{\eta}}\;\mathcal{K}^{{\rm L}, (1)}_{\phi_a}\,,\vspace*{0.2cm}\\
\hspace*{-0.2cm}a^{(1)} = a\,, & \hspace*{-0.2cm} a^{(2)} = a^{(1)} + {\delta\tilde{\eta}}\;b^{(1)}\,,\vspace*{0.2cm}\\
\hspace*{-0.2cm}b^{(1)} = b\,, & \hspace*{-0.2cm} b^{(2)} = b^{(1)} + {\delta\tilde{\eta}}\;\mathcal{K}^{{\rm L}, (1)}_a\,,\vspace*{0.2cm}
\end{array}\right. \hspace*{-0.9cm} ~~\Longrightarrow~~ \hspace*{-0.3cm}
\left\lbrace
\begin{array}{rcl}
\hspace*{-0.2cm}\tilde{\phi}_{a,+0}\hspace*{-0.2cm} &=&\hspace*{-0.2cm} \tilde{\phi}_a^{(1)} +\hspace*{-0.1cm} {1\over2}\delta\tilde{\eta}\left[\left(\mathcal{G}_{ab}^{(1)}\right)^{-1}\hspace*{-0.2cm}\tilde{\pi}^{(1)}_{b}+\left(\mathcal{G}_{ab}^{(2)}\right)^{-1}\hspace*{-0.2cm}\tilde{\pi}^{(2)}_{b}\right]\,,\vspace*{0.2cm}\\
\hspace*{-0.2cm}a_{+0} \hspace*{-0.2cm}&=&\hspace*{-0.2cm} a^{(1)} + \hspace*{-0.1cm}{1\over2}\delta\tilde{\eta}\left[b^{(1)}+b^{(2)}\right]\,,\vspace*{0.2cm}\\
\hspace*{-0.2cm}\tilde{\pi}_{a,+0}\hspace*{-0.2cm}&=&\hspace*{-0.2cm}\tilde{\pi}^{(1)}_{a}+\hspace*{-0.1cm}{1\over2}\delta\tilde{\eta}\left[\mathcal{K}^{{\rm L}, (1)}_{\phi_a}+\mathcal{K}^{{\rm L}, (2)}_{\phi_a}\right]\,,\vspace*{0.2cm}\\
\hspace*{-0.2cm}b_{+0}\hspace*{-0.2cm}&=&\hspace*{-0.2cm}b^{(1)}+\hspace*{-0.1cm}{1\over2}\delta\tilde{\eta}\left[\mathcal{K}^{{\rm L}, (1)}_{a}+\mathcal{K}^{{\rm L}, (2)}_{a}\right]\,.\vspace*{0.2cm}
\end{array}\right.\nonumber\\
\end{eqnarray}
Similarly, the approximation order can be improved to 4th order or $n$th order by using the equivalent adapted versions of the algorithms of Eqs.~(\ref{eq:RK4algorithm_1}) and (\ref{eq:RKLSalgorithm_1}) respectively.

\subsection{Working example: Ricci reheating}\label{sec:ricci-reheating-example}

Applications of non-canonical scalar field interactions have been studied in many setups. For instance, in preheating, the presence of non-minimal kinetic terms has been investigated both analytically and with the use of lattice simulations in~\cite{Lachapelle:2008sy,Child:2013ria,Rahmati:2014cwa,Li:2019ncw,Brandenberger:2019njw,Adshead:2023nhk,Adshead:2024ykw,Huang:2024amu,Adshead:2025gka}. Alternatively, preheating scenarios with a scalar field non-minimally coupled to gravity 
have been considered in~\cite{Ema:2016dny,Bassett:1997az,Tsujikawa:1999jh,Tsujikawa:1999iv,Fu:2019qqe,Garcia-Bellido:2008ycs,Figueroa:2016dsc,Sfakianakis:2018lzf,Bezrukov:2020txg,DeCross:2015uza,DeCross:2016fdz,DeCross:2016cbs,Nguyen:2019kbm,vandeVis:2020qcp,Watanabe:2006ku,Opferkuch:2019zbd,Dimopoulos:2018wfg,Bettoni:2021zhq,Figueroa:2021iwm,Figueroa:2024asq,Figueroa:2024yja}. As an example, we discuss here the case of \emph{Ricci reheating}~\cite{Figueroa:2016dsc,Opferkuch:2019zbd,Laverda:2023uqv,Figueroa:2024asq}, where the post-inflationary dynamics of the inflaton do not undergo coherent oscillations. Instead, the Universe transitions from quasi-de Sitter (qdS) to a stiffer phase with $w>1/3$ (often close to the maximum value in kination domination, $w\simeq 1$). During this stage, the (background) Ricci scalar,
\begin{equation}
\bar R = 6\left(2H^2+\dot H\right) \,,
\end{equation}
becomes negative. Thus, a spectator scalar field $\chi$ with non-minimal coupling to gravity of the form $\tfrac12\,\xi R\,\chi^2$, with $\xi>0$, acquires a tachyonic effective mass $\langle m_{\rm eff}^2 \rangle \simeq \xi \bar R <0$. Long-wavelength $\chi$-modes grow until
backreaction (self-interactions and/or gravity) shuts off the instability.

As a working example, we consider a minimally coupled inflaton $\phi$ that drives the background dynamics initially, and a non-minimally coupled spectator scalar field $\chi$,
\begin{equation}
\mathcal{L} = -\tfrac12(\partial\phi)^2 - V_{\rm inf}(\phi) - \tfrac12(\partial\chi)^2 - \tfrac12\,\xi R\,\chi^2 - V(\chi)\,.
\end{equation}
To realize Ricci reheating,
we employ a \emph{monotonic plateau-to-zero} potential,
\begin{align}
V_{\rm inf}(\phi) &= \frac{V_0}{2}\!\left[1-\tanh\!\left(\frac{\beta\,\phi}{m_p}\right)\right]\,,
\label{eq:Vinf-plateau-kination}
\end{align}
where $V_0$ sets the inflationary energy scale and $\beta$ controls how rapidly the potential falls.
For large negative $\phi$, the potential is nearly flat ($V\simeq V_0$), sustaining inflation.
As $\phi$ passes through zero, $V_{\rm inf}$ smoothly decreases to zero over a field range
$\Delta\phi \sim m_p/\beta$. Once the potential energy becomes negligible, the field's kinetic energy
dominates and the background enters a \emph{kination} phase with equation of state $w\simeq1$.
During this stage,
\begin{equation}
\bar R(t) = 6\!\left(2H^2+\dot H\right) = -\,\frac{2}{3t^2} < 0 \qquad (w=1)\,,
\label{eq:R-kination}
\end{equation}
so that $\chi$ acquires a negative curvature mass term, providing the characteristic tachyonic mass of Ricci reheating.

The lattice evolution follows the Jordan-frame equations summarized in \cref{sec:NMC}, recall \cref{eq:EOMpi,eq:pia,eq:Rnew}. To set the initial fluctuations we assume the Bunch-Davies vacuum and evolve the modes using the canonical field $\sigma \equiv a\,\chi$ in conformal time
($\alpha=1$), whose linear mode equation reads
\begin{equation}
\sigma_k'' + \Big[k^2 + a^2\big(\xi-\tfrac{1}{6}\big)R\Big]\,\sigma_k = 0\,.
\label{eq:sigma-eq}
\end{equation}
Note that the potential term $V(\chi)$ is neglected at this stage, as we assume the initial fluctuations to be small. This allows us to determine the initial power spectrum and initialize a lattice simulation before the tachyonic growth saturates or before self-interactions become relevant. 

\begin{figure}[t]
  \centering
  \includegraphics[width=0.495\textwidth]{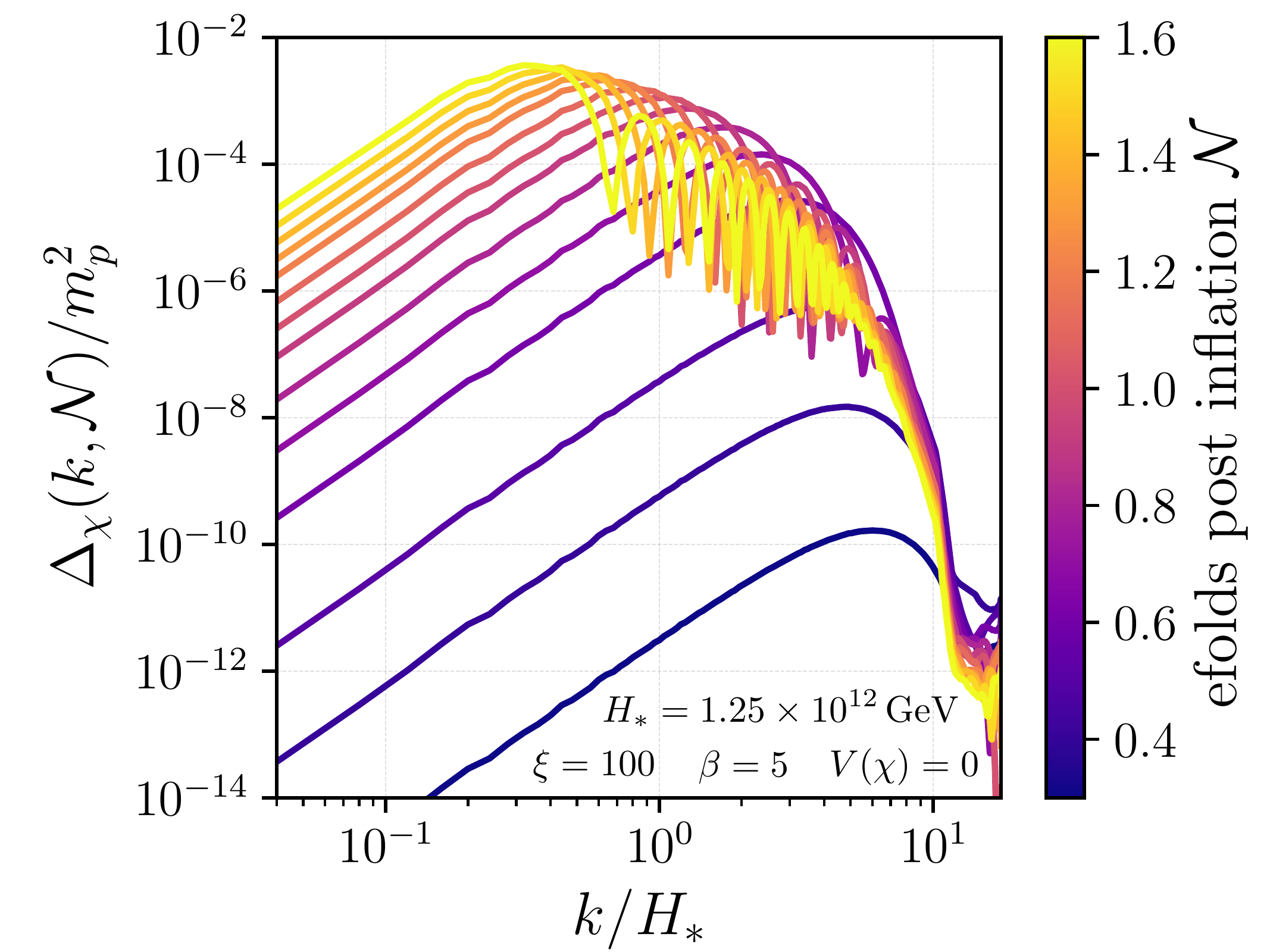}
  \includegraphics[width=0.495\textwidth]{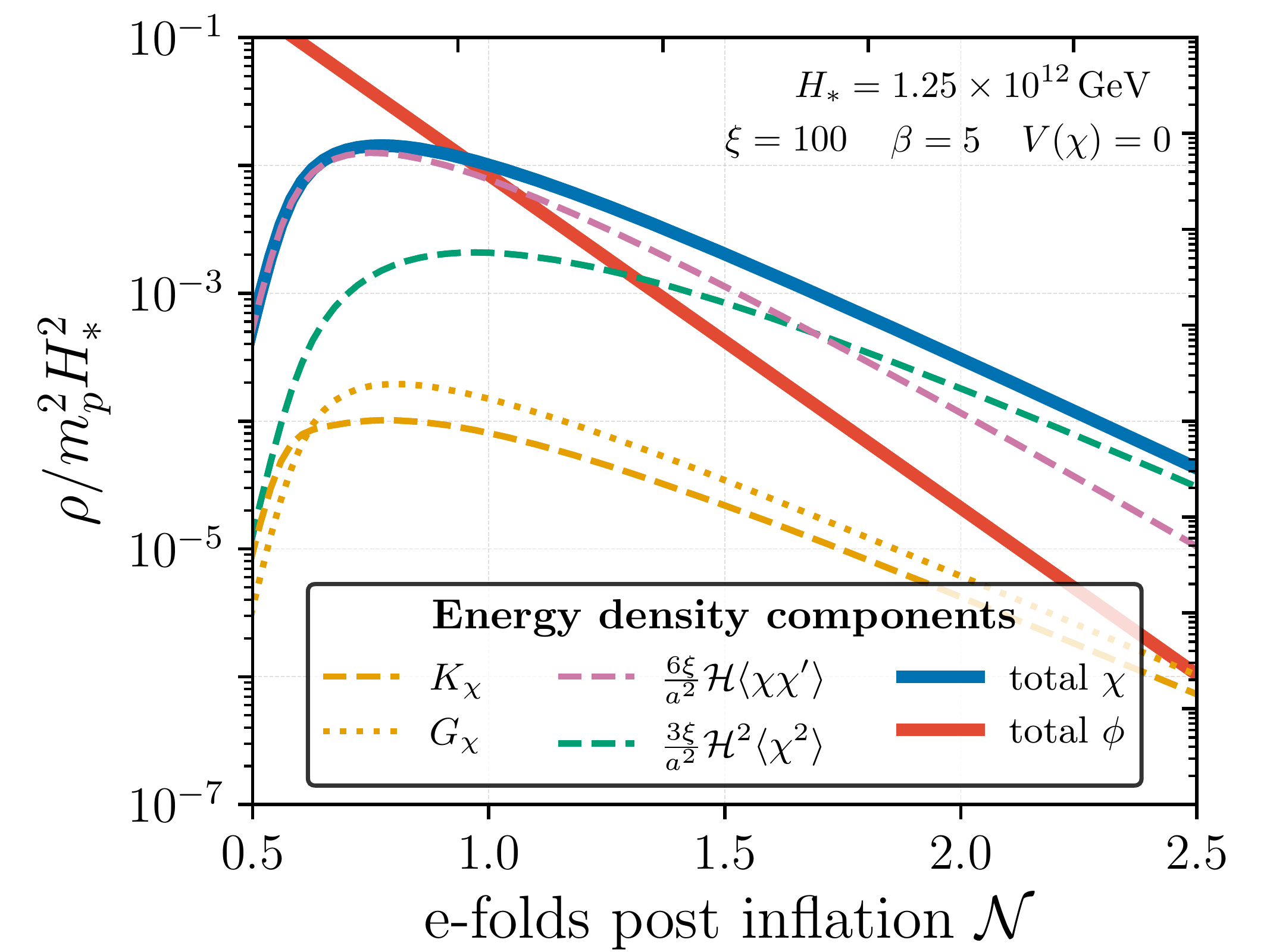}
  \caption{\textbf{Left:} Evolution of $\Delta_\chi(k,\mathcal{N})/m_p^2$ at several $\mathcal{N}>\mathcal{N}_*$. Amplification occurs for modes inside the curvature-induced tachyonic band $k/a \lesssim \sqrt{\xi|R|}$; in the $k/H_*$ normalization, this is $k/H_* \lesssim \kappa_c(\mathcal{N})$ with $\kappa_c(\mathcal{N})\equiv a\sqrt{\xi|R|}/H_*$ (\emph{during exact kination} $\kappa_c\propto e^{-2\mathcal{N}}$, so we see the peak drift to lower $k/H_*$).
\textbf{Right:} Energy–density evolution showing transfer from the inflaton background to the non–minimally coupled spectator. The growth saturates as the energy density in $\chi$ becomes comparable to that in $\phi$. Also shown as dashed/dotted lines are the components of $\chi$'s energy density.}
  \label{fig:RR-spec-energydensity}
\end{figure}

We start the lattice a short time after inflation ends,
when the background has entered the kination phase but before the tachyonic growth has saturated (e.g. at
$\mathcal{N}\equiv\ln a \sim 0.2$-$0.5$). For the lattice results below, time evolution is performed with the RK2 algorithm given above, {\it c.f.}~Eq.~(\ref{eq:RKLSalgorithmNMC_2}). Here we use $400$ points per dimension and a fine temporal step $H_\star\,\Delta t \simeq 2.5\times10^{-3}$, which is required to resolve the ultraviolet tail of the $\chi$-spectrum if self-interactions become important. Throughout each simulation we monitor the lattice Friedmann constraint using the volume-averaged quantities
$\{a(\mathcal{N}),\langle\rho_\phi\rangle,\langle\rho_\chi\rangle\}$ obtained from the evolution equations, and verify that it remains satisfied to better than $10^{-3}$ at all times.

In \cref{fig:RR-spec-energydensity} we show results for the evolution of the power spectra of the NMC scalar field (left) and the different contributions to the total energy density (right). To understand the dynamics of the field, we rewrite \cref{eq:sigma-eq} governing the linear regime of the spectator modes, using cosmic time $t$,
\begin{equation}
\ddot\chi_k + 3H\,\dot\chi_k + \left(\frac{k^2}{a^2} + \xi \bar R + V_{,\chi,\chi}\right)\chi_k = 0\,,
\label{eq:chik-cosmic}
\end{equation}
where $V_{,\chi,\chi}\equiv \partial^2V/\partial\chi^2$. During kination and for $V_{,\chi,\chi}\approx0$, the effective frequency
is approximately $\omega_k^2 \simeq k^2/a^2 + \xi \bar R$, so all modes with $k/a \lesssim (\xi\,|\bar R|)^{1/2}$ grow tachyonically. Numerically we see (i) exponential growth while $\bar R<0$, (ii) a moving instability cutoff in $k$, and (iii) saturation as backreaction turns on. The key result here is that by solving this system on the lattice we are able to see the back-reaction of the produced $\chi$ particles onto the background, eventually leading to a radiation-scaling Universe once the energy density in $\chi$ dominates over that in the inflaton $\phi$, as plot in the right panel of Fig.~\ref{fig:RR-spec-energydensity}. While in this example we have not included interactions in $V(\chi)$, these can be straightforwardly added to the lattice simulation to study their effect on the saturation of the tachyonic instability and the final equation of state of the Universe, see~\cite{Figueroa:2024asq}.

\section{Axion interactions}
\label{sec:Axion}
~~~~The {\it triangular} or {\it chiral} anomaly leads to important phenomenology in particle physics. For instance, in Abelian gauge theories it controls the decay of the neutral pion into photons~\cite{Adler:1969gk,Bell:1969ts}, whilst in non-Abelian gauge theories, it is crucial for resolving the $U_A(1)$ problem~\cite{Veneziano:1979ec,Witten:1979vv}, and in the electroweak theory, it leads to baryon and lepton non-conservation~\cite{tHooft:1976rip,tHooft:1976snw}. The chiral anomaly refers to the fact that axial currents are not conserved at the quantum level, due to the existence of non-trivial vacua in gauge theories. In QCD, the presence of such anomaly and the expectation that it should lead to a certain level of CP violation, contradicts observations, raising up the Strong CP problem~\cite{Peccei:1977hh}. One possible solution, called the Peccei-Quinn (PQ) mechanism~\cite{Weinberg:1977ma,Wilczek:1977pj,Peccei:1977hh,Peccei:1977ur}, includes an additional global $U(1)$
~symmetry in the Standard Model (SM) that is spontaneously broken at energy scales way above the QCD scale. As a consequence, a weakly interacting and long-lived pseudoscalar particle emerges, the \textit{axion}, which becomes slightly massive at the QCD phase transition.

In cosmology, the {\it QCD axion}, and more generally {\it axion-like particles} (ALPs), are long-time sought dark matter (DM) candidates~\cite{Preskill:1982cy,Abbott:1982af,Dine:1982ah}, which are however difficult to detect due to their feeble coupling to the SM~\cite{Kim:1979if,Shifman:1979if,Zhitnitsky:1980tq,Dine:1981rt}. The axion DM paradigm~\cite{Marsh:2015xka,Sikivie:2020zpn,Irastorza:2021tdu} can lead to very different scenarios depending on whether the PQ symmetry breaking occurs before or after the end of inflation. In the former case, the standard misalignment mechanism~\cite{Preskill:1982cy,Abbott:1982af,Dine:1982ah} of a homogeneous axion field is the dominant framework for axion production. This is not the only possibility, however. Several alternative misalignment mechanisms have been proposed, {\it e.g.}~kinetic \cite{Co:2019jts,Chang:2019tvx,Co:2019wyp,Domcke:2020kcp,Co:2020jtv,Harigaya:2021txz,Chakraborty:2021fkp,Kawamura:2021xpu,Co:2021qgl,Co:2021lkc,Gouttenoire:2021wzu,Gouttenoire:2021jhk,Fonseca:2019ypl,Madge:2021abk,Eroncel:2022vjg,Eroncel:2022efc} or trapped misalignment~\cite{Jeong:2022kdr,DiLuzio:2021gos,DiLuzio:2024fyt} are among the most widely studied examples. Lattice simulations \cite{Morgante:2021bks,Berges:2019dgr,Chatrchyan:2020pzh,Chatrchyan:2023cmz,Ratzinger:2020oct,Weiner:2020sxn,Fasiello:2025ptb,Co:2025jnj} are becoming increasingly important in this context, as the intrinsic non-linearities of the axion potential play a crucial role in obtaining reliable dark-matter predictions. In the post-inflationary symmetry-breaking scenario, by contrast, axionic string and string-wall networks are formed, becoming the leading production mechanism. The evolution of axion strings and axion radiation is governed by highly non-linear classical equations of motion, for which only lattice simulations can provide reliable numerical predictions. To date it remains an intense research area, with ongoing debate and studies based on lattice simulations~\cite{Yamaguchi:1998gx,Yamaguchi:1999wt,Yamaguchi:2000fg,Yamaguchi:2002sh,Hiramatsu:2010yu,Hiramatsu:2012gg,Kawasaki:2014sqa,Fleury:2015aca,Lopez-Eiguren:2017dmc,Klaer:2017qhr,Klaer:2017ond,Gorghetto:2018myk,Kawasaki:2018bzv,Vaquero:2018tib,Buschmann:2019icd,Klaer:2019fxc,Hindmarsh:2019csc,Gorghetto:2020qws,Gorghetto:2021fsn,Buschmann:2021sdq,Benabou:2023ghl,Saikawa:2024bta,Kim:2024wku,Correia:2024cpk,Correia:2025nns}.  LCT for the simulation of networks topological defects are presented in \cref{sec:DefectsV}, while in this section we focus on axion interactions in the context of inflation. 

Cosmic inflation can also incorporate ALPs in models generically referred to as {\it axion inflation}. In this context, 
the inflaton is identified with an ALP that enjoys a shift symmetry, 
ensuring in this way the stability of the inflaton potential against radiative corrections. 
Since the original proposal~\cite{Freese:1990rb}, numerous models based on this framework have been developed, see {\it e.g.}~\cite{Adams:1992bn, Dimopoulos:2005ac, Easther:2005zr, Bachlechner:2014hsa,Bachlechner:2014gfa, McAllister:2008hb, Silverstein:2008sg,Marchesano:2014mla,Kappl:2015esy,Arkani-Hamed:2003xts,Kim:2004rp,Kaloper:2008fb,Abe:2014xja}. Interaction of ALPs with gauge sectors, Abelian or non-Abelian, can be of special interest in cosmology. These take place in the form of derivative couplings, as imposed by shift symmetry. The lowest dimension gauge invariant operator that one can write is the dimension-5 {\it Chern-Simons} coupling $\phi F \tilde F$, which is a chiral operator. By \textit{axion interactions} we will refer from now on to this type of interaction. 

In axion inflation, a chiral excitation of the gauge sector leads to exponentially high occupation numbers of the gauge field, eventually backreacting onto the inlaton sector (and on the expanding background). As a consequence, steeper inflaton potentials are allowed compared to the absence of axion interactions. The Abelian case of axion inflation~\cite{Anber:2006xt,Anber:2009ua,Barnaby:2010vf,Adshead:2013qp,Cheng:2015oqa} is the most studied proposal, particularly for a hidden $U(1)$ gauge sector. A characteristic feature of these scenarios is that 
an exponential growth of one of the gauge field chiralities is developed during inflation, leading to a rich phenomenology with distinctive observational signatures: non-Gaussian perturbations~\cite{Barnaby:2010vf,Barnaby:2011qe,Barnaby:2011vw,Cook:2011hg,Pajer:2013fsa,Dimastrogiovanni:2018xnn}, chiral GW backgrounds~\footnote{The production of GWs is also very efficient in the $SU(2)$ case~\cite{Maleknejad:2016qjz,Dimastrogiovanni:2016fuu,Mirzagholi:2020irt,Watanabe:2020ctz,Fujita:2022jkc,Badger:2024ekb}.}~\cite{Sorbo:2011rz, Barnaby:2011qe, Cook:2013xea, Adshead:2013qp, Bartolo:2016ami,Maggiore:2019uih,LISACosmologyWorkingGroup:2022jok,Bastero-Gil:2022fme,Garcia-Bellido:2023ser}, efficient preheating \cite{Adshead:2015pva,Cuissa:2018oiw,Adshead:2019igv, Adshead:2019lbr,Adshead:2023mvt}, successful magnetogenesis~\cite{Garretson:1992vt, Anber:2006xt, Adshead:2016iae, Durrer:2023rhc} and realization of the baryon asymmetry mechanism~\cite{Giovannini:1997eg,Anber:2015yca,Fujita:2016igl,Kamada:2016eeb,Maleknejad:2016dci,Jimenez:2017cdr,Cado:2022evn}. Alternative couplings of the ALP with non-Abelian sectors and their associated phenomenology, have also been extensively studied in {\it e.g.}~Gauge-flation~\cite{Maleknejad:2011sq,Maleknejad:2011jw,Maeda:2013daa} and Chromo-Natural inflation~\cite{Adshead:2012kp,Adshead:2013nka} models, as well as couplings to fermions in~\cite{Adshead:2015kza,Adshead:2018oaa,Domcke:2018eki,Domcke:2019qmm,Cado:2022pxk}.

The exponential amplification of an Abelian gauge field through the Chern–Simons coupling has also motivated studies beyond the context of inflation. For instance, the audible axion model \cite{Machado:2018nqk,Machado:2019xuc,Ratzinger:2020oct} explores the standard misalignment scenario for ALPs, potentially leading to observable gravitational wave (GW) backgrounds. Similar mechanisms of dark sector amplification have also been considered in attempts to resolve the Hubble tension via the decay of an ultralight axion~\cite{Gonzalez:2020fdy,Weiner:2020sxn}.

While axion interactions have been studied under various levels of approximation, see {\it e.g.}~\cite{Cheng:2015oqa,Notari:2016npn,DallAgata:2019yrr,Domcke:2020zez,Gorbar:2021rlt,Peloso:2022ovc,Iarygina:2023mtj,Galanti:2024jhw,Alam:2024fid,Dimastrogiovanni:2024lzj,Dimastrogiovanni:2025snj} for recent works, only LCT have proven capable of fully capturing the truly local non-linear dynamics. The complete locality provided by lattice simulations has been employed to re-assess the effects of axion interactions during inflation~\cite{Caravano:2021bfn,Caravano:2022epk,Figueroa:2023oxc,Caravano:2024xsb,Sharma:2024nfu,Figueroa:2024rkr} and during preheating~\cite{Cuissa:2018oiw,Adshead:2019lbr,Adshead:2019igv,Adshead:2023mvt}, 
and in the context of both the audible axion model~\cite{Ratzinger:2020oct} and decaying ultralight axions~\cite{Weiner:2020sxn}. 

\subsection{Axion Chern-Simons interaction with a $U(1)$ gauge field}

We review  
now the continuum description of an ALP coupled to a dark $U(1)$ gauge sector, setting up the theoretical framework for later discretization of the system. We consider the lowest dimensional interaction term that preserves shift-symmetry with a gauge field, which in our notation reads
\begin{equation}\label{eqn:InteractionPhiFFdual}
    \mathcal{L}_{\rm int} = - \frac{1}{4 \Lambda} \phi F_{\mu \nu} \Tilde{F}^{\mu \nu} \, ,
\end{equation}
with $\Lambda$ representing an energy scale, $F_{\mu \nu}$ the field strength [{\it c.f.}~Eq.~(\ref{eq:AbCovDerivCont})] of a $U(1)$ gauge field $A_{\mu}$, and and $\Tilde{F}_{\mu \nu} \equiv \frac{1}{2}\epsilon_{\mu\nu\rho\sigma}F^{\rho\sigma}$ the dual counterpart,
with $\epsilon_{\mu\nu\rho\sigma}$ the completely antisymmetric Levi-Civita tensor, which in a curved spacetime takes the form $\epsilon_{0123}= 1/\sqrt{-g}$. 
Let us consider the action $S_{\rm tot} =  S_{\rm g} + S_{\rm m}$, where $S_{\rm g} \equiv {1\over2}m_p^2 \int dx^4 \sqrt{-g} \,R$ is the standard Hilbert-Einstein action for gravity, and the action of the matter fields is given by
\begin{eqnarray}\label{eqn:AxionInteractiosAction}
S_{\rm m} = -\int {\rm d}x^4 \sqrt{-g}\left[\frac{1}{2}\partial_\mu \phi\partial^\mu\phi+V(\phi) +\, \frac{1}{4}F_{\mu\nu}F^{\mu\nu} - \frac{\alpha_{\Lambda}}{4}\frac{\phi}{m_p} \FFdual \right]\, ,
\end{eqnarray}
with $\alpha_{\Lambda} \equiv m_p/\Lambda$. This action is invariant under local U(1) transformations where $A_{\mu} \rightarrow A_{\mu} + \partial_{\mu} \alpha(x)$, with $\alpha(x)$ any real function. It is also shift symmetric $\phi \rightarrow \phi +c$, with $c$ an arbitrary real constant. The potential $V(\phi)$, however, is expected to explicitly break the shift symmetry. In a spatially flat FLRW background, {\it c.f.}~Eq.~(\ref{eq:FLRWmetric}), varying the above action~(\ref{eqn:AxionInteractiosAction}) leads to the EOM
\begin{eqnarray}
& &\ddot{\phi} = -3H\dot{\phi}+\frac{1}{a^2}\nabla^2\phi-\frac{\partial V(\phi)}{\partial \phi}+\frac{\alpha_\Lambda}{a^3 m_p}\vec{E}\cdot\vec{B}\,,\label{eqn:eom1}\\
& &\dot{\vec{E}} = -H\vec{E}-\frac{1}{a^2}\vec{\nabla}\times\vec{B}-\frac{\alpha_\Lambda}{a m_p}\left(\dot{\phi}\vec{B}-\vec{\nabla}\phi\times\vec{E}\right),\label{eqn:eom2}\\
& &\vec{\nabla}\cdot\vec{E} \,\, = -\frac{\alpha_{\Lambda}}{am_p}\vec{\nabla}\phi\cdot\vec{B}\,,\quad\quad\quad{\rm [Gauss~Constraint]}\label{eqn:Gauss}
\end{eqnarray}
with $\vec E$ and $\vec B$ the electric and magnetic fields
defined in Eq.~(\ref{eq:ElectricMagneticDefs}), and where we have chosen to work in the temporal gauge $A_0=0$.

Assuming for greater generality that self-consistent expansion is sourced by both the ALP and gauge field, the Friedmann equations read
\begin{eqnarray}
& &\ddot{a}=-\frac{a}{3m_p^2}\big( 2E_{ K}-E_{ V}+E_{ EM} \big)\,,\label{eqn:ddaAxion}\\ 
& &H^2=\frac{1}{3m_p^2}\big(E_{ K}+E_{ G}+E_{ V}+E_{ EM}\big)\,,\quad\quad\quad{\rm [Hubble~Constraint]}\label{eqn:HubbleAxion}
\end{eqnarray}
where the different homogeneous energy density contributions are given by the expressions  
\begin{eqnarray}\label{eqn:energyDensityTerms}
E_{ K} \equiv \frac{1}{2}\langle\dot{\phi}^2\rangle\; , \quad E_{ G} \equiv \frac{1}{2a^2}\langle(\vec\nabla\phi)^2\rangle\; , \quad
E_{ V} \equiv \langle V(\phi) \rangle\;, \quad E_{ EM} \equiv \frac{1}{2a^4}\langle a^2\vec E^2+\vec B^2\rangle \;,
\end{eqnarray}
with $\langle ... \rangle$ denoting volume average over sufficiently large regions to encompass all relevant wavelengths of the fields. Here $E_K$, $E_G$, $E_V$ denote, respectively, the kinetic, gradient and potential energy densities of the axion, while EM refers to the electromagnetic energy density associated to $A_\mu$. 

As mentioned in the introduction of this Section, the tachyonic excitation of the gauge field and the chiral nature of the Chern-Simons coupling~(\ref{eqn:InteractionPhiFFdual}), are very prominent features of the expected dynamics. This can be illustrated by considering a homogeneous axion rolling down its potential, ignoring for the time being the backreaction from the gauge sector. The EOM for the gauge field can then be re-written under these simplifying assumptions as follows
\begin{eqnarray}\label{eq:linAxion}
\left( \frac{d^2}{d\tau^2} - \nabla^2 + \frac{\alpha_{\Lambda}}{m_p} \; \frac{d \phi}{d\tau}  \;\vec{\nabla} \times \right) \vec{A}(\tau,{\bf x}) = 0 \; ,
\end{eqnarray}
where we introduced conformal time $d\tau \equiv dt/a(t)$ for convenience. As Eq.~(\ref{eq:linAxion}) is linear, it is natural to introduce a mode-decomposition os the gauge field as
\begin{eqnarray}
\vec A(\tau,{\bf x}) = \sum_{\lambda} \int \frac{{\rm d}^3k}{(2\pi)^3}  A^\lambda(\tau,{\bf k})\vec{\varepsilon}^{\,\lambda}(\hat{\bf k})  e^{i{\bf k}\cdot{\bf x}}\,,
\label{eq:FTransform}
\end{eqnarray}
where the vectors $\lbrace \vec{\varepsilon}^{\,+}(\hat{\bf k})\, ,\vec{\varepsilon}^{\,-}(\hat{\bf k})\rbrace$ form a {\it chiral basis}, satisfying
\begin{eqnarray}
\hat {\bf k}\cdot\vec\varepsilon^{\,\lambda}(\hat{\bf k})=0\,,\quad
\hat{\bf k} \times \vec \varepsilon^{\,\lambda}(\hat{\bf k})=-i\lambda \vec\varepsilon^{\,\lambda}(\hat{\bf k})\,,
\quad
\varepsilon_i^{\,\lambda}(\hat{\bf k})^*=\varepsilon_i^{\,\lambda}(-\hat{\bf k})\,,\quad \vec \varepsilon^{\,\lambda'}\hspace*{-1mm}(\hat{\bf k})\cdot\vec\varepsilon^{\,\lambda}(\hat{\bf k})^* = \delta_{\lambda\lambda'}\,.
\label{eqn:polarisationvectors}
\end{eqnarray}
Promoting now the Fourier amplitudes into a quantum operator by means of standard annihilation and creation operators, $A^\lambda(\tau,{\bf k}) \rightarrow \hat A^\lambda(\tau,{\bf k}) \equiv \hat a_{{\bf k}}\mathcal{A}^\lambda(\tau,{\bf k})$ + $\hat a_{-{\bf k}}^\dag\mathcal{A}^{\lambda}(\tau,-{\bf k})^{*}$, with $[\hat a_{{\bf k}}, \hat a_{{\bf k}'}^{\dagger}] \equiv \delta^{(3)}({\bf k}-{\bf k}')$,\footnote{In this line, the hat notation indicates quantum operators and should not be confused with the use of hats to denote unit vectors throughout the paper.} the mode functions $\mathcal{A}^{\pm}(\tau,{\bf k})$ then follow the EOM
\begin{eqnarray}
& &\left[\partial_\tau^2+\left(k^2\mp k\frac{\alpha_{\Lambda}}{m_p}\frac{d\phi}{d\tau}\right)\right] \mathcal{A}^\pm(\tau,{\bf k})=0\,,\label{eq:linA}
\end{eqnarray}
making manifest the parity breaking nature of (\ref{eqn:InteractionPhiFFdual}), as the $\mp$ sign differentiates the behavior of the two chiralities, $\mathcal{A}^+$ and $\mathcal{A}^-$. For modes $k<\frac{\alpha_{\Lambda}}{m_p}\frac{d\phi}{d\tau}$, one of the polarizations (which one depends on the sign of $d\phi/d\tau$), 
 experiences a tachyonic instability. 
For example, in axion inflation 
it is customary to introduce the instability parameter
\begin{equation}
    \xi \equiv \frac{|\dot{\phi}|}{2 H \Lambda }\,,
\end{equation}
so that Eq.~\eqref{eq:linA} is re-written as
\begin{equation}
\left(\partial_\tau^2+k^2\pm sign(\dot\phi)\frac{2 k \xi}{|\tau|}\right) \mathcal{A}^\pm(\tau,{\bf k})=0\,,\label{eq:linA2}
\end{equation}
where the slow-roll condition $\tau \simeq -1/(aH)$ has been used. Assuming that during inflation, deep inside the Hubble scale, the modes are in the Bunch-Davies vacuum, an approximated expression for the amplitude of the tachyonic modes can be found for $\xi > \mathcal{O}(1)$. Choosing (without loss of generality) $sign(\dot\phi) = -1$, this solution reads~\cite{Anber:2009ua}
\begin{equation}
\mathcal{A}^+(\tau,{\bf k}) \simeq \frac{1}{\sqrt{2k}}\left(\frac{k}{2\xi aH}\right)^{1/4} e^{\pi\xi - 2\sqrt{2\xi k/(aH)}}\,,~~~~k|\tau| \ll 2\xi\,,\label{eqn:amplifiedAplus}
\end{equation}
based on the assumption that during slow-roll $\xi$ can be considered (approximately) constant. Expression~(\ref{eqn:amplifiedAplus}) exhibits the exponential growth of a mode that experiences a tachyonic instability. On the other hand, the other polarization $\mathcal{A}^-(\tau,{\bf k})$ remains close to the Bunch-Davies vacuum solution. 

\subsection{Lattice formulation of axion interactions}
\label{subsec:axionLattice}

Simulating models with an interaction as in Eq.~(\ref{eqn:InteractionPhiFFdual}), requires evolution schemes beyond those used in canonical scalar-gauge field interactions presented in {\tt The Art-I}~\cite{Figueroa:2020rrl}, as the evolution kernel of the gauge field prevents the use of symplectic methods. Furthermore, the discretization of $F\tilde{F}$ needs special care in order to preserve its continuum properties, namely the fact that it is a total derivative, $F\tilde F = \partial^\mu K_\mu$, with $K_\mu$ the {\it Chern-Simons} current. In order to see this in detail, let us first write the contribution to the action one can build from Eq.~(\ref{eqn:InteractionPhiFFdual}), in terms of electric $\vec{E}$ and magnetic $\vec{B}$ fields [{\it c.f.}~Eq.~(\ref{eq:ElectricMagneticDefs})], as
\begin{equation}\label{eq:axCouplingCont}
    S_{\rm int}= \int \text{d}t \text{d}^3x \frac{\alpha_{\Lambda}}{m_p}\phi\,\vec{E}\cdot\vec{B}\;.
\end{equation}
Following 
Ref.~\cite{Figueroa:2017qmv}, one should write a lattice operator $(\vec{E}\cdot \vec{B})_{\rm Latt}$ that respects the natural discrete symmetries of the lattice, {\it i.e.}~spatial cubic symmetry and time-reversal symmetry. As electric and magnetic fields on a lattice can be thought as living at semi-integer   lattice sites / time steps 
\begin{equation}\label{eq:standardLatticeElecAndMagSite}
E_i \equiv  E_i(t+\hat{0}/2,\textbf{n}+\hat{\imath}/2)\;,~~~
B_i \equiv B_i(t,\textbf{n}+\hat{\jmath}/2+\hat{k}/2)\,,
\end{equation} 
one should not use the naive operator $(\vec{E}\cdot \vec{B})_{\rm Latt} = \sum_i E_iB_i$. Instead, Ref.~\cite{Figueroa:2017qmv} proposed the action
\begin{equation}\label{eq:axCouplingLAT}
     S^{\rm L}_{\rm int}=\frac{\alpha_{\Lambda}}{m_p}   \sum_{n_0,\textbf{n}} \delta t \delta x^3  \phi \sum_i E^{(2)}_{i}(B^{(4)}_{i}+B^{(4)}_{i,+0})\;,
\end{equation}
based on {\it improved} electric and magnetic field operators that live at the lattice sites, defined as
\begin{eqnarray}\label{eq:improvedLatticeElecAndMag}
    \begin{array}{rcl}
    E^{(2)}_{i}(t+\hat{0}/2,\textbf{n})&\equiv&\frac{1}{2}(E_i+E_{i,-i})\;,\\
    B^{(4)}_{i}(t,\textbf{n})&\equiv&\frac{1}{4}(B_i+B_{i,-j}+B_{i,-k}+B_{i,-j-k})\;,
    \end{array}
\end{eqnarray}
where $B_i^{(4)}$ is the well-known {\it clover} definition of a
magnetic field on a site.
The choice of $B^{(4)}$ and $E^{(2)}$ is introduced in order to respect the cubic symmetry, while the
symmetrization between present- and forward-time $B^{(4)}$ fields is set by time-reflection symmetry.
In addition to reproducing 
the continuum at order $\mathcal{O}(\delta x_\mu^2)$, Eq.~(\ref{eq:axCouplingLAT}) leads to a correct lattice version of the Bianchi identities, and to a total derivative form on the lattice\footnote{The statement that $(\vec{E}\cdot \vec{B})_{\rm Latt}$ can be expressed exactly as a total derivative in Abelian lattice gauge theories goes back to~\cite{Moore:1996qs,Moore:1996wn}, where the motion of Chern-Simons number in $SU(2)$ and $SU(3)$ gauge at high temperatures under a chemical potential was investigated, improving previous non-Abelian studies~\cite{Ambjorn:1990pu}. While no explicit details were given for the Abelian case in~\cite{Moore:1996qs,Moore:1996wn}, Ref.~\cite{Figueroa:2017qmv} presented all proofs and formulas necessary for numerical simulations.} as $(\vec{E}\cdot \vec{B})_{\rm Latt} \propto \nabla_\mu^+ K^\mu$. 

A technical difficulty that emerges 
when using the operator~(\ref{eq:axCouplingLAT}), is that due to the semi-sum $(B^{(4)}_{i}+B^{(4)}_{i,+0})$, 
the lattice equations of motion 
are not explicit in time~\cite{Figueroa:2017qmv}. This requires the use of an implicit evolution scheme, with multiple iterative sub-steps within each time-step $\delta t$. While such implicit schemes have been proposed for evolution in Minkowski~\cite{Figueroa:2017qmv} and in a comoving grid with self-consistent expansion~\cite{Cuissa:2018oiw}, in this review we follow instead the scheme introduced in~\cite{Figueroa:2023oxc}. The latter proposed a variation of the above lattice prescription, allowing for an explicit evolution scheme,  while still preserving all the relevant desired properties at the lattice level, such as the Bianchi identities and the shift symmetry.

Following Ref.~\cite{Figueroa:2023oxc} we use a {\it hybrid} lattice formulation, where one discretizes only spatial derivatives, while still treats the temporal coordinate as a continuous variable (recall discussions in Sect.~\ref{subsec:Algorithms}). To formalize this idea, we re-write the previous lattice action demanding only cubic symmetry, so that
\begin{equation}\label{eq:axCouplingHybLAT}
     S^{\rm L}_{\rm int} = \frac{\alpha_{\Lambda}}{m_p}  \int dt \sum_{\textbf{n}} \delta x^3 \phi \sum_i E^{(2)}_{i}B^{(4)}_{i}\;,
\end{equation}
with $E^{(2)}_{i}$ and $B^{(4)}_{i}$ following the definitions in Eq.~(\ref{eq:improvedLatticeElecAndMag})   but with continuous time. Accordingly, we write a hybrid lattice action for the whole system as
\begin{eqnarray}\label{eq:matterDiscreteAxionCouplingActionExplicit}
	S_{\rm m}^{\rm L} = \delta x^3\sum_{\textbf{n}}\int \hspace*{-1mm}dt\, a^3 \hspace*{-0.5mm}\left\{ \frac{(\dot \phi)^2}{2} - \sum_i \frac{(\nabla^+_i\phi)^2}{2a^2} - V(\phi) + \sum_i \frac{1}{2a^2}\left(E_i^2 - a^{-2}B_i^2\right)  + \frac{\alpha_\Lambda\phi}{m_pa^3} \sum_i E_i^{(2)}B_i^{(4)} \right\}, 
\end{eqnarray}
and upon variation, we obtain the following set of discrete equations, 
\begin{eqnarray}\label{eq:explicitEOMscalar}
	&&\ddot{\phi} =  -3H\dot{\phi} + \frac{1}{a^2} \sum_i \nabla_i^-\nabla_i^+ \phi - \frac{dV(\phi)}{d\phi} + \frac{\alpha_\Lambda}{a^3m_p} \sum_i E_i^{(2)}B_i^{(4)} \;,\\
    \label{eq:explicitEOMgauge}
	 &&\dot{E}_i = -HE_i - \frac{1}{a^2} \sum_{j,k} \epsilon_{ijk} \nabla_j^- B_k - \frac{\alpha_\Lambda}{2am_p} \left(\dot{\phi} B_i^{(4)} + \dot{\phi}_{+i}B^{(4)}_{i,+i} \right) \\
	&& \hspace*{5.3cm}+ \frac{\alpha_\Lambda}{4am_p} \sum_\pm \sum_{j,k} \epsilon_{ijk}  \left( \left[ (\nabla_j^\pm \phi) E_{k,\pm j}^{(2)} \right]_{+i} +  \left[ (\nabla_j^\pm \phi) E_{k,\pm j}^{(2)}  \right]   \right),\nonumber \\
    \label{eq:explicitGaussLaw}
     &&\sum_i \nabla_i^- E_{i} = -\frac{\alpha_{\Lambda}}{2am_{p}} \sum_{\pm} \sum_i
\left( \nabla_i^\pm \phi \right) B_{i,\pm i}^{(4)}\,~~~~{\rm [Gauss~Constraint]}\;.
\end{eqnarray}
We note that all terms of each equation live at the same lattice site, and that these discrete equations approximate the spatial continuum counterpart expressions to order $O(\delta x^2)$. The form of Eqs.~(\ref{eq:explicitEOMscalar})-(\ref{eq:explicitEOMgauge}) makes manifest that explicit evolution schemes are suitable to solve them. The only requisite is to use an integration method that allows for conjugate momenta in the evolution kernels: {\it i.e.}~non-symplectic integrators, like the explicit-in-time {\it Runge-Kutta} (RK) solvers introduced in Sect.~\ref{subsubsec:NonSymplecticInt}. In~\cite{Figueroa:2023oxc,Figueroa:2024rkr,Lizarraga:2025aiw}, RK integrators of order 2 and 3 have been actually implemented and shown to work successfully for this system of equations. 

It is convenient to generalize the equations to $\alpha$-time $d\eta = a^{-\alpha} dt$, and rewrite them in terms of \textit{program variables} $\tilde\eta \equiv \omega_*\eta, \tilde x^i \equiv \omega_* x^i, \tilde{\phi} \equiv \phi / f_*,  \tilde{A}_i \equiv A_i/\omega_*,  \tilde{E}_i \equiv E_i/\omega_*^2, \tilde{B}_i \equiv B_i/\omega_*^2$, {\it c.f.}~Eq.~(\ref{eq:GaugeProgramVar}), with $f_*,\,\omega_*$ some convenient constants of dimension $+1$. In particular,  
we obtain matter field kernels as 
\begin{eqnarray}
	&&\mathcal{K}^{\rm L}_{\phi}[a,\tilde{\phi},\tilde{A}_j,a',\tilde{\pi}_{\phi},\tilde{E}_j] =  -(3-\alpha)\frac{a'}{a}\tilde{\pi}_{\phi} + a^{2(\alpha-1)} \sum_i \tilde{\nabla}_i^-\tilde{\nabla}_i^+ \tilde{\phi}\label{eq:explicitEOMscalarKernelAlpha} \\ 
    &&\hspace*{6.3cm}- a^{2\alpha}\frac{d\tilde{V}(\tilde{\phi})}{d\tilde{\phi}} + \left(\frac{\omega^2_{*}}{f_{*}m_p}\right)\alpha_\Lambda a^{\alpha-3}\sum_i \tilde{E}_i^{(2)}\tilde{B}_i^{(4)} \;,\nonumber\\ 
	 && \mathcal{K}^{\rm L}_{A_i}[a,\tilde{\phi},\tilde{A}_j,a',\tilde{\pi}_{\phi},\tilde{E}_j] = (\alpha-1)\frac{a'}{a}\tilde{E}_i - a^{2(\alpha-1)} \sum_{j,k} \epsilon_{ijk} \tilde{\nabla}_j^- \tilde{B}_k  \\ \label{eq:explicitEOMgaugeKernelAlpha} &&\hspace*{4.3cm}- \left(\frac{f_{*}}{m_p}\right) \frac{\alpha_\Lambda a^{\alpha-1}}{2}\left(\tilde{\pi}_{\phi} \tilde{B}_i^{(4)} + \tilde{\pi}_{\phi,+i}\tilde{B}^{(4)}_{i,+i} \right) \nonumber\\  &&\hspace*{4.3cm}+ \left(\frac{f_{*}}{m_p}\right)\frac{\alpha_\Lambda a^{\alpha-1}}{4} \sum_\pm \sum_{j,k} \epsilon_{ijk}  \left\{ \left[ (\tilde{\nabla}_j^\pm \tilde{\phi}) \tilde{E}_{k,\pm j}^{(2)} \right]_{+i} +  \left[ (\tilde{\nabla}_j^\pm \tilde{\phi}) \tilde{E}_{k,\pm j}^{(2)}  \right]   \right\} \;,\nonumber 
\end{eqnarray}
where we have introduced $\tilde{\pi}_{\phi}=\tilde{\phi}'$ and $\tilde{E}_i=\tilde{A}'_i$ as conjugate momenta, and $\tilde{E}_i^{(2)}$ is defined in terms of the latter. 
This allows to formulate the second order differential Eqs.~(\ref{eq:explicitEOMscalar})-(\ref{eq:explicitEOMgauge}) as a coupled system of first-order differential equations,
\begin{equation}
\left\lbrace
\begin{array}{rcl}\label{eq:coupledFirstDiffEqAxion}
\tilde{\pi}_\phi' &=& \mathcal{K}^{\rm L}_{\phi}[a,\tilde{\phi},\tilde{A}_j,a',\tilde{\pi}_{\phi},\tilde{E}_j]\,, \\[5pt]
\tilde{E}_i' &=&\mathcal{K}^{\rm L}_{A_i}[a,\tilde{\phi},\tilde{A}_j,a',\tilde{\pi}_{\phi},\tilde{E}_j]\,, \\[5pt]
\tilde{\phi}'&=&\tilde{\pi}_\phi\,,\\[5pt]
\tilde{A}_i'&=&\tilde{E}_i\,,
\end{array}
\right.
\end{equation}
the solution of which satisfies (to machine precision) the Gauss constraint
\begin{equation}\label{eq:explicitGaussLawAlpha}
	\sum_i \tilde{\nabla}_i^- \tilde{E}_i = - \left[\frac{f_{*}}{m_p}\right]\frac{\alpha_\Lambda a^{\alpha-1}}{2} \sum_\pm \sum_i (\tilde{\nabla}_i^\pm \tilde{\phi}) \tilde{B}_{i,\pm i}^{(4)} \; .
\end{equation}

Additionally, if one considers self-consistent evolution as sourced by the ALP and the gauge field, we need to write the evolution kernel for the scale factor, which in program variables reads
\begin{eqnarray}\label{eq:explicitScaleFactorKernelAlpha}
    &&\mathcal{K}_{a}^{\rm L}[a,\tilde{E}^{\phi}_{K},\tilde{E}^{\phi}_{G},\tilde{E}^{\phi}_{V},\tilde{E}^{A}_{K},\tilde{E}^{A}_{G}]\\
    &&\hspace*{2.3cm}=\left(\frac{f_{*}}{m_{\text{p}}}\right)^2\frac{a^{2\alpha+1}}{3}\left[(\alpha-2)\tilde{E}^{\phi}_{K}+\alpha\tilde{E}^{\phi}_{G}+(\alpha+1)\tilde{E}^{\phi}_{V}+(\alpha-1)\tilde{E}^{A}_{K}+(\alpha-1)\tilde{E}^{A}_{G}\right]\,,\nonumber
\end{eqnarray}
with the energy density terms given by
\begin{equation}\label{eq:energyDensityTermsAlpha}
\begin{gathered}
\tilde{E}^{\phi}_{K} \equiv \frac{1}{2a^{2\alpha}}\left\langle\tilde{\pi}^2_{\phi}\right\rangle\; , \quad \tilde{E}^{\phi}_{G} \equiv \frac{1}{2a^2}\Big\langle(\tilde{\nabla}^{\hspace{-0.5mm}+}\hspace{-0.5mm}\tilde{\phi})^2\Big\rangle\; , \quad \tilde{E}^{\phi}_{V} \equiv \left\langle \tilde{V}(\tilde{\phi})\right\rangle\;, 
\\
\tilde{E}^{A}_{K} \equiv \left(\frac{\omega_{*}}{f_{*}}\right)^2\frac{1}{2a^{2(\alpha+1)}} \Big\langle\tilde{E}^2_i\Big\rangle \;, \quad \tilde{E}^{A}_{G} \equiv \left(\frac{\omega_{*}}{f_{*}}\right)^2\frac{1}{2a^4} \Big\langle\tilde{B}^2_i\Big\rangle \;,
\end{gathered}
\end{equation}
Turning the evolution of the scale factor into a system of coupled first-order differential equations,
\begin{equation}\label{eq:coupledFirstDiffEqExpansion}
\left\lbrace
\begin{array}{rcl}
b' &=&  \mathcal{K}^{\rm L}_{a}[a,\tilde{E}^{\phi}_{K},\tilde{E}^{\phi}_{G},\tilde{E}^{\phi}_{V},\tilde{E}^{A}_{K},\tilde{E}^{A}_{G}]\,, \\[5pt]
a'&=&b\,,
\end{array}
\right.
\end{equation}
This comes together with the lattice version of the Hubble constraint, which needs to be satisfied by the evolution algorithm to the desired precision,
\begin{equation}\label{eq:HubbleConstraintAlpha}
      b^2 = \left(\frac{f_{*}}{m_p}\right)^2\frac{a^{2(\alpha+1)}}{3}\left(\tilde{E}^{\phi}_{K}+\tilde{E}^{\phi}_{G}+\tilde{E}^{\phi}_{V}+\tilde{E}^{A}_{K}+\tilde{E}^{A}_{G}\right).
\end{equation}

In the case of self-consistent expansion, one must therefore simultaneously solve Eqs.~(\ref{eq:coupledFirstDiffEqAxion}) and~(\ref{eq:coupledFirstDiffEqExpansion}). While the Gauss constraint in Eq.~(\ref{eq:explicitGaussLawAlpha}) is preserved to machine precision during the evolution of the matter fields, the Hubble constraint in Eq.~(\ref{eq:HubbleConstraintAlpha}) is only expected to be preserved at $\mathcal{O}(d\tilde\eta^n)$, with $n = 2, 3, ...$ depending on the evolution algorithm used. As said before, we need non-symplectic time integration algorithms such as Runge-Kutta (RK) methods, like those introduced in Sect.~\ref{subsubsec:NonSymplecticInt}. As an example, we present here explicitly a RK2 algorithm for this system (based on modified-Euler), 
\begin{eqnarray}
\hspace{-0.75cm}
\left.
\begin{array}{ll}
\tilde{\phi}^{(1)} = \tilde{\phi}\,, & \tilde{\phi}^{(2)} = \tilde{\phi}^{(1)} + \delta\tilde{\eta}\tilde{\pi}^{(1)}_{\phi}\,,\vspace*{0.2cm}\\
\tilde{\pi}^{(1)}_{\phi} = \tilde{\pi}_{\phi} & \tilde{\pi}^{(2)}_{\phi} = \tilde{\pi}^{(1)}_{\phi} + {\delta\tilde{\eta}}\mathcal{K}^{{\rm L}, (1)}_{\phi}\,,\vspace*{0.2cm}\\
\tilde{A}^{(1)}_i=\tilde{A}_i & \tilde{A}^{(2)}_i = \tilde{A}^{(1)}_i + {\delta\tilde{\eta}}\tilde{E}^{(1)}_i\,,\vspace*{0.2cm}\\
\tilde{E}^{(1)}_{i}=\tilde{E}_i\,, & 
\tilde{E}^{(2)}_{i} = \tilde{E}^{(1)}_{i} + {\delta\tilde{\eta}}\mathcal{K}^{{\rm L}, (1)}_{A_i}\,,\vspace*{0.2cm}\\
a^{(1)} = a\,, & a^{(2)} = a^{(1)} + {\delta\tilde{\eta}}b^{(1)}\,,\vspace*{0.2cm}\\
b^{(1)} = b\,, & b^{(2)} = b^{(1)} + {\delta\tilde{\eta}}\mathcal{K}^{{\rm L}, (1)}_{a}\,,\vspace*{0.2cm}
\end{array}
\right\rbrace ~~~\Longrightarrow~~~
\left\lbrace
\begin{array}{rcl}
\tilde{\phi}_{+0}&=& \tilde{\phi}^{(1)} + {1\over2}\delta\tilde{\eta}\left[\tilde{\pi}^{(1)}_{\phi}+\tilde{\pi}^{(2)}_{\phi}\right]\,,\vspace*{0.2cm}\\
\tilde{A}_{i,+0} &=& \tilde{A}^{(1)}_i + {1\over2}\delta\tilde{\eta}\left[\tilde{E}^{(1)}_{i}+\tilde{E}^{(2)}_{i}\right]\,,\vspace*{0.2cm}\\
a_{+0} &=& a^{(1)} + {1\over2}\delta\tilde{\eta}\left[b^{(1)}+b^{(2)}\right]\,,\vspace*{0.2cm}\\
\tilde{\pi}_{\phi, +0}&=&\tilde{\pi}^{(1)}_{\phi}+{1\over2}\delta\tilde{\eta}\left[\mathcal{K}^{{\rm L}, (1)}_{\phi}+\mathcal{K}^{{\rm L}, (2)}_{\phi}\right]\,,\vspace*{0.2cm}\\
\tilde{E}_{i,+0}&=&\tilde{E}^{(1)}_{i}+{1\over2}\delta\tilde{\eta}\left[\mathcal{K}^{{\rm L}, (1)}_{A_i}+\mathcal{K}^{{\rm L}, (2)}_{A_i}\right]\,,\vspace*{0.2cm}\\
b_{+0}&=&b^{(1)}+{1\over2}\delta\tilde{\eta}\left[\mathcal{K}^{{\rm L}, (1)}_{a}+\mathcal{K}^{{\rm L}, (2)}_{a}\right]\,,\vspace*{0.2cm}
\end{array}\right.\nonumber\\
\end{eqnarray}
where $\mathcal{K}^{{\rm L}, (l)}_{a} = \mathcal{K}^{{\rm L}}_{a}[a^{(l)},\tilde{E}^{\phi,(l)}_{K},\tilde{E}^{\phi,(l)}_{G},\tilde{E}^{\phi,(l)}_{V},\tilde{E}^{A,(l)}_{K},\tilde{E}^{A,(l)}_{G}]$, $\mathcal{K}^{{\rm L}, (l)}_{\phi} = \mathcal{K}^{{\rm L}}_{\phi}[a^{(l)},\tilde{\phi}^{(l)},\tilde{A}_j^{(l)},b^{(l)},\tilde{\pi}^{(l)}_{\phi},\tilde{E}_j^{(l)}]$ and $\mathcal{K}^{{\rm L}, (l)}_{A_i} = \mathcal{K}^{{\rm L}}_{A_i}[a^{(l)},\tilde{\phi}^{(l)},\tilde{A}_j^{(l)},b^{(l)},\tilde{\pi}^{(l)}_{\phi},\tilde{E}_j^{(l)}]$, with $l=1,2$. Alternatively, one can easily adapt as well for this problem the higher order integrator RK4 presented in Eq.~(\ref{eq:RK4algorithm_1}), or the low-storage RK methods encapsulated by Eq.~(\ref{eq:RKLSalgorithm_1}). 

\noindent In an inflationary context, rather than $\alpha$-time it is convenient to use e-folding as the time variable, 
\begin{align}
    d\mathcal{N}=d\ln{a}=Hdt\;.
\end{align}
In this case, the field kernels read
\begin{eqnarray}    
	&&\hspace*{-1cm}\mathcal{K}^{\rm L}_{\mathcal N,\phi}[a,\tilde H,\tilde{\phi},\tilde{A}_j,\tilde{\pi}_{\phi},\tilde{E}_j]= -\tilde{\pi}_{\phi} + \frac{1}{\tilde H}\left\{a^{-2} \sum_i \tilde{\nabla}_i^-\tilde{\nabla}_i^+ \tilde{\phi}\right. \label{eq:explicitEOMscalarKernelEfold}\\
    &&\hspace*{7.3cm}\left.- \frac{d\tilde{V}(\tilde{\phi})}{d\tilde{\phi}} + \left[\frac{\omega^2_{*}}{f_{*}m_p}\right]\alpha_\Lambda a^{-3}\sum_i \tilde{E}_i^{(2)}\tilde{B}_i^{(4)}\right\} \; ,\nonumber\\ 
	 &&\hspace*{-1cm}\mathcal{K}_{\mathcal N,A_i}^{\rm L}[a,\tilde H,\tilde{\phi},\tilde{A}_j,\tilde{\pi}_{\phi},\tilde{E}_j] = -\tilde{E}_i+\frac{1}{\tilde H}\left\{ - a^{-2} \sum_{j,k} \epsilon_{ijk} \tilde{\nabla}_j^- \tilde{B}_k \right. \label{eq:explicitEOMgaugeKernelEfold} \\&&\hspace*{4.3cm}- \left[\frac{f_{*}}{m_p}\right] \frac{\alpha_\Lambda}{2}a^{-1}\left(\tilde{\pi}_{\phi} \tilde{B}_i^{(4)} + \tilde{\pi}_{\phi,+i}\tilde{B}^{(4)}_{i,+i} \right)  \nonumber \\
	&&\hspace*{4.3cm} \left. + \left[\frac{f_{*}}{m_p}\right]\frac{\alpha_\Lambda}{4}a^{-1} \sum_\pm \sum_{j,k} \epsilon_{ijk}  \left\{ \left[ (\tilde{\nabla}_j^\pm \tilde{\phi}) \tilde{E}_{k,\pm j}^{(2)} \right]_{+i} +  \left[ (\tilde{\nabla}_j^\pm \tilde{\phi}) \tilde{E}_{k,\pm j}^{(2)}  \right]   \right\}\right\} \; , \nonumber
\end{eqnarray}
with $\tilde \pi_{\phi}=\dot{\tilde \phi}$ and $\tilde{E}_i=\dot{\tilde A}$ the conjugate momenta (defined via derivatives with respect to cosmic time). The dynamics can then be described as a  system of coupled first-order differential equations, which takes the form 
\begin{equation}
\left\lbrace
\begin{array}{rcl}\label{eq:coupledFirstDiffEqAxionNefolding}
\frac{d\tilde{\pi}_\phi}{d\mathcal{N}} &=& \mathcal{K}^{\rm L}_{\mathcal N,\phi}[a,\tilde H,\tilde{\phi},\tilde{A}_j,\tilde{\pi}_{\phi},\tilde{E}_j]\,,\\[5pt]
\frac{d\tilde{E}_i}{d\mathcal{N}} &=& \mathcal{K}_{\mathcal N,A_i}^{\rm L}[a,\tilde H,\tilde{\phi},\tilde{A}_j,\tilde{\pi}_{\phi},\tilde{E}_j] \,, \\[5pt]
\frac{d\tilde{\phi}}{d\mathcal{N}}&=&\tilde{\pi}_\phi/\tilde H\,,\\[5pt]
\frac{d\tilde{A}_i}{d\mathcal{N}}&=&\tilde{E}_i/\tilde H\,,
\end{array}
\right.
\end{equation}
with $\tilde H \equiv H/\omega_*$ and where the scale factor is given by $a = a_{*} e^{\mathcal{N} - \mathcal{N}_{*}}$, by construction, with $a_{*} \equiv a(\mathcal{N}_{*})$ the scale factor value at some initial e-folding time $\mathcal{N}_{*}$. 

For self-consistent expansion, the Hubble parameter is evolved as
\begin{equation}
        \frac{d\tilde H}{d\mathcal{N}} = \mathcal{K}_{H}[a,\tilde H,\tilde{E}^{\phi}_{K},\tilde{E}^{\phi}_{G},\tilde{E}^{A}_{K},\tilde{E}^{A}_{G}] \equiv -\left(\frac{f_{*}}{m_{p}}\right)^2\frac{1}{3\tilde H}\left[3\tilde{E}^{\phi}_{K}+\tilde{E}^{\phi}_{G}+2\tilde{E}^{A}_{K}+2\tilde{E}^{A}_{G}\right]\, ,\label{eq:explicitEOMhubbleKernelEfold}
\end{equation}
with the energy components defined as in Eq.~(\ref{eq:energyDensityTermsAlpha}) for cosmic time. The evolution of the universe is therefore controlled in this case by a first order differential equation, instead of a second order equation as for $\alpha$-time (which was then split into two coupled first order equations). Moreover, note that, independently of this change of the time variable, the constraints remain unchanged, simply using Eqs.~\ref{eq:explicitGaussLawAlpha} and \ref{eq:HubbleConstraintAlpha} in cosmic time to be consistent with the definition of the conjugate momenta.

\subsection{Chirality on the lattice}

Here we consider methods for filtering either chirality of the gauge field on the lattice. This is particularly useful if we want to set the initial conditions of the two chiralities separately, or simply to measure the spectra of a given chirality. We discuss first how to construct a chiral basis in the continuum, and then specialize the case to the lattice. Then, we explain how a projector can be constructed that allows to extract the different chiralities of a gauge field.

We begin by noting that a gauge field\footnote{Recall that we work in the temporal gauge, so $A_0 = 0$, and $A_\mu = (0,\vec A)$.} $\vec{A}(\textbf{k})$ can always be expressed in Fourier space as a linear combination of a Cartesian unitary vector basis
\begin{equation}\label{eq:cartesianBasisVector}
    \vec A \equiv A_1\hat e_1 + A_2\hat e_2 + A_3\hat e_3\;,
\end{equation}
but also, equivalently, it can be decomposed in a chiral basis, as
\begin{equation}\label{eq:chiralnBasisVector}
    \vec A \equiv A^{+}\vec\varepsilon^{\,+} + A^{-}\vec\varepsilon^{\,-} + A^{\parallel}\hat{\textbf{k}}\;.
\end{equation}
We can extract the chiral components from the Cartesian basis at any moment, simply by means of $A^{\pm} = \vec{\varepsilon}^{\, \pm} \cdot \vec{A}$, and the longitudinal mode from $A^{\parallel} = \hat{\textbf{k}} \cdot \vec{A}$. 

To construct the chiral basis, we need to build an orthonormal basis around the momentum vector $\textbf{k} = (k_1, k_2, k_3)$, which in spherical coordinates $(\theta, \phi)$ takes the form
\begin{equation}\label{eq:reciprocalPositionVector}
   {\bf k} = k(\sin \theta \cos \varphi, \sin \theta \sin \varphi, \cos \theta)\,. 
\end{equation}
with $k = \sqrt{k_1^2 + k_2^2 + k_3^2}$. Without loss of generality, we consider the unit vector $\hat{e}_{3}$ as a reference direction to construct a basis of states orthogonal to $\hat{\textbf{k}}=\textbf{k}/k$. We define a plane orthogonal to the $\hat{\bf{k}}$ generated by the orthonormal vectors $(\vec{u}, \vec{v})$,
\begin{equation}\label{eq:orthogonalPlanePostionVector}
    \begin{array}{rcl}
\vec{v} (\hat{\bf k})&=&\frac{\hat{e}_3 \times \hat{\bf k}}{|\hat{e}_3 \times \hat{\bf k}|} = (-\sin \varphi, \cos \varphi, 0)\;, \\[10pt]
\vec{u}(\hat{\bf k}) &=& \vec{v} \times \hat{\bf k} =(\cos \theta \cos \varphi, \cos \theta \sin \varphi, - \sin \theta) \; .
    \end{array}
\end{equation}
With these, we then construct the orthonormal chiral vectors as
\begin{equation}\label{eq:chiralVectorCont}
\vec{\varepsilon}^{\,\pm} = \frac{\vec{u} \pm i \vec{v}}{\sqrt{2}} \; .
\end{equation}
Together with $\hat{\bf{k}}$, they form the chiral basis $\{\hat{\textbf{k}}, \vec{\varepsilon}^{\, +}, \vec{\varepsilon}^{\,-}\}$ of Eq.~(\ref{eq:chiralnBasisVector}). 

To construct equivalent chiral vectors on a lattice, we need to work with lattice momenta, recall~Sect.~\ref{subsec:LatticeMomentum}. Considering forward/backward lattice derivatives living in between lattice sites, the required lattice momentum, {\it c.f.}~Eq.~(\ref{eq:latticeMomentum0}), is
\begin{equation}\label{eq:latticeMomentum}
     k_{{\rm L},i} = 2\frac{\sin (\pi \tilde{n}_i / N)}{ \delta x} \, .
\end{equation}
Thus, we simply need to construct the angles $\{\theta, \varphi\}$ that define $\{\vec{u}, \vec{v}\}$ in expression (\ref{eq:orthogonalPlanePostionVector}), but in this case 
with respect to $\textbf{k}_{\rm L}$. The corresponding lattice angles $\{\theta_{\rm L}, \varphi_{\rm L}\}$ can then be simply defined as
\begin{equation}\label{eq:anglesLatticeChiral}
    \begin{array}{rcl}
     \sin \theta_{\rm L} = \frac{\sqrt{k_{{\rm L},1}^2 + k_{{\rm L},2}^2}}{\sqrt{k_{{\rm L},1}^2 + k_{{\rm L},2}^2 +k_{{\rm L},3}^2 }}\,, \quad 
    \sin \varphi_{\rm L} = \frac{k_{{\rm L},2}}{\sqrt{k_{{\rm L},1}^2 + k_{{\rm L},2}^2}}\, , \\
    \cos \theta_{\rm L} = \frac{k_{{\rm L},3}}{\sqrt{k_{{\rm L},1}^2 + k_{{\rm L},2}^2 +k_{{\rm L},3}^2 }}\, , \quad 
    \cos \varphi_{\rm L} = \frac{k_{{\rm L},1}}{\sqrt{k_{{\rm L},1}^2 + k_{{\rm L},2}^2}}\, ,
    \end{array}
\end{equation}
and with them we can construct the chiral vectors on the lattice
\begin{equation}\label{eq:chiralVectorLAT}
    \vec{\varepsilon}_{\rm L}^{\,\pm} = \frac{\vec{u}_{\rm L} \pm i \vec{v}_{\rm L}}{\sqrt{2}} \;, 
\end{equation}
with $\vec{u}_{\rm L}, \vec{v}_{\rm L}$ simply defined analogously as in~(\ref{eq:orthogonalPlanePostionVector}), but in terms of $\{\theta_{\rm L}, \varphi_{\rm L}\}$.

To determine the different chiralities of the gauge field, it is convenient to construct
a chiral projector. To do this, we start from the chiral operator in the continuum
\begin{equation}\label{eq:ChiralOperator}
    \Sigma_{ij}(\hat{\textbf{k}})\equiv -i\epsilon_{ijl}\hat k_l\,,
\end{equation}
which has eigenvalues $+1$, $-1$ and $0$ for the triad of eigenvectors $\{\vec{\varepsilon}^{\, +}, \vec{\varepsilon}^{\, -}, \hat{\textbf{k}}\}$, in this same order. If we rename the latter as $\vec{\varepsilon}^{\,(+1)} \equiv \vec{\varepsilon}^{\, +} $, $\vec{\varepsilon}^{\,(-1)} \equiv \vec{\varepsilon}^{\, -}$, and $\vec{\varepsilon}^{\,(0)} \equiv \hat{\textbf{k}}$, the chiral operator essentially satisfies
\begin{equation}\label{eq:ChiralOperatorOnChiralVector}
    \Sigma_{ij}\varepsilon_j^{\,(\sigma)} = \sigma \varepsilon_i^{\,(\sigma)}\,,~~~\sigma = -1,0,+1.
\end{equation}
Since a vector field $\vec{A}(\textbf{k})$ can be written as in Eq.~(\ref{eq:chiralnBasisVector}), this implies that if we apply the chirality operator to it, we obtain
\begin{equation}\label{eq:ChiralOperatorOnVector}
    \Sigma_{ij}(\hat{\textbf{k}})A^{(\sigma)}_j(\hat{\textbf{k}}) = \sigma A^{(\sigma)}_i(\hat{\textbf{k}})\,,~~~ \sigma = -1,0,+1\,,
\end{equation}
where $A^{(+1)} \equiv A^{+}$, $A^{(-1)} \equiv A^{-}$, and $A^{(0)} \equiv A^{\parallel}$. With the chiral operator, we can then construct the helicity projector
\begin{equation}\label{eq:hlicityProjector}
    \Pi^{\pm}_{ij}(\hat{\textbf{k}}) \equiv \frac{1}{2}\left[(\Sigma^2(\hat{\textbf{k}}))_{ij} \pm \Sigma_{ij}(\hat{\textbf{k}})\right]
  = \frac{1}{2}\left(P_{ij}(\hat{\textbf{k}}) \pm  \Sigma_{ij}(\hat{\textbf{k}})\right)=\frac{1}{2}\left( \delta_{ij} - \frac{k_{i}k_{j}}{k^2} \mp \frac{i}{k}\epsilon_{ijk}k_{k}\right)\,,
\end{equation}
where in the second line we have used $(\Sigma^2(\hat{\textbf{k}}))_{ij} \equiv \Sigma_{il}(\hat{\textbf{k}})\Sigma_{lj}(\hat{\textbf{k}}) = P_{ij}(\hat{\textbf{k}})$, with $P_{ij}(\hat{\textbf{k}}) \equiv \delta_{ij} - \hat k_i\hat k_j$ the transverse projector. The helicity projector, fulfills the same properties (\ref{eqn:polarisationvectors}) as the chiral vectors:
by applying it to the different components of the vector field, we obtain that
\begin{equation}\label{eq:hlicityProjectorOnVectorComp}
    \begin{array}{rcl}
       \Pi^{\lambda}_{ij}(\hat{\textbf{k}})A_j^{\parallel}(\hat{\textbf{k}}) &=& 0\;,\\
       \Pi^{\lambda}_{ij}(\hat{\textbf{k}})A_j^{\lambda'}(\hat{\textbf{k}}) &=& {1\over2}(1+\lambda\lambda')A_i^{\lambda'}(\hat{\textbf{k}}) \equiv A_i^{\lambda}(\hat{\textbf{k}})\delta_{\lambda\lambda'}\;,
    \end{array}
\end{equation}
where $\lambda = \pm$. Thus, with this projector, we can filter either helicity
\begin{equation}\label{eq:hlicityProjectorOnVector}
    A^{\pm}_i(\hat{\textbf{k}}) \equiv \Pi^{\pm}_{ij}(\hat{\textbf{k}})A_j(\hat{\textbf{k}})\,.
\end{equation}

On the lattice, the helicity projector is simply built analogously using the lattice momenta (\ref{eq:latticeMomentum}), so that we define
\begin{equation}\label{eq:hlicityProjectorLat}
    \Pi^{{\rm L},\lambda}_{ij}(\tilde{\textbf{n}}) = \frac{1}{2}\left( \delta_{ij} - \frac{k_{\text{L},i}k_{\text{L},j}}{k_{\text{L}}^2} -\lambda \frac{i}{k_{\text{L}}}\epsilon_{ijk}k_{\text{L},k}\right)\; .
\end{equation}
For a vector on the lattice, it performs the desired filtering of its chiral components simply through
\begin{equation}\label{eq:hlicityProjectorOnVectorLat}
    A_j^\lambda(\tilde{\textbf{n}}) = \Pi^{{\rm L},\lambda}_{ij}(\tilde{\textbf{n}})A_j(\tilde{\textbf{n}})\,.
\end{equation}

\begin{table}[t]
\begin{center}
\begin{tabular}{|c||c|c|}
 \hline
 $\alpha_{\Lambda}$& $N$ & $k_{\rm IR}/m$ \\
 \hline
 \hline 
 12   &   320  & 0.1932  \\
 14   &   480  & 0.1932  \\
 18   &   3072 & 0.1932  \\
 \hline
\end{tabular}
\end{center}
\caption{Benchmark coupling  representative of the different backreaction regimes of axion inflation, together with the lattice parameters used to simulate them, see Fig.~\ref{fig:EnergyComponentsofRegimesAxionInflation}.}
\label{tab:AxionInlationSimulations}
\end{table}

\subsection{Working example: Axion inflation with $V(\phi) = {1\over2}m^2\phi^2$}

In the introduction to this section, we stress the importance of properly addressing the backreaction of the gauge field onto the background dynamics, which can only be done properly through lattice simulations. As an example, here we present some of the results of our recent lattice simulations of the specific case of axion inflation with $V(\phi) = {1\over2}m^2\phi^2$, as presented in \rrcite{Figueroa:2023oxc,Figueroa:2024rkr,Lizarraga:2025aiw}. As explained, the dynamics of this system are characterized by Eqs.~(\ref{eqn:eom1}),~(\ref{eqn:eom2}) and~(\ref{eqn:ddaAxion}) subject to the constrains in Eqs.~(\ref{eqn:Gauss}) and~(\ref{eqn:HubbleAxion}). Such system of equations is formulated in the lattice following the prescription presented in Sect.~\ref{subsec:axionLattice}, which respects the shift symmetry and the Bianchi identities at the lattice level. The full set of lattice equations to be evolved is given by \cref{eq:explicitEOMscalar,eq:explicitEOMgauge,eq:explicitGaussLaw}. Given the choice of quadratic inflaton potential, we choose the program variables as $f_*=\omega_*=m$, with the mass fixed to $m=1.5 \times 10^{13}$ GeV in order to satisfy CMB constraints. Finally, initial conditions are set according to what is explained in Sec.~\ref{subsubsec:ICfromGeneralPS}, either starting from the Bunch–Davies vacuum solution directly on the lattice, or introducing an excited spectrum obtained from the homogeneous linear dynamics (simulated without the need of a lattice) before backreaction of the gauge field becomes relevant.

The level of backreaction to be developed in the system is controlled by the coupling $\alpha_{\Lambda}$, which measures the strength of the inflaton–gauge interaction. We distinguish three regimes based on lattice simulations \cite{Figueroa:2024rkr}: {\it weak backreaction} for $\alpha_{\Lambda} \lesssim 13.1$, where the trajectory of slow roll inflation is altered only after the end of inflation, and the production of gauge field during inflation is almost negligible; {\it mild backreaction} for $13.1\lesssim \alpha_{\Lambda}\lesssim14.3$, where inflation initially ends at the same moment as in the slow roll regime, but backreaction then kicks-in so that the system re-enters into a second period of inflation for a few e-folds; and finally {\it strong backreaction} for $\alpha_{\Lambda} \gtrsim 14.3$, where backreaction happens during the inflationary period deviating drastically the dynamics from the slow roll trajectory, delaying the end of inflation by several e-folds (the stronger the coupling, the more additional e-folds). Here we show the results of simulations for three benchmark cases representative of each of the three regimes. The corresponding lattice parameters used for the simulations are summarized in Table~\ref{tab:AxionInlationSimulations}. 

In order to characterize the different regimes, we follow the evolution of the slow-roll parameter $\epsilon_{H} \equiv -\dot{H}/H^2$, which has a definition in terms of the energy components of the system as
\begin{eqnarray}
    \epsilon_{H} = 1 + \frac{2E_{ K}-E_{ V} +E_{ EM} }{E_{\rm tot}} \, ,
\end{eqnarray}
where definitions of the energy components are given in Eq.~\eqref{eq:HubbleConstraintAlpha}, and $E_{\rm tot}$ is the total energy of the system. The  end of inflation can be identified as the last time when the condition $\epsilon_H = 1$ occurs. 

\begin{figure}[!t]
    \centering
    \begin{minipage}{\textwidth} 
    \centering
        \includegraphics[width=\textwidth]{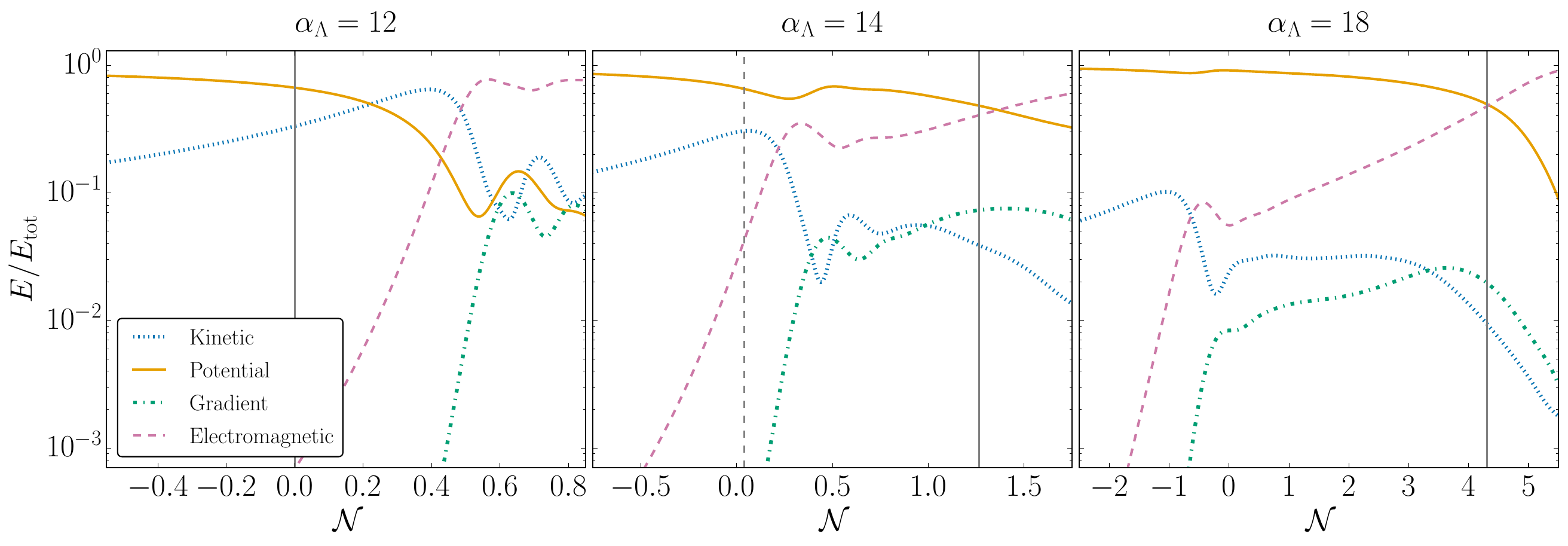}
    \end{minipage}
    
    \caption{
       Evolution of the kinetic (dotted), potential (solid), gradient (dashed-dotted) and electromagnetic (dashed) energy components for $\alpha_{\Lambda} = 12$, $\alpha_{\Lambda} = 14$ and $\alpha_{\Lambda} = 18$. The vertical grey lines indicate the moments at which $\epsilon_H = 1$ is reached from $\epsilon_H < 1$. We use dashed lines when this occurs more than once, in order to distinguish them from the solid line, which corresponds to the final crossing.}
    \label{fig:EnergyComponentsofRegimesAxionInflation}
\end{figure}

In Fig.~\ref{fig:EnergyComponentsofRegimesAxionInflation} we follow the evolution of the different energy components for each of the benchmark cases. For the case $\alpha_{\Lambda} = 12$, the backreaction of the gauge sector is not relevant during inflation so the end of inflation is not delayed. For $\alpha_{\Lambda}=14$, the backreaction is not strong enough to affect inflation before $\mathcal{N}=0$, so the original end according to slow-roll is not delayed. However, backreaction becomes relevant afterwards, causing the system to re-enter the regime $\epsilon_H < 1$ and hence continues with another period of inflation, which only ends eventually once the electromagnetic energy of the gauge field overtakes the inflaton potential energy. Finally, for $\alpha_{\Lambda}=18$, the gauge electromagnetic energy surpasses the inflaton kinetic energy early on, and sustains a new regime of {\it (electro)magnetic slow-roll}~\cite{Figueroa:2023oxc,Figueroa:2024rkr}. This phase ends once the gauge field energy surpasses the inflaton potential energy, leaving the universe fully reheated. From determining the last moment at which $\epsilon_H = 1$ (solid vertical gray line), we find no delay for $\alpha_{\Lambda}=12$, a delay of $\Delta \mathcal{N} \sim 1.3$ e-foldings for $\alpha_{\Lambda}=14$ (the dashed vertical line marks the first time at which $\epsilon_H=1$, after which the system re-enters the inflationary regime), and a delay of $\Delta \mathcal{N} \sim 4$ e-foldings for $\alpha_{\Lambda}=18$. It is important to remark that these results are already obtained for lattice simulations that use a proper separation between IR and UV scales. This needs to be increased the bigger the coupling, to ensure all relevant scales of the problem are captured and so the results are convergent (this is, insensitive to changes of the IR and UV cutoffs). In particular, the benchmark case here for the strong backreaction requires above 3000 points/dimension, so these are very demanding simulations, see~\cite{Figueroa:2024rkr} for further details.

\begin{figure}[t!]
    \centering
    \begin{subfigure}{0.49\textwidth}
        \includegraphics[width=\linewidth]{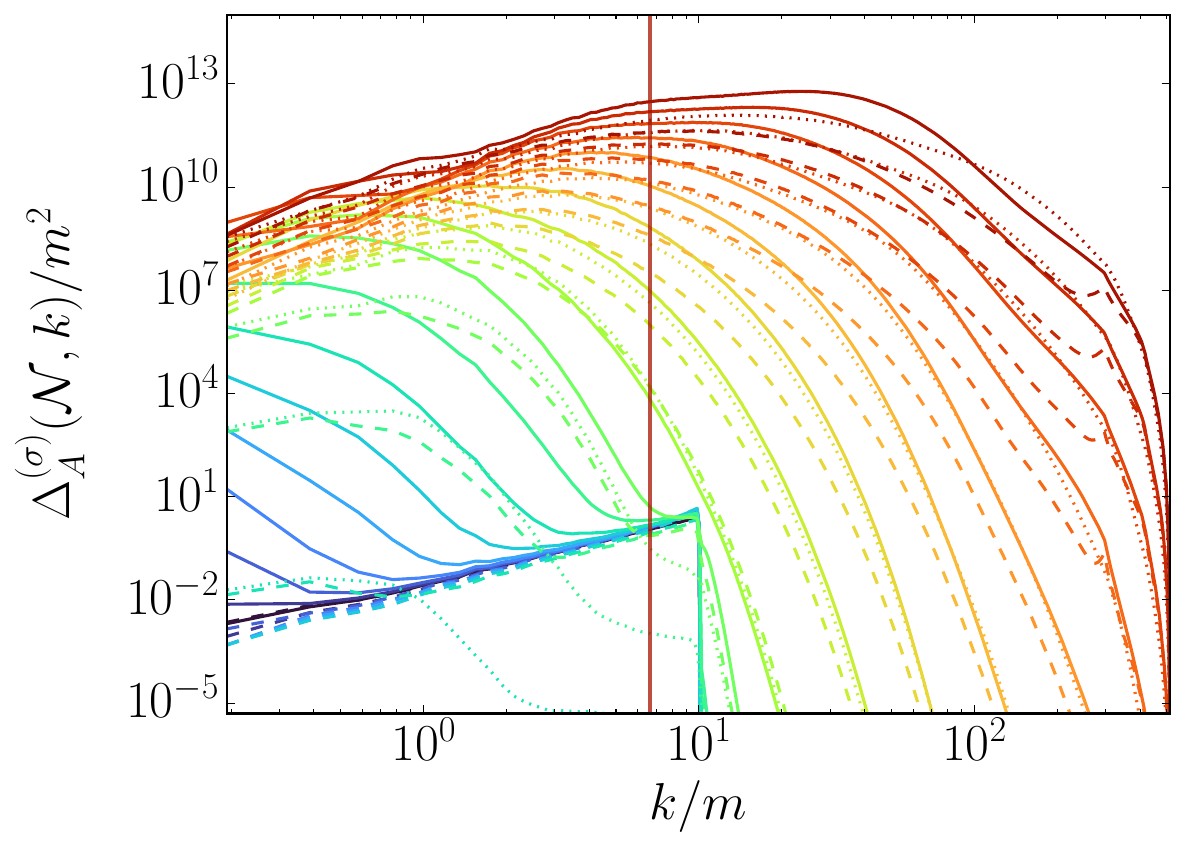} 
    \end{subfigure}
    \begin{subfigure}{0.49\textwidth}
        \includegraphics[width=\linewidth]{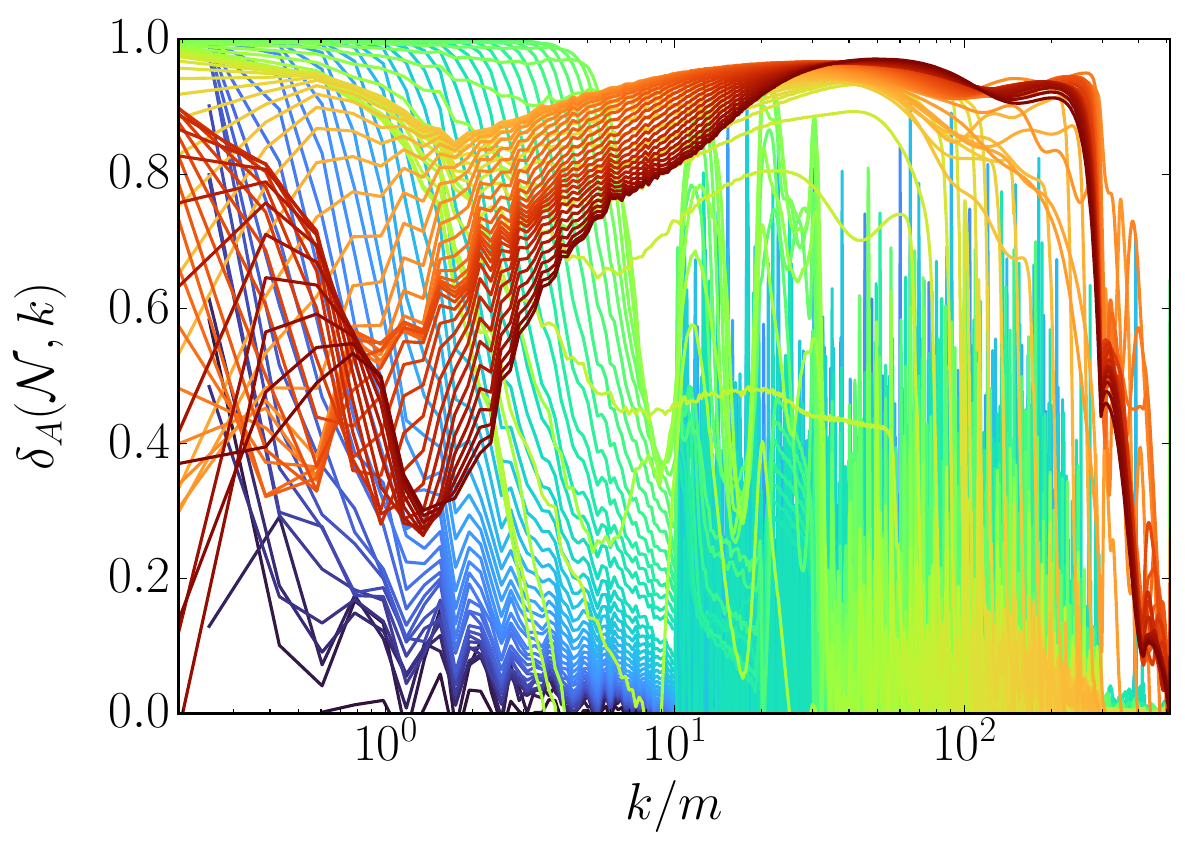} 
    \end{subfigure}
    \caption{Evolution of the gauge field for $\alpha_{\Lambda}=18$, from earlier to later times, represented by colder and warmer colors, respectively. We represent the period from $\mathcal{N} = -4.5$ to $\mathcal{N} = 4$ e-foldings. \textbf{Left:} Evolution of power spectra for each chirality and the longitudinal mode, plotted every 0.5 e-folds, for $\sigma = +$ (solid), $-$ (dashed), $\parallel$ (dotted),  \textbf{Right:} chiral imbalance $\delta_{A}(\mathcal{N},k)$ defined in Eq.~(\ref{eq:chiralImb}) plotted every 0.1 e-folds.}
    \label{fig:GaugeSpectraComp}
\end{figure}

In Fig.~\ref{fig:GaugeSpectraComp} we show the evolution of local properties of the gauge field for the strong backreaction case of $\alpha_{\Lambda}=18$. The left panel displays the evolution until the end of inflation of the power spectra $\Delta_A(\mathcal{N},k)$ for the gauge field chiralities and the longitudinal mode, labeled by $\sigma = {+, -, \parallel}$.  In the linear regime, only the $(+)$ mode is amplified, while the $(-)$ component remains in the vacuum state, and the longitudinal $(\parallel)$ remains vanishing. As non-linearities and inhomogeneities grow, mode–mode interactions excite the $(-)$ and $(\parallel)$ components. This can be understood by analyzing the equations of motion in Fourier space, where mode convolutions mix different chiralities, exciting them only if the inflaton field develops gradients (see Section V.D of~\cite{Figueroa:2024rkr} for further details). On another note, a key feature is the transfer of power from IR to UV scales during the extra e-folds, driven by non-linear effects. This highlights the need for a large UV coverage to achieve convergence in the strong backreaction regime. Morever, another notable feature is that the peak of excitation follows the comoving Hubble radius but remains essentially sub-horizon at the end (vertical line).

Finally, the right panel of Fig.~\ref{fig:GaugeSpectraComp} shows the evolution of the chiral imbalance, defined as
\begin{eqnarray}\label{eq:chiralImb}
    \delta_A(\mathcal{N},k) \equiv \frac{\Delta^{(+)}_{A}(\mathcal{N},k)-\Delta^{(-)}_{A}(\mathcal{N},k)}{\Delta^{(+)}_{A}(\mathcal{N},k)+\Delta^{(-)}_{A}(\mathcal{N},k)}\,.
\end{eqnarray}
Initially, when only the $(+)$ mode is excited, the imbalance reaches $\delta_A \simeq 1$ for the amplified modes. As non-linearities grow, the excitation of the $(-)$ component reduces $\delta_A$ in a scale-dependent manner, approaching a partial chiral balance on certain scales, mostly in the IR region. Note that this behavior is a characteristic feature of the fully local non-linear dynamics, which only lattice simulations can capture, while homogeneous-backeaction approaches~\cite{Cheng:2015oqa,Notari:2016npn,DallAgata:2019yrr,Domcke:2020zez,Sobol:2019xls,Gorbar:2021rlt,Durrer:2023rhc,Durrer:2024ibi,vonEckardstein:2023gwk} fail to grasp. Even so, it is worth mentioning that at the end of the inflationary period, on the most UV scales where the peak of the power spectrum is located, the (+) mode still dominates, nearly preserving the typical chiral imbalance.

\section{Cosmic defects} 
\label{sec:DefectsV}
In this section we introduce cosmic defects~\cite{Kibble:1976sj,Kibble:1980mv,Vilenkin:1984ib,Vilenkin:2000jqa} and their simulation on a lattice. Defects are stable macroscopic field theory configurations that originate in theories with spontaneously symmetry breaking of a group $G$ down to a smaller group $H$. More concretely, they originate when the vacuum manifold $\cM\cong G/H$, has a non-trivial homotopy group, $\pi_n(\cM)\neq \mathcal{I}$, which characterizes the topological properties of the defect. Defects can be {\it topological} configurations, corresponding to well-defined regions in space where the underlying scalar fields are trapped far from the vacuum manifold. This is the case {\it e.g.} of domain walls, strings, or monopoles~\cite{Kibble:1976sj,Vilenkin:1981kz,Vilenkin:1984ib,Vachaspati:1984dz,Barriola:1989hx,Hindmarsh:1994re,Vilenkin:2000jqa,Durrer:2001cg}. Defects can also be {\it non-topological} configurations confined within the vacuum manifold, characterized by gradients that interpolate in between distant regions. This is the case of textures or self-ordering scalar fields, see for example~\cite{Turok:1991qq,Jaffe:1993tt,Durrer:1998rw,Durrer:2001cg,Fenu:2009qf,Garcia-Bellido:2010qjz,Figueroa:2010zx,Fenu:2013tea,Durrer:2014raa}.

In the early universe, cosmic defects may originate after a phase transition via the so-called Kibble mechanism~\cite{Kibble:1976sj}. After spontaneous symmetry breaking,  causally disconnected patches of the universe are expected to fall into different elements of the vacuum manifold. When these patches enter into causal contact at a later time, cosmic defects form at the interface between the patches to guarantee continuity of the fields. As this process happens randomly everywhere in the universe, defects do not form isolated, but are instead part of a network of defects\footnote{In the case of cosmic strings, a network of these can also emerge in String Theory scenarios, if fundamental strings are stretched out to cosmological scales~\cite{Dvali:2003zj,Copeland:2003bj}.}. These networks are expected to evolve until reaching a scaling regime, where the number of defects per Hubble patch is constant.\footnote{In the case of global cosmic strings, logarithmic corrections to the scaling regime have been argued~\cite{Gorghetto:2018myk,Gorghetto:2020qws,Buschmann:2021sdq,Saikawa:2024bta,Benabou:2024msj}, though some of the largest simulations to date (2025) are compatible with standard scaling~\cite{Correia:2024cpk}.} Once in the scaling regime, the statistical properties of the network are expected to be independent of the microphysics, {\it i.e.}~of the underlying field theory.

Cosmic defects are expected to leave an observable cosmological imprint on the universe, due to the large amount of energy they contain and their extended lifetime. These could be, for example, signatures on the cosmic microwave background~\cite{Garcia-Bellido:2010qjz,Fenu:2013tea,Ade:2013xla,Durrer:2014raa,Lizarraga:2014xza,Charnock:2016nzm,Lizarraga:2016onn,Lopez-Eiguren:2017dmc,Figueroa:2010zx,Ringeval:2010ca,Regan:2014vha} or the formation of a background of gravitational waves~\cite{Vilenkin:1981bx,Vachaspati:1984gt,Damour:2000wa,Damour:2001bk,Damour:2004kw,Figueroa:2012kw,Hiramatsu:2013qaa,Blanco-Pillado:2017oxo,Auclair:2019wcv,Gouttenoire:2019kij,Figueroa:2020lvo,Gorghetto:2021fsn,Chang:2021afa,Yamada:2022aax,Yamada:2022imq,Servant:2023mwt,Servant:2023tua,Dimitriou:2025bvq}, which could be detected by ongoing or future detectors. As the field dynamics governing the evolution of the defects is highly non-linear, any detailed study requires of lattice simulations. 

\subsection{Continuum formulation of cosmic defects}

As discussed above, the macroscopic properties of the defects only depend on the topological properties of the underlying field theory, and are expected to be independent of the microphysics details. Thus, it is customary to use field theories with a minimal field content to study the dynamics of cosmic defects on a lattice. In the following, we introduce simple models that are typically used to study the formation and evolution of cosmic defects, mainly focusing on topological ones such as cosmic strings or domain walls, though we will also briefly comment on non-topological defects. In general, we will consider theories consisting of $N$ real scalar fields, with a symmetry-breaking potential of the form,
\begin{equation}\label{eq:defects:potential}
V_N[\{\phi_a\}]=\frac{\lambda}{4}\left(\sum_{a=1}^N\phi_a^2-v^2\right)^2\,.
\end{equation}
Through this section, for simplicity, we will work in conformal time $d\tau \equiv dt/a$ and denote $f'=\text{d}f/\text{d}\tau$.

\subsubsection{Cosmic strings}\label{sec:defects:CosmicStringstheory}
 
Cosmic strings are one-dimensional structures that arise in theories with a broken $U(1)$ symmetry, in which the vacuum manifold has the topology of a circle, $\cM= S^1\cong U(1)$, and a non-trivial first homotopy group, $\pi_1(S^1)=\mathbb{Z}$. Cosmic strings arise in a myriad of different high-energy theories, and can be classified in different types depending on their properties. The simplest type of them are the so-called global and local strings, on which we focus in this review. 

{\bf Global cosmic strings --.} These are characterized by long-range interactions between string segments and the presence of both massive and massless particles that can be emitted from the strings. They are predicted, among others, by axion-like dark matter models~\cite{Svrcek:2006yi,Arvanitaki:2009fg}, including those that aim to solve the strong CP problem of QCD~\cite{Weinberg:1977ma,Wilczek:1977pj,Peccei:1977hh,Peccei:1977ur}. The dynamics of global strings have been widely studied on the lattice~\cite{Correia:2024cpk,Hindmarsh:2021vih,Hindmarsh:2019csc,Gorghetto:2018myk,Gorghetto:2020qws,Gorghetto:2021fsn,Saikawa:2024bta,Eggemeier:2019khm,Vaquero:2018tib,Matsunami:2019fss,Baeza-Ballesteros:2023say,Buschmann:2021sdq,Benabou:2024msj,Correia:2025nns}, see also~\cite{Klaer:2019fxc,Klaer:2017ond,Klaer:2017qhr} for simulations of high-tension global strings in an low-energy effective theory.

On the lattice, the dynamics of global cosmic strings are usually studied using a simple model with a single complex scalar field,
\begin{equation}\label{eq:defects:actionglobal}
S_\text{global}=- \int \text{d}^4 x \sqrt{-g}\,\Big\{(\partial_\mu\varphi)^*\partial^\mu\varphi+V_2[\varphi]\Big\}\,,
\end{equation}
where the scalar potential is given by \cref{eq:defects:potential} for $N=2$, after identifying $\varphi=(\phi_1+i\phi_2)/\sqrt{2}$.

This theory is invariant under global $U(1)$ transformations, $\varphi\rightarrow\text{e}^{i\gamma}\varphi$, with $\gamma$ a constant value, and presents two phases: a symmetric phase with $\langle \varphi \rangle = 0$, in which the $U(1)$ symmetry is realized, and a spontaneously broken phase with $\langle |\varphi|^2\rangle =v^2/2$, characterized by both massless ($\theta$) and massive ($\chi$) modes, the latter having a mass $m_\chi=\sqrt{2\lambda}v$. The symmetry breaking pattern leads to a vacuum manifold with the topology of a circle, $\cM= S^1$, allowing for the formation of global cosmic strings, characterized by a typical core radius $r_\text{c}\sim 1/m_\chi$. The evolution of global strings is controlled by the dynamics of the underlying field theory. For the particular model under consideration, the complex scalar field evolves following the equation of motion,
\begin{equation}
(a^2 \varphi')'-a^2\nabla^2\varphi=-2a^4\lambda\varphi\left(|\varphi|^2-\frac{v^2}{2}\right)\,.
\end{equation}

{\bf Local cosmic strings --.} These are characterized by short range interactions between string segments, that decay exponentially fast with the distance. They are predicted in a variety of grand unified theories~\cite{Copeland:2009ga,Copeland:2011dx}, and are also expected in models with a broken local $U(1)$ symmetry. Local strings have also been widely studied on the lattice~\cite{Hindmarsh:2021mnl,Hindmarsh:2017qff,Daverio:2015nva,Baeza-Ballesteros:2024otj,Saurabh:2020pqe,Blanco-Pillado:2023sap}, usually embedded in Abelian-Higgs model, consisting of a single complex scalar field coupled to a $U(1)$ gauge field,
\begin{equation}\label{eq:defects:actionlocal}
S_\text{local} = -\int \text{d}^4x \sqrt{-g}\left\{(D^A_\mu\varphi)^*(D_A^\mu\varphi)+\frac{1}{4}F_{\mu\nu}F^{\mu\nu}+V_2[\varphi]\right\}\,,
\end{equation}
where the covariant derivative and the field strength tensor are defined in \cref{eq:AbCovDerivCont}, and the potential is the same as for global strings. The Abelian-Higgs action is invariant under gauge transformations
\begin{equation}
\varphi(x) \rightarrow \displaystyle\text{e}^{i\alpha(x)}\varphi(x)\,,~~~
A_\mu(x) \rightarrow
\displaystyle A_\mu(x) + \partial_\mu \alpha(x)\,,
\end{equation}
where $\alpha(x)$ is a function of the spacetime coordinates, and we denote $e=g_A Q_A$ in this section for simplicity. As in the global case, two phases are present in this theory: an unbroken high-energy phase, $\langle \varphi\rangle =0$, and a broken phase at lower energies, $\langle |\varphi|^2\rangle = v^2/2$. The latter is characterized by a non-trivial vacuum manifold, $\cM = S^1$, and thus cosmic strings may form after the phase transition. The low-energy phase possesses two types of massive modes, a scalar field with mass $m_\chi=\sqrt{2\lambda} v$, and a vector field with mass $m_A=ev$, respectively. These two fields characterize the scalar and magnetic core radii of the resulting strings, respectively. A widespread case of study, due to its simplicity, is the so-called \textit{critical case}, $2\lambda=e^2$, in which $m_\chi=m_A$, and hence the strings have a unique core radius. 

The evolution of local strings is governed by the dynamics of the underlying fields. In the temporal gauge, $A_0=0$, they follow the equations of motion,
\begin{equation}
\left.\begin{array}{rcl}
(a^2\varphi')'-a^2 D_A^2\varphi & = & \displaystyle -2a^{4}\lambda\varphi\left(|\varphi|^2-\frac{v^2}{2}\right)\,,\\[10pt]
A_i''-\nabla^2A_i + \nabla_j \nabla_i A_j & = & \displaystyle 2a^{2}e\text{Im}[\varphi^*D_i\varphi]\,.
\end{array}\right.
\end{equation}
Here, one can check that the dynamics of the fields only depend on the ratio $\beta=e^2/2\lambda$, with the critical case corresponding to $\beta=1$. This requires rescaling both the gauge fields and the coordinates, and is shown explicitly later in \cref{subsubsec:LocalStrings}.

\subsubsection{Domain walls}\label{sec:defects:DWtheory}
Domain walls are two-dimensional structures that arise in theories with a broken discrete symmetry that leads to a true vacuum with disconnected sectors. The simplest example is the case in which the vacuum manifold has two elements, $\cM\cong \mathbb{Z}_2$, for which $\pi_2(\cM)=\mathbb{Z}_2$, allowing for the formation of domain walls that separate regions that lie on different vacua. This simple model has been subject to multiple lattice studies, see e.g.~\cite{Press:1989yh,Garagounis:2002kt,Hiramatsu:2010yz, Hiramatsu:2013qaa,Ferreira:2023jbu,Dankovsky:2024zvs,Heilemann:2025iwv,Notari:2025kqq,Blasi:2025tmn}.  

The dynamics of domain walls are typically studied on the lattice using a simple model containing a single real scalar field, with action,
\begin{eqnarray}\label{eq:defects:actionDW}
S_\text{DW}=-\int \text{d}^4 x \sqrt{-g}\,\Big\{\frac{1}{2}\partial_\mu \phi \partial^\mu \phi+V_1[\phi]\Big\}\,,
\end{eqnarray}
with the potential given in \cref{eq:defects:potential} for $N=1$. This action is invariant under $\mathbb{Z}_2$ reflections in field space, $\phi\rightarrow -\phi$.
As before, the theory admits two different phases: a symmetric phase at high energies, $\langle \phi \rangle =0$, and a spontaneously broken phase at low energies, $\langle \phi\rangle = \pm v$, with $\cM=\{-v,v\}\cong \mathbb{Z}_2$. The phase transition may lead to the formation of domain walls with width $w_\text{DW}^{-1}\sim\sqrt{2\lambda} v$. In the model above, the scalar field forming the domain walls follows the equation of motion
\begin{equation}
(a^2 \phi')'-a^2\nabla^2\phi=-a^4\lambda\phi(\phi^2-v^2)\,.
\end{equation}

However, early-universe models leading to the formation of domain wall networks are constrained due to the scaling of their energy density with the expansion of the universe, $\rho_{\rm dw} \sim \sigma H \sim a^{-2}$, with 
\begin{equation}
    \sigma \equiv \int_{-v}^{+v}\sqrt{2V_1(\phi)}
\end{equation}
 the domain wall tension, {\it i.e.}~the mass per unit surface. This leads to an overclosed universe and a modification of the standard expansion history for $\sigma \gtrsim (1 {\rm MeV})^3$~\cite{Zeldovich:1974uw,Lazanu:2015fua}. For larger tensions, the existence of domain walls is only viable if they completely annihilate at sufficiently early times. One mechanism to achieve that, typically considered in the literature, consists in explicitly breaking the exact $\mathbb{Z}_2$ symmetry by introducing a small \textit{bias} term in the potential, such as
\begin{equation}\label{eq:defects:potentialDWbias}
V[\phi]= 
V_1[\phi]+qv\phi^3\ ,
\end{equation}
where $q \ll 1$ is a time-independent dimensionless constant. Such term introduces a small difference of potential energy $\Delta V \simeq 2 q (1+9 q^2)^{3/2}$ between both vacua while maintaining the local maximum of the potential at $\phi=0$. Domain walls may still form after a phase transition, but the existence of a vacuum pressure makes them eventually decay, leaving all the field in the lowest-energy configuration. In this biased case, the equation of motion for the field reads,
\begin{equation}
(a^2 \phi')'-a^2\nabla^2\phi=-a^4\lambda\phi(\phi^2-v^2)-3a^4qv\phi^2\,.
\end{equation}
The collapse of domain walls due to the presence of a bias potential term has been widely studied on the lattice~\cite{Larsson:1996sp,Correia:2014kqa,Correia:2017aqf,Correia:2018tty,Kitajima:2023cek,Kitajima:2023kzu, Ferreira:2024eru, Cyr:2025nzf, Notari:2025kqq,Babichev:2025stm, Dankovsky:2025pjg}. The domain walls can also annihilate if e.g.~one of the potential minima is slightly more populated than the other, see \cite{Larsson:1996sp}.

\subsubsection{Other defects}

As a last example, we consider a general theory with $N$ singlet scalar fields, $\phi_a$, with $a=1...N$, and action
\begin{equation}
S=-\int\text{d}^4x\sqrt{-g}\left\{\frac{1}{2}\sum_{a=1}^N\partial_\mu\phi_a\partial^\mu\phi^a+V_N[\{\phi_a\}]\right\}\,,
\end{equation}
with the potential given in \cref{eq:defects:potential}. This action is invariant under a global $O(N)$ symmetry, which is spontaneously broken to $O(N-1)$ at low energies, leading to the formation of different types of topological and non-topological defects, depending on the value of $N$. In particular, $N=1$ and $N=2$ correspond to the cases of domain walls and cosmic strings covered above, respectively, while $N=3$ represents the case of scalar monopoles, with homotopy group $\pi_0(\cM)=\mathbb{Z}$~\cite{Vachaspati:2016abz}. Finally, values of $N\geq 4$ lead to the formation of non-topological defects called textures~\cite{Figueroa:2020lvo}. In all cases, the evolution of the defects is governed by the underlying field dynamics, which follow the equations of motion,
\begin{equation}
(a^2\phi_a')'-a^2\nabla^2\phi_a=-a^4\lambda \phi_a\left(\sum_{b=1}^N \phi_b^2 - v^2\right)\,.
\end{equation}

Finally, we note that other 
types of topological and non-topological defects, might have also been produced in the early universe~\cite{Vachaspati:1997rr,Achucarro:1999it,Vachaspati:2000cq}, and have been studied as well on the lattice. The list includes, but is not reduced to, magnetic monopoles~\cite{Vachaspati:2016abz,Hindmarsh:2025vxh}, $U(1)\times U(1)$ multi-tension strings \cite{Urrestilla:2007yw,Lizarraga:2016hpd,Correia:2022spe}, current-carrying strings~\cite{Correia:2022spe,Correia:2024wsq}, semi-local strings~\cite{Achucarro:2013mga,Lopez-Eiguren:2017ucu,Achucarro:2019blr},  necklaces \cite{Hindmarsh:2016dha,Hindmarsh:2018zch}, and electroweak dumbbells~\cite{Patel:2023sfm,Patel:2023ybi}. Other examples are the case of domain walls in theories with multiple disconnected vacua~\cite{Wu:2022tpe,Wu:2022stu,Liu:1996ea}, and string-bounded domain walls that originate in theories of axion strings~\cite{Zhitnitsky:1980tq,Dine:1981rt,Kim:1986ax}. While these will not be explicitly covered in this review, the lattice techniques we present here can be straightforwardly extended to them.

\subsection{Lattice simulations of cosmic defects}\label{defects:latticesimulations}

Simulating the dynamics of cosmic topological defects on the lattice relies on the same canonical techniques as presented in {\tt The Art-I}~\cite{Figueroa:2020rrl}. Apriori, one could simulate a phase transition and the creation of the network of defects, and wait until they reach scaling. However, such a simulation would in general be affected by two important limitations:
\begin{itemize}
\item As the width of the defects is constant in physical units, there is a \textbf{loss of resolution} when simulating in comoving coordinates, limiting the time extend of the simulation. 
\item The approach to the scaling regime can be very slow, so not enough simulation time may be available before it is reached. In addition, the duration of lattice simulations of cosmic defects is also softly limited by the half-light-crossing-time of the lattice, after which causality issues have been found to affect the scaling dynamics of the network. 
\end{itemize} 

Instead of simply relying on canonical techniques, one typically deals with the two above problems by introducing some modifications of the simulation of defects. Namely, the use of initial conditions that allow to speed up the reach of scaling, and the application of resolution-preserving techniques. In addition, it is also common to monitor defect-specific observables to keep track of the properties of the defect network. In the following, we explain in detail the lattice techniques used to achieve a scaling regime in the simulation of defects, and introduce as well some relevant observables. We focus on the case of cosmic strings and domain walls, though the techniques presented here can be easily extended to other types of defects. Finally, we note that in simulations of cosmic defect networks, one typically assumes a fixed evolution for the scale factor, $a\equiv a(\tau)$. 

\subsubsection{Simulation of global strings}
\label{subsubsec:GlobalStrings}

Simulating the dynamics of global cosmic strings generated from the theory described by \cref{eq:defects:actionglobal}, follows the canonical approach described in Sect.~4 of {\tt The Art-I}~\cite{Figueroa:2020rrl}, based on the the symplectic methods summarized in this review in \cref{sec:LatticeTechniques}, applied in this occasion to complex scalar fields. For the model~(\ref{eq:defects:actionglobal}), it is customary to define dimensionless program variables, {\it c.f.}~ Sect.~\ref{subsubsec:ProgramVariables}, using the following scales: the $vev$ of the scalar field $\fstar=v$, and the mass scale\footnote{The exact mass scale of the massive mode is actually $m_{\chi} = \sqrt{2\lambda}v$, but we opt to take $\omegastar=\sqrt{\lambda} v$ just to capture the parametric dependence, avoiding to drag $\mathcal{O}(1)$ numbers in the program variables' definition.} of the massive mode $\omegastar=\sqrt{\lambda} v$. Then, a hybrid approach to the lattice equations of motion for the complex field, discretizing space but maintaining a continuous time variable,  leads to
\begin{equation}
(a^2 \tilde{\varphi}')' - a^2 \tilde{\nabla}^-_i\tilde{\nabla}^+_i \tilde{\varphi}=-2 a^4 \tilde{\varphi} \left(|\tilde{\varphi}|^2 -\frac{1}{2}\right)\,,
\end{equation}
where time derivatives are now taken with respect to conformal time in program units, $\tilde{f}'=\text{d}\tilde{f}/\text{d}\tilde{\tau}$. Defining the conjugate momentum of the field as
\begin{equation}
\tilde{\pi}_\varphi = a^2 \tilde{\varphi}'\,,
\end{equation}
allows to characterize the dynamics by a system of first-order differential equations,
\begin{equation}\label{eq:defects:globalscheme}
\left\lbrace\begin{array}{rcl}
	\tilde{\pi}_\varphi' & = & \cK_\varphi[a, \tilde{\varphi}]\,,\\[5pt]
	\tilde{\varphi}' & = & a^{-2}\tilde{\pi}_\varphi\,.
\end{array}\right.
\end{equation}
where we have introduced the complex scalar kernel,
\begin{equation}\label{eq:defects:globalkernel}
\cK_\varphi[a,\tilde{\varphi}]=a^2\tilde{\nabla}^-_i\tilde{\nabla}^+_i \tilde{\varphi}-2a^4\tilde{\varphi}\left(|\tilde{\varphi}|^2-\frac{1}{2}\right)\,.
\end{equation}
These equations can be solved using any symplectic integration scheme\footnote{Technically, the equations can also be solved by non-symplectic integrators, like those described in \cref{subsubsec:NonSymplecticInt}. Such schemes are however not optimal for the stability of the solution, given the conservative Kernel~(\ref{eq:defects:globalkernel}). }, such as leapfrog, velocity- or position-verlet algorithms, or higher-order Yoshida integrators, as summarized in \cref{subsubsec:SymplecticInt}. 

In principle, one could simulate directly the phase transition and allow the network to form naturally, either by starting from a field on the false vacuum with some fluctuations, or by using a time-dependent potential (as {\it e.g.}~in thermal field theory in an expanding universe). This leads however to a very slow approach to scaling, see Ref.~\cite{Correia:2024cpk} for a recent discussion on this, and references therein. Instead, it is more common to use a multi-step process that leads to a more rapid approach to scaling that we now describe. It is in this sense that simulations of cosmic defect networks in scaling, require some non-canonical initial procedure. 

As a first step, we can think of initializing the complex scalar field with Gaussian random fluctuations that follow a power spectrum $\cP_{\phi_i}(k)$ peaked around a characteristic length $\ell_\text{str}$. This scale plays the role of a correlation length, and controls the initial density of the resulting string network. One choice used in the literature is
\begin{equation}\label{eq:initialPSstrings}
\cP_{\phi_i}(k) = \frac{k^3 v^2 \ell_\text{str}^3}{\sqrt{2\pi}}\text{exp}\left(-\frac{1}{2}k^2\ell^2_\text{str}\right)\,,
\end{equation}
as this ensures that the initial conditions obey $\langle |\tilde{\varphi}|^2 \rangle =1/2$, so that the system lies on the broken phase, with randomly distributed phases. The conjugate momentum is initialized to zero.

The field configuration resulting from this procedure is typically too energetic. To get rid of this excess of energy we could of course let the field slowly evolve in an expanding background, so that the energy density dilutes according to standard canonical dynamics. An alternative more rapid approach is to perform a {\it dissipative} evolution, achieved by artificially introducing a friction term $\Gamma_D\phi'$ in the EOM, with $\Gamma_D$ a diffusion rate. In the limit where such friction term dominates, this corresponds to evolving the field according to a first-order heat-like equation on a fixed background,
\begin{equation}\label{eq:defects:diffusionglobal}
\tilde \Gamma_D\tilde{\varphi}' = \tilde{\cK}_\varphi[a(\tilde{\tau}_0), \tilde{\varphi}]\,.
\end{equation}
where the scale factor is frozen to a value of reference set at the initial time $\tau_0$ of the simulation, and $\tilde \Gamma_D \equiv \Gamma_D / \omega_{*}$. One typically fixes $\tilde \Gamma_D = 1$, as this is equivalent to re-absorbing such rate in the time variable, which during diffusion is just an auxiliary variable unrelated to physical time. The above equation with $\tilde \Gamma_D = 1$, can be solved numerically with a non-symplectic evolution scheme, such as forward Euler,
\begin{equation}\label{eq:diffusionschemeglobal}
\begin{array}{rcl}
\text{IC} & : & \{\tilde{\varphi}\}\text{ at }\tilde{\tau}_0\,,\\[4pt]
\tilde{\varphi}_{+0} & = & \tilde{\varphi} + \delta\tilde{\tau}_\text{diff}   \tilde{\cK}_\varphi[a(\tau_0), \tilde{\varphi}]\,,
\end{array}
\end{equation}
or a higher-order Runge-Kutta integrator. We note that one should use a different time step than that used for the standard evolution, as the {\it Courant} stability condition for the diffusion equation is different and reads $\delta \tilde{\tau}_\text{diff}\leq 3\delta\tilde{x}^2$. Diffusing the fields for $\Delta\tilde\tau_{\rm diff} \sim \cO(10)$ units leads typically to well defined networks~\cite{Hindmarsh:2019csc,Hindmarsh:2021vih,Correia:2024cpk,Correia:2025nns}. 

\begin{figure}[!t]
    \centering
    \begin{minipage}{\textwidth} 
    \centering
        \includegraphics[width=0.75\textwidth]{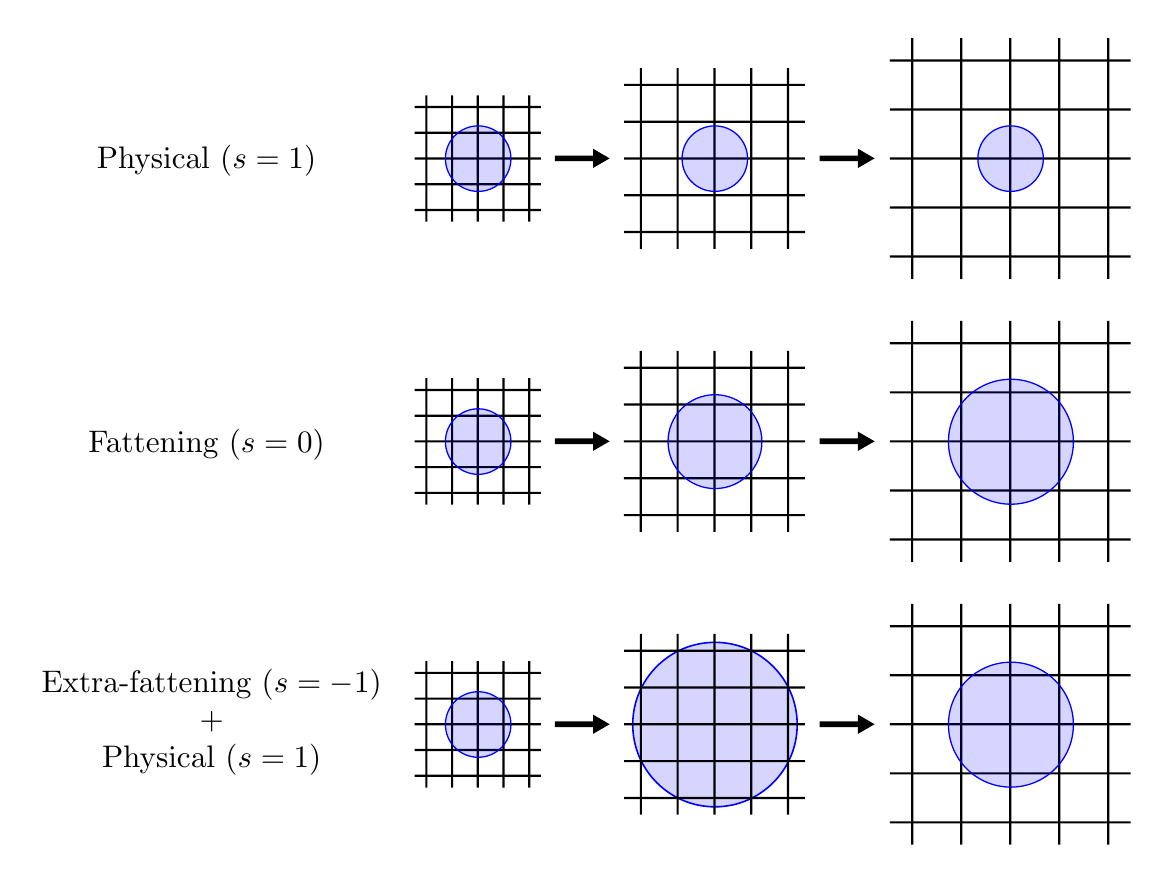}
    \end{minipage}
    \caption{
        Schematic representation of the loss-of-resolution problem for cosmic strings (top row) and the two main resolution-preserving techniques used in simulations: fattening (middle row) and the use of an initial phase of extra-fattening followed by physical evolution (bottom row). The blue circle represents the transverse section of the string.}
    \label{fig:defects:lossresolution}
\end{figure}

After the string network has been generated, the dynamics could be simulated according to the equations of motion in \cref{eq:defects:globalscheme}, waiting until the scaling regime is reached, before extracting physical measurements. However, when working in an expanding background, this evolution comes with a drawback: the radius of the string core is constant in real space, so the comoving width of the strings decreases with time as
\begin{equation}
w_\text{c}(\tau)=w_0\frac{a_0}{a(\tau)}\,,
\end{equation}
where $w_0 = w_\text{c}(\tau_0)$ and we define the scale factor at the start of the simulation as $a_0\equiv a(\tau_0)$. This is schematically represented in the first row of \cref{fig:defects:lossresolution}. To prevent this loss of resolution, which can lead to large discretization effects~\cite{Baeza-Ballesteros:2023say,Pierobon:2023ozb}, it is common to apply a resolution preserving approach known as {\it (extra-)fattening}~\cite{Press:1989yh,Moore:2001px,Bevis:2006mj}. The idea is to use modified equations of motion for the fields
\begin{equation}\label{eq:defects:globalfatteningeom}
(a^2 \tilde{\varphi}')' - a^2 \tilde{\nabla}_i^-\tilde{\nabla}_i^+ \tilde{\varphi} = -2 a^4 \left(\frac{a}{a_0}\right)^{2(s-1)} \tilde{\varphi} \left(|\tilde{\varphi}|^2 -\frac{1}{2}\right)\,,
\end{equation} 
where the dimensionless constant $s$ controls the evolution of the (comoving) string width, which now scales as
\begin{equation}
w_\text{c}(\tau)= w_0\left(\frac{a_0}{a}\right)^s\,.
\end{equation}
Different choices of $s$ correspond to different resolution-preserving techniques, as discussed below. Alternatively, the modification can be regarded as promoting the scalar self-coupling constant $\lambda$ to a time-dependent parameter,
\begin{equation}\label{eq:defects:globalstringsfatteninglambderedefinition}
\lambda(\tau) = \left(\frac{a}{a_0}\right)^{2(s-1)} \lambda_0\,,
\end{equation}
with $\lambda_0=\lambda(\tau_0)$ the value at the start of the simulation, corresponding to the value used to define the program variables. We note that early proposals of resolution-preserving approaches~\cite{Press:1989yh} proposed to also modify the coefficient in front of the friction term in \cref{eq:defects:globalfatteningeom} in order to preserve the scaling of the fluctuations around the vacuum, $\langle|\varphi|-v/\sqrt{2}\rangle\propto a^{-3/2}$, as is discussed later for domain walls in \cref{eq:defects:DWfatteningeom}. This requirement, however, has been argued to have negligible impact on the simulations~\cite{Moore:2001px} and is no longer considered.

The modified dynamics in \cref{eq:defects:globalfatteningeom} can be solved similarly as in the physical case. After defining the conjugate momentum $\tilde{\pi}_\varphi=a^2\tilde{\varphi}$, we re-write the EOM as
\begin{equation}
\left\lbrace\begin{array}{rcl}
	\tilde{\pi}_\varphi' & = & \cK_\varphi^\text{fat}[a, \tilde{\varphi};s]\,,\\[10pt]
	\tilde{\varphi}' & = & a^{-2}\tilde{\pi}_\varphi\,,
\end{array}\right.
\end{equation}
where the new kernel is
\begin{equation}\label{eq:defects:kernelextrafattening}
\cK_\varphi^\text{fat}[a,\tilde{\varphi};s]=a^2\tilde{\nabla}_i^-\tilde{\nabla}_i^+ \tilde{\varphi}-2a^4 \left(\frac{a}{a_0}\right)^{2(s-1)}\tilde{\varphi}\left(|\tilde{\varphi}|^2-\frac{1}{2}\right)\,.
\end{equation}
These equations can then be solved with canonical numerical techniques as discussed in \cref{subsubsec:SymplecticInt}.

Different choices of $s$ correspond to distinct resolution-preserving techniques. A commonly used scheme is the so-called {\it extra-fattening}, where the fields (previously diffused) are evolved with $s=-1$, so that their comoving core grows in time. This is followed by a phase of standard evolution ($s=1$) in which the comoving width of the strings decreases again. The first phase can be considered as part of the preparation of the initial condition, so that the strings are sufficiently well resolved on the lattice when scaling is finally reached during the physical-evolution phase. Typically, the extra-fattening phase is set to last for $\Delta\tau_\text{efat}=\sqrt{\tau_0 \tau_\text{end}}$, where $\tau_\text{end}$ is the end-time of the simulation (typically set to match the half-crossing time). In this way, the comoving width at $\tau_\text{end}$ equals the width before the start of the extra-fattening. 

Another scheme is the so-called {\it fattening}, which corresponds to using $s=0$. In this case, the comoving width of the strings stays constant, as the physical width is actually growing. This is usually applied after diffusion, kept for the full duration of the simulation. Both extra-fattening and the fattening techniques are schematically represented in the middle and bottom rows of~\cref{fig:defects:lossresolution}, respectively. A more sophisticated resolution-preserving approach is the use of adaptative mesh-refinement techniques, based on the use of a lattice time-dependent irregular spacing, which is adapted locally to a finer resolution in those regions close to the string cores. For results on global cosmic strings using these methods, see~\cite{Drew:2019mzc,Drew:2022iqz,Drew:2023ptp,Buschmann:2021sdq,Benabou:2024msj}.

\begin{figure}[!h]
    \centering
        \includegraphics[width=0.32\textwidth]{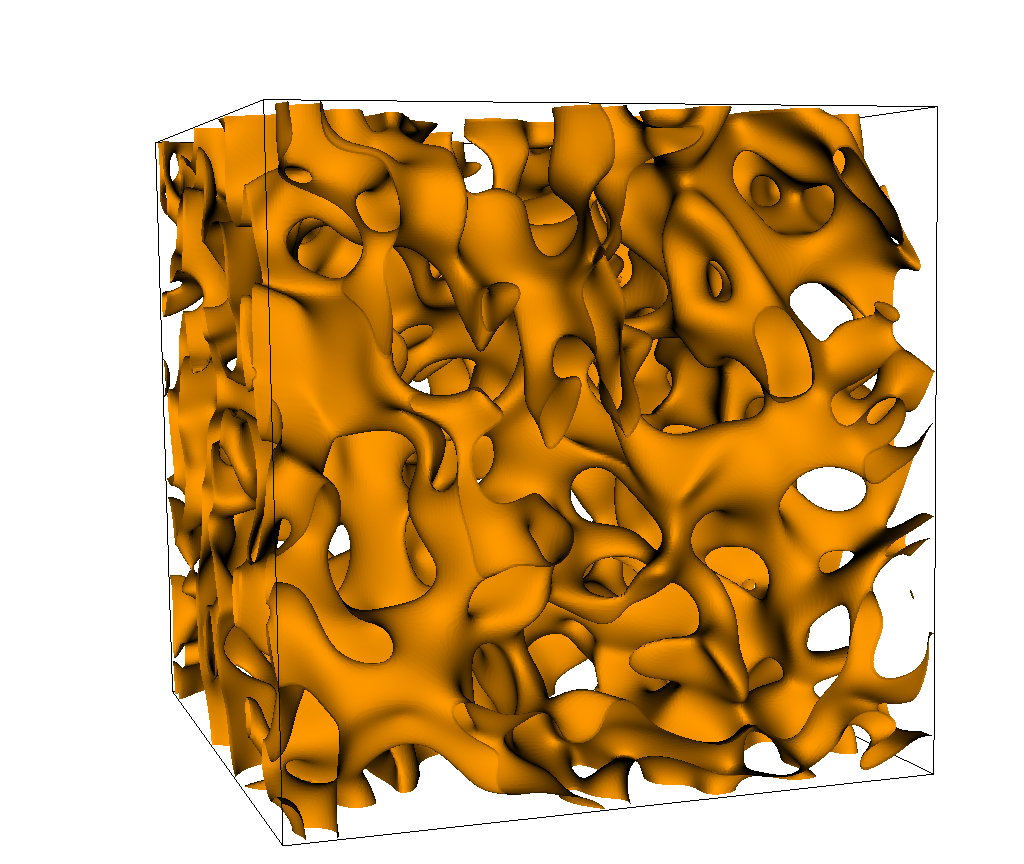}
       \includegraphics[width=0.32\textwidth]{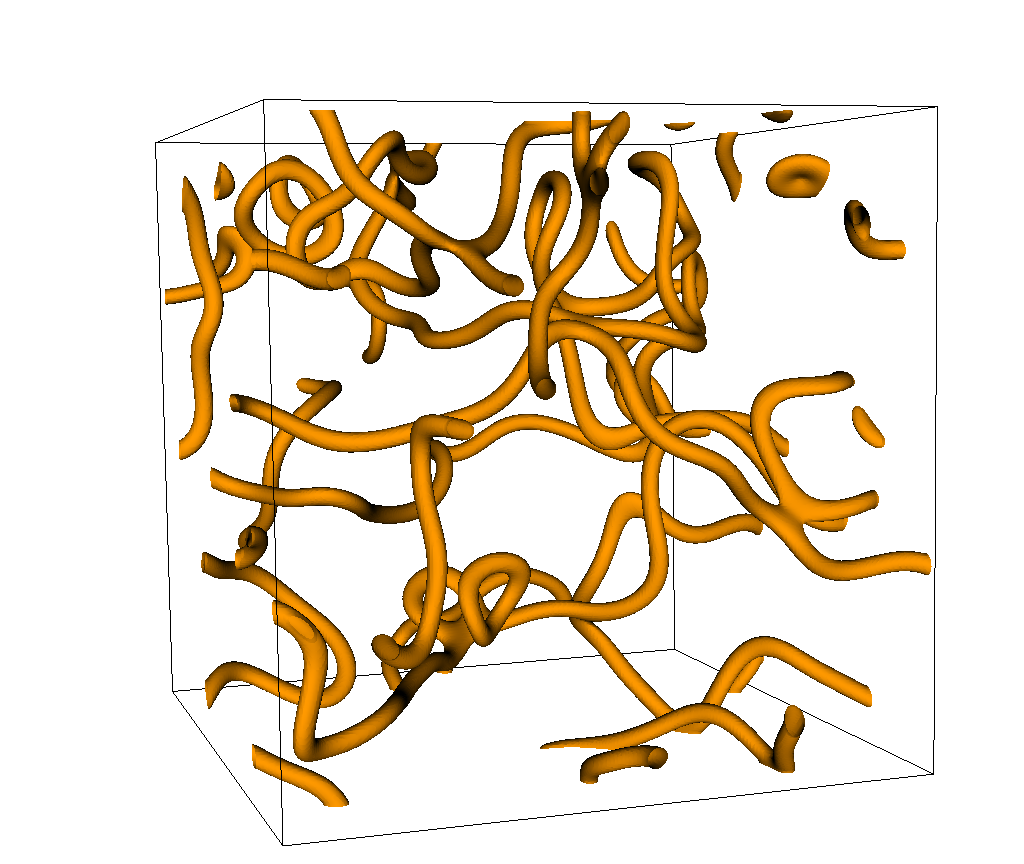}\
        \includegraphics[width=0.32\textwidth]{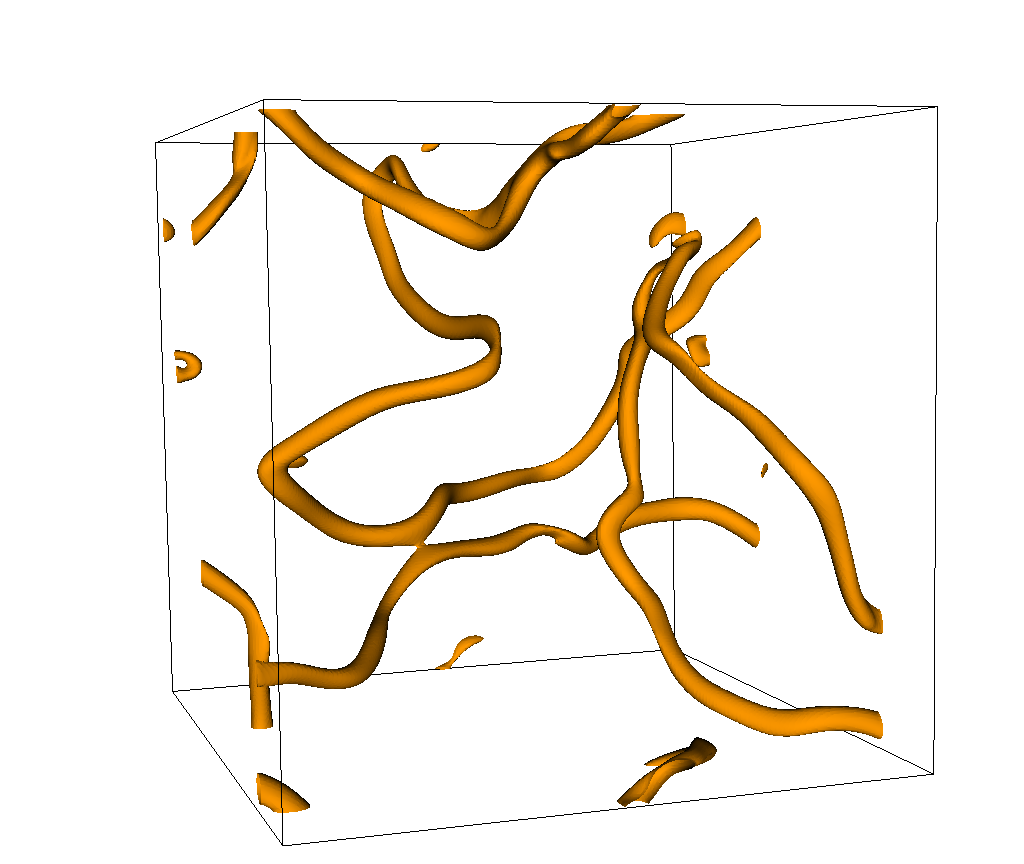}
    \caption{Snapshots of $|\varphi^2|=0.25v^2$ isosurfaces of from a simulation of global cosmic strings generated with $a(\tau)=\tilde{\tau}/70$, $\tilde{\ell}_\text{str}=5$ and $\tilde{\tau}_\text{end}=92$, in a lattice with $\tilde{L}=64$, $\delta\tilde{x}=0.25$. The snapshots correspond to the random initial conditions (left), the configuration after the diffusion evolution (center), and the network after the extra-fattening phase (right).
         }
    \label{fig:defects:snapshots}
\end{figure}

In \cref{fig:defects:snapshots} we represent, as an example, snapshots of a network of global strings simulated on a lattice with $\tilde{L}=64$, $\delta\tilde{x}=0.25$ using a dissipation phase followed by extra-fattening, with $a(\tau)=\tilde{\tau}/70$, $\tilde{\ell}_\text{str}=5$ and $\tilde{\tau}_\text{end}=92$. The snapshots represent the isosurfaces with $|\varphi|^2=0.1v^2$ at some relevant moments of the simulation: the initial configuration of the simulation generated using \cref{eq:initialPSstrings} (left), the field configuration after the diffusion phase (center), and the network after the end of extra-fattening (right).

\begin{figure}[!t]
    \centering
    \begin{minipage}{0.47\textwidth} 
    \centering
    \textbf{Global strings}\\[4pt]
        \includegraphics[width=\textwidth]{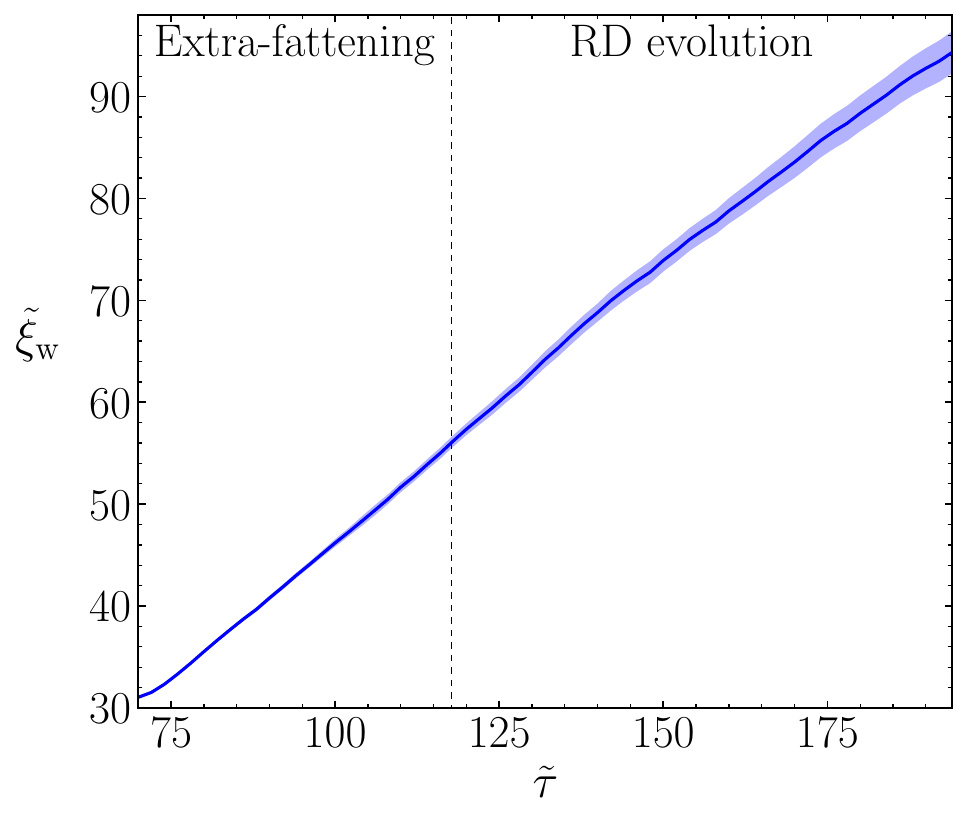}
    \end{minipage}
    \begin{minipage}{0.47\textwidth} 
    \centering
    \textbf{Local strings}\\[4pt]
        \includegraphics[width=\textwidth]{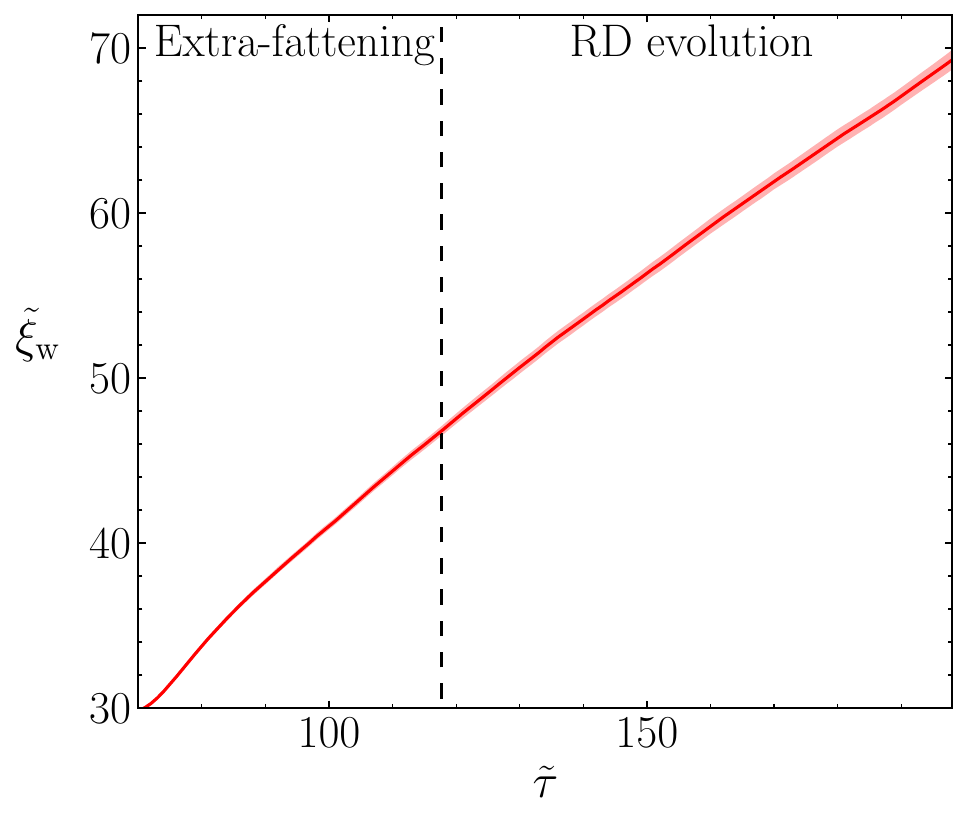}
    \end{minipage}
    
    \caption{
        Time-evolution of the comoving mean string separation of a network of global (left) and local (right) strings, averaged over multiple independent realizations. The initial conditions are generated as described in \cref{subsubsec:GlobalStrings,subsubsec:LocalStrings}, respectively, using $\tilde{\ell}_\str=15$ and $\tilde{\tau}_\text{diff}=20$ units of diffusion,  and simulations are performed in periodic lattices with $N=1024$ and $\delta\tilde{x}=0.25$. The vertical dotted line indicates the end of the extra-fattening phased, followed by physical evolution in a radiation-dominated (RD) background, and the shaded region represents one standard deviation.}
    \label{fig:defects:stringevolution}
\end{figure}

In simulations of cosmic strings we use observables that keep track of the string dynamics. Of particular interest is to measure the {\it winding} around each lattice plaquette. For a plaquette with vertices $\{\mathbf{n}, \mathbf{n}+\hat{\imath}, \mathbf{n}+\hat{\imath}+\hat{\jmath}, \mathbf{n}+\hat{\jmath}\}$, we define the winding number as~\cite{Vachaspati:1984dz,Rajantie:1998vv},
\begin{equation}\label{eq:defects:globalwinding}
W_{ij}(\mathbf{n})=\frac{1}{2\pi}\left(Y_i(\mathbf{n})+Y_j(\mathbf{n}+\hat{\imath})-Y_i(\mathbf{n}+\hat{\jmath})-Y_j(\mathbf{n})\right)\,,
\end{equation}
where 
\begin{equation}
Y_i(\mathbf{n})=\left[\theta(\mathbf{n})-\theta(\mathbf{n}+\hat{\imath})\right]_\pi\,,
\end{equation}
is the phase variation along each link. Here $\theta(\mathbf{n})$ is the phase of the complex scalar field and $[\theta]_\pi$ sets $\theta$ in the range $-\pi <\theta \leq\pi$. A non-zero winding around a plaquette indicates that that plaquette is pierced by the core of a string, thus giving an estimate of the location of the string cores in a simulation. This estimate can be refined using the values of the fields at the plaquette vertices, see for example~\cite{Fleury:2015aca,Drew:2019mzc}, for different estimation procedures. It is also worth mentioning that the definition in \cref{eq:defects:globalwinding} only allows to measure winding numbers as $\pm 1$. Higher winding number would require of a generalized definition using multiple plaquettes\footnote{Strings with higher winding numbers in scaling networks are however not found in lattice simulations initialized by the procedures explained above.}. The total number of pierced plaquettes, $N_\text{w}$, can be used to obtain an estimate of the total length of the strings in the lattice frame,
\begin{equation}\label{eq:defects:totallengthstrings}
L_\text{w}=\frac{2}{3}\delta x N_\text{w}\,.
\end{equation}
Here, the $2/3$ factor takes into account the {\it Manhattan} effect~\cite{Fleury:2015aca}, i.e., the fact that after entering some lattice cell, strings may leave it through any of the other five plaquettes. As an example, we represent in the left panel of \cref{fig:defects:stringevolution} the evolution of the mean-string separation, $\xi=\sqrt{(aN)^3/L_\text{w}}$, averaged over 10 realizations of global string networks evolving in radiation domination, with $a(\tau)=\tau/\tau_0$ and $\tilde{\tau}_0=70$. Networks are generated using the initialization procedure described above, {\it i.e.}~applying diffusion + extra-fattening, with $\tilde{\ell}_\str=15$ and $\tilde{\tau}_\text{diff}=20$ units of diffusion, in lattices with $N=1024$ and $\delta\tilde{x}=0.25$. 

It is also interesting to characterize the energy components of the strings. To capture only the energy close to the string cores, it is convenient to use a weight function centered on the string cores~\cite{Hindmarsh:2019csc,Hindmarsh:2021vih,Correia:2024cpk}. An option, for example, is to use a weight based on the potential~\cite{Hindmarsh:2021vih,Baeza-Ballesteros:2023say},
\begin{equation}\label{eq:defects:weightfunction}
W[\varphi]=\frac{4V[\varphi]}{\lambda v^4}\,\Theta\left(\frac{v^2}{2}-|\varphi|^2\right)\,,
\end{equation}
which is normalized such that $W[0]=1$ at the center of the string core ($\varphi = 0$). Here $\Theta$ is the Heviside function that prevents the inclusion of field values $|\varphi|^2>v^2/2$. Using this weight, we can define the total kinetic, gradient and potential energies of the strings, through
\begin{equation}
\begin{array}{rcl}
E_{K,\str} & = & \displaystyle\frac{\delta x^3}{a^3}\sum_{\mathbf{n}} W[\varphi(\mathbf{n})]|\pi_\varphi(\mathbf{n})|^2\,,\\[10pt]
 E_{G,\str} & = &\displaystyle a \delta x^3\sum_{\mathbf{n}}\sum_i W[\varphi(\mathbf{n})] |\partial_i\varphi(\mathbf{n})|^2\,,\\[10pt]
  E_{V,\str} & = &\displaystyle \delta x^3\sum_{\mathbf{n}} W[\varphi(\mathbf{n})] V[\varphi(\mathbf{n})]\,,
\end{array}
\end{equation}
where the total energy of the network is
\begin{equation}
E_\str=E_{K,\str}+E_{G,\str}+E_{V,\str}\,.
\end{equation}

The weighted energy components can additionally be used to obtain estimates of the proper length of the strings and their mean squared velocity~\cite{Correia:2024cpk,Hindmarsh:2021vih,Hindmarsh:2019csc},
\begin{equation}
L_\text{str} 
= \displaystyle\frac{1}{a}\frac{E_\text{str}+f_\text{V}E_{\mathcal{L},\text{str}}}{\mu(1+f_\text{V})}\,,
~~~~
v_\text{rms}^2 
= \displaystyle\frac{E_\text{str}+E_{\mathcal{L},\text{str}}}{E_\text{str}+f_\text{V}E_{\mathcal{L},\text{str}}}\,,
\end{equation}
where 
\begin{equation}
E_{\mathcal{L},\text{str}}\equiv \delta x \sum_{\bm{n}} W[\varphi(\bm{n})]\mathcal{L}(\bm{n})=E_{K,\text{str}}-E_{G,\text{str}}-E_{V,\text{str}}\,
\end{equation}
is the weighted Lagrangian 
of the global string, and $\mu$ and $f_\text{V}$ are the weighted tension and fraction of potential energy of an infinite straight string solution~\cite{Vilenkin:2000jqa}. For cosmic strings with $\pm1$ winding number and described by the model \cref{eq:defects:actionglobal}, using the weight function in \cref{eq:defects:weightfunction} they take the values
\begin{equation}
\mu = 1.7824 v^2\,,~~~~
f_\text{V} = 0.3689\,.
\end{equation}
We note that it has been argued that one can reduce discretization effects by computing these coefficients at finite lattice spacing~\cite{Correia:2024cpk}.

\subsubsection{Simulation of local strings}
\label{subsubsec:LocalStrings}

Lattice simulations of local cosmic strings follow similar procedures as simulations of global strings. The physical field dynamics are simulated with canonical methods, but it is typically complemented with dedicated initialization and resolution-preserving techniques, as in the global case. We typically define program variables, {\it c.f.}~Sect.~\ref{subsubsec:ProgramVariables}, using $\fstar=v$ and $\omegastar=\sqrt{\lambda} v$. 
The equations of motion in the temporal gauge, $A_0=0$, read in program units 
\begin{equation}\label{eq:defects:localeomsprogramvariables}
\left.\begin{array}{rcl}
(a^2\tilde{\varphi}')'-a^2\tilde{D}_{A,i}^-\tilde{D}_{A,i}^+\tilde{\varphi} & = & \displaystyle-2a^4\tilde{\varphi}\left(|\tilde{\varphi}|^2-\frac{1}{2}\right)\,,\\[10pt]
\tilde A_i''-\tilde{\nabla}_j^-\tilde{\nabla}_j^+ \tilde{A}_i + \tilde{\nabla}_j^-\tilde{\nabla}_i^+ \tilde{A}_j & = &\displaystyle 2 a^2 e \left(\frac{\fstar}{\omegastar}\right)^2 \text{Im}[\tilde{\varphi}^*\tilde{D}^+_i \tilde{\varphi}]\,.
\end{array}\right.
\end{equation}
The Gauss constraint, on the other hand, reads
\begin{equation}\label{eq:defects:gausslawprogramvariables}
\begin{array}{rcl}
\tilde{\nabla}_i \tilde{A}_i' & = & \displaystyle 2 a^2 e \left(\frac{\fstar}{\omegastar}\right)^2 \text{Im}[\tilde{\varphi}^*\tilde{\varphi}']\,. 
\end{array}
\end{equation}
At this point, it is simple to check that the dynamics only depend on the ratio $\beta=e^2/2\lambda$, as claimed in \cref{sec:defects:CosmicStringstheory}. If we rescale the dimensionless gauge fields by defining $\hat{\tilde{A}}_\mu = \tilde{A}_\mu / e$ and substitute the values of $\fstar$ and $\omegastar$, the dependence on $\beta$ becomes explicit on the the equation of motion of the gauge fields,
\begin{equation}
   \begin{array}{rcl}
\hat{\tilde {A}}_i''-\tilde{\nabla}_j^-\tilde{\nabla}_j^+ \hat{\tilde{A}}_i + \tilde{\nabla}_j^-\tilde{\nabla}_i^+ \hat{\tilde{A}}_j & = &\displaystyle  a^2 \beta \text{Im}[\tilde{\varphi}^*\hat{\tilde{D}}^+_i \tilde{\varphi}]\,,
\end{array} 
\end{equation}
with $\hat{\tilde{D}}_i^+=\tilde{\nabla}_i^+-i\hat{\tilde{A}}_i$, while the equation of motion of the scalar field is independent of $\lambda$ and $e^2$. One can similarly check that the Gauss' constraint is also only dependent on the ratio $\beta$.

To simulate the dynamics of a network of local cosmic strings following these equations, one defines the conjugate momenta
\begin{equation}
\begin{array}{rcl}
\tilde{\pi}_\varphi & \equiv & a^2 \tilde{\varphi}'\,,\\[5pt]
 (\tilde{\pi}_A)_i & \equiv & A_i'\,,
 \end{array}
\end{equation}
so that we can rewrite the equations of motion as a system of first-order differential equations,
\begin{equation}\label{eq:defects:eomfirstorderlocal}
\left\lbrace\begin{array}{rcl}
\tilde{\pi}_\varphi' &=& \cK_\varphi[a,\tilde{\varphi}, \tilde{A}]\,, \\[5pt]
(\tilde{\pi}_A)_i' &=&\cK_{A_i}[a,\tilde{\varphi}, \tilde{A}]\,, \\[5pt]
\tilde{\varphi}'&=&a^{-2}\tilde{\pi}_\varphi\,,\\[5pt]
\tilde{A}_i'&=&(\tilde{\pi}_A)'_i\,,
\end{array}\right.
\end{equation}
where we have introduced the kernels,
\begin{equation}
\begin{array}{rcl}
\cK_\varphi[a,\tilde{\varphi},\tilde{A}_j]& =&  \displaystyle a^2\tilde{D}_{A,i}^-\tilde{D}_{A,i}^+\tilde{\varphi} -2a^4\tilde{\varphi}\left(|\tilde{\varphi}|^2-\frac{1}{2}\right)\,,\\[15pt]
\cK_{A_i}[a,\tilde{\varphi},\tilde{A}_j] & = & \displaystyle\tilde{\nabla}_j^-\tilde{\nabla}_j^+ \tilde{A}_i -  \tilde{\nabla}_j^-\tilde{\nabla}_i^+ \tilde{A}_j +2 a^2 e \left(\frac{\fstar}{\omegastar}\right)^2 \text{Im}[\tilde{\varphi}^*\tilde{D}_i \tilde{\varphi}]\,.
\end{array}
\end{equation}
These equations can be solved using any symplectic integrator from \cref{subsubsec:SymplecticInt}.

As in the case of global strings, lattice simulations of local cosmic strings can be complemented by a suitable initialization and a resolution-preserving approach. A widely used option is to initialize the complex scalar field with Gaussian random fluctuations that follow a power spectrum as in \cref{eq:initialPSstrings}, while the gauge field and all conjugate momenta are set to zero. To eliminate the excess energy in the initial configuration, the fields are evolved through a dissipation phase, which also allows a magnetic field to be formed inside the strings. Similarly as for the global case, one uses first-order diffusion equations like
\begin{equation}\label{eq:defects:diffusionlocal}
\left\lbrace\begin{array}{rcl}
\tilde{\varphi}' & = & \displaystyle\tilde{\cK}_\varphi[a(\tilde{\tau}_0), \tilde{\varphi}, \tilde{A}_j]\,,\\[5pt]
\tilde{A}_i' & = & \displaystyle\tilde{\cK}_{A_i}[a(\tilde{\tau}_0), \tilde{\varphi}, \tilde{A}_j]\,,
\end{array}\right.
\end{equation}
which can be solved with any non-symplectic evolver. For example, the forward Euler algorithm reads\vspace*{-4mm}
\begin{equation}
\begin{array}{l}
\text{IC}:  \displaystyle\{\tilde{\varphi},\tilde{A}_j\}\text{ at }\tilde{\tau}_0\,,\\[6pt]
\left\lbrace\begin{array}{rcl}
\tilde{\varphi}_{+0} & = &\displaystyle \tilde{\varphi} + \delta\tilde{\tau}_\text{diff}   \tilde{\cK}_\varphi[a(\tau_0), \tilde{\varphi}, \tilde{A}_j]\,,\\[10pt]
\tilde{A}_{i,+0} & = &\displaystyle \tilde{A}_i + \delta\tilde{\tau}_\text{diff} \tilde{\cK}_{A_i}[a(\tau_0), \tilde{\varphi}, \tilde{A}_j]\,.
\end{array}\right.
\end{array}
\end{equation}
As in the global case, we note that the time step to solve the diffusion equations is different from that of standard physical evolution, as the Courant stability condition for a diffusion equation as those above is different, see the discussion around~Eqs.~(\ref{eq:defects:diffusionglobal}) and (\ref{eq:diffusionschemeglobal}).

Once the network is generated after the dissipation phase, it can be evolved using canonical techniques. However, when working on an expanding background, the same loss-of-resolution problem as for global strings happens. To prevent this, similar resolution-preserving approaches can be taken, based on simulating a modified version of the equations of motion~\cite{Press:1989yh, Moore:2001px, Bevis:2006mj},
\begin{equation}\label{eq:defects:localfatteningeom}
\left.\begin{array}{rcl}
(a^2 \tilde{\varphi}')'  & = &a^2 \tilde{\nabla}_i^-\tilde{\nabla}_i^+ \tilde{\varphi} + \displaystyle -2 a^4 \left(\frac{a}{a_0}\right) ^ {2(s-1)}\tilde{\varphi} \left(|\tilde{\varphi}|^2 -\frac{1}{2}\right)\,,\\[10pt]
\displaystyle \left[\left(\frac{a}{a_0}\right)^{-2(s-1)}\tilde{A}_i'\right]' & = &\displaystyle \left(\frac{a}{a_0}\right)^{-2(s-1)}\left[\tilde{\nabla}_j^-\tilde{\nabla}_j^+\tilde{A}_i - \tilde{\nabla}_j^-\tilde{\nabla}_i^+ \tilde{A}_j \right] +  2 a^2 e \left(\frac{\fstar}{\omegastar}\right)^2 \text{Im}[\tilde{\varphi}^* \tilde{D}_i \tilde{\varphi}]\,,
\end{array}\right.
\end{equation}
complemented by a modified Gauss constraint
\begin{equation}
\tilde{\nabla}_i \tilde{A}_i'=2 a^2\left(\frac{a}{a_0}\right)^{2(s-1)} e \left(\frac{\fstar}{\omegastar}\right)^2 \text{Im}[\tilde{\varphi}^*\tilde{\varphi}']\,.
\end{equation}
As for the global case, $s=-1$ and $s=0$ correspond to the so-called extra-fattening and fattening, respectively---see the discussion below \cref{eq:defects:kernelextrafattening} for a more in depth discussion of these two possibilities. We recall that the $s=1$ case corresponds to the physical evolution---see \cref{eq:defects:eomfirstorderlocal}.

To numerically solve these equations, it is convenient to generalize the definition of the conjugate momentum associated to the gauge field, 
\begin{equation}
(\tilde{\pi}_A)_i=\left(\frac{a}{a_0}\right)^{-2(s-1)}\tilde{A}_i'\,,
\end{equation}
so that the equations of motion are rewritten as
\begin{equation}
\left\lbrace\begin{array}{rcl}
\tilde{\pi}_\varphi' &=&\displaystyle \cK^\text{fat}_\varphi[a,\tilde{\varphi}, \tilde{A}; s]\,, \\[10pt]
(\tilde{\pi}_A)_i' &=&\displaystyle\cK^\text{fat}_{A_i}[a,\tilde{\varphi}, \tilde{A};s]\,, \\[10pt]
\tilde{\varphi}'&=&\displaystyle a^{-2}\tilde{\pi}_\varphi\,,\\[10pt]
\tilde{A}_i'&=&\displaystyle \left(\frac{a}{a_0}\right)^{2(s-1)}(\tilde{\pi}_A)'_i\,.
\end{array}\right.
\end{equation}
where we have introduced modified kernels,
\begin{equation}
\begin{array}{rcl}
\cK^\text{fat}_\varphi[a,\tilde{\varphi},\tilde{A}_j;s] & = &\displaystyle  a^2\tilde{D}_{A,i}^-\tilde{D}_{A,i}^+\tilde{\varphi} -2 a^4 \left(\frac{a}{a_0}\right) ^ {2(s-1)}\tilde{\varphi}\left(|\tilde{\varphi}|^2-\frac{1}{2}\right)\,,\\[10pt]
\cK_{A_i}^\text{fat}[a,\tilde{\varphi},\tilde{A}_j;s] & = &\displaystyle \left(\frac{a}{a_0}\right)^{-2(s-1)}\left[\tilde{\nabla}_j^-\tilde{\nabla}_j^+\tilde{A}_i - \tilde{\nabla}_j^-\tilde{\nabla}_i^+ \tilde{A}_j \right] +2 a^2 e \left(\frac{\fstar}{\omegastar}\right)^2 \text{Im}[\tilde{\varphi}^*\tilde{D}_i \tilde{\varphi}]\,.
\end{array}
\end{equation}

It is worth mentioning that, after a redefinition of the gauge fields, $A_\mu\rightarrow A_\mu / e$, these modified equations of motion can be seen as the equations of motion of a modified theory in which the scalar and gauge coupling constants are promoted to time-dependent parameters,
\begin{equation}\label{eq:lambdaAndGaugeScaling}
\begin{array}{rcl}
\lambda(\tau) & = & \displaystyle \left(\frac{a}{a_0}\right) ^ {2(s-1)}\lambda_0\,,\\[10pt]
 e^2(\tau) & = & \displaystyle \left(\frac{a}{a_0}\right) ^ {2(s-1)}e^2_0\,,
\end{array}
\end{equation}
with $\lambda_0=\lambda(\tau_0)$ and $e_0=e(\tau_0)$. Note these transformations (\ref{eq:lambdaAndGaugeScaling}) cannot be applied directly in the action \cref{eq:defects:actionlocal} before redefining the gauge fields as indicated above. Without the redefinition of the gauge field, these time-dependent parameters would need to be complemented by the addition of extra terms in the action to reproduce the equations of motion in \cref{eq:defects:localfatteningeom}.

Finally, we discuss  dedicated observables for local strings. First, we note that we can determine the location of the string cores from the position of the pierced plaquettes. These are characterized by a non-zero winding number, defined as in \cref{eq:defects:globalwinding}, where now the contribution from each link is defined in a gauge invariant manner,
\begin{equation}
Y_i(\mathbf{n})=\left[e\delta x A_i(\mathbf{n})+\theta(\mathbf{n})-\theta(\mathbf{n}+\hat{\imath})\right]_\pi-e\delta x A_i(\mathbf{n})\,.
\end{equation}
As in the global case, the total number of pierced plaquettes can be used to obtain an estimate of the total length of the strings on the lattice frame, see \cref{eq:defects:totallengthstrings}. In the right panel of \cref{fig:defects:stringevolution} we represent an example of the evolution of the mean-string separation, $\xi=\sqrt{(aN)^3/L_\text{w}}$, of local string networks evolving in radiation domination, with $a(\tau)=\tau/\tau_0$ and $\tilde{\tau}_0=70$, averaged over 20 independent realizations. The initial conditions are generated with $\tilde{\ell}_\str=15$ and $\tilde{\tau}_\text{diff}=20$ units of diffusion, in lattices with $N=1024$ and $\delta\tilde{x}=0.25$, including an extra-fattening phase.

One can also measure the different string energy components. Using a weight function $W[\varphi]$, like the one introduced in \cref{eq:defects:weightfunction}, we can define the kinetic, gradient, potential, electric and magnetic energy components of the strings, respectively, as
\begin{equation}
\begin{array}{rcl}
E_{K,\str}&=&\displaystyle\frac{\delta x^3}{a^3}\sum_{\mathbf{n}} W[\varphi(\mathbf{n})] |\pi_\varphi(\mathbf{n})|^2\,,\\[10pt]
 E_{G,\str} & =&\displaystyle a\delta x^3\sum_{\mathbf{n}}\sum_i W[\varphi(\mathbf{n})]|D_i\varphi(\mathbf{n})|^2\,,\\[10pt]
E_{V,\str}&=&\displaystyle a^3\delta x^3 \sum_{\mathbf{n}} W[\varphi(\mathbf{n})] V[\varphi(\mathbf{n})]\,,\\[10pt]
 E_{E,\str} & =&\displaystyle\frac{a}{2}\sum_{{\mathbf{n}}}W[\varphi(\mathbf{n})] \sum_i E_i^2(\mathbf{n})\,,\\[10pt]
 E_{B,\str} & = &\displaystyle \frac{a}{2}\sum_{\mathbf{n}} W[\varphi(\mathbf{n})] \sum_i B_i^2(\mathbf{n})\,,
\end{array}
\end{equation}
where we have used the definitions of the electric and magnetic fields from \cref{eq:ElectricMagneticDefs}. From here, we can define also the total weighted energy and Lagrangian of the local strings,
\begin{equation}
\begin{array}{rcl}
E_\str & = & E_{K,\str}+E_{G,\str}+E_{V,\str}+E_{\text{E},\str}+ E_{B,\str}\,,\\[5pt]
E_{\mathcal{L},\str} & = & E_{K,\str}-E_{G,\str}-E_{V,\str}+E_{\text{E},\str}- E_{B,\str}\,,
\end{array}
\end{equation}
which can be used to obtain estimators of the proper comoving length of the strings and the mean-squared velocity, as~\cite{Hindmarsh:2017qff}
\begin{equation}
L_\text{str} = \displaystyle \frac{1}{a}\frac{E_\text{str}-\Delta f E_{\mathcal{L},\text{str}}}{\mu(1+\Delta f)}\,,
~~~~~
v_\text{rms}^2  = \displaystyle \frac{E_\text{str}+E_{\mathcal{L},\text{str}}}{E_\text{str}-\Delta f E_{\mathcal{L},\text{str}}}\,.
\end{equation}
Here $\mu$ is the weighted tension of the static string and $\Delta f=f_\text{B}-f_\text{V}$ is the difference between the fraction of magnetic and potential energy of the static solution~\cite{Nielsen:1973cs}, respectively. Using the weight in \cref{eq:defects:weightfunction} and the Nielsen-Olsen vortex static solution  from the theory defined in \cref{eq:defects:actionlocal} with $\pm1$ winding number, one finds
\begin{equation}
\mu = 1.4415v^2\,,~~~~
f_\text{B} = 0.2047\,,~~~~
f_\text{V} = 0.2056\,.
\end{equation}
It is worth noting that in the case of local strings, alternative estimators of the proper length and mean squared velocity can also be defined, see {\it e.g.}~\rcite{Hindmarsh:2017qff}.

\subsubsection{Simulation of domain walls}
\label{subsubsec:DW}

Lattice simulations of domain walls can also benefit from similar techniques to those discussed for cosmic strings. The dynamics of domain walls can be simulated by evolving the equations of motion of the underlying scalar field. For the model introduced in \cref{eq:defects:actionDW} with biased potential given by  \cref{eq:defects:potentialDWbias}, one typically defines dimensionless program variables ({\it c.f.}~Sect.~\ref{subsubsec:ProgramVariables}) using two scales, $\fstar=v$ and $\omegastar=\sqrt{\lambda} v$, so that
and the equations of motion on the lattice read
\begin{equation}\label{eq:defects:eomsDW}
(a^2\tilde{\phi}')'-a^2\tilde{\nabla}_i^-\tilde{\nabla}_i^+\phi=-a^4\left[\tilde{\phi}(\tilde{\phi}^2-1)+3\frac{q}{\lambda}\tilde{\phi}^2\right]\,.
\end{equation}
The case $q=0$ recovers the unbiased domain walls, but from now one we will assume $q \neq 0$. This equation of motion depends on the single parameter $g=q/\lambda$, and can be simulated using canonical techniques presented in {\tt The Art-I}~\cite{Figueroa:2020rrl}, summarized in \cref{subsubsec:SymplecticInt}. If one defines the conjugate momentum
\begin{equation}\label{eq:defects:diffusionDW}
\tilde{\pi}_\phi=a^2\tilde{\phi}'\,,
\end{equation}
the equation of motion can be rewritten as a system of first order differential equations,
\begin{equation}\label{eq:defects:DWscheme}
\left\lbrace\begin{array}{rcl}
	\tilde{\pi}_\phi' & = &\displaystyle \cK_\phi[a, \tilde{\phi}]\,,\\[5pt]
	\tilde{\phi}' & = &\displaystyle a^{-2}\tilde{\pi}_\phi\,.
\end{array}\right.
\end{equation}
where we have introduced the scalar kernel,
\begin{equation}
\cK_\phi[a,\tilde{\phi}]=a^2\tilde{\nabla}_i^-\tilde{\nabla}_i^+ \tilde{\phi}-a^4\left[\tilde{\phi}(\tilde{\phi}^2-1)+3\frac{q}{\lambda}\tilde{\phi}^2\right]\,.
\end{equation}

As in the case of cosmic strings, in order to ensure that the network of domain walls reaches scaling as soon as possible, some procedure can be introduced before one solves the above EOM with canonical techniques. Typically, simulations of domain walls are initialized using random initial conditions with $\langle\phi\rangle=0$, introducing fluctuations (typically characterized by a power spectrum, but not necessarily) that destabilize the vacuum, thus emulating a phase transition. After this, a dissipative phase can be implemented, ensuring the network reaches the scaling regime faster, as in the case of strings. This can be achieved similarly as with strings, by using a first-order diffusion equation with fixed background, which reads
\begin{equation} \label{eq:difusDW}
\tilde{\phi}' = \tilde{\cK}_\phi[a(\tilde{\tau}_0), \tilde{\phi}]\,.
\end{equation}
This equation can be solved using any non-symplectic integrator, as for example the forward Euler algorithm, 
\begin{equation}
\begin{array}{rcl}
\text{IC} & : & \{\tilde{\phi}\}\text{ at }\tilde{\tau}_0\,,\\[10pt]
\tilde{\phi}_{+0} & = & \tilde{\phi} + \delta\tilde{\tau}_\text{diff}   \tilde{\cK}_\varphi[a(\tau_0), \tilde{\phi}]\,.
\end{array}
\end{equation}
To obtain a domain wall network in the scaling regime, it is typically enough to apply such diffusion during few units of program time after the formation of the domain walls, see Appendix B.2 of \cite{Notari:2025kqq}.

As in the case of cosmic strings, domain walls also suffer from a loss of ultraviolet resolution at late times. Although not always used, one approach to overcome this problem is the use of a fattening technique, similarly as in the case of cosmic strings. Such technique was actually originally proposed for domain walls in~\cite{Press:1989yh}, and it amounts to rewriting the scalar field equation of motion as follows,
\begin{equation}\label{eq:defects:DWfatteningeom}
\tilde{\phi}''+\gamma \frac{a'}{a}\phi'-\tilde{\nabla}_i^-\tilde{\nabla}_i^+\phi=-a^\beta\left[\tilde{\phi}(\tilde{\phi}^2-1)\right]\,,
\end{equation}
with $\beta$ and $\gamma$ real numbers, and where we have particularized to the unbiased case $q=0$. For $\beta = \gamma =2$, we recover the physical case. As proposed in \cite{Press:1989yh}, one can prevent the loss of domain wall resolution by fixing instead $\beta = 0$, which keeps their comoving width constant. Additionally, after such a change, the Hubble damping localizes the scalar field into the minima of the potential as $\langle \phi - v \rangle_{\rm rms} \propto  a^{- \gamma/2 - \beta/4}$, so if one fixes $\beta = 0$, one must also fix $\gamma = 3$ in order to recover the physical result $\langle \phi - v \rangle_{\rm rms} \propto  a^{-3/2}$.

An important observable when studying domain walls is the estimator of their total area, $A$. This can be obtained by counting the number of links on the lattice that cross by the domain walls, that is, the number on links across which the scalar field changes sign. This number, needs then to be corrected by a weight function that takes into account the inclination of the domain walls. Numerically, this is implemented as~\cite{Press:1989yh}
\begin{equation}
\displaystyle A=a^2\delta x^2 \sum_{\mathbf{n}}\sum_{i} \delta_{\mathbf{n},i}\frac{|\nabla \phi(\mathbf{n})|}{\displaystyle\sum_{j}|\nabla_j\phi(\mathbf{n})|}\,,
\end{equation}
where $\delta_{n,i}=1$ if $\phi(\mathbf{n})\phi(\mathbf{n}+\hat{\imath})<0$, and 0 otherwise.

\section{Initial conditions}
\label{sec:InitialConditions}

In this section, we shift our attention to procedures for initializing fields on a lattice. A common practice to set up the initial conditions for simulations of scalar fields has traditionally consisted of introducing their homogeneous modes and, on top of these, small fluctuations characterized by a quantum vacuum spectrum. In scalar-gauge theories, gauge-field amplitudes are typically set to zero, and once the scalar spectrum of quantum fluctuations is set, the longitudinal mode of the electric field is then enforced to respect the Gauss's law. We refer to this initialization procedure in scalar(-gauge) theories as canonical initial conditions, see Sect.~7 of {\tt The Art~I}~\cite{Figueroa:2020rrl} for a detailed discussion. In this section, we explore initial conditions that differ from such prescription. There are, in fact, various types of  non-canonical initial conditions, which we classify according the main ingredient that characterizes them:\vspace*{0.2cm}

\indent {\bf 1. Arbitrary power spectra.} This corresponds to initial conditions determined by a spectrum of fluctuations that has already been excited well above the vacuum. For example, in many scenarios of inflation or (p)reheating, a finite range of modes (that are initially in vacuum) are noticeably excited out of the vacuum due to some instability mechanism ({\it e.g.}~parametric resonance, tachyonic growth, etc), while the dynamics of the system remains linear. To save simulation time, it is convenient to run the linear dynamics separately (through non-lattice methods), and then use the spectrum of excited modes obtained from the linear regime as an initial input on the lattice, just before the dynamics become non-linear. We discuss how to create field fluctuations according to an arbitrary power spectrum in Sect.~\ref{subsec:ArbitrarySpectrum}.\vspace*{0.2cm}

{\bf 2. Algorithmic methods.} These methods are characterized by the application of some algorithm(s) that `evolve' the fields into a desirable configuration that is used as an initial condition. For example, in scalar-gauge theories, as an alternative to imposing the Gauss's constraint to the longitudinal mode of the electric field, we may impose completely unconstrained fluctuations to gauge fields, and then remove the unwanted transverse modes by a {\it minimization procedure}~\cite{Ambjorn:1990pu,Moore:1996qs}. We can also impose thermal initial conditions in a system via some {\it Langevin dynamics}, while exactly preserving the Gauss's law~\cite{Krasnitz:1995xi}. We discuss some representative cases in Sect.~\ref{subsubsec:ICalgorithmsProb}. Another 
example of algorithmic initialization is used in simulations of cosmic defects, in which we want to obtain an initial network which reaches the {\it scaling regime} in a small fraction of the simulation duration. The initialization methods used in this case start with some random initial procedure that mimics symmetry breaking, and then usually incorporates a {\it diffusion} method and/or some resolution-preserving technique, such as {\it extra-fattening}. As these methods have been covered in detail in Sect.~\ref{defects:latticesimulations}, we simply summarize them in Sect.~\ref{subsubsec:ICalgorithmsDefects}.\vspace*{0.2cm}

{\bf 3. Spatial configurations.} In some cases we need to set up very specific field configurations in coordinate space, according to a certain spatial profile. This is typically the case in first order phase transition scenarios, where scalar field bubble configurations need to be introduced at the nucleation points, following given spatial-profile recipes. It is also the case of topological defects, such as monopoles, strings, or domain walls, if we want to study them individually (as opposed to following their collective network dynamics), or if we want to focus on particular defect structures, such as kinks or cusps on a string. While quite different among themselves, all these cases are characterized by the fact that there exist 
solutions (either analytical or numerical) to the spatial profile of the desired configurations, which we need to introduce on the lattice whilst respecting periodic boundary conditions. We briefly discuss some examples of this type of initialization in Sect.~\ref{subsec:ICspatialProfile}.

\subsection{Initial condition set by a general spectrum}
\label{subsec:ArbitrarySpectrum}

In this section, we discuss how to set up initial field configurations based on some arbitrary power spectrum. 
In Sect.~\ref{subsubsec:GeneralPS} we discuss some aspects regarding the general properties of power spectra. We explain next, in Sect.~\ref{subsubsec:ICexamplesGeneralPS},  how to initialize field fluctuations
characterized by an external power spectrum. Finally, in Sect.~\ref{subsubsec:ICfromGeneralPS} we comment on some particular examples from scenarios presented in this review, such as those presented in Sects.~\ref{sec:NMCoupled_scalars} and~\ref{sec:Axion} of this review. 

\subsubsection{Properties of a power spectrum}
\label{subsubsec:GeneralPS}

There are two general aspects that characterize a power spectrum, namely its dependency on the scale factor normalization, and its mass dimensions. Let us inspect first how it transforms under
a re-scaling of the scale factor $a(\eta) \longrightarrow C\cdot a(\eta)$, with $C > 0$ an arbitrary positive constant. Reproducing Eq.~\eqref{eq:continuumPS} for convenience, we write 
\begin{eqnarray}\label{eq:continuumPSII}
\langle {\tt f}^2 \rangle \equiv \int {\rm d}\log k\;\Delta_{\tt f} (k)\,,~~~{\rm where}~~~ \Delta_{\tt f} \equiv {k^3\over 2\pi^2}\mathcal{P}_{\tt f}\,, ~~~{\rm and}~~~\langle {\tt f}_{\bf k} {\tt f}_{{\bf k}^{\prime}}^* \rangle = (2\pi)^3 \mathcal{P}_{\tt f}(k) \delta (\mathbf{k}-\mathbf{k^{\prime}})\,.
\end{eqnarray}
We know that, by construction, the expectation value $\langle {\tt f}^2 \rangle$ 
does not depend on the scale factor normalization. As the measure ${\rm d} \log k = {\rm d} k / k$ is independent of $C$, it follows that the power spectrum $\Delta_{{\tt f}}$ must be invariant under scale factor re-scalings. Then, as the physical momenta $p$ is independent of the scale factor, we have $k = a p \propto C$, and so $\mathcal{P}_{{\tt f}} \propto \Delta_{\tt f}/k^3 $ is forced to scale as $\propto C^{-3}$ under such transformation.

Second, we consider the mass dimension, which we indicate with square brackets {\it e.g.} $[m_p]=+1$.  If ${\tt f}({\bf x})$ represents the amplitude of a scalar field or (of the spatial component of) a gauge field, the expectation value $\langle {\tt f}^2 \rangle$ has mass dimension $[ \langle {\tt f}^2 \rangle] = +2$. As $[d\log k] = 0$, it follows that $[\Delta_{\tt f}] = +2$. This reflects an important property: the power spectrum $\Delta_{\tt f}(k)$ of a function $\tt f({\bf x})$, always inherits the dimensions of the squared function ${\tt f}^2$. As $[k^3] = +3$, it follows that $[\mathcal{P}_{\tt f}] = -1$. 

Having these two points in mind, we can write a general parametrization of $\mathcal{P}_{\tt f}(k)$ as
\begin{eqnarray}\label{eqn:VarianceGeneralForm}
    \mathcal{P}_{\tt f}(k) = \mathcal{P}_{\tt f} (k/a;a) \equiv \frac{1}{a^3} \mathcal{F}(k/a) \,,
\end{eqnarray}
where $\mathcal{F}(k/a)$ is a function of massdimension $[\mathcal{F}] = -1$, which characterizes the variance of ${\tt f}({\bf x})$ and depends only on the modulus of the physical momentum $p=k/a$. We write therefore
\begin{eqnarray}
    \langle {\tt f}^2 \rangle = \int {\rm d} \log k \; \frac{(k/a)^3}{2\pi^2} \mathcal{F}(k/a)\;.
\end{eqnarray}
A dimensionless version of the power spectrum in program units can also be written down, defining $\widetilde{\mathcal{F}} = E_* \mathcal{F}$, with $E_*$ some energy scale of dimensions $[E_*] = +1$. This leads to
\begin{eqnarray}\label{eq:f2programUnits}
    \langle \tilde{{\tt f}}^2 \rangle = \left(\dfrac{\omega_*^3 }{ E_* f_*^2}\right)\int {\rm d} \log \kappa \; \frac{(\kappa/a)^3}{2\pi^2} \widetilde{\mathcal{F}}(\kappa/a)\,,
\end{eqnarray}
with the dimensionless pre-factor collapsing to $(\omega_*/f_*)^2$ for the standard choice $E_* = \omega_*$.

Analogously, the power spectrum $\Delta_{\dot{{\tt f}}}$ of the time derivative of a function $\dot {\tt f}$, defined through
\begin{eqnarray}
   \langle {\dot{{\tt f}}}^2 \rangle= 
   \int {\rm d} \log k \, \Delta_{\dot{{\tt f}}} (k)\;,
\end{eqnarray}
is also invariant under re-scalings of the scale factor, given that the left-hand side of the above equation is also invariant under such transformation. For broader generality we write
\begin{eqnarray}
    \langle {\tt f}'^2 \rangle =  \int {\rm d} \log k \, \Delta_{{\tt f}'} (k)\;,
\end{eqnarray}
where ${\tt f}' \equiv \partial {\tt f}/\partial \eta$, with $\eta$ the $\alpha$-time. It follows immediately that $\Delta_{{\tt f}'} \equiv a^{2\alpha}\Delta_{\dot{{\tt f}}}$, requiring $\Delta_{{\tt f}'}$ to transform as 
$\propto C^{2\alpha}$ in order to keep $\Delta_{\dot{{\tt f}}}$ invariant. Identifying ${\tt f}({\bf x})$ again with a scalar field or (the spatial component of) a gauge field, leads to identify the mass dimensions $[ \langle {\tt f}'^2 \rangle] = +4$. As $[{\rm d} \log k] = 0$ and $[k^3] = +3$, it then follows that $[\Delta_{{\tt f}'}] = +4$ and $[\mathcal{P}_{{\tt f}'}] = +1$. We thus write an analogous parametrization as before,
\begin{eqnarray}\label{eqn:Varianceof fdotGeneralForm}
    \mathcal{P}_{{\tt f}'}(k) = \frac{a^{2\alpha}}{a^3} \mathcal{G}(k/a) \;,
\end{eqnarray}
with $\mathcal{G}(k/a)$ a function of mass dimension $[\mathcal{G}] = +1$, which characterizes the variance of ${\tt f}'$, and depends only on the modulus of the physical momentum $p=k/a$. This leads to
\begin{eqnarray}
    \langle {\tt f}'^2 \rangle = \int {\rm d} \log k \; a^{2\alpha} \frac{(k/a)^3}{2\pi^2} \mathcal{G}(k/a)\;.
\end{eqnarray}
As before, a dimensionless version in program units can be introduced by writing  $\widetilde{\mathcal{G}} = \mathcal{G}/M_*$, with $M_*$ a mass scale with dimensions $[M_*] = +1$. This leads to
\begin{eqnarray}\label{eq:fprime2programUnits}
    \langle \tilde{{\tt f}'}^2 \rangle = a^{2\alpha} \left(\dfrac{\omega_* M_*}{f_*^2}\right)\int {\rm d}\log \kappa \; \frac{(\kappa/a)^3}{2\pi^2} \widetilde{\mathcal{G}}(\kappa/a)\,,
\end{eqnarray}
with the dimensionless pre-factor again collapsing to $(\omega_*/f_*)^2$ for the choice $M_* = \omega_*$. It is important to note that contrary to Eq.~(\ref{eq:f2programUnits}), which is invariant under re-scalings $a(\eta) \rightarrow C\cdot a(\eta)$, Eq.~(\ref{eq:fprime2programUnits}) instead transforms as $\propto C^{2\alpha}$, due to the $a^{2\alpha}$ factor in front. Only for cosmic time ($\alpha = 0$),  Eq.~(\ref{eq:fprime2programUnits}) does become invariant under re-scalings of the scale factor. 

\subsubsection{Initial conditions from a general power spectrum}
\label{subsubsec:ICfromGeneralPS}

We describe now how to set up the initial condition of a field ${f}({\bf n})$ on the lattice, when its fluctuations are determined by an arbitrary power spectrum $\Delta_{\tt f}(k) \equiv {k^3}\mathcal{P}_{\tt f}(k)/2\pi^2$, where $\mathcal{P}_{\tt f}(k)$ is the variance of its continuous counterpart field in Fourier space, ${\tt f}({\bf k})$, {\it c.f.}~Eq.~\eqref{eq:continuumPSII}. While it is customary to set the spectrum $\Delta_{\tt f}(k)$ to mimic quantum vacuum fluctuations in most traditional applications, this does not need to be the case. For example, in scenarios where the dynamics of the field fluctuations are initially linear, one could first simulate the evolution of these fluctuations in a one-dimensional grid of momentum magnitudes, and then plug the resulting power spectrum on the lattice just before the dynamics becomes non-linear. 

To set the field fluctuations in Fourier space following a given power spectrum, one needs to sample from a Gaussian distribution
with vanishing mean and variance given by the power spectra. To see this, let us reproduce first the lattice counterpart of a continuum power spectrum, {\it c.f.}~Eq.~(\ref{eq:discretePST1}), 
\begin{eqnarray}\label{eq:discretePST2}
\Delta_{f}(k(|{\bf \tilde{n}}|)) \equiv {k(\tilde{\bf n})\over 2\pi}\frac{\delta x}{N^5} \#_{R(\tilde{\bf n})} \big\langle \big|{f}(\tilde{\bf n})\big|^2\big\rangle_{R(\tilde{\bf n})} 
= \frac{k^3(\tilde {\bf n})}{2\pi^2}\;{\Upsilon_{|\tilde{\bf n}|}}\;\left(\frac{\delta x}{N}\right)^3 \big\langle \big|{f}(\tilde{\bf n})\big|^2\big\rangle_{R(\tilde{\bf n})}\,.
\end{eqnarray}
Here $\langle  ...  \rangle_{R(\tilde{\bf n})}$ represents an angular average over a spherical shell $R(\tilde{\bf n})$ of radius $|\tilde{\bf n}^{\prime}|\in \big[|\tilde{\bf n}|,|\tilde{\bf n}|+ \Delta\tilde{n}\big)$, with $\Delta\tilde{n}$ a given radial binning, and $\#_{R(\tilde{\bf n})} $ the bin {\it multiplicity}, {\it i.e.}~the number of lattice sites within each spherical shell. In the second equality we have re-introduced the quantity in Eq.~(\ref{eq:Upsilon})
\begin{eqnarray}\label{eq:UpsilonII}
\Upsilon_{|\tilde{\bf n}|} \equiv \frac{\#_{R(\tilde{\bf n})}}{4\pi|\tilde{\bf n}|^2}\;,
\end{eqnarray} 
which controls whether one is using {\tt Type-I} ($\#_{R(|\tilde{\bf n}|)}$ exact, $\Upsilon_{|\tilde{\bf n}|} \neq 1$) or {\tt Type-II} ($\#_{R(|\tilde{\bf n}|)} = 4\pi|\tilde{\bf n}|^2$, $\Upsilon_{|\tilde{\bf n}|} = 1$) power spectra, see the discussion in Sect.~\ref{subsec:PS}. 
 
Denoting the average from statistical sampling of the Gaussian distribution
as $\langle ... \rangle_{\rm stat}$, and identifying $\langle  ...  \rangle_{R(\tilde{\bf n})} \longrightarrow \langle ... \rangle_{\rm stat}$, Eq.~(\ref{eq:discretePST2})
implies that the variance of the  
field fluctuations 
$\langle |{f}(\tilde{\bf n})|^2 \rangle_{\rm stat}$ on the lattice is
\begin{eqnarray}
     \langle |{f}(\tilde{\bf n})|^2 \rangle_{\rm stat} = \frac{k^2(|{\bf \tilde{n}}|) {N}^5}{\pi  \delta x  \#_{R(\tilde{\bf n})}} \,\mathcal{P}_{{\tt f}} (k) = 
     {1\over \Upsilon_{|\tilde{\bf n}|}}\left(\frac{{N}}{\delta \tilde{x}}\right)^3 \mathcal{P}_{\tt f}(k)\,, 
\end{eqnarray}
which in program units and using Eq.~\eqref{eqn:VarianceGeneralForm}
becomes
\begin{eqnarray}\label{eq:varF}
\langle |\tilde {f}(\tilde{\bf n})|^2 \rangle_{\rm stat} = \left(\frac{\omega_*^3}{f_*^2 E_*}\right)\frac{\kappa^2(|\tilde{\bf n}|)  {N}^5}{\pi  \delta \tilde x \#_{R(\tilde{\bf n})}}{\widetilde{\mathcal{F}}(\kappa/a)\over a^3}
= \left(\frac{\omega_*^3}{f_*^2 E_*}\right)
{1\over \Upsilon_{|\tilde{\bf n}|}}\left(\frac{{N}}{\delta \tilde{x}}\right)^3 {\widetilde{\mathcal{F}}(\kappa/a)\over a^3}\,.
\end{eqnarray}
The analogous expression for the time derivative variance reads
\begin{eqnarray}\label{eq:varDF}
     \langle |\tilde {f}'(\tilde{\bf n})|^2 \rangle_{\rm stat} = \left(\frac{\omega_* E_*}{f_*^2 }\right)\frac{\kappa^2(|\tilde{\bf n}|)  {N}^5}{\pi  \delta \tilde x \#_{R(\tilde{\bf n})}}{\widetilde{\mathcal{G}}(\kappa/a)\over a^{3-2\alpha}}
= \left(\frac{\omega_* E_*}{f_*^2 }\right)  
{1\over \Upsilon_{|\tilde{\bf n}|}}\left(\frac{{N}}{\delta \tilde{x}}\right)^3 {\widetilde{\mathcal{G}}(\kappa/a)\over a^{3-2\alpha}} \;. 
\end{eqnarray}

\noindent  
We can therefore write general expressions for the variances on the lattice as
\begin{eqnarray}\label{eq:varFandDFgeneral}
\langle |\tilde {f}(\tilde{\bf n})|^2 \rangle_{\rm stat} \equiv {\mathcal{C}_*\over \Upsilon_{|\tilde{\bf n}|}} \left(\frac{{N}}{\delta \tilde{x}}\right)^3 {\widetilde{\mathcal{F}}(\kappa/a)\over a^3}\,,~~~~ \langle |\tilde { f}'(\tilde{\bf n})|^2 \rangle_{\rm stat} \equiv {\mathcal{D}_*\over \Upsilon_{|\tilde{\bf n}|}} \left(\frac{{N}}{\delta \tilde{x}}\right)^3 {\widetilde{\mathcal{G}}(\kappa/a)\over a^{3-2\alpha}}
\end{eqnarray}
where
\begin{eqnarray}
\mathcal{C}_* \equiv \left\lbrace 
\begin{array}{cl}
\left(\frac{\omega_*^3}{f_*^2 E_*}\right) & ,~[\,{\rm General}\,]\vspace*{2mm}\\
\left(\frac{\omega_*}{f_*}\right)^2 & ,~[\,E_* = \omega_*\,]
\end{array}\right.\,,
\quad\quad\quad 
\mathcal{D}_* \equiv \left\lbrace 
\begin{array}{cl}
 \left(\frac{\omega_* M_*}{f_*^2}\right) & ,~[\,{\rm General}\,]\vspace*{2mm}\\
\left(\frac{\omega_*}{f_*}\right)^2 & ,~[\,M_* = \omega_*\,]
\end{array}\right.\,.
\end{eqnarray}
In practice, we initialize a field ${\tilde f}(\tilde{\bf n}) = {\tilde R}(\tilde{\bf n}) + {\tilde I}(\tilde{\bf n})$, and its time-derivative ${\tilde f}'(\tilde{\bf n}) = {\tilde R}'(\tilde{\bf n}) + {\tilde I}'(\tilde{\bf n})$, on each point $\tilde {\bf n}$ of the reciprocal lattice, by drawing independent random realizations of their real and imaginary parts, from Gaussian distributions with vanishing mean and variance given by ${1\over2}$ times the expressions in~(\ref{eq:varFandDFgeneral}), respectively. Alternatively, we can also
initialize the fields ${\tilde f}(\tilde{\bf n}) = |{\tilde f}(\tilde{\bf n})|e^{i\alpha(\tilde {\bf n})}$ and ${\tilde f}'(\tilde{\bf n}) = |{\tilde f}'(\tilde{\bf n})|e^{i\beta(\tilde {\bf n})}$, by taking $\alpha({\bf \tilde{n}})$ and $\beta({\bf \tilde{n}})$ as two random independent phases drawn from a uniform distribution over the range $[0, 2\pi)$, and drawing the values of $|{\tilde f}(\tilde{\bf n})|$ and $|{\tilde f}'(\tilde{\bf n})|$ from a {\it Rayleigh} distribution with expected square amplitudes given by (\ref{eq:varFandDFgeneral}). The two procedures are equivalent.

\subsubsection{Examples of initial conditions}
\label{subsubsec:ICexamplesGeneralPS}

We now present some applications of the procedure outlined in the previous section. We first discuss the standard case of initial conditions set from quantum vacuum fluctuations. In the case of scalar or gauge fields, 
these are characterized by the expectation values (variance)
\begin{eqnarray}\label{eqn:scalarvacuumPS}
    \mathcal{P}_{\tt f} (k) = \frac{1}{a^3} \frac{1}{2\omega_k}~~{\rm and}~~ \mathcal{P}_{{\tt f}'} (k) = \frac{1}{a^{3-2\alpha}} {\omega_k\over 2}\,, ~~~~~{\rm with}~~~  \omega_k \equiv \left[(k/a)^2 + m_{\tt  f}^2\right]^{1/2}\,,
\end{eqnarray}
where $m_{\tt f}$ is the mass of the field. From here we extract 
\begin{eqnarray}
\mathcal{F}(k/a) \equiv \frac{1}{2}\left[(k/a)^2 +  m_{ \tt f}^2\right]^{-1/2}\,,~~~~{\rm and}\,~~~\mathcal{G}(k/a) \equiv \frac{1}{2}\left[(k/a)^2 + m_{\tt f}^2\right]^{1/2}\,,
\end{eqnarray}
or, equivalently, in program variables,
\begin{eqnarray}
\widetilde{\mathcal{F}}(\kappa/a) \equiv \frac{1}{2}\left[(\kappa/a)^2 +  {\tilde m}_{\tt f}^2\right]^{-1/2}\,,~~~{\rm and}~~~~ \widetilde{\mathcal{G}}(\kappa/a) \equiv \frac{1}{2}\left[(\kappa/a)^2 + {\tilde m}_{\tt f}^2\right]^{1/2}\,,
\end{eqnarray}
where ${\tilde m}_{\tt f} \equiv m_{\tt f}/\omega_*$. In order to set up a spectrum mimicking quantum vacuum fluctuations, one just needs to plug these expressions in Eq.~(\ref{eq:varFandDFgeneral}), and sample the phase and the modulus from the relevant uniform and Rayleigh distributions, respectively. This is valid for both scalar fields and the components (in a cartesian or chiral basis) of gauge fields.

As a second example, we consider the physics case discussed in Sect.~\ref{sec:ricci-reheating-example}. In that example, the non-minimal coupled field to gravity is initialized on the lattice at some time, namely e-fold $N_0$, with some fluctuations that do not follow the canonical power spectrum of vacuum fluctuations. This power spectrum is determined from solving the homogeneous linear dynamics of the modes $\sigma_k$ introduced in Eq.~\eqref{eq:sigma-eq}. Deep in inflation, when all lattice modes are well inside the horizon and the evolution is adiabatic, the amplitude of the modes $\sigma_k$ is initially given by the Bunch-Davies (Minkowski) vacuum $\sigma_k \simeq e^{-ik/aH}/\sqrt{2k}$. From here, we evolve Eq.~\eqref{eq:sigma-eq} on the background $(a(\mathcal{N}),R(\mathcal{N}))$ until the starting time of the lattice $\mathcal{N}_0$, chosen before non-linearities become relevant. The amplitude $\sigma_k$ and time derivative $\sigma_k'$ evaluated at time $\mathcal{N}_0$ are used to build the field amplitude $\chi$ and its derivative, $  \chi(\mathcal{N}_0,\mathbf{k})=\sigma(\mathcal{N}_0,\mathbf{k})/a(\mathcal{N}_0)$ and $\chi'(\mathcal{N}_0,\mathbf{k})=(\sigma'(\mathcal{N}_0,\mathbf{k})-\mathcal{H}(\mathcal{N}_0)\,\sigma(\mathcal{N}_0,\mathbf{k}))/a(\mathcal{N}_0)$. From these, we can build the functions $\mathcal{\tilde F}(\kappa/a(\mathcal{N}_0))$ and $\mathcal{\tilde G}(\kappa/a(\mathcal{N}_0))$ that determine the variance of the distribution from which the amplitudes of the scalar field fluctuations are sampled. For an extended description of this initialization procedure, see~\cite{Figueroa:2021iwm,Figueroa:2024yja,Figueroa:2024asq}. 

Finally, we also consider the system discussed in Sect.~\ref{sec:Axion}. There we presented the evolution of Abelian gauge fields, which initially lie in a vacuum configuration and are later excited through an axion-like coupling. In order to initialize the gauge field, we chose to decompose its amplitude ($A_i$) and time derivative ($E_i=\dot{A}_i$) in a chiral basis, {\it c.f.}~Eq.~\eqref{eq:chiralnBasisVector}, identifying the chiral (transverse) components $A^{\pm}, E^{\pm}$ and the longitudinal modes $ A^{\parallel}, E^{\parallel}$. As we want to initially satisfy the Gauss's constraint in the absence of scalar field fluctuations, we demand to have only transverse vectors, {\it i.e.}~$\vec{k}\cdot\vec{A}=\vec{k}\cdot\vec{E}=0$, or equivalently, we initially set the parallel component of the fields to zero, {\it i.e.}~$A^{\parallel} = 0$, $E^{\parallel} = 0$. The procedure to set the initial conditions for each remaining chiral components is equivalent to that of a scalar field. This means we write for each chiral amplitude the following
\begin{eqnarray}
A^{\pm}(k) &\simeq& f^{\pm}_k e^{ik/(aH)}\,,
\label{eqn:ABDcosmic}\\
E^{\pm}(k) &\simeq& -i g_k^{\pm} e^{ik/(aH)}\,,\label{eqn:EBDcosmic}
\end{eqnarray}
where $\tau \simeq -1/(aH)$ is the conformal time deep inside inflation, and $f^+_k, f^-_k, g^+_k, g^-_k$ are the independent amplitude realizations of a Gaussian random field with zero mean and variance given by the Bunch--Davies vacuum configuration, whose variance functions are given by $\mathcal{\widetilde F}_{\rm BD} (\kappa /a) = a/(2\kappa)$ and $\mathcal{\widetilde G}_{\rm BD}(\kappa/a) = \kappa/(2a)$. Note that it is vital to set the phase difference between the amplitude and its time derivative exactly to $\pi/2$, otherwise the modes will evolve incorrectly. 

As an alternative, it is possible to initialize the vector fields with one of the chiralities in an excited state, while the other remains in vacuum. For this, we first solve the linearized equation of motion given in Eq.~\eqref{eq:linA} in cosmic time $t$, and the amplitude and its time derivative evaluated at some moment are used to build the functions $\mathcal{\widetilde{ F}}_{\rm exc}$ and $\mathcal{\widetilde{ G}}_{\rm exc}$. For example, if one chooses to excite the positive chirality, the amplitude and time derivative of each chirality are given by 
\begin{eqnarray}
&\tilde A^{+}(\kappa)& \simeq {\rm F}^{+}_k \, e^{i\kappa/(aH)}\,,\\  
&\tilde A^{-}(\kappa)& \simeq {f}^{-}_k \, e^{i\kappa/(aH)}\,, \label{eqn:ABDcosmicExcited}\\
&\tilde E^{+}(\kappa)& \simeq -\,{\rm G}_k^{+} \, e^{i [\kappa/(aH) - \pi/2 \Theta(\kappa)]}\,, \\  
&\tilde E^{-}(\kappa)& \simeq -i \,{g}_k^{-} \, e^{i \kappa/(aH)}\,,\label{eqn:EBDcosmicExcited}
\end{eqnarray}
with $f_k^-$ and $g_k^-$ the amplitudes sampled from the vacuum configuration, and ${\rm F}_k^+$ and ${\rm G}_k^+$ are sampled from a distribution with variance given by $\langle |\tilde {A}_{\rm exc}(\tilde{\bf n})|^2 \rangle_{\rm stat}=\mathcal{\widetilde{ F}}_{\rm exc}$ and $\langle |\tilde {E}_{\rm exc}(\tilde{\bf n})|^2 \rangle_{\rm stat}=\mathcal{\widetilde{ G}}_{\rm exc}$. Finally, the function $\Theta(\kappa)$ is a parametrization of the transition between the excited branch and the vacuum branch of the power spectrum, extracted from the solution of the linearized mode's equation,  see~\cite{Figueroa:2024rkr} for a detailed explanation of this initialization procedure. 

\subsection{Initial conditions via algorithms}
\label{subsec:ICalgorithms}

In some circumstances, we need to apply some algorithm(s) that ‘evolve’ fields into a desired configuration, which is then used as the initial condition for our lattice simulation. This is different from the circumstance discussed in Sect.~\ref{subsubsec:ICexamplesGeneralPS}, where fluctuations are created according to a defined power spectrum, as the resulting initial field fluctuations follow a distribution that cannot be described simply by a power spectrum. In Sect.~\ref{subsubsec:ICalgorithmsProb} we discuss how to obtain a field configuration $\{f({\bf n})\}$ that samples a given non-Gaussian probability distribution $P(f)$, as often done {\it e.g.}~when studying fields close to thermal equilibrium. In Sect.~\ref{subsubsec:ICalgorithmsDefects}, on the other hand, we comment on initialization procedures to create defect networks close to scaling, albeit more briefly, as the considered methods have already been described described in detail in Sect.~\ref{sec:DefectsV}.

\subsubsection{Initial condition from probability distributions}\label{subsubsec:ICalgorithmsProb}

Here we consider methods to obtain field configurations $\{f({\bf n})\}$ that represent samples drawn from a given probability distribution $P(f)$. This is the typical situation when studying the classical real-time dynamics of fields close to thermal equilibrium, such as is done for the computation of the $SU(2)$ sphaleron rate, whose most precise prediction was given in~\cite{DOnofrio:2014rug}. In this case, the appropriate probability distribution is nothing else than the Boltzmann distribution $\mathcal{P}(f)\propto\exp(-\beta H)$, with $\beta$ an inverse temperature and $H$ the Hamiltonian of the system of interest\footnote{Another particularly interesting example, which has nothing to do with dynamics, is to consider the distribution $\exp(-S_E)$ with $S_E$ an Euclidean action. This allows for directly computing the Euclidean path integral of field theories numerically and is at the core of lattice QCD simulations, see for instance \cite{Gattringer:2010zz}.}. In this subsection, we review the procedure to produce such configurations. 

The type of algorithms needed to simulate random processes are usually referred to as {\it Monte-Carlo} algorithms. They all rely on variations of the same central idea: under specific conditions, random walks equilibrate and the equilibrium state can be described by a probability distribution. Turning this idea around makes it a very powerful tool: by designing appropriate random walks in field space, we can generate field configurations that follow the desired distribution. 

To understand how this is turned into a practical algorithm, it is in practice enough to consider~\textit{Markov chains}~\cite{Gattringer:2010zz}. These are random walks whose transition probability between two specific field configurations $A$ and $B$ depends only on $A$ and $B$ themselves, and not on the history or the random walk. We can represent the Markov chain schematically as 
\begin{equation}
f_1\xrightarrow{T(f_1|f_2)} f_2\xrightarrow{T(f_2|f_3)}  f_3\to \cdots\to f_n\to\cdots
\end{equation}
with $f_1$ the initial configuration and $T$ a transition probability that only depends on the current state of the chain and the next proposed state, but not on anything that happened previously.  In order to have a chain that reaches equilibrium, the transitions $T$ need to be such that the probability of arriving to a given state is equal to the probability of getting out of that state, {\it i.e.}~$\sum_f T(f'|f)P(f) = \sum_{f}T(f|f') P(f')$. Otherwise, the distribution would not be stationary. The most straightforward way to ensure this property is to impose the \textit{detailed balance} condition 
\begin{equation}
T(f'|f)P(f) = T(f|f') P(f')\,,
\end{equation}
which makes sure that the weighted probability of going from a state $f$ to $f'$ is the same as going from $f'$ to $f$. To summarize, to generate configurations in thermal equilibrium, one starts from an initial configuration, usually chosen as a random or a vacuum configuration, and updates it using some Monte-Carlo algorithm. Once the chain equilibrates after sufficient updates, all newly generated configuration are distributed according to the desired thermal distribution, and so can be used as a starting point for our lattice simulations.

Monte-Carlo algorithms thus create Markov chains that equilibrate to the correct distribution probability. However, choices of the transition probability lead to different behaviors of the chain (for instance, some equilibrate faster than others). We present next the Metropolis algorithm, the ``original" and most simple variation of this idea. We also comment on other variations and point the reader to relevant literature. \vspace*{-0.3cm}\\

{\bf Metropolis algorithm}. The Metropolis algorithm allows to generate a Markov chain of configurations that follow the correct Boltzmann distribution. At every step, starting from a configuration $f$, one proposes random updates to the fields $\{f+\delta f\}$, accepting them with a probability $\min(1,\exp(-\beta\Delta H))$, where $\Delta H$ is the change in the Hamiltonian after the update. If the new configuration is accepted, it is added to the Markov chain, and used as the starting point for the next Metropolis update. Instead if it is rejected, one duplicates the last configuration $f$ in the Markov chain, and restarts the update procedure. The proof that this transition probability satisfies detailed balance with the Boltzmann distribution as an equilibrium distribution is straightforward, and can be found {\it e.g.~}in~\cite{Gattringer:2010zz}. In the most relevant cases, the theory under consideration is local, and the updates can be made locally. This is crucial in order to have a practical algorithm, as any global update would lead to changes in the Hamiltonian that scale with the volume, leading to exponentially small update probabilities. 

In order to be concrete, we show how the Metropolis algorithm works in the case of a scalar field $\psi$ on a $3$-dimensional lattice, characterized by the following Hamiltonian
\begin{equation}
H=\sum_{\mathbf{n}}\left[\frac12  \dot{\phi}^2 +\frac{1}{2{\delta{x}^2}}\sum_{i=1}^3 \left({\phi}_{+i}-\phi\right)^2 + V(\phi)\right] \, .
\end{equation}
Given some configuration $f$, the idea of the algorithm is to updated, at each step, the field at a single position, $\phi(\mathbf{n)}_{new}=\phi(\mathbf{n)}+\delta\phi(\mathbf{n)}$, and accept or reject the update before moving to the next position. Here $\delta\phi(\mathbf{n)}$ a random uniform number within the interval $[-\epsilon,\epsilon]$, with $\epsilon$ some positive real number. The change in the Hamiltonian due to this update is easy to compute and, crucially, is almost local: it depends only on $\phi(\mathbf{n})$ and its nearest neighbors $\phi(\mathbf{n}+\hat{\imath})$, $i = 1,2,3$, so that
\begin{align}
\Delta H\equiv H_{\phi(\mathbf{n)}_{\rm new}} - H_{\phi(\mathbf{n)}} &= -\frac{1}{{\delta{x}^2}}\sum_{i}\left(\phi_{+i}+\phi_{-i}\right)\left(\phi(\mathbf{n)}_{\rm new}-\phi(\mathbf{n)}\right)+ \frac{3}{{\delta{x}^2}}\left(\phi(\mathbf{n})^2_{\rm new}-\phi(\mathbf{n})^2\right)\notag \\
&+ V(\phi(\mathbf{n})_{\rm new})- V(\phi(\mathbf{n})) \ . \label{eq:deltaHMetro}
\end{align}
~~~~After update a single site, the updated configuration $\phi(\mathbf{n})_{\rm new}$ is accepted with probability \\ $\min(1,\exp(-\beta\Delta H))$. If rejected, one instead keeps the original field. In other words, if the Hamiltonian decreases, the field change is always accepted. If it increases, the field change is accepted with a probability that is exponentially suppressed by the increase in the Hamiltonian. The acceptance probability depends on the parameter $\epsilon$ which sets the size of the typical field proposal. In practice, this parameter is usually tuned to get an acceptance probability of about $50\%$, reaching an appropriate compromise between acceptance probability and a fast exploration of field space (a too small $\epsilon$ leads to the update being  almost always accepted while almost not moving in field space, while a too large $\epsilon$'s leads to vanishing acceptance probabilities).

The main advantage of the Metropolis algorithm is its simplicity and flexibility. It is not particularly ``efficient", as its capability to explore the phase space is rather limited. Ideally we would like to consider the fields drawn from the Markov chain as independently distributed variables. In practice, however, fields are auto-correlated~\cite{Wolff:2003sm}, and we need to wait for a certain number of updates for the fields to `decorrelate'. For example, for the example of the scalar field described above, rather than adding the field configuration to the Markov chain after each step of Metropolis, one usually completes a full sweep of the lattice before appending to the set of configurations. 

One of the main aims of designing more efficient algorithms is to reduce the autocorrelation as much as possible. One possible alternative is the use of \textit{over-relaxation}, based on deterministic (``microcanonical") updates that do not change the value of the Hamiltonian, but map the field to different areas in the phase space. This needs to be done in combination with standard Metropolis steps to maintain ergodicity and explore field configurations with different energies. The practical implementation of this technique, however, depends on the specific theory. For example, we refer the reader to~\cite{Fodor:1994ih, Kajantie:1995kf} for its application in the context of the electroweak theory. 

The Metropolis accept-reject step itself is often also surpassed by more efficient algorithms, which are however theory specific. For example, one common method is the {\it Heatbath} algorithm, which relies on the fact that if local fields can be sampled directly from the local probability distribution $\exp(-\beta\Delta H)$, the accept-reject step is not necessary. This is sometimes possible and leads to vast improvements. See~\cite{Kajantie:1995kf} for an example in the electroweak sector, or~\cite{Gattringer:2010zz} for pure $SU(N)$ theories.

Finally, another method which is most often used in Lattice QCD for its flexibility and efficiency is the Hybrid Monte-Carlo approach. In this case, the random update is replaced by a deterministic trajectory of the field in an auxiliary theory. By adding fiducial degrees of freedom, the fields can be made to follow equations of motion that produce new configurations that are in local equilibrium. For a detailed explanation we refer again to~\cite{Gattringer:2010zz}. 

\subsubsection{Initial condition for scaling  networks of cosmic defects}
\label{subsubsec:ICalgorithmsDefects}

Another type of algorithmic initializations are the procedures used to simulate networks of cosmic defects. In these scenario, one aims to study defect networks that evolve in the {\it scaling regime}, as once this is reached, the macroscopic properties of the network are expected to be independent of the UV details of the underlying field theory. The main challenge is that a network of defects originated after a phase transition may take too long to reach this regime in a lattice simulation, before resolution-loss problems of the defects become manifest on the lattice or the simulation results are affected by causality-violation problems due to the periodicity of the lattice~\cite{Hindmarsh:2019csc,Hindmarsh:2021vih,Correia:2024cpk}. As we have discussed in detail these problems and their solution in Sect.~\ref{sec:DefectsV}, here we just give a brief account of the methods employed. 

The main goal of this initialization is to use a set of artificial procedures to initially bring the network of defects close to the scaling regime, before the standard physical evolution is applied. Typically, one first considers some random initialization method that mimics the symmetry breaking process of the underlying scalar field(s). For example, for cosmic strings, we create (following the method in Sect.~\ref{subsubsec:ICfromGeneralPS}) scalar field fluctuations with the power spectrum in Eq.~(\ref{eq:initialPSstrings}). The latter depends on a length scale $\ell_\text{str}$, which is a tuneable parameter that controls the initial density of the resulting network. For other defects, similar methods mimicking symmetry breaking can be employed, the details of which are not particularly relevant. Whichever the random process initially used, the resulting configuration is often evolved afterwards with a dissipative method, either using directly a diffusion equation~\cite{Hindmarsh:2019csc,Hindmarsh:2021vih,Correia:2024cpk}, or artificially increasing the Hubble friction in the EOM~\cite{Notari:2025kqq}. This dissipates the excess energy from the symmetry breaking process more efficiently than canonical evolution in an expanding background, rapidly leading to well-formed and localized defect configurations in the network. After this, the network is expected to arrive much faster to the scaling regime via canonical evolution. The diffusion equations are presented in \cref{eq:defects:diffusionglobal,eq:defects:diffusionlocal,eq:difusDW}, for global strings, local strings, and domain walls, respectively.

Due to the natural loss of resolution of defect cores during physical evolution, it is also common to complement the above dissipative process with a subsequent resolution-preserving method. The most commonly used method is the so-called {\it fattening} technique~\cite{Press:1989yh,Bevis:2006mj,Moore:2001px}, by which the comoving width of the defects is artificially maintained constant, ensuring the resolution of the defect width till the end of the simulation. Alternatively, one can use the {\it extra-fattening} method, where the fields (previously diffused) are initially evolved with a set of equations that allow the comoving core radius to grow proportionally to the scale factor. This is then followed by a phase of standard evolution in which the comoving width of the strings decreases again, so that the defect resolution at the end of the simulation is the same as it was at the beginning. The extra-fattening phase can be considered as part of the preparation of the initial condition, so that the defects are sufficiently well resolved on the lattice when scaling is finally reached during physical evolution. The form of the equations of motion of the fields during these special phases is presented for global and local cosmic strings in \cref{eq:defects:globalfatteningeom,eq:defects:localfatteningeom}, respectively, and depends on a single parameter $s$ that controls the regime: extra-fattening ($s = -1$), fattening ($s = 0$), and physical evolutions ($s = 1$). An analogous fattening method for domain walls has also been introduced in Eq.~\ref{eq:defects:DWfatteningeom}.

\subsection{Initial condition via spatial profiles}
\label{subsec:ICspatialProfile}

Finally, we comment briefly on the use of spatial field configurations as initial conditions. This is the case in which all fields and their conjugate momenta are initially set to certain values on all lattice sites, according to a given spatial profile, which can be known either analytically or numerically. Setting up a profile requires typically a multi-step procedure, as we describe below for some  examples. After setting up the final step, fields can  be evolved from then onward, according to the corresponding evolution kernels of the system.

A common case where this initialization technique is used, is to obtain isolated topological defects, typically aiming at understanding particular structures or configurations of these. The most studied case is that of cosmic strings, where a spatial profile can be constructed initially from the known solution of an infinite static defect, such as the Nielsen-Olsen vortex in the local case~\cite{Nielsen:1973cs}. To this semi-analytic solution, several modifications might still be applied, depending on the desired final configuration. In particular, once static solutions are set on the lattice, operations of interest that can be applied, range from boosting the strings to obtain a non-static configuration with non-zero conjugate momentum~\cite{Matsunami:2019fss,Saurabh:2020pqe,Baeza-Ballesteros:2023say,Baeza-Ballesteros:2024otj}, to multiplying the profiles by another defect solution to obtain a configuration with multiple defects~\cite{Matsunami:2019fss,Saurabh:2020pqe,Baeza-Ballesteros:2023say,Baeza-Ballesteros:2024otj}, or shifting them in the transverse direction, to obtain a string with a fixed structure~\cite{Drew:2019mzc,Drew:2022iqz,Drew:2023ptp}, such as a sine-like profile. 
In the latter case, in which the initial conjugate momenta is set to zero, it is often common to perform a short 
phase of diffusion afterwards, to minimize the effects of the transformations.

Introducing a spatial profile as an initial condition, might require the modification of the profile itself, in order to accommodate periodic boundary conditions on the lattice. In the case of strings, periodic boundaries enforce the total topological charge associated to strings to be zero, forbidding solutions with a single infinite string. One typically solves this issue by introducing a pair of infinite straight strings with opposite winding numbers, ensuring that the total topological charge vanishes. Even when considering initial conditions with topological charge zero, we may still need to modify the field profiles far from the defect cores, to ensure periodicity on the lattice~\cite{Matsunami:2019fss,Saurabh:2020pqe,Baeza-Ballesteros:2023say,Baeza-Ballesteros:2024otj}. Alternatively, one can use  dissipative (Sommerfeld) boundary conditions in some direction, allowing in this way for a non-zero topological charge~\cite{Drew:2019mzc,Drew:2022iqz,Drew:2023ptp}.

It is also worth mentioning another method that has been used for the initialization of local static strings of arbitrary shape~\cite{Hindmarsh:2021mnl,Hindmarsh:2017qff,Baeza-Ballesteros:2024otj}, which relies on the use of the compact discretization of the gauge fields. In this technique, one first sets the magnetic flux of those plaquettes that should contain the string core to $2\pi$, while setting the scalar field everywhere to the vacuum with a fixed phase. After a short phase of diffusion, one ends up with a definite string configuration with approximately the chosen shape. We note, however, that this procedure leads to the presence of a ``frozen'' magnetic flux. While this flux has been argued not affect the dynamics of the strings, a more systematic study of the presence of this remanent structure is still missing. 

Another example of a system in which one may want to set initial conditions via spacial profiles are first order phase transitions. In these processes  scalar-field bubbles nucleate by quantum tunneling through a potential barrier. In lattice simulations, bubbles are typically introduced by hand at the nucleation points, following a given spacial profile. 

One of the most commonly used bubble profiles is the so-called {\it critical profile}, which corresponds to
the most likely field configuration of the bubbles (critical bubbles are those just energetic enough to expand and avoid collapse). The
profile of critical bubbles is invariant under four-dimensional
Euclidean rotations~\cite{Coleman:1977py}, {\it i.e.}~it obeys an
$\mathrm{O}(4)$ symmetry, which allows to write the bubble profile
as a function of a single variable, $\rho = (\tau_\text{E}^2 + r^2)^{1/2}$, with $r$ the spatial distance from the bubble center and $\tau_\text{E}\equiv it$ the Euclidean time. In the thin wall limit ,the scalar field profile of a critical bubble is characterized by a simple $tanh$-ansatz~\cite{Cutting:2018tjt,Cutting:2020nla}, which depends on the radius $R_c$ of the critical bubble, and a width $\delta_c$ interpolating between the unbroken and the broken phases (hence controlling the thickness of the critical bubble wall). Alternatively, some lattice simulations consider nucleated bubbles by inserting a scalar field configuration with a Gaussian profile~\cite{Hindmarsh:2013xza,Hindmarsh:2015qta,Hindmarsh:2017gnf}, with
radii slightly larger than the critical bubble radius $R_c$. More precise bubble profiles can be obtained using~{\tt CosmoTransitions}~\cite{ Wainwright:2011kj}, a publicly available package  that finds the temperature-dependent phase minima of a given model, its critical temperatures, and the actual nucleation temperatures and tunneling profiles of the transition. 

We note that in a thermal phase transition, once the bubbles are created, these expand and perturb the surrounding medium -- the fluid --, which represents particle species coupled to the scalar field responsible for the phase transition. One needs therefore to incorporate such scalar-fluid coupling in the simulations~\cite{Hindmarsh:2013xza,Hindmarsh:2015qta,Hindmarsh:2017gnf}. An in-depth discussion on lattice techniques applied to {\it scalar-gauge-fluid} (SGF) dynamics, will be presented in an upcoming review, {\tt The Art-III}, currently under preparation as part of our monograph series on LCT.

\section{Scalar dynamics in $d+1$ dimensions} \label{sec:Sims2D}
So far we have focused on simulations of field theories on lattices with three spatial dimensions, which we refer to as simulations in $(3+1)$-dimensions, where the second number refers to the temporal component. The aim of this section is to generalize previous formalisms to lattice simulations in $(d+1)$-dimensions with an arbitrary number of spatial dimensions $d$, with particular emphasis on the $d=2$ case, due to its relevance in the literature. We restrict ourselves to singlet scalar field theories, {\it i.e.}~we do not consider the simulation of scalar-gauge theories or metric perturbations in $d \neq 3$ dimensions.

Simulating scalar fields on lattices with $d \neq 3$ spatial dimensions may be of interest for two main reasons. First of all, one might want to simulate a field theory intrinsically defined in $d \neq 3$ dimensions, with scalar fields propagating {\it e.g.}~in a (2+1)-dimensional FLRW spacetime, $ds^2 = a^2(\eta) (-d\eta^2 + \delta_{ij} dx_i dx_j)$ with $i=1,2$. In this case, the corresponding equations of  motion in $(2+1)$-dimensions, can be discretized and solved with adequate evolution algorithms, just as much as we have done for the theories in $(3+1)$-dimensions presented in previous sections. Alternatively, one might be interested in simulating a field theory defined in three spatial dimensions, but implemented in a lattice of lower dimensionality, with the aim of approximating the three-dimensional dynamics at a reduced computational cost. This is justified as long as there is statistical isotropy in the system, both at the level of the initial conditions and the dynamics, as it is typically the case for scalar field interactions. Reducing the dimensionality, for example from $d=3$ to $d=2$, can be quite beneficial, accelerating the simulation by a factor $\sim N$, with $N \approx 10^2 - 10^3$ in most applications. This enable investigations that require very long simulation times, very large lattices, or extensive scans over model parameters. The ability to accurately capture three-dimensional dynamics using lower-dimensional simulations, must be however assessed on a case-by-case basis. 

In this section we focus on the second type of circumstance, this is, the case in which we want to simulate $(3+1)$-dimensional physics on a $(d+1)$-dimensional lattice, particularly with $d = 2$ or $d = 1$. In \cref{sec:2dim-theory}, We generalize the definitions and properties of a lattice to an arbitrary number of spatial dimensions $d$, and also discuss the setting of initial field fluctuations on such a lattice. As an example, results from $(2+1)$-dimensional simulations of a preheating model are presented in \cref{sec:2dim-examples}, where we additionally compare them with the results from fully fledged $(3+1)$-dimensional simulations of the same model, showing significant agreement of the results.

\subsection{Lattice definition and properties in $d+1$ dimensions}\label{sec:2dim-theory}

We begin this section by generalizing the basic definitions and concepts of a $d$-dimensional lattice, similarly as what we did in {\it c.f.}~\cref{sec:LatticeTechniques} for the case $d=3$. We consider a cubic $d$-dimensional lattice with $N$ points per dimension, side length $L$, lattice spacing $\delta x \equiv L/N$, and periodic boundary conditions. The points of such a lattice can thus be labeled by the vector of integers, 
\begin{eqnarray}
    {\bf n} = (n_1, \dots,  n_d), ~~~~{\rm with}~~
    n_i = 0, \dots , N-1  \,,~~~{\rm for}~~ i  = 1,\dots,d\,,
\end{eqnarray}
with ${\bf x} = {\bf n}\,\delta x$ representing comoving spatial coordinates on the lattice. Associated to such a position space lattice, we can define a \textit{reciprocal lattice} representing momentum coordinates, whose sites can be tagged by the following vector,
\begin{eqnarray}
    \tilde{\bf n} = (\tilde n_1, \dots, \tilde n_d), ~~~~{\rm with}~~
    \tilde n_i = -\frac{N}{2}+1, -\frac{N}{2}+2, ... ,-1,0,1, ... , \frac{N}{2} - 1, \frac{N}{2}  \,,~~~  i  = 1,\dots,d\,.
\end{eqnarray}
The discrete Fourier transform (DFT) in $d$ spatial dimensions  of a lattice function $f$ is defined as follows,
\begin{eqnarray}
    f({\bf n}) \equiv {1\over N^d}\sum_{\tilde {\bf n}} e^{-i{2\pi\over N} {\bf \tilde n n}} f({\bf \tilde n}) ~~~~ \Longleftrightarrow ~~~~  f({\bf \tilde n}) \equiv \sum_{\bf n} e^{+i{2\pi\over N} {\bf n \tilde n} }f({\bf n})\,,\label{eq:DiscreteFTdim} 
\end{eqnarray}
which recovers the usual expression from~\cref{eq:FTdiscrete} for $d=3$. Similarly, Eq. (59) in \texttt{The Art-I} can be generalized as follows,
\be
\sum_{\bf n} e^{i{2\pi\over N} {\bf n \tilde n} }=N^d\delta_{0,{\bf \tilde n}}\,.\label{eq:DFTidentityDims}
\ee
The minimum and maximum momentum per dimension resolved by the $d$-dimensional cubic lattice are the same as in the $d=3$ case, i.e.~$k_{\rm IR} = 2 \pi /L$ and $k_{\rm UV}=(N/2) k_{\rm IR}$ respectively, {\it c.f.}~Eq.~(\ref{eq:IRandUVmodes}). However, the maximum momentum $k_{\rm max}$ that can be captured, which corresponds to the diagonal of the lattice, is now
\begin{eqnarray}\label{eq:kmax}
    k_{\rm max}=\sqrt{k_{1,\rm UV}^2+...+k_{d,\rm UV}^2}=\frac{\sqrt{d}}{2}Nk_{\rm IR}=\sqrt{d}\frac{\pi}{\delta x}\,.
\end{eqnarray}

Let us think now of the simulation of a scalar field propagating in a $(3+1)$-dimensional FLRW spacetime, but on a lattice of $(d+1)$-dimensions. The discretization of the scalar field equations of motion and their integration by evolution algorithms is analogous to the $d=3$ case. The only  difference is that now one needs to remember that the Laplacian term in the equation of motion, Eq.~\eqref{eq:singlet-eom}, sums over $d$ spatial directions, i.e.~$\nabla^2 \phi = \sum_{i=1}^{d} \partial^2 \phi / \partial x_i^2$. A similar change applies to the gradient component of the scalar field's energy density, ${G}_{\phi} = \frac{1}{2 a^2} \sum_{i=1}^d (\partial_i \phi)^2$, {\it c.f.}~Eq.~\eqref{eq:energy-contrib}. Note, in addition, that we aim at simulating the $(3+1)$-dimensional model in a lattice of lower dimensionality, and so we use the form of the equations of motion in three spatial dimensions, as given in Eq.~\eqref{eq:singlet-eom}.

We also discuss how the power spectrum of the scalar fields is defined on a $d$-dimensional lattice, which is relevant for the measuring of spectral observables and to set-up initial conditions following some Gaussian distribution of fluctuations---see \cref{subsec:ArbitrarySpectrum} for an in depth discussion of the initialization procedure. In a $(d+1)$-dimensional theory, the ensemble average of a discrete function $f$ is substituted by the following $d$-dimensional volume average,
\be
\langle { f}^2 \rangle_V = \frac{\delta x^d}{L^d}\sum_{\bf n} { f}^2({\bf n}) = {1\over N^{2d}}\sum_{\tilde{\bf n}} \left| { f} (\tilde{\bf n})\right|^2 \,,
\ee
where we have used the Fourier transform given by Eq.~\eqref{eq:DiscreteFTdim} and the identity~\ref{eq:DFTidentityDims} (note that we recover~\cref{eq:Averagef2} for $d=3$). Following a similar derivation as that presented in~\cref{sec:LatticeTechniques}, this expression can be further developed by decomposing the sum over in radial and angular parts. It follows that
\be\label{eq:dDimf2Ave}
\langle {f}^2 \rangle_V = \frac{1}{2 \pi} {\delta x\over N^{2d-1}} \sum_{|\tilde{\bf n}|}\Delta\log k(\tilde{\bf n})~k(|\tilde {\bf n}|) ~\#_{|\bf \tilde{n}| }^{(d)} ~\big\langle \big|{f}(\tilde{\bf n})\big|^2\big\rangle_{R(\tilde{\bf n})} \ ,  
\ee
where $k(|\tilde {\bf n}|) \equiv k_{\rm IR}|\tilde {\bf n}|$, and 
\be \langle ( ... ) \rangle_{R(\tilde{\bf n})} \equiv \frac{1}{\#_{|\bf \tilde{n}| }^{(d)}}\sum_{\tilde{\bf n}^{\prime}\in R(\tilde{\bf n})}( ... )\,,\ee
represents an angular average over a $(d-1)$-dimensional `spherical' shell $R(\tilde{\bf n})$ of radii $|\tilde{\bf n}^{\prime}| \in \big[|\tilde{\bf n}|,|\tilde{\bf n}|+ \Delta\tilde{n}\big)$, with $\Delta\tilde{n}$ a given radial binning, and $\#_{|\bf \tilde{n}| }^{(d)}$ the \textit{multiplicity}, {\it i.e.}~the number of lattice points contained within such shell. The exact values of the multiplicities depend on the radius and width of each shell. In the case of the canonical choice $\Delta\tilde{n} = 1$, they can be approximated as 
\begin{itemize}\label{eq:multiplicity_approx}
    \item $d=1: \,\,\,\#_{|\bf \tilde{n}| }^{(1)} = 2$, 
    \item $d=2: \,\,\,\#_{|\bf \tilde{n}| }^{(2)} \simeq 2 \pi |\tilde{\bf n}| = L\, k(|\tilde {\bf n}|)$,
    \item $d=3: \,\,\,\#_{|\bf \tilde{n}| }^{(3)} \simeq 4 \pi  |\tilde{\bf n}|^2 = \frac{L^2}{\pi} k^2(|\tilde {\bf n}|)$,
    \item ...
    \item $d = ...: \,\,\,\#_{|\bf \tilde{n}| }^{(d)} \simeq \Omega_d  |\tilde{\bf n}|^{d-1} = {L^{d-1}\over 2^{d-2}\pi^{(d-2)/2}\Gamma(d/2)}k^{d-1}(|\tilde {\bf n}|)$\,, ~~~$\Omega_d \equiv {2\pi^{d/2}\over \Gamma(d/2)}$ {\small ($d$-dimensional {\it solid angle})}\,,
\end{itemize}
where $\Gamma(x)$ is the Gamma function. These expressions approximate\footnote{Notice that for $d=1$ the multiplicity is not an approximation as there are exactly two lattice points per canonical bin.} quite well the exact multiplicities at most momenta scales except at very infrared and ultraviolet modes, see~\cref{subsec:PS} for more details. 

Using these results, we can introduce  a generalization of the power spectrum of scalar fluctuations $\Delta_{f}(k)$ to a $d$-dimensional lattice. As discussed in~\cref{sec:LatticeTechniques}, a function ${\tt f}(\bf x)$ defined in the continuum is represented on a lattice by a discrete function $f({\bf n})$, which takes the same value as the continuum function, 
{\it i.e.}~$f({\bf n}) \equiv {\tt f}({\bf x} = {\bf n} \, \dx)$. By identifying the volume average on the lattice, $\langle f^2 \rangle_V$, with the statistical ensemble average in the continuum, $\langle {\tt f}^2 \rangle$, {\it c.f.}~\cref{eq:continuumPS}, we can infer a lattice equivalent of the power spectrum of a field in 3-dimensions. Generalizing to $d$-spatial dimensions, we write
\be
\langle {\tt f}^2 \rangle = \int d\log k~\Delta_{\tt f}(k)~~, ~~~\Delta_{\tt f}(k) \equiv {\Omega_d k^d\over (2\pi)^d}\mathcal{P}_{\tt f}(k)~~,~~~ \langle {\tt f}_{\bf k} {\tt f}_{{\bf k}^{\prime}}^* \rangle = (2\pi)^d \mathcal{P}_{\tt f}(k) \delta (\mathbf{k}-\mathbf{k^{\prime}})\,,~ \label{eq:dDimsPS}
\ee
and comparing this definition against Eq.~(\ref{eq:dDimf2Ave}), we obtain a lattice-power spectrum as
\begin{eqnarray}
\Delta_{f} (k(|\tilde {\bf n}|)) &=&  {1\over 2\pi}{\delta x \over N^{2d-1}} k(|\tilde {\bf n}|) \#_{|\bf \tilde{n}|}^{(d)} ~\big\langle \big|{f}(\tilde{\bf n})\big|^2\big\rangle_{R(\tilde{\bf n})}\\
&=&  {\#_{|\bf \tilde{n}|}^{(d)}k^d(|\tilde {\bf n}|)\over (2\pi)^d |{\bf \tilde{n}}|^{d-1}} ~\left(\frac{\delta x}{N}\right)^d ~\big\langle \big|{f}(\tilde{\bf n})\big|^2\big\rangle_{R(\tilde{\bf n})}\\
&\equiv& \frac{k^d(\tilde {\bf n})}{2^{d-1}\,\pi^{d/2}\,\Gamma(d/2)} \,\Upsilon_{|{\bf \tilde n}|}^{(d)} ~\left(\frac{\delta x}{N}\right)^d ~\big\langle \big|{f}(\tilde{\bf n})\big|^2\big\rangle_{R(\tilde{\bf n})}\,,\label{eq:dDimDelta}
\end{eqnarray}
which reduces to Eq.~\eqref{eq:discretePST1}
for $d=3$, as it should. Here $\Upsilon_{|{\bf \tilde n}|}^{(d)}$ is a generalization of the $d=3$ factor defined in Eq.~\eqref{eq:Upsilon}, given by
\begin{eqnarray}
    \Upsilon_{|{\bf \tilde n}|}^{(d)} \equiv {\#_{|\bf \tilde{n}|}^{(d)}\over \Omega_d |{\tilde{\bf n}}|^{d-1}} = \frac{\#_{|\bf \tilde{n}|}^{(d)}\Gamma(d/2)}{2\,\pi^{d/2} \, |{\tilde{\bf n}}|^{d-1}}\,.
\end{eqnarray}
By inverting Eq.~(\ref{eq:dDimDelta}), we obtain the following expression for the field variance, which is used to initialize the scalar fluctuations
\begin{eqnarray}\label{eq:nDimVariance}
\big\langle \big|{f}(\tilde{\bf n})\big|^2\big\rangle_{R(\tilde{\bf n})} = \frac{2^{d-1}\,\pi^{d/2}\,\Gamma(d/2)}{k^d(\tilde {\bf n})} \, \frac{1}{ \Upsilon_{|{\bf \tilde n}|}^{(d)}} \, \left(\frac{N}{\delta x }\right)^d \, \Delta_{f}(k(|\tilde {\bf n}|))\equiv \frac{1}{ \Upsilon_{|{\bf \tilde n}|}^{(d)}} \, \left(\frac{N}{\delta x }\right)^d \, \mathcal{P}_{f}(k(|\tilde {\bf n}|)) \,.
\end{eqnarray}
For $d=1,2$ and $3$, this becomes
\begin{align}
    &\hspace{-0.39cm} \bullet \,\,d=1: \,\,\,\,\,\,\big\langle \big|{f}(\tilde{\bf n})\big|^2\big\rangle_{R(\tilde{\bf n})}  = \frac{\pi}{k(\tilde {\bf n})} \, \frac{1}{ \Upsilon_{|{\bf \tilde n}|}^{(1)}} \, \left(\frac{N}{\delta x }\right) \, \Delta_{f}(k(|\tilde {\bf n}|)) \equiv \frac{1}{ \Upsilon_{|{\bf \tilde n}|}^{(1)}} \, \left(\frac{N}{\delta x }\right) \, \mathcal{P}_{f}(k(|\tilde {\bf n}|)) \ , \\
    &\hspace{-0.39cm}  \bullet \,\,d=2: \,\,\,\,\,\,  \big\langle \big|{ f}(\tilde{\bf n})\big|^2\big\rangle_{R(\tilde{\bf n})}  = \frac{2\,\pi}{k^2(\tilde {\bf n})} \, \frac{1}{ \Upsilon_{|{\bf \tilde n}|}^{(2)}} \, \left(\frac{N}{\delta x }\right)^2 \, \Delta_{f}(k(|\tilde {\bf n}|)) \equiv \frac{1}{ \Upsilon_{|{\bf \tilde n}|}^{(2)}} \, \left(\frac{N}{\delta x }\right)^2 \, \mathcal{P}_{f}(k(|\tilde {\bf n}|))\ ,\label{eq:var2d}\\
    &\hspace{-0.39cm} \bullet \,\,d=3: \,\,\,\,\,\,  \big\langle \big|{ f}(\tilde{\bf n})\big|^2 \big\rangle_{R(\tilde{\bf n})}  = \frac{2\,\pi^{2}}{k^3(\tilde {\bf n})} \, \frac{1}{ \Upsilon_{|{\bf \tilde n}|}^{(3)}} \, \left(\frac{N}{\delta x }\right)^3 \, \Delta_{f}(k(|\tilde {\bf n}|))  \equiv \frac{1}{ \Upsilon_{|{\bf \tilde n}|}^{(3)}} \, \left(\frac{N}{\delta x }\right)^3 \, \mathcal{P}_{f}(k(|\tilde {\bf n}|))\ . \label{eq:var3d}
\end{align}

With this in mind, it is straightforward to initialize the fields following a Gaussian distribution of fluctuations at each point of the reciprocal lattice, with the variance given by the above expressions for any chosen power spectrum. As in the case of $d=3$ dimensions, to achieve this one must take into account that in momentum space, $f(\tilde{\bf n})$ is a complex field. The Gaussian distribution can thus be obtained by letting the absolute value $|f(\tilde{\bf n})|$  follow a Rayleigh distribution with variance given by Eq.~(\ref{eq:nDimVariance}), and arbitrary phase drawn from a uniform distribution within $[0,2\pi)$, as explained at the end of~\cref{subsubsec:ICfromGeneralPS}.

\subsection{Working example: preheating in (2+1)-dimensions}  \label{sec:2dim-examples}

In the following, we illustrate the ability of simulations in $d=2$ spatial dimensions to mimic $(3+1)$-dimensional physics, by studying an example of post-inflationary preheating dynamics. For this purpose we consider a model consisting of two singlet scalar fields, an inflaton field $\phi$ and a `daughter' field $\chi$, with the following potential,
\be\label{Chap2D:Potential}
V(\phi,\chi)=V_{\rm inf}(\phi)+V_{\rm int}(\phi,\chi)=\frac{1}{2}\Lambda^4 \, {\rm tanh}^2\left( \frac{\phi}{M}\right)+\frac{1}{2}g^2\phi^2\chi^2, \ee
where $\Lambda$ and $M$ are two mass scales, and $g$ is a dimensionless coupling constant. The inflationary potential $V_{\rm inf}(\phi)$ is inspired by the class of $\alpha$-attractor T-models~\cite{Kallosh:2013hoa}, while the interaction part $V_{\rm int}(\phi,\chi)$ serves as a portal coupling to reheat the universe.

Before presenting the numerical results, we describe how inflation and preheating proceed in this model. The inflaton potential is quadratic around the minimum $\phi=0$, while it flattens for field amplitudes $|\phi|\gg M$ towards the plateau $V_{\rm inf} \rightarrow \Lambda^4/2$. The transition between these two regimes is determined by the inflection points $\phi_{\rm i}/M =\pm {\rm arcsinh}(\sqrt{1/2}) \simeq \pm 0.658$. Slow-roll inflation occurs when the homogeneous inflaton $\bar \phi\equiv \phi(t)$ lies at large field amplitudes (we take $\phi > 0$ without loss of generality). The inflaton slowly rolls towards the minimum of the potential, triggering the end of inflation approximately at the field amplitude $\phi_* \equiv (M/2) {\rm arcsinh}\left( \sqrt{8}\mpl /M\right)$, when the first potential slow-roll parameter obeys $\varepsilon_V(\phi_*)=\mpl^2 V'^2/V^2\equiv1$. CMB observations \cite{Planck:2018jri} constrain the values of the two mass scales $\Lambda$ and $M$. The observed amplitude of the scalar perturbations yields a constraint $\Lambda\equiv \Lambda (M,N_k)$, with $N_k$ the number of e-folds of inflation, while the upper bound of the tensor-to-scalar ratio, $r<0.036$~\cite{BICEP:2021xfz}, imposes an experimental upper bound $M\lesssim 8.5\mpl$, see e.g.~\cite{Antusch:2022mqv,Figueroa:2024yja}.

At the end of inflation, the energy budget is completely dominated by the homogeneous inflaton $\bar{\phi}$, which starts oscillating around the minimum of its potential. The oscillation frequency is given by its effective mass,
\be\label{Chap2D:OscFrequ}
m^2(\phi)\equiv \frac{\partial^2 V_{\rm inf}}{\partial \phi^2}=\frac{\Lambda^4}{M^2}\left({\rm sech}^4 \left( \frac{\phi}{M} \right) - 2\, {\rm sech}^2 \left( \frac{\phi}{M} \right){\rm tanh}^2 \left( \frac{\phi}{M} \right)\right)=\frac{\Lambda^4}{M^2}-\frac{4\Lambda^4\phi^2}{M^4}+ ... , \,
\ee
which sets the typical time scale of the preheating dynamics. The amplitude of the oscillating inflaton decreases due to the expansion of the universe, and after few oscillations it becomes $|\phi| \ll M$. The oscillation frequency is then approximately given by the first term in the last identity of Eq.~(\ref{Chap2D:OscFrequ}), which can be identified with the mass of the inflaton, $m_\phi\equiv \Lambda^2/M$. During the oscillatory regime, the inflaton's amplitude decays as $\bar{\phi}(t)\propto t^{-1}$, and the oscillation-averaged equation of state is given by $\bar{w}=0$~\cite{Turner:1983he}. 

The initial oscillatory dynamics gives rise to a process called \textit{parametric resonance}~\cite{Kofman:1994rk}. Through the inflaton-daughter interactions, the oscillations of the homogeneous inflaton excites fluctuations $\delta \tilde{\chi}_k$ of the daughter field, up to a given cutoff. 
The daughter field modes grow expontially, until they start backreacting on the dynamics of the inflaton, leading to the decay of the inflaton and the termination of parametric resonance. Once the system enters fully into this non-linear backreaction regime, re-scattering effects between the inflaton and the daughter field lead to populate higher momentum modes,  broadening of the initial spectrum due to parametric resonance.

The initial oscillations of the inflaton suggest the following choice of rescaling parameters, see Eq.~\eqref{eq:GaugeProgramVar}: $\alpha=0$, $f_*=\phi_*$ and $\omega_*=\Lambda^2/M$. The potential in Eq.~\eqref{eq:ProgramPotMultiScalar} can then be written in program variables as follows,
\begin{align}\label{Chap2D:Potential2}
\tilde{V}(\tilde{\phi},\tilde{\chi}) \equiv \frac{1}{f_*^2 \omega_*^2}V(\tilde{\phi},\tilde{\chi}) &=\frac{1}{2}\left( \frac{M}{\phi_*}\right)^2 {\rm tanh}^2\left( \frac{\phi_*\tilde{\phi}}{M}\right)+\frac{1}{2}q_*\tilde{\phi}^2\tilde{\chi}^2 = \frac{1}{2}\tilde{\phi}^2+...+\frac{1}{2}q_*\tilde{\phi}^2\tilde{\chi}^2 \, ,
\end{align}
where we have also defined the \textit{resonance parameter} $q_*\equiv g^2\phi_*^2 /\omega_*^2$, and the last identity is the approximation around the minimum of the potential. The dynamics of the system are described on the lattice by the corresponding discretized field EOM, given in Eqs.~\eqref{eq:EOMScalar-Discr_Hybrid} and \eqref{eq:EOMScaleFactor-Discr_Hybrid}, which we have evolved with the velocity-verlet scheme (see Sect.~\ref{subsubsec:SymplecticInt}). 
\begin{figure}[t]
   \centering
    \includegraphics[width=7cm]{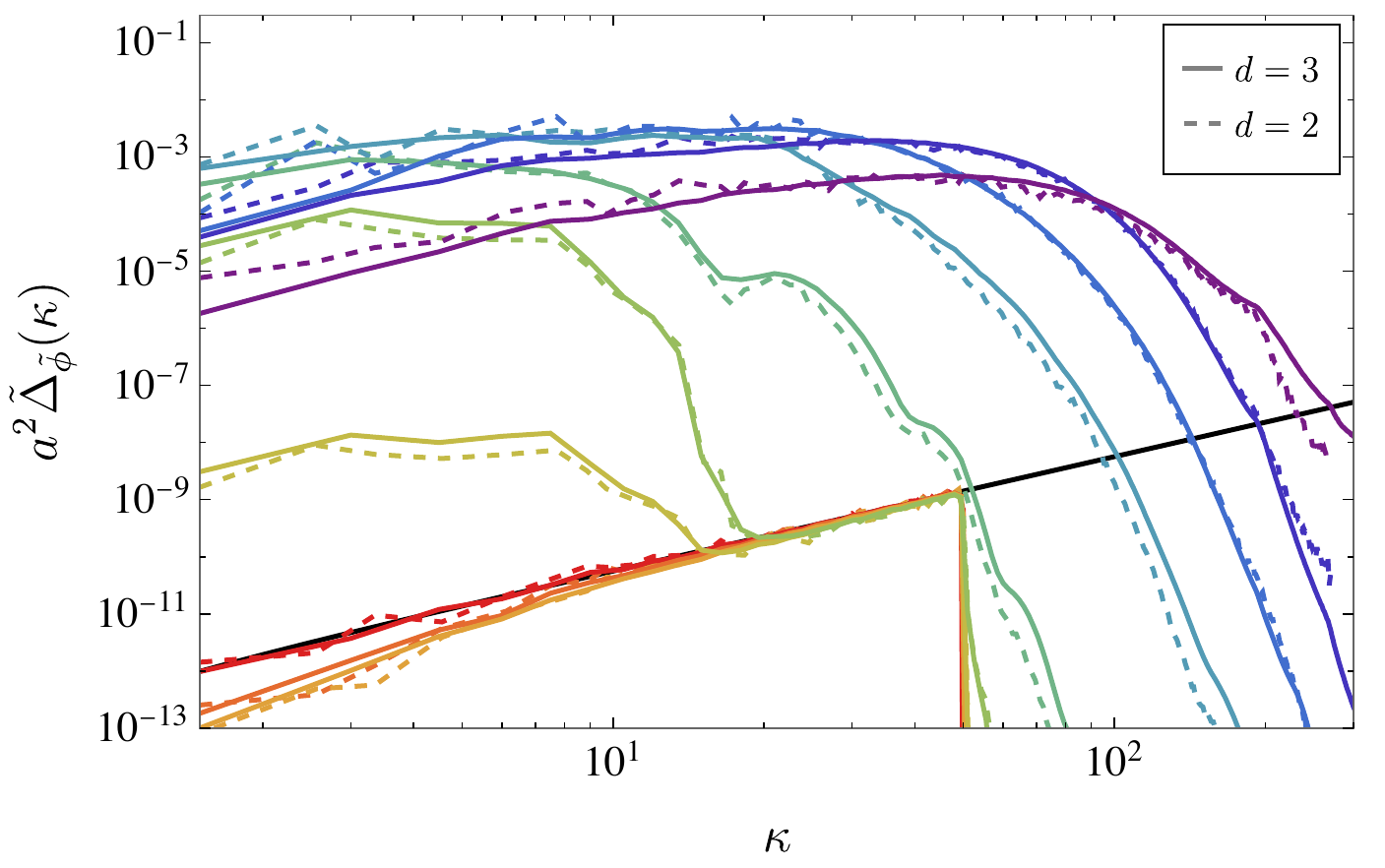} \hspace{0.2cm}
    \includegraphics[width=7cm]{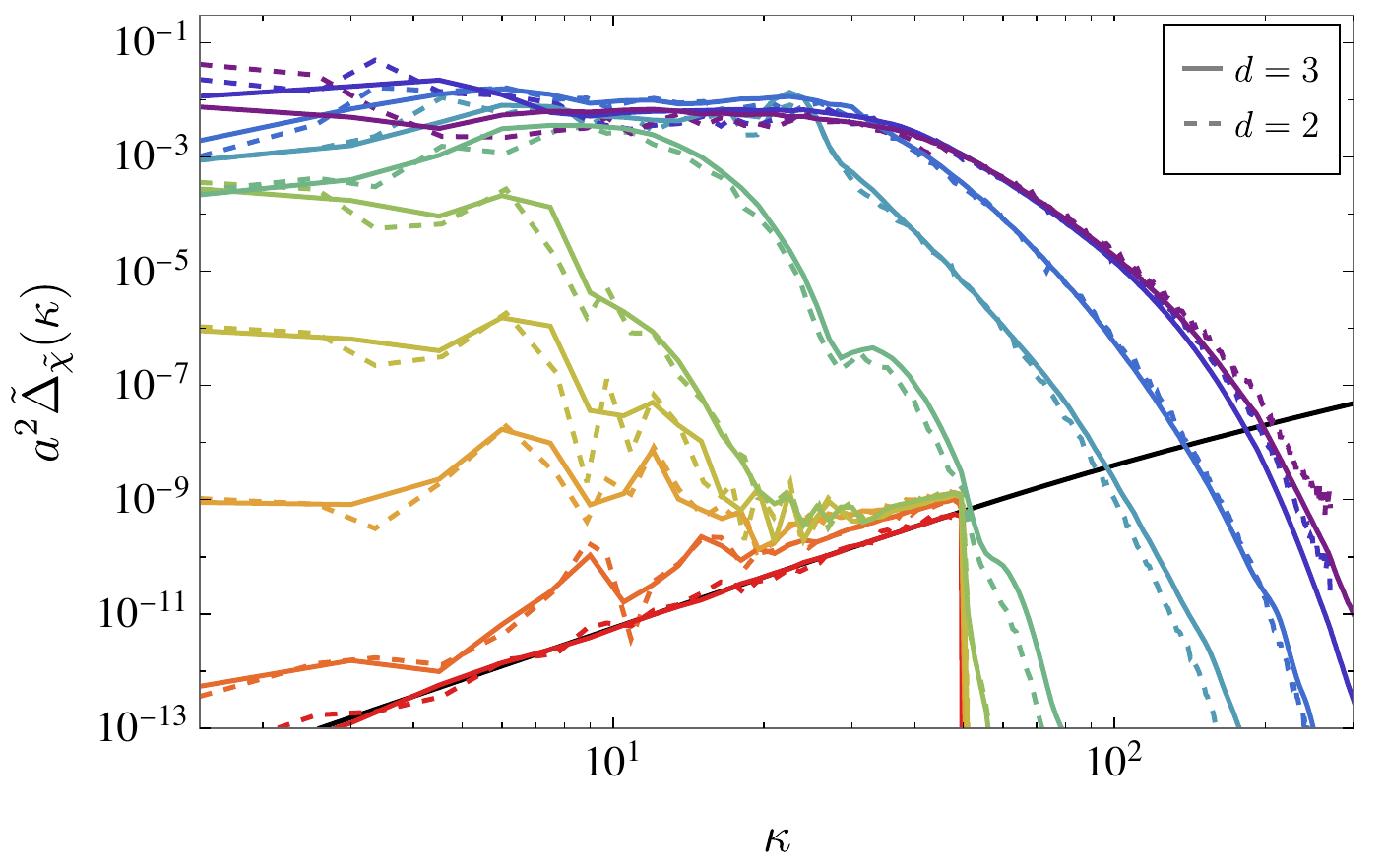} \vspace{0.5cm} 
    
    \includegraphics[width=7cm]{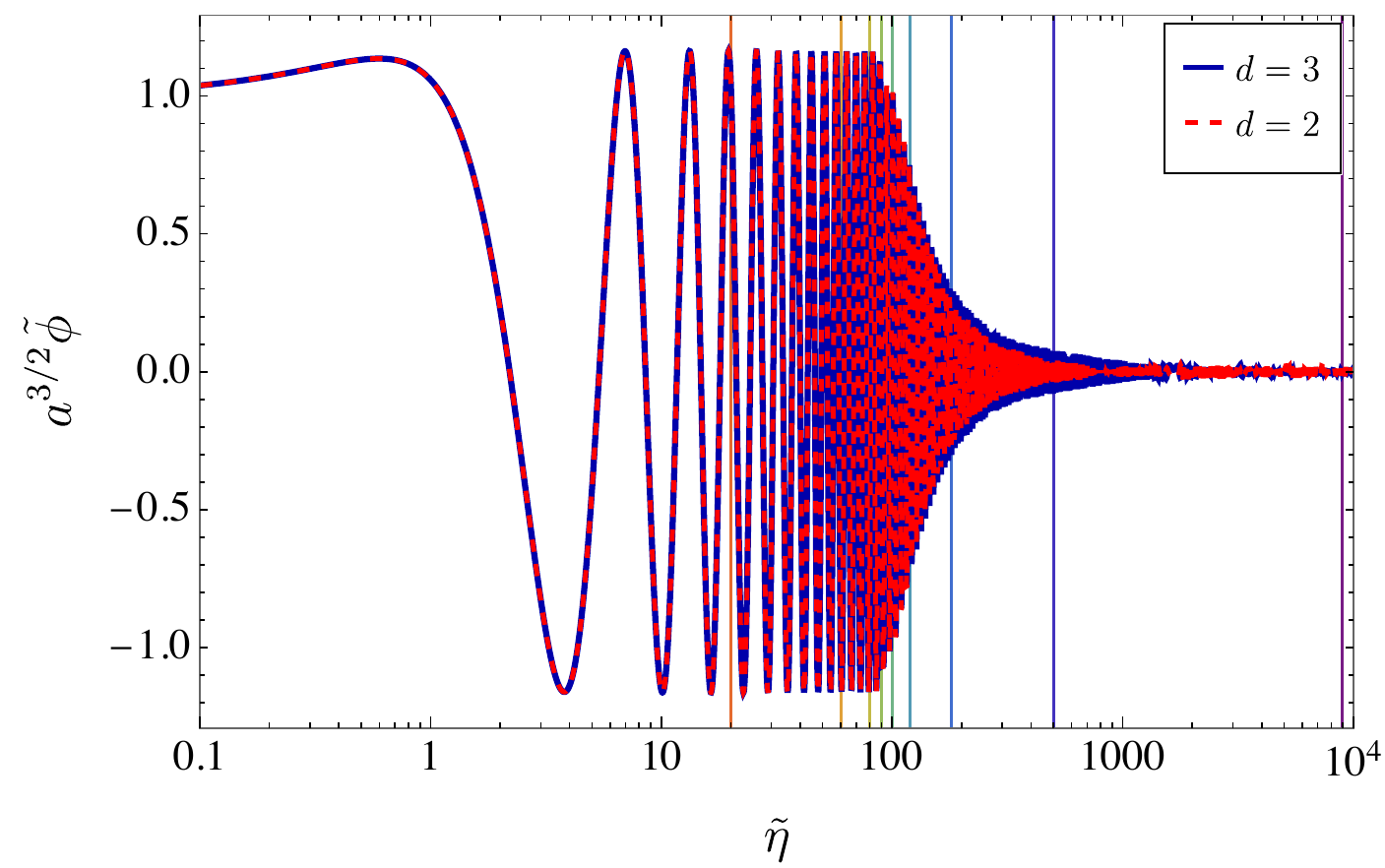} \hspace{0.2cm}
    \includegraphics[width=7cm]{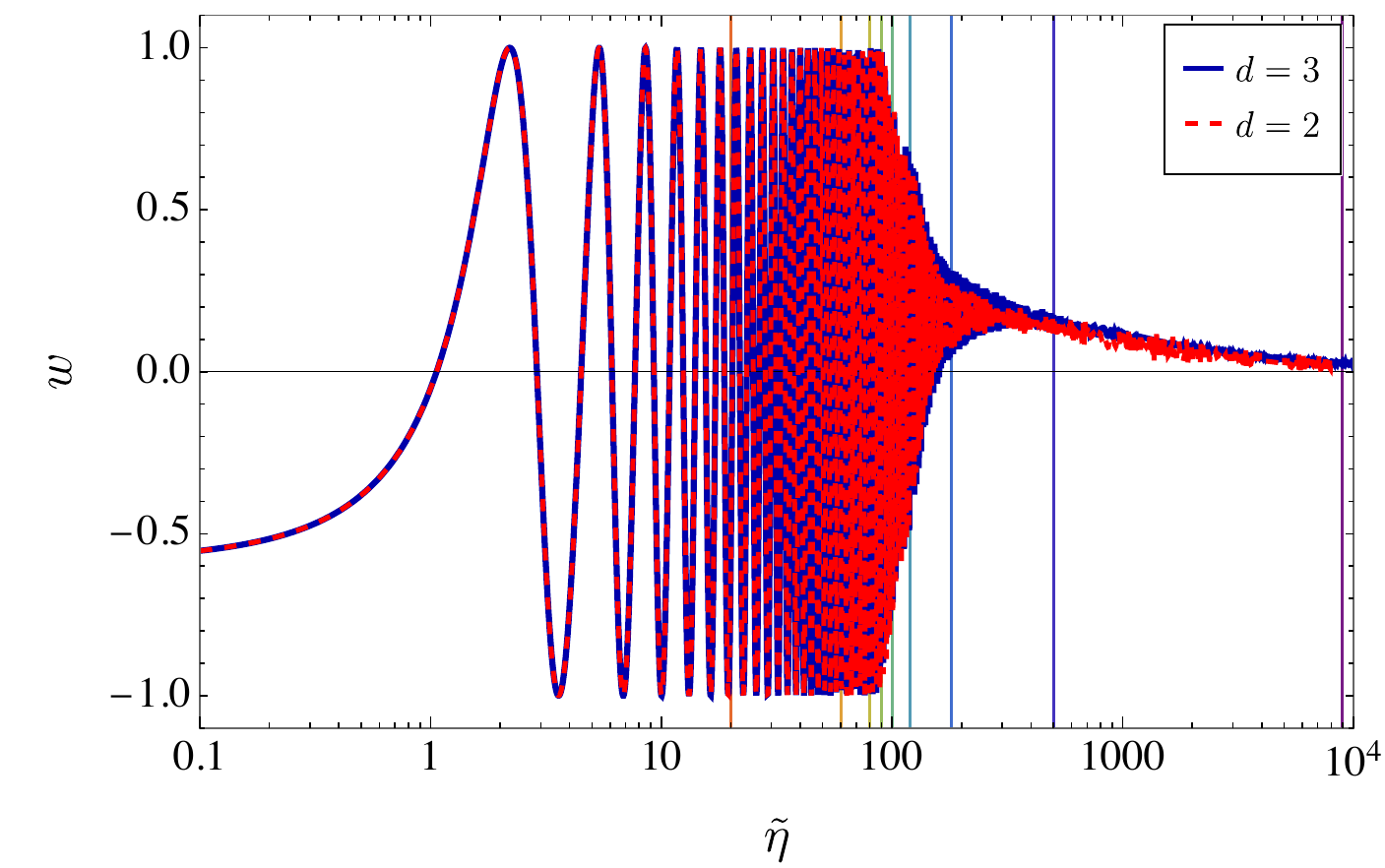} \vspace{0.3cm}
    \caption{Comparison of results from simulations of the model described in Sect.~\ref{sec:2dim-examples} in 3+1 (solid lines) and 2+1 dimensions (dashed) for the case $M=5 \mpl$, $N_k=60$ and $q_*=10^4$. {\bf Top}: Inflaton (top-left) and daughter field spectra (top-right) rescaled by $a^2 (\tilde{\eta})$ as a function of $\kappa \equiv k /\omega_*$. They are shown at different time instances, evolving from red at $\tilde{\eta}=0$ to purple at $\tilde{\eta}=10^4$. The coloured vertical lines in the lower panels indicate the times at which the spectra have been taken (the first spectrum at $\tilde{\eta}=0$ is not indicated). The black solid line shows the analytical continuum spectrum of initial fluctuations. {\bf Bottom}: Evolution of the rescaled inflaton field $a^{3/2} \tilde{\phi}$ (bottom-left) and equation of state $w$ (bottom-right). Blue corresponds to the 3+1 and red to the 2+1 dimensional simulation.} \label{fig:spectra-comparison1}
\end{figure}

\vspace*{0.5cm}
{\bf Comparison of 3+1 and 2+1 dimensional lattice simulations}. We have simulated the preheating stage of this model in both $(2+1)$- and $(3+1)$-dimensions for the choice $M=5 m_p$, $N_k=60$ (which fixes $\Lambda^4=9,84\times10^{-10}\mpl^4$) and $q_*=10^4$. We initialize the system at the end of inflation with homogeneous values $\bar{\phi}_*=1.35\mpl$ and $\dot{\bar{\phi}}_*=-4.28\times10^{-6}\mpl^2$, as well as $\bar{\chi}_*=\dot{\bar{\chi}}_*=0$, on top of which we impose initial quantum vacuum fluctuations given by Eq.~\eqref{eqn:scalarvacuumPS}, up to a UV cutoff $\kappa_c=50$. We have set the lattice parameters to $N=256$ and $\kappa_{\rm IR}=1.5$. Let us remark that since both simulations are obtained for the same values of $N$ and $\kappa_{\rm IR}$, the maximum momentum captured in each lattice $\kappa_{\rm max}$, see Eq.~\eqref{eq:kmax}, is larger by a factor $\sqrt{3/2}$ for $d=3$ compared to the $d=2$ case. 

The results of our comparison are presented in Fig.~\ref{fig:spectra-comparison1}, where solid lines corresponding to the (3+1)-dimensions case, and the dashed to the (2+1)-dimensional simulation. The top panels in each figure show the field spectra $\tilde{\Delta}_{{\phi}}(\kappa)$ and $\tilde{\Delta}_{{\chi}}(\kappa)$ at different times, where $\tilde{\Delta}_{f} \equiv \Delta_{f} /f_*^2$. The initial discretized power spectra captures well the initial spectrum of vacuum fluctuations in the continuum, depicted by the black line. The evolution of the spectra of both fields is approximated well by the (2+1)-dimensional simulation, which includes the excitation of modes during the initial resonance stage and also the occupation of higher modes during the following non-linear regime. In particular, the position and amplitude of the peaks in $\tilde{\Delta}_{{\phi}}(\kappa)$ and $\tilde{\Delta}_{{\chi}}((\kappa))$ coincide in both simulations. We do however observe slightly stronger fluctuations in the $d=2$ spectra (specifically for lower $\kappa$ values) due to the lower number of lattice points contained in the momentum bins, especially in the most infrarred ones. 

The lower panels show the volume-averaged inflaton amplitude $\tilde{\phi}$ and equation of state $w\equiv \bar{p}/\bar{\rho}$ (which determines the evolution of the scale factor). We can see here as well that the (3+1)-dimensional (blue) evolution is approximated well by the (2+1)-dimensional (red) case. This includes the initial homogeneous phase, which is matched almost exactly, the onset of backreaction, and the evolution during the non-linear stage, where the equation of state returns to $w=0$. 

This similarity between simulations in $(2+1)$- and $(3+1)$-dimensions is actually also found for other inflaton potentials of the form $V_{\rm inf} \propto {\rm tanh}^p(|\phi|/M)$ for $p > 2$, as long as $M\gg \mpl$~\cite{Antusch:2021aiw}, and also for other types of interactions~\cite{Antusch:2025ewc}. However, the applicability of lower dimensional simulations needs to be checked on a case by case basis, and will not always work. For example, model realizations of Eq.~\eqref{Chap2D:Potential} with $M\ll \mpl$ give rise to the formation of oscillons~\cite{Gleiser:1993pt,Copeland:1995fq,Amin:2011hj}, which are massive, long-lived, compact field configurations. In these cases one observes that $(2+1)$-dimensional simulations do not approximate well the $3+1$ dimensional ones. In general, whenever compact 3-dimensional configurations are expected to form in random locations in the 3-dimensional model (oscillons in this case, but could be also topological defects in other scenarios), reducing the dimensionality of the simulations will not be a good strategy (typically because statistical isotropy is already broken to some extent, but also because reducing the dimensionality of space might prevent directly the formation of the original compact objects).

\section{Gravitational waves}
\label{sec:GW}
We focus now on how to implement the dynamics of gravitational waves (GWs) on a lattice. We review first in Sect.~\ref{subsec:GWcont} the definition of GWs and their dynamics in the continuum. In Sect.~\ref{subsec:GWlattice},  we present lattice analogues of tensor metric perturbations representing GWs, a discretized version of the equations of motion of GWs in an expanding background, and a proper lattice definition of the energy density power spectrum of a GW background (GWB). 

\begin{mdframed}
{\bf Note -.} This section on GWs is unrelated to the non-canonical aspects (interactions, initial conditions, and dimensionality) discussed in the previous Sections~\ref{sec:NMCoupled_scalars} - \ref{sec:Sims2D} of this monograph. We take the opportunity to present here the theoretical basis of our implementation(s) of GW dynamics on a lattice, as previous to this manuscript, our methods for simulating GWs had been only presented in our publicly available 
\href{http://cosmolattice.net/technicalnotes}{\color{blue}\it \CL -- Technical Notes}, which are neither published, nor available in the ArXiv. By presenting the methods in this review, proper credit will be given to our implementation of GW dynamics on a lattice. As a novelty, we include here a new algorithm that improves on the memory requirements of previous methods. 
\end{mdframed}

\subsection{Gravitational waves in the continuum}\label{subsec:GWcont}

Gravitational waves (GWs) are identified as tensor perturbations $h_{ij}$, {\it i.e.}~transverse and traceless fields, on top of a FLRW background
\begin{equation}
\text{d}s^2 = -a^{2\alpha}(\eta)\text{d}\eta^2 + a^2(\eta)(\delta_{ij} + h_{ij})\text{d}x^i\text{d}x^j\:, \quad\quad {\rm with}\quad\quad\partial_i h_{ij}=0\,,~h_{ii}=0\,.
\end{equation}
The dynamics of GWs is linear, and expressed in $\alpha$-time, is described by an EOM of the form~\cite{Caprini:2018mtu}
\begin{equation}\label{eqn:EoMGWs}
      {h}_{ij}'' + (3-\alpha) \frac{a'}{a} h_{ij}' - a^{-2(1-\alpha)}\nabla^2 h_{ij} = \dfrac{2 }{m_p^2 a^{2(1-\alpha)}}\Pi_{ij}^{\text{TT}}\,, 
\end{equation}
where the source of the GWs  is the transverse-traceless (TT) part of the anisotropic tensor $\Pi_{ij}^\text{TT}$, which obeys the transversality and tracelessness conditions, $\partial_i \Pi_{ij}^\text{TT} = \Pi_{ii}^\text{TT}=0$, $\forall \,{\bf{x}}, \eta$. The anisotropic stress tensor $\Pi_{ij}$, describes the deviation of the total energy momentum tensor $T_{\mu\nu}$, from the form of a perfect fluid. The spatial components, relevant for the production of GWs, read
\begin{equation}\label{eq:gws:Pitensor}
   \Pi_{ij} \equiv T_{ij} - \bar pg_{ij}\:,
\end{equation}
with $\bar p$ the homogeneous background pressure and $g_{ij}$ the spatial-spatial part of the perturbed FLRW metric. Below, we define an \textit{effective} anisotropic stress tensor ${\Pi}^{\text{eff}}_{ij}$, which contains only the parts of $\Pi_{ij}$ that do not vanish after projecting to their TT-part, {\it i.e.}~verifying $(\Pi_{ij}^{\rm eff})^{\rm TT} = \Pi_{ij}^{\rm TT}$. For a general theory with Lagrangian as in~\cref{eq:lagrangian}, including canonically normalized real singlet ($\lbrace \phi_b \rbrace$), $U(1)$-charged complex ($\lbrace \varphi_b \rbrace$), and $[SU(N) \times U(1)]$-charged doublet ($\lbrace \Phi_b \rbrace$) scalar fields, as well as the corresponding Abelian and non-Abelian gauge fields, $A_{\mu}$ and $B_{\mu} \equiv B_{\mu}^c T_c$, respectively, it takes the form~\cite{PhDthesisFigueroa} 
 \begin{eqnarray}\label{eq:PiEff}
 \begin{array}{rcl}
{\Pi}^{\text{eff}}_{ij} & =&\displaystyle  \partial_i {\phi}_{b}\: \partial_j {\phi}_{b} + 2\, \text{Re}\left[(D^A_i {\varphi}_{b})^*\:(D^A_j {\varphi}_{b})\right] + 2\, \text{Re}\left[(D_i {\Phi}_{b})^\dagger\:(D_j {\Phi}_{b})\right]   \\[10pt] 
& & \displaystyle \hspace*{1.59cm}- \left(a^{-2\alpha}E_i E_j + a^{-2} B_i B_j\right) \hspace*{0.25cm}- \left(a^{-2\alpha} E_i^c E_j^c + a^{-2}  B_i^c B_j^c\right)\,,
\end{array}
\end{eqnarray}
with $E_j$ and $B_j$ the Abelian 
electric and magnetic fields, in this same order, and $E_j^c$ and $B_j^c$ the non-Abelian counterparts, as defined in~\cref{eq:ElectricMagneticDefs}. 

\begin{mdframed}
{\bf Important Note -.} The form of Eq.~(\ref{eq:PiEff}) remains valid in the case of an axion-like field interacting with gauge fields (either Abelian or non-Abelian), as {\it e.g.}~considered in Sect.~\ref{sec:Axion}. It is also applicable in the case of (the physical evolution of) topological defects, such as domain walls or cosmic strings, as described in Sect.~\ref{sec:DefectsV}. The form of ${\Pi}^{\text{eff}}_{ij}$, however, needs to be modified to account for the non-canonical interactions considered in Sect.~\ref{sec:NMCoupled_scalars}, like in the case of non-minimally coupled scalars to gravity (Sect.~\ref{sec:NMC}), or scalars with non-minimal kinetic terms (Sect.~\ref{sec:NMKfields}). It is also important to note also that strictly speaking the above description of GWs only applies to $(3+1)$-dimensions, as there is no proper notion of GWs in the $(2+1)$- or $(1+1)$-dimensional spaces considered in Sect.~\ref{sec:Sims2D}. 
\end{mdframed}

Obtaining the TT part of a tensor in coordinate space corresponds to a non-local operation. In Fourier space, on the other hand, it is easy to construct a projector that filters out only the TT degrees of freedom of a given tensor. The GW source can be written as
\begin{equation}\label{eq:gws:TTprojectionpi}
    \Pi_{ij}^\text{TT}({\bf k},\eta) = \Lambda_{ij,lm}({\hat{\bf k}}) \Pi_{lm}^{\rm eff} ({\bf k},\eta)\,,
\end{equation}
where $\Lambda_{ij,lm}({\hat{\bf k}})$ is a projection operator defined as~\cite{Carroll:2004st,Baumann_2022}
\begin{eqnarray}
\Lambda_{ij,lm}({\hat{\bf k}}) \equiv P_{il}({\hat{\bf k}})  P_{jm}({\hat{\bf k}}) - \dfrac12 P_{ij}({\hat{\bf k}}) P_{lm}({\hat{\bf k}}) \,, \quad\quad \text{with}\quad\quad P_{ij}(\hat{\bf k})= \delta_{ij} - {\hat{ k}}_i {\hat{k}}_j\,, 
\label{eqn:ProjectorTT}
\end{eqnarray}
with ${\hat{ k}}_i \equiv {k_i/ k}$. Thanks to the fact that $P_{ij}$ is transverse ($P_{ij} \hat{k}_j = 0$) and  idempotent ($P_{ij}P_{jm} = P_{im}$), one can easily see that the TT conditions in Fourier space, $k_i\Pi_{ij}^{\rm TT}({\bf {k}},\eta)=\Pi_{ii}({\bf {k}},\eta)^{\rm TT} = 0$, are satisfied at any time. 

The transverse-traceless projection described by \cref{eq:gws:TTprojectionpi} is needed to construct the source of the GW degrees of freedom to solve \cref{eqn:EoMGWs}. This projection, however, represents an expensive computational operation, given that it is non-local in position space. In lattice simulations, the determination of the source of the GWs would require to go back-and-forth to Fourier space in every evolution step, making very inefficient the simulation of GW emission on a lattice. 

In Ref.~\cite{Garcia-Bellido:2007fiu} a workaround was proposed to overcome this computational problem: noting that  $\Pi_{ij}^\text{TT}({\bf k},\eta)$ in Fourier space is just a linear combination of the components of the full tensor $\Pi_{ij}^{\rm eff}({\bf k},\eta)$, {\it c.f.}~Eq.~(\ref{eq:gws:TTprojectionpi}), and that the (particular) solution to Eq.~(\ref{eqn:EoMGWs}) is linear in $\Pi_{ij}^{\rm TT}$, one can write the truly TT tensor perturbations ({\it i.e.}~the GWs) as
\begin{equation}\label{eqn:ustohs}
    h_{ij}({\bf k},\eta) = \Lambda_{ij,lm}(\hat{{\bf k}}) u_{lm}({\bf k},\eta)\,,
\end{equation}
where $u_{ij}({\bf k},\eta)$ is the Fourier transform of 6 unphysical degrees of freedom ($\lbrace u_{11}, u_{12}, u_{13}, u_{22}, u_{23}, u_{33}\rbrace$) contained in the symmetric tensor $u_{ij}({\bf x},\eta)$, which are 
the solution to the linear equation
\begin{equation}
     {u}_{ij}'' + (3-\alpha) \frac{a'}{a} u_{ij}' - a^{-2(1-\alpha)}\nabla^2 u_{ij} = \dfrac{2 }{m_p^2 a^{2(1-\alpha)}}\Pi_{ij}^\text{eff}\,.\label{eq:gws:usEoM}
\end{equation}
This is analogous to Eq.~(\ref{eqn:EoMGWs}), but 
without TT projection of the source. Contrary to Eq.~(\ref{eqn:EoMGWs}), Eq.~(\ref{eq:gws:usEoM}) can actually be evolved in configuration space for as long as we want, without having to go back-and-forth to and from Fourier space. Only when we desire to obtain the physical degrees of freedom truly representing GWs, $h_{ij}$, we Fourier-transform the  solution to \cref{eq:gws:usEoM}, $u_{ij}({\bf x},\eta) \rightarrow u_{ij}({\bf k},\eta)$, and apply Eq.~(\ref{eqn:ustohs}). This procedure, in particular, needs to be performed whenever one wants to measure the energy density of a GWB in a simulation.

As a novelty in this review, we propose to push even further the previous workaround, by exploiting that GWs are traceless. We can work instead with only 5 unphysical degrees of freedom, given by a symmetric traceless tensor $v_{ij}$, which obeys 
\begin{equation}
v_{33} = -(v_{11}+v_{22})\,.
\end{equation}
These fields ($\lbrace v_{11}, v_{12}, v_{13}, v_{22}, v_{23}\rbrace$) are sourced by the traceless part of the effective anisotropic stress tensor,
\begin{equation}\label{eq:gws:vsEoM}
     {v}_{ij}'' + (3-\alpha) \frac{a'}{a} v_{ij}' - a^{-2(1-\alpha)}\nabla^2 v_{ij} = \dfrac{2 }{m_p^2 a^{2(1-\alpha)}}\left[\Pi_{ij}^\text{eff}-\frac{1}{3}\delta_{ij}\Pi_{kk}^\text{eff}\right]\,,
 \end{equation}
and can be used to reconstruct at any time the physical GWs fields by TT projection as 
\begin{equation}\label{eqn:vstohs}
    h_{ij}({\bf k},\eta) = \Lambda_{ij,lm}(\hat{{\bf k}}) v_{lm}({\bf k},\eta)\,.
\end{equation}
The advantage of using 5 degrees of freedom described by $v_{ij}$ versus the 6 degrees of freedom contained in $u_{ij}$, is that we save a sixth of the computational memory required to represent GWs. The disadvantage is that the source in Eq.~(\ref{eq:gws:vsEoM}), compared to the source in Eq.~(\ref{eq:gws:usEoM}), requires the extra computation of the term $-{1\over3}\delta_{ij}\Pi_{kk}^{\rm eff}$ at every lattice site (at every time step), which may be time consuming. It depends therefore on the situation whether simulating 5 or 6 degrees of freedom is more convenient, depending on what is more precious, saving memory or simulation time.

\noindent The energy density of a GW background (GWB) is given by~\cite{Caprini:2018mtu}
\begin{eqnarray}
\label{eqn:GWenergydensity}
\rho_{\rm GW}(\eta) &=& \dfrac{m_p^2}{4a^{2\alpha}}\dfrac{1}{V}\int_{V} \hspace*{-1mm}\text{d}^3{\bf x} \,\,{h}'_{ij}({\bf x},\eta){h}'_{ij}({\bf x},\eta)\label{eq:gws:rho1}\\
&\simeq& \dfrac{m_p^2}{4 a^{2\alpha} V}\int_{V} \dfrac{\text{d}^3{\bf k}}{(2\pi)^3} \,{h}'_{ij}({\bf k},\eta){h}^{\prime*}_{ij}({\bf k},\eta)\label{eq:gws:rho2} \\ 
&\equiv& \int \dfrac{\text{d} \rho_{\rm GW}}{\text{d} \log k} \text{d} \log k\:,\label{eq:gws:rho3}
\end{eqnarray}
where the spatial average over a volume $V$ is assumed to encompass all relevant wavelengths of the perturbations $h_{ij}$, and in the second line we have we have taken the limit $kV^{1/3} \gg 1$, so that $\int_{V} d{\bf x}e^{-i{\bf x}({\bf k}-{\bf k'})} \longrightarrow (2\pi)^3\delta^{(3)}({\bf k}-{\bf k'})$. The energy density power spectrum (per logarithmic interval) is then defined as
\begin{equation}
  \left(\dfrac{\text{d} \rho_{\rm GW}}{\text{d} \log k}\right)(k,\eta) = \dfrac {m_p^2 k^3 }{8\pi^2a^{2\alpha} V} \int  \dfrac{\text{d} \Omega_k}{4\pi} \,{h}'_{ij}({\hat{\bf k}},k,\eta) {h}^{\prime*}_{ij}({\hat{\bf k}},k,\eta) \: ,
\end{equation}
with $\text{d}\Omega_k$ a solid angle measure in momentum-space. 

For stochastic processes the volume average can be replaced by an ensemble average $\langle...\rangle$ over the random realizations of the tensor fluctuations, 
\begin{eqnarray}\label{eqn:GWPS}
    \rho_{\rm GW}(\eta) &=& \dfrac{m_p^2}{4a^{2\alpha}} \langle{h}'_{ij}({\bf x},\eta) {h}^{\prime*}_{ij}({\bf x},\eta)\rangle \nonumber \\
  &=& \dfrac{m_p^2}{4a^{2\alpha}}\int \dfrac{\text{d}^3{\bf k}}{(2\pi)^3}\dfrac{\text{d}^3{\bf k'}}{(2\pi)^3}  e^{-i {\bf x}({\bf k} - {\bf k'})} \cdot \langle{h}'_{ij}({\bf k},\eta) {h}^{\prime*}_{ij}({\bf k},\eta)\rangle \nonumber \\
  &\equiv& \dfrac{m_p^2}{8\pi^2a^{2\alpha}}\int\dfrac{\text{d}k}{k} k^3 P_{h'}(k,\eta)\:,
\end{eqnarray}
where we have introduced the power spectrum of the time derivative of $h_{ij}$ in the third line
as
\begin{equation}\label{eqn:GWenergydensitydef}
     \langle {h}'_{ij}({\bf k},\eta){h}'_{ij}({\bf k'},\eta) \rangle = (2\pi)^3 P_{h'}(k,\eta)\delta^{(3)}({\bf k} - {\bf k'})\,.
\end{equation}
Comparing Eq.~(\ref{eq:gws:rho3}) and~(\ref{eqn:GWPS}), leads to a form of the energy density power spectrum of a GWB, which normalized to the critical energy density $\rho_\text{c} \equiv 3a^{-2\alpha}\mathcal{H}^2 m_p^2 = 3H^2 m_p^2$, reads
\begin{equation}
    \Omega_\text{GW} \equiv \frac{1}{\rho_\text{c}}\dfrac{\text{d} \rho_{\rm GW}}{\text{d} \log k}\ = \dfrac{k^3}{24\pi^2\mathcal{H}^2}  P_{h'}(k,\eta) = \dfrac{k^3}{24\pi^2a^{2\alpha}H^2}  P_{h'}(k,\eta)\,.
\end{equation}
This last quantity is usually known as the fractional GW energy density power spectrum.

\subsection{Gravitational waves on the lattice}
\label{subsec:GWlattice}
 
On the lattice, the dynamics of GWs are studied using the unphysical fields introduced in the previous section. These are evolved according to a discretized versions of either Eq.~(\ref{eq:gws:usEoM}) or (\ref{eq:gws:vsEoM}), which can be solved following the canonical techniques presented in {\tt The Art-I}, reviewed in \cref{subsec:Algorithms}. Without loss of generality we discuss now the case of the $v$-fields, while analogous techniques would apply to the $u$-fields.

To solve the EOM of the unphysical $v_{ij}$ fields, we introduce program variables as described in \cref{eq:GaugeProgramVar}. It is convenient to also rescale the unphysical fields as
\begin{equation}\label{eq:gws:programvars}
\quad\tilde{v}_{ij}=\left(\dfrac{m_p}{f_*}\right)^2v_{ij}\:.
\end{equation}
Numerically, $\tilde{v}$-fields are evolved by defining the conjugate momenta, ${(\tilde\pi_{{v}})}_{ij} = a^{3 - \alpha} {\tilde{v}}_{ij}'$, which allows to rewrite Eq.~(\ref{eq:gws:vsEoM}) as a system of first order differential equations,
\begin{equation}\label{eq:systemGWs}
    \left\{\begin{array}{rcl}
         \tilde{v}_{ij}'&=&\displaystyle a^{\alpha-3}{(\Tilde{\pi}_{v})}_{ij} \:, \\[10pt]
         {(\tilde\pi_{{v}})}_{ij}' &=&\displaystyle  a^{1+\alpha}\tilde{\nabla}^2 \tilde{v}_{ij} + 2a^{1+\alpha}\left[\tilde{\Pi}^{\text{eff}}_{ij}-\frac{1}{3}\delta_{ij}\tilde{\Pi}_{kk}^\text{eff}\right]\:,
    \end{array}\right.
\end{equation}
where the effective anisotropic stress tensor is expressed in program units as
\begin{equation}
\begin{array}{rcl}
     \Tilde{\Pi}^{\text{eff}}_{ij} = \displaystyle  \frac{\Pi_{ij}^{\text{eff}}}{\omega_*^2 f_*^2} & = & \tilde{\nabla}_i \Tilde{\phi}_b\: \tilde{\nabla}_j \Tilde{\phi}_b + 2 \text{Re}\{(\Tilde{D}^A_i \Tilde{\varphi}_b)^*\:(\Tilde{D}^A_j \Tilde{\varphi}_b)\} + 2 \text{Re}\{(\Tilde{D}_i \Tilde{\Phi}_b)^*\:(\Tilde{D}_j \Tilde{\Phi}_b)\}\\[10pt]
     & & \displaystyle - \left(\frac{\omega_*^2}{f_*^2}\right)\left[a^{-2\alpha}\tilde{{E}}_{i}\:\tilde{{E}}_{j} + a^{-2} \tilde{{B}}_i \: \tilde{{B}}_j+ a^{-2\alpha}\tilde{{E}}^c_{i}\:\tilde{{E}}_{j} + a^{-2} \tilde{{B}}^c_i \: \tilde{{B}}^c_j\right]\,.
     \end{array}
\end{equation}
This system of first-order differential equations can be solved using any symplectic integrator, as leapfrog, velocity-Verlet, position-Verlet, or Yoshida. Alternatively they could also be solved by non-symplectic integrators, like Runge-Kutta, but given that the EOM are relativistic wave-like equations, there would be no particular advantage on this. The $u$-field EOM would look analogous to Eq.~(\ref{eq:systemGWs}), but simply making the replacement $\Big[\tilde{\Pi}^{\text{eff}}_{ij}-\frac{1}{3}\delta_{ij}\tilde{\Pi}_{kk}^\text{eff}\Big] \longrightarrow  \tilde{\Pi}^{\text{eff}}_{ij}$ in the kernel of ${(\tilde\pi_{{u}})}_{ij}$. 

As explained in the previous section, the physical degrees of freedom representing GWs, $h_{ij}$, can be recovered at any time during the simulation by means of a TT projection. In particular, after transforming the fields to Fourier space, one projects to obtain the TT component via
\begin{equation}\label{eq:gws:hLattUoU}
    h_{ij}({\tilde{\bf n}},\eta) = \Lambda^{\rm L}_{ij,lm} ({\tilde{\bf n}}) v_{lm}({ \tilde{\bf n}},\eta)\,,
\end{equation}
where $\Lambda^{\rm L}_{ij,lm}$ is a lattice version of the TT projector introduced in Eq.~(\ref{eqn:ProjectorTT}), analogously defined as~\cite{Figueroa:2011ye}
\begin{eqnarray}\label{eq:gws:ProjectorTTlatt}
    \Lambda_{ij,lm}^{\rm L}({\tilde{\bf n}}) \equiv P^\text{L}_{il}({\tilde{\bf n}})  P^{\text{L}*}_{jm}({ \tilde{\bf n}}) - \dfrac12 P^\text{L}_{ij}({ \tilde{\bf n}}) P^{\text{L}*}_{lm}({\tilde{\bf n}})\,, 
\quad\quad
   P^\text{L}_{ij}({\tilde{\bf n}})= \delta_{ij} - \dfrac{{ k}({\tilde{\bf n})}_{{\rm L},i} { k}({\tilde{\bf n}})_{{\rm L},j}^*}{k_{\rm L}^2({\tilde{\bf n}})}\,, 
    \end{eqnarray}
where ${\bf k}_{{\rm L}}(\tilde{\bf n})$ is the {\it lattice momentum} corresponding to some lattice derivative operator, {\it c.f.}~Sect.~\ref{subsec:LatticeMomentum}. The lattice momenta can be either real or complex, and thus also the TT projector, depending on the choice of lattice derivative. For the neutral derivative given in Eq.~(\ref{eq:neutrald}), for example, we define real projectors as 
\begin{eqnarray}\label{eq:realP}
\Lambda_{ij,lm}^{\text{L},0} = P^{\text{L},0}_{il}P^{\text{L},0}_{jm} - \dfrac{1}{2}P^{\text{L},0}_{ij}P^{\text{L},0}_{lm} \: ,~~~P_{ij}^{\text{L},0} = \delta_{ij} - \dfrac{k^{0}_{\text{L},i}k^{0}_{\text{L},j}}{|{\bm k}^0_{\text{L}}|^2} \: ,
\end{eqnarray}
where we omit the arguments of the projectors for legibility.
On the other hand, the projectors become complex for the lattice momenta ${\bf k_\text{L}^\pm}$ corresponding to the forward/backward derivatives given in Eq.~(\ref{eq:forwardbackwardd}), with 
\begin{eqnarray}\label{eq:complexP}
\Lambda_{ij,lm}^{\text{L},\pm} = P^{\text{L},\pm}_{il}{P_{jm}^{\text{L},\pm*}} - \dfrac{1}{2}P_{ij}^{\text{L},\pm}P_{lm}^{\text{L},\pm*}\,,~~~ P_{ij}^{\text{L},\pm} = \delta_{ij} - \dfrac{(k_{\text{L},i}^\pm)^*k_{\text{L},j}^\pm}{|{\bm k}_{\text{L}}^\pm|^2}\,.
\end{eqnarray}

Lattice TT projectors obey a series of properties similar to their continuum counterpart. For example, in the case of the complex projectors for forward/backward derivatives, they obey the following properties
\begin{eqnarray}
\begin{array}{ll}
    \text{1}) ~\sum_{i}k_{\text{L},i}^\pm P^{\text{L},\pm}_{ij} = 0\:, \quad \quad\quad  &\text{2)}~ \sum_{i}(k_{\text{L},i}^\pm)^{*}P^{\text{L},\pm}_{ij} \neq 0\:,\\[15pt]
    \text{3)}~\sum_{j}k_{\text{L},j}^\pm P^{\text{L},\pm}_{ij} \neq 0 \:,\quad \quad \quad &\text{4)}~\sum_{j}(k_{\text{L},j}^\pm)^*P^{\text{L},\pm}_{ij} = 0 \:,\\[15pt]
    \text{5)}~ {P^{\text{L},\pm}_{ij}}^* = P^{\text{L},\pm}_{ji}\:, \quad \quad \quad
    &\text{6)}~P^{\text{L},\pm}_{ij}(-{\tilde{\bf n}}) = P^{\text{L},\pm}_{ji}({\tilde{\bf n}})\:,  \\[15pt]
     \text{7)}~P^{\text{L},\pm}_{ij}P^{\text{L},\pm}_{jk} = P^{\text{L},\pm}_{ik}\:, \quad \quad \quad
     &\text{8)}~ P^{\text{L},\pm}_{ij}P^{\text{L},\pm}_{ki} \neq P^{\text{L},\pm}_{ik} \:,
\end{array}
\end{eqnarray}
the most relevant of them are the idempotence of the projector (property 7) and their hermiticity (property 5). These properties also imply that the projector is transverse with respect to the lattice derivative associated to the particular choice of lattice momentum used to define it (property 1). Note that the real projector obeys a similar set of properties, except for the fact that it is symmetric instead of hermitian. A proof of all these properties can be found in~Ref.~\cite{Figueroa:2011ye}. 
 
After projecting to the physical  degrees of freedom $h_{ij}$, one can use these to measure the energy density of a GWB on a lattice. The GW energy density power spectrum is computed with the discrete counterpart of Eq.~(\ref{eq:gws:rho1}),
\begin{eqnarray}\label{eq:gws:GWEDlat}
    \rho_{\rm GW} (\eta) &=& \dfrac{m_p^2}{4 a^{2\alpha}N^3} \sum_{\bf n} {h}'_{ij}({\bf n}, \eta){h}'_{ij}({\bf n}, \eta)\nonumber \\
    &=& \dfrac{m_p^2}{4a^{2\alpha}} \dfrac{1}{N^6} \sum_{\tilde{\bf  n}} {h}'_{ij}({ \tilde{\bf n}}, \eta)h^{\prime*}_{ij}({ \tilde{\bf n}}, \eta)\nonumber \\
    &=& \dfrac{m_p^2}{4a^{2\alpha}} \dfrac{1}{N^6} \sum_{l}\sum_{{\tilde{\bf n}} \epsilon R(l)}{h}'_{ij}({\tilde{ \bf n}}, \eta)h^{\prime*}_{ij}({\tilde{\bf n}}, \eta)\:,
\end{eqnarray}
where in the second line we have applied the discrete Fourier transform on the two $h$-fields, and used $\Sigma_{\bf n} e^{ik_{\rm IR}dx{\bf n}({\tilde{\bf n}}-{\tilde{\bf n}'})} = N^3\delta_{\tilde{\bf n} \tilde{\bf n}'}$. Finally, 
in the last line we have split the summation over spherical bins. As explained in~Sect.~\ref{subsec:PS}, the construction of the power spectrum depends on the different ways of counting the bin's multiplicity 
$\#_l$, {\it i.e.}~the number of modes that fit within each bin. Here we follow Ref.~\cite{Figueroa:2011ye}, which considers bins centered at ${k}(l) = k_{\rm IR}l$ with regular widths $\Delta k \equiv k_{\rm IR}$, and approximates the number of points within the $l$-th bin as $ \#_l \simeq 4\pi l^2$, where $l = 1, 2, 3, ...$. In such a case, one obtains
\begin{eqnarray}
   \rho_{\rm GW} (\eta) &=& \dfrac{m_p^2}{4 a^{2\alpha}N^6} \sum_{l} 4\pi l^2
    \langle{h}'_{ij}({\tilde{\bf n}}, \eta){h}^{\prime*}_{ij}({\tilde{\bf n}}, \eta)\rangle_{R(l)} \nonumber\\
    &=& \sum_{l} \left\{  \dfrac{m_p^2 \delta x^6 }{8\pi^2a^{2\alpha} L^3} k^3(l)
    \langle{h}'_{ij}({\tilde{\bf n}}, \eta){h}^{\prime*}_{ij}({\tilde{\bf n}}, \eta)\rangle_{R(l)} \right\}\Delta \log k \, . \label{eq:latticeGWenergy}
\end{eqnarray}
with $\langle...\rangle_{R(l)}$ denoting average over all the modes $\tilde{\bf n}$ within the $l$-th spherical shell, and $\Delta \log k \equiv k_{\rm IR}/k(l)$. From here, we define the GW energy density power spectrum on the lattice as
\begin{equation}\label{eq:gws:discretePS}
  \left(\dfrac{\text{d} \rho_{\rm GW}}{\text{d} \log k}\right) (l) =  \dfrac{m_p^2 k(l)^3}{8\pi^2 a^{2\alpha}  L^3} \left<\left[\delta x^3 {h}'_{ij}(|\Tilde{\textbf{n}}|,\eta)\right]\left[\delta x^3 {h}'_{ij}(|\Tilde{\textbf{n}}|,\eta)\right]^*\right>_{R(l)} \: .
\end{equation}
Given the approximation used for the $\#_{l}$, the above expression corresponds to a {\tt Type\,II} lattice spectrum, 
in analogy to the definitions of scalar fields' power spectra discussed in Sect.~\ref{subsec:PS}. Normalizing by the critical energy density, we finally write
\begin{eqnarray}
\Omega_{\rm GW} ({\tilde{\bf n}}, \eta) = \dfrac{1}{\rho_\text{c}}\dfrac{m_p^2 k^3(l)}{8\pi^2 a^{2\alpha}} \dfrac{\delta x^3}{N^3} \left<\left[ {h}'_{ij}(\Tilde{\textbf{n}},\eta)\right]\left[ {h}'_{ij}(\Tilde{\textbf{n}},\eta)\right]^*\right>_{R({\tilde{\bf n}})} \hspace*{3.5cm}[{\tt Type~II}]\,.
\end{eqnarray}
Of course, an analogous {\tt Type I} lattice version of the GWB spectrum can also be easily constructed as~\cite{GWmodule:2022,GWmodule:2023} 
\begin{eqnarray}
\Omega_{\rm GW} ({\tilde{\bf n}}, \eta) =\displaystyle \dfrac{1}{\rho_\text{c}} \dfrac{m_p^2 k(l) }{8\pi a^{2\alpha} } \dfrac{\delta x}{N^5} \#_{l} \left<\left[ {h}'_{ij}({ \tilde{\bf n}},\eta)\right]\left[ h'_{ij}({ \tilde{\bf n}},\eta)\right]^*\right>_{R(l)} \hspace*{3.5cm}[{\tt Type~I}]\,.
\end{eqnarray}

In light of both {\tt Type I} and {\tt Type II} expressions, we note that the object of interest to construct in the lattice is the bilinear product ${h}'_{ij}({\tilde{\bf n}}){h}_{ij}^{\prime *}({ \tilde{\bf n}})$. This can be computed directly from the conjugate momenta of the $v$-fields (or alternatively from the $u$-fields), 
\begin{equation}
    {h}'_{ij}=\dfrac{\omega_*}{a^{3-\alpha}}\Lambda_{ij,lm}^\text{L}(\pi_{v})_{lm}=\dfrac{\omega_*}{a^{3-\alpha}}\left(\dfrac{f_*}{m_p}\right)^2\Lambda_{ij,lm}^\text{L}(\tilde{\pi}_{v})_{lm}\:.
\end{equation}
Then, the bilinear product can be written as a linear combination of two type of traces 
\begin{equation}\label{eq:gws:Traces}
        {h}'_{ij}{h}_{ij}^{\prime*} = \text{Tr}({\tt P}\,{\tt v}'\,{\tt P}\,{\tt v}^{\prime*}) - \dfrac{1}{2} \text{Tr}({\tt P}\,{\tt v}')\text{Tr}({\tt P}\,{\tt v}^{\prime*}) \: ,
\end{equation}
where ${\tt {v}}'$ and ${\tt P}$ are matrices with elements $({\tt v}')_{ij} = {v}'_{ij}$ and $({\tt P})_{ij}=P^\text{L}_{ij}$. \Cref{eq:gws:Traces} is valid for both real and complex valued projectors. Numerically, this can be explicitly implemented by defining the matrix products $w_{ij} \equiv P^\text{L}_{ik}{v}'_{kj}$ and $\tilde{w}_{ij}\equiv P^\text{L}_{ik}{v}^{\prime*}_{kj}$, and using them to determine the trace values from
\begin{eqnarray}
\text{Tr}({\tt P}\,{\tt v}')\hspace*{-.2cm}&=&\hspace*{-.2cm} w_{11} + w_{22} + w_{33}\:\\
   \text{Tr}({\tt P}\,{\tt v}^{\prime*}) \hspace*{-.2cm}&=&\hspace*{-.2cm} \tilde{w}_{11} + \tilde{w}_{22} + \tilde{w}_{33}\:\\
     \text{Tr}({\tt P}\,{\tt v}'\,{\tt P}\,{\tt v}^{\prime*}) \hspace*{-.2cm}&=&\hspace*{-.2cm} w_{11}\tilde{w}_{11} +w_{22}\tilde{w}_{22} + w_{33}\tilde{w}_{33}+ w_{12}\tilde{w}_{21}  \\
      \hspace*{-.2cm}& &\hspace*{-.2cm} + w_{21}\tilde{w}_{12} + w_{13}\tilde{w}_{31} + w_{31}\tilde{w}_{13} + w_{23}\tilde{w}_{32} + w_{32}\tilde{w}_{23}\:\nonumber
\end{eqnarray}
In the real case, these computations can be shortened since $\tilde{w}_{ij}=w^*_{ij}$.

Finally, we also present lattice observables to monitor the {\it transversality} and {\it tracelesness} of the $h_{ij}({\bf {n}},\eta)$ field in a simulation. In particular, we propose to track the lattice-volume average of the following dimensionless ratios,
\begin{eqnarray}\label{eq:transversality}
\delta(\eta) &\equiv&\displaystyle \dfrac{\langle \nabla^\text{L}_i h_{ij}({\bf {n}},\eta)\rangle}{\langle S^\text{L}_i h_{ij}({\bf {n}},\eta\rangle}\:,\hspace*{3.5cm} [{\tt Transversality~indicator}]\\[10pt]
\label{eq:tracelessness}
\lambda(\eta) &\equiv&\displaystyle \frac{\left\langle \sum_i |h_{ii}({\bf {n}},\eta)|\right\rangle}{\left\langle |\sum_i h_{ii}({\bf {n}},\eta|\right\rangle}\:.\hspace*{3.5cm}[{\tt Tracelessness~indicator}]
\end{eqnarray}
Here $\nabla^{\rm L}$ is the discretized spatial derivatives, {\it c.f.}~Sect.~\ref{subsec:LatticeMomentum}, associated to the lattice momentum used to construct the TT projector, while $S^{\rm L}_i$ are defined as 
\begin{eqnarray}
     S^{0}_i h_{ij} \equiv \displaystyle  \dfrac{h_{ij}({\bf n}+\hat{\imath},\eta)+h_{ij}({\bf n}-\hat{\imath},\eta)}{2\delta x}\:,~~~~
    S^{\pm}_i h_{ij} \equiv\displaystyle \dfrac{ h_{ij}({\bf n}\pm\hat{\imath},\eta)+ h_{ij}({\bf n},\eta)}{\delta x} \:.
\end{eqnarray}
Both the transversality and tracelesness indicators are expected to be very small, typically down to machine precision. One typically observes $\delta, \lambda \sim \mathcal{O}(10^{-16})-\mathcal{O}(10^{-15})$, though the details depend on the scenario, see~\cite{Figueroa:2011ye}
 for explicit examples.

\section{Summary and outlook}

~~~~~~The use of numerical techniques to study the non-linear field dynamics of early Universe scenarios has increased considerably in recent years, such that a new field---{\bf Lattice Cosmology}---is emerging on its own merits. {\bf Lattice Cosmology Techniques} (LCT) have established themselves as a clear pathway towards understanding the early Universe, thanks to their ability to describe its phenomenology and their potential to provide accurate predictions (of paramount necessity in the current era of precision observational cosmology). The basic framework of LCT was first summarized in our dissertation {\tt The Art-I}~\cite{Figueroa:2020rrl}, which introduced the necessary concepts and methods to simulate scalar-singlet and scalar-gauge theories in an expanding Universe on a lattice. The current monograph, which we refer to as {\tt The Art-II}, is a compendium of non-canonical aspects of field theories, which extends the application of LCT to situations beyond the canonical cases covered in {\tt The Art-I}. 

In this monograph, we discuss lattice implementations of {\bf (A)} non-canonical interactions, including non-minimal scalar field couplings to gravity, $\phi^2R$, non-minimal scalar kinetic theories described by $\mathcal{G}_{ab}(\lbrace\phi_c\rbrace)\partial_\mu\phi_a\partial^\mu\phi_b$, and axion-like particle (ALP) interactions with Abelian gauge fields, $\phi F_{\mu\nu}\tilde F^{\mu\nu}$; {\bf (B)} methods to set up special initial configurations, including the generation of cosmic defect networks ({\it e.g.}~cosmic strings and domain walls) close to the scaling regime, field configurations based on arbitrary power spectra or spatial profiles, and other algorithmic techniques, such as {\it e.g.}~those required to set up thermal initial conditions; and {\bf (C)} scalar field dynamics in $(d + 1)$-dimensions, with $d \neq 3$, with particular emphasis on $d = 2$ spatial dimensions. Unrelated to non-canonical circumstances, we also discuss implementation(s) of gravitational wave (GW) dynamics on a lattice. {\tt The Art-II} is just the second part of a series of dissertations that are expected to further appear in the near future, extending the application of LCT to fluid-scalar-gauge dynamics ({\tt The Art-III}), and to the interaction between matter fields and gravitational degrees of freedom ({\tt The Art-IV}).

We now summarize briefly the content of this monograph. In Section~\ref{sec:LatticeTechniques}, we present a recapitulation of {\tt The Art-I}~\cite{Figueroa:2020rrl}. In Section~\ref{sec:NMCoupled_scalars}, we consider lattice formulations of non-canonical interactions of scalar fields, either non-minimally coupled to gravity via $\phi^2 R$, or with non-minimal kinetic terms, $\mathcal{G}_{ab}\partial_{\mu}\phi_a \partial^{\mu}\phi_b$. As an example, we discuss some results of the {\it Ricci reheating} scenario. In Section~\ref{sec:Axion}, we present a lattice formulation of the interaction of an axion-like particle (ALP) with an Abelian gauge sector, through the coupling $\phi F\tilde F$. We put particular care into building a proper lattice representation of $F\tilde F$ as a total derivative, and discuss the notion of chirality on the lattice. As a working example, we discuss some results from the non-linear dynamics of the {\it axion inflation} scenario. In Section~\ref{sec:DefectsV}, we discuss how the creation of cosmic defects can be studied on a lattice. In particular, for cosmic strings and domain walls, we discuss techniques to accelerate the achievement of {\it scaling} in a network and to address the loss-of-resolution of the defect cores due to the expansion of the universe, and also introduce specific observables for each type of defect. In Section~\ref{sec:InitialConditions}, we discuss techniques for the creation of a variety of initial conditions on a lattice. We present methods to create field configurations based on arbitrary external power spectra, probabilistic methods to create {\it e.g.}~thermal field fluctuations, and discuss cases where the initial condition is given by known spatial profiles. In Section~\ref{sec:Sims2D}, we present techniques for performing scalar field simulations in an arbitrary number of spatial dimensions, and discuss how this can be used to reduce the cost of three-dimensional simulations by mimicking the same dynamics in lower-dimensional lattices. As an example, we compare the results from three- and two-dimensional simulations for a simple preheating scenario. Finally, in Section~\ref{sec:GW}, we review our approach to producing and evolving gravitational waves (GWs) on a lattice. In particular, we discuss variants of a technique used to measure the power spectrum of transverse–traceless tensor modes from the evolution of auxiliary fields sourced by a general source that is not transverse–traceless. 

The techniques presented in this dissertation represent the theoretical basis for the non-canonical field theory aspects (interactions, initial conditions, dimensionality) and GW dynamics implemented in \href{http://www.cosmolattice.net}{\color{blue} \tt ${\mathcal C}$osmo${\mathcal L}$attice~v2.0}, which will be released in 2026, soon after the publication of this monograph. The presentation of techniques in this document should also be understood as a theoretical review on non-canonical LCT, complementing our previous review {\tt The Art\,-\,I} on canonical cases. 
Most of the physics and technical aspects that we present here in {\tt The Art\,-\,II} have already been  tested in our private branch of \CLns. The incorporation of non-canonical field-theory aspects in {\tt v2.0} of the code will maintain the same framework as in previous versions, in which physics and code-specific features are separated, allowing the user to write expressions in a serial manner while parallelization and performance are never sacrificed. \CL {\tt v2.0} will continue to be a publicly available, user-friendly, and modular C++ MPI-based code. As a novelty, users will also be able to run it on GPUs, in addition to CPUs. Visit our website \href{https://cosmolattice.net/}{\color{blue}https://cosmolattice.net/} for further details.

To conclude, we comment on further applications of LCT to circumstances beyond those considered in {\tt The Art-I} or {\tt The Art-II}. Among these, we find of great interest:
\begin{itemize}
\item The study of first-order phase transitions through the nucleation of scalar field bubbles. In the thermal case, bubbles grow and merge while surrounded by a plasma coupled to the scalar field~\cite{Guth:1981uk,Steinhardt:1981ct,Witten:1984rs}. Due to the growth and collision of the bubbles~\cite{Hogan:1986dsh,Kosowsky:1991ua,Huber:2008hg,Caprini:2009fx,Kosowsky:1992rz}, the plasma becomes excited in the form of sound waves, which may also lead to shocks and turbulent motions. All these phenomena, in particular, are very efficient sources of GWs~\cite{Kosowsky:2001xp,Hindmarsh:2013xza,Hindmarsh:2015qta,Weir:2017wfa,Caprini:2019egz,Schwaller:2015tja,Caprini:2015zlo}, enabling the possibility to observe such events with upcoming GW detectors. A detailed assessment of the specific non-linear dynamics of a first-order phase transition and its GW signatures, however, is only possible through lattice simulations that simultaneously incorporate fluid dynamics coupled to the corresponding scalar and/or gauge field sectors. Furthermore, the use of a hydrodynamic approach may also become relevant in the study of reheating~\cite{Elia:2025auj}. A review on lattice techniques applied to {\it scalar-gauge-fluid} (SGF) dynamics, {\tt The Art-III}, is currently under preparation as part of our monograph series on LCT.

    \item The production of gravitational perturbations, which can be very efficient in many early Universe scenarios~\cite{Khlebnikov:1997di,Easther:2006gt,Easther:2006vd,GarciaBellido:2007af,Dufaux:2007pt,Dufaux:2008dn,Dufaux:2010cf,Zhou:2013tsa,Bethke:2013aba,Bethke:2013vca,Antusch:2016con,Antusch:2017flz,Antusch:2017vga,Liu:2018rrt,Figueroa:2017vfa,Fu:2017ero,Caprini:2018mtu,Lozanov:2019ylm,Adshead:2019lbr,Adshead:2019igv,Armendariz-Picon:2019csc}, as well as their impact on the background dynamics~\cite{Bassett:1998wg,Bassett:1999mt,Bassett:1999ta,Finelli:2000ya,Chambers:2007se,Bond:2009xx,Imrith:2019njf,Musoke:2019ima,Giblin:2019nuv,Martin:2020fgl}. On the one hand, the production of large scalar perturbations can lead to the formation of primordial black holes, see {\it e.g.}~\cite{Cotner:2019ykd,Martin:2019nuw,GarciaBellido:1996qt,Green:2000he,Hidalgo:2011fj,Torres-Lomas:2014bua,Suyama:2004mz,Suyama:2006sr,Cotner:2018vug}. On the other, gravitational backreaction may be relevant to describe systems where gravitational perturbations are efficiently produced, as {\it e.g.}~in axion inflation~\cite{Linde:2012bt,Adshead:2023mvt}. In this context,
    following the evolution of gravitational perturbations and determining their potential impact on the overall dynamics is crucial to assess the observability of these type of inflationary scenarios. A review on lattice techniques applied to gravitational interactions of matter fields, {\tt The Art-IV}, is under preparation as part of our monograph series on LCT.

    \item The possibility of adding fermions in the simulations. As the notion of `classical fields' does not exist for fermions, a straightforward discretization and evolution of the Dirac equation is of limited use, as it is able to capture only linear dynamics of fermion excitations. Nevertheless, it was realized in Ref.~\cite{Aarts:1998td} that it is possible to study the real-time dynamics of fermions in a semi-classical formulation of the out-of-equilibrium Schwinger–Keldysh framework; see also Refs.~\cite{Kasper:2014uaa,Buividovich:2015jfa,Mace:2019cqo}. It is possible to combine the lattice implementation of Ref.~~\cite{Aarts:1998td} with the `low-cost' fermions introduced in Ref.~~\cite{Borsanyi:2008eu}. For instance, in Refs.~\cite{Saffin:2011kc,Saffin:2011kn,Mou:2013kca,Mou:2015aia}, the out-of-equilibrium dynamics of classical scalar fields coupled to quantum fermions was successfully simulated. These simulations are rather costly in terms of memory, and only small lattices have been considered to date.

    \item The simulation of interacting  fields including fully-quantum effects. The classical approach exploited by LCT is expected to fail if the phenomena of interest involve low occupation numbers, like in thermalisation processes or late-time field evolution~\cite{Aarts:2001yn,Arrizabalaga:2004iw, Berges:2004yj,Berges:2015kfa}, or whenever purely quantum-mechanical phenomena are expected, such as instantons or quantum tunneling effects~\cite{Hertzberg:2020tqa,Tranberg:2022noe}. In this respect, the two-particle irreducible (2PI) effective action~\cite{Cornwall:1974vz,Berges:2004yj,Berges:2015kfa} provides a compelling method for studying quantum field dynamics, see {\it e.g.}~\cite{Kainulainen:2021eki,Kainulainen:2022lzp,Tranberg:2023uzs,Kainulainen:2024etd,Tranberg:2024klz}. Including 2PI techniques, however, has a large associated cost. Mainly, it requires a diagrammatic expansion (typically on small couplings, but possibly also on the inverse number of fields) which must be truncated, as well as to include a renormalization procedure, so that counter-terms are introduced to absorb infinities. In addition, the resulting equations of motion are very costly to integrate numerically, as they have memory kernels, which need to be truncated in time. The dynamics of simple scalar field systems have been investigated with appropriate expansion schemes at leading order, next-to-leading order and even next-to-next-to-leading order, finding good convergence towards the classical result as the order of the truncation increases, whenever the dynamics develop large occupation numbers~\cite{Tranberg:2023uzs,Kainulainen:2024etd,Tranberg:2024klz,Tranberg:2025lto}. Alternatively, methods based on tensor networks have been used recently~\cite{Budd:2026rmw} to study the real-time dynamics of interacting quantum fields in (1+1) dimensions. 
\end{itemize}

As a final remark, we note that analogous methods to LCT have also been used in applications of other areas of high energy physics. 
For example, classical-statistical simulations have been used to study the Abelian~\cite{Buividovich:2015jfa,Buividovich:2016ulp,Figueroa:2017hun,Figueroa:2019jsi,Mace:2019cqo,Mace:2020dkp} and non-Abelian~\cite{Akamatsu:2015kau} dynamics associated to the chiral anomaly, or to compute quantities such as the sphaleron-rate \cite{Philipsen:1995sg,Ambjorn:1995xm,Arnold:1995bh,Arnold:1996dy,Arnold:1997yb,Moore:1997sn,Bodeker:1998hm,Moore:1998zk,Moore:1999fs,Bodeker:1999gx,Arnold:1999uy, Tang:1996qx,Ambjorn:1997jz,Moore:2000mx,DOnofrio:2012phz,DOnofrio:2015gop}. They have also been used to study spectral quantities~\cite{Boguslavski:2018beu,Schlichting:2019tbr} and some other properties of the quark-gluon plasma~\cite{Laine:2009dd,Laine:2013lia,Panero:2013pla,Boguslavski:2020tqz}.

\label{sec:discussion}

\section*{Acknowledgments}

We thank the Institute of Basic Sciences and the local organizers of the \href{https://indico.ific.uv.es/event/8110/}{\color{blue}CosmoLattice School 2025} (Mohammad Ali Gorji, Dong-Won Jung, and Masahide Yamaguchi) for their kind hospitality during our stay in Daejeon, South Korea. We are also thankful to the many students that attended the school, for the stimulant atmosphere they created and the great many questions they asked. 

The work of DGF (0000-0002-4005-8915) is supported by the grants CIPROM/2022/69, EUR2022-134028, PROMETEO/2021/083, and PID2023-148162NB-C22. JL (0000-0002-1198-3191) and AU (0000-0002-0238-8390) acknowledge support from Eusko Jaurlaritza IT1628-22 and by the PID2024-156016NB-I00 grant funded by MCIN/AEI/10.13039/501100011033/ and by ERDF: “A way of making Europe”. In particular, AU gratefully acknowledges the support from the University of the Basque Country grant PIF20/151. FT (0000-0003-1883-8365) is supported by a \textit{Beatriu de Pinós} fellowship (2022-BP-00063) from the Ministry of Research and Universities of the Government of Catalonia, and also acknowledges support from grants PID2022-137268NB-C52 from the Spanish Ministry of Science and Innovation, 2021-SGR-00872 from AGAUR, and CEX2024-001451-M funded by MICIU/AEI/10.13039/501100011033. The work of NL was supported by the Czech Science Foundation, GAČR, Project No. 24-13079S. A.F. is supported by the Deutsche Forschungsgemeinschaft (DFG, German Research Foundation) through the Emmy Noether Programme Project No. 545261797. KM acknowledges
support from the Swiss National Science Foundation
(project number P500PT-214466). 

Computations for this work were performed on the local SOM clusters at the Instituto de Física Corpuscular (IFIC), including {\tt Graviton}, with funding contributions from EUROPA EXCELENCIA-2022 Grant No.~EUR2022-134028 and the European Union NextGenerationEU (PRTR-C17.I01) and Generalitat Valenciana Grant No.~ASFAE/2022/020, the ARINA and Solaris clusters at the University of
the Basque Country (UPV/EHU), the Hyperion cluster
from the DIPC Supercomputing Center, the MareNostrum 5 cluster at Barcelona Supercomputing Center (BSC), the FinisTerrae III cluster at Centro de Supercomputación de Galicia (CESGA) and the Lluis Vives and Tirant II clusters at the University of
Valencia (UV). We acknowledge the EuroHPC Joint Undertaking for awarding this project access to the EuroHPC supercomputer LUMI, hosted by CSC (Finland) and the LUMI consortium through a EuroHPC Regular Access call.

\newpage
\appendix
\section*{Appendix: Coefficient tables for higher order integrators}

We provide in Table \ref{tab:VVnCoeffs} the weights of the time steps required in the construction of higher orders of the velocity Verlet algorithms, see Section \ref{subsubsec:SymplecticInt}. Furthermore, we show in Table \ref{tab:RKlsCoefficients} some of the specific coefficients for the low-storage Runge–Kutta schemes discussed in Sect.~\ref{subsubsec:NonSymplecticInt}, following Refs.~\cite{Carpenter1994Thirdorder2R,Carpenter1994Fourthorder2R,Bazavov:2021pik,Bazavov:2025dzo,Bazavov:2025exj}.

\begin{table}[H]
    \centering
\begin{tabular}{|c|c|c|c|}
\hline
    Name & Order & $w_i = \frac{\delta t_i}{\delta t}$ & $q$  \\
     \hline
    $VV4$ & $O(\delta \eta^4)$ & \begin{tabular}{c}
          $w_1 = w_3 = 1.351207191959657771818$  \\
          $w_2 = -1.702414403875838200264$
     \end{tabular} & 3 \\
     \hline
     $VV6$ & $O(\delta t\eta^6)$ &  \begin{tabular}{c}
          $w_1 = w_7 = 0.78451361047755726382$  \\
          $w_2 = w_6 =  0.23557321335935813368$ \\
          $w_3 = w_5 = -1.1776799841788710069$ \\
          $w_4 = 1.3151863206839112189$
     \end{tabular} & 7 \\
     \hline
     $VV8$ & $O(\delta \eta^8)$ & \begin{tabular}{c}
          $w_1 = w_{15} = 0.74167036435061295345$  \\
          $w_2 = w_{14} = -0.40910082580003159400$  \\
          $w_3 = w_{13} = 0.19075471029623837995$  \\
          $w_4 = w_{12} = -0.57386247111608226666$  \\
          $w_5 = w_{11} = 0.29906418130365592384$  \\
          $w_6 = w_{10} = 0.33462491824529818378$  \\
          $w_7 = w_9 = 0.31529309239676659663$  \\
          $w_8 = -0.79688793935291635402$
     \end{tabular} & 15 \\
     \hline
      $VV10$ &  $O(\delta \eta^{10})$ & \begin{tabular}{c}
          $w_1 = w_{31} = -0.48159895600253002870$  \\
          $w_2 = w_{30} = 0.0036303931544595926879$  \\
          $w_3 = w_{29} = 0.50180317558723140279$  \\
          $w_4 = w_{28} = 0.28298402624506254868$  \\
          $w_5 = w_{27} = 0.80702967895372223806$  \\
          $w_6 = w_{26} = -0.026090580538592205447$  \\
          $w_7 = w_{25} = -0.87286590146318071547$  \\
          $w_8 = w_{24} = -0.52373568062510581643$  \\
          $w_9 = w_{23} = 0.44521844299952789252$  \\
          $w_{10} = w_{22} = 0.18612289547097907887$  \\
          $w_{11} = w_{21} = 0.23137327866438360633$  \\
          $w_{12} = w_{20} = -0.52191036590418628905$  \\
          $w_{13} = w_{19} = 0.74866113714499296793$  \\
          $w_{14} = w_{18} = 0.066736511890604057532$  \\
          $w_{15} = w_{17} = -0.80360324375670830316$  \\
          $w_{16} = 0.91249037635867994571$  \\
     \end{tabular} & 31 \\
     \hline
\end{tabular}
    \caption{Weights of the time steps required to construct higher-order velocity Verlet algorithms. A given algorithm requires $q$ iterations. The coefficients are symmetric, in each case, with respect to the intermediate $\omega_i$ parameter. Note that we reported here only the algorithms of a given order with the minimal number of steps. For others, see Ref.~\cite{Kahan:1997:CCR}. }
    \label{tab:VVnCoeffs}
\end{table}

\begin{table}[H]
    \centering
\begin{tabular}{|c|c|c|c|}
\hline
    Name & Order & Stages & Coefficients  \\
     \hline
    RK(2,2) & $O(\delta \eta^2)$ & $s=2$ & \begin{tabular}{l}
          $A_1 = 0$  \\
          $A_2 = -1$\\
          $B_1 = 1$\\
          $B_2 = 1/2$        
     \end{tabular} \\
     \hline
     RK(3,3) & $O(\delta \eta^3)$ & $s=3$ &  \begin{tabular}{l}
          $A_1 = 0$  \\
          $A_2=-5/9$ \\
          $A_3=-153/128$ \\
          $B_1=1/3$ \\
          $B_2=15/16$ \\
          $B_3=8/15$ 
     \end{tabular} \\
     \hline
     RK(3,4) & $O(\delta \eta^3)$ & $s=4$ &  \begin{tabular}{l}
          $A_1 = 0$  \\
          $A_2=-0.7825460361923583$ \\
          $A_3=-2.042914325731225$ \\
          $A_4=-1.799337253940777$ \\
          $B_1=0.06688758201974097$ \\
          $B_2=2.876554598956719$ \\
          $B_3=0.5534657361343982$ \\
          $B_4=0.391273018096179$ 
     \end{tabular} \\
     \hline
RK(4,5) & $O(\delta \eta^4)$ & $s=5$ &  \begin{tabular}{l}
          $A_1 = 0$  \\
          $A_2=-567301805773/1357537059087$ \\
          $A_3=-2404267990393/2016746695238$ \\
          $A_4=-3550918686646/2091501179385$ \\
          $A_5=-1275806237668/842570457699$ \\
          $B_1=1432997174477/9575080441755$ \\
          $B_2=5161836677717/13612068292357$ \\
          $B_3=1720146321549/2090206949498$ \\
          $B_4=3134564353537/4481467310338$ \\
          $B_5=2277821191437/14882151754819$ 
     \end{tabular} \\
     \hline
\end{tabular}
    \caption{Specific coefficients for various low-storage Runge–Kutta schemes, including their order and number of stages, as discussed in Sect.~\ref{subsubsec:NonSymplecticInt} and based on Refs.~\cite{Carpenter1994Thirdorder2R,Carpenter1994Fourthorder2R,Bazavov:2021pik,Bazavov:2025dzo,Bazavov:2025exj}. Other coefficients and order/stages can also be found in these references. }
    \label{tab:RKlsCoefficients}
\end{table}

 \newpage

\begin{multicols}{2}
\footnotesize
\bibliography{manual,intro,defects,AxionInteractions,2Dsection,GWsection,NMscalars,summary,initCond}

@article{Kallosh:2013hoa,
    author = "Kallosh, Renata and Linde, Andrei",
    title = "{Universality Class in Conformal Inflation}",
    eprint = "1306.5220",
    archivePrefix = "arXiv",
    primaryClass = "hep-th",
    doi = "10.1088/1475-7516/2013/07/002",
    journal = "JCAP",
    volume = "07",
    pages = "002",
    year = "2013"
}

@article{Planck:2018jri,
    author = "Akrami, Y. and others",
    collaboration = "Planck",
    title = "{Planck 2018 results. X. Constraints on inflation}",
    eprint = "1807.06211",
    archivePrefix = "arXiv",
    primaryClass = "astro-ph.CO",
    doi = "10.1051/0004-6361/201833887",
    journal = "Astron. Astrophys.",
    volume = "641",
    pages = "A10",
    year = "2020"
}

@article{BICEP:2021xfz,
    author = "Ade, P. A. R. and others",
    collaboration = "BICEP, Keck",
    title = "{Improved Constraints on Primordial Gravitational Waves using Planck, WMAP, and BICEP/Keck Observations through the 2018 Observing Season}",
    eprint = "2110.00483",
    archivePrefix = "arXiv",
    primaryClass = "astro-ph.CO",
    doi = "10.1103/PhysRevLett.127.151301",
    journal = "Phys. Rev. Lett.",
    volume = "127",
    number = "15",
    pages = "151301",
    year = "2021"
}

@article{Turner:1983he,
    author = "Turner, Michael S.",
    title = "{Coherent Scalar Field Oscillations in an Expanding Universe}",
    reportNumber = "EFI-83-29-CHICAGO",
    doi = "10.1103/PhysRevD.28.1243",
    journal = "Phys. Rev. D",
    volume = "28",
    pages = "1243",
    year = "1983"
}

@article{Adler:1969gk,
    author = "Adler, Stephen L.",
    title = "{Axial vector vertex in spinor electrodynamics}",
    doi = "10.1103/PhysRev.177.2426",
    journal = "Phys. Rev.",
    volume = "177",
    pages = "2426--2438",
    year = "1969"
}

@article{Bell:1969ts,
    author = "Bell, J. S. and Jackiw, R.",
    title = "{A PCAC puzzle: $\pi^0 \to \gamma \gamma$ in the $\sigma$ model}",
    doi = "10.1007/BF02823296",
    journal = "Nuovo Cim. A",
    volume = "60",
    pages = "47--61",
    year = "1969"
}

@article{Veneziano:1979ec,
    author = "Veneziano, G.",
    title = "{U(1) Without Instantons}",
    reportNumber = "CERN-TH-2651",
    doi = "10.1016/0550-3213(79)90332-8",
    journal = "Nucl. Phys. B",
    volume = "159",
    pages = "213--224",
    year = "1979"
}

@article{Witten:1979vv,
    author = "Witten, Edward",
    title = "{Current Algebra Theorems for the U(1) Goldstone Boson}",
    reportNumber = "HUTP-79/A014",
    doi = "10.1016/0550-3213(79)90031-2",
    journal = "Nucl. Phys. B",
    volume = "156",
    pages = "269--283",
    year = "1979"
}

@article{tHooft:1976rip,
    author = "'t Hooft, Gerard",
    editor = "Shifman, Mikhail A.",
    title = "{Symmetry Breaking Through Bell-Jackiw Anomalies}",
    reportNumber = "PRINT-76-0254 (HARVARD)",
    doi = "10.1103/PhysRevLett.37.8",
    journal = "Phys. Rev. Lett.",
    volume = "37",
    pages = "8--11",
    year = "1976"
}

@article{tHooft:1976snw,
    author = "'t Hooft, Gerard",
    editor = "Shifman, Mikhail A.",
    title = "{Computation of the Quantum Effects Due to a Four-Dimensional Pseudoparticle}",
    reportNumber = "PRINT-76-0551 (HARVARD)",
    doi = "10.1103/PhysRevD.14.3432",
    journal = "Phys. Rev. D",
    volume = "14",
    pages = "3432--3450",
    year = "1976",
    note = "[Erratum: Phys.Rev.D 18, 2199 (1978)]"
}

@article{Preskill:1982cy,
    author = "Preskill, John and Wise, Mark B. and Wilczek, Frank",
    editor = "Srednicki, M. A.",
    title = "{Cosmology of the Invisible Axion}",
    reportNumber = "HUTP-82-A048, NSF-ITP-82-103",
    doi = "10.1016/0370-2693(83)90637-8",
    journal = "Phys. Lett. B",
    volume = "120",
    pages = "127--132",
    year = "1983"
}

@article{Abbott:1982af,
    author = "Abbott, L. F. and Sikivie, P.",
    editor = "Srednicki, M. A.",
    title = "{A Cosmological Bound on the Invisible Axion}",
    reportNumber = "PRINT-82-0695 (BRANDEIS)",
    doi = "10.1016/0370-2693(83)90638-X",
    journal = "Phys. Lett. B",
    volume = "120",
    pages = "133--136",
    year = "1983"
}

@article{Dine:1982ah,
    author = "Dine, Michael and Fischler, Willy",
    editor = "Srednicki, M. A.",
    title = "{The Not So Harmless Axion}",
    reportNumber = "UPR-0201T",
    doi = "10.1016/0370-2693(83)90639-1",
    journal = "Phys. Lett. B",
    volume = "120",
    pages = "137--141",
    year = "1983"
}

@article{Kim:1979if,
    author = "Kim, Jihn E.",
    title = "{Weak Interaction Singlet and Strong CP Invariance}",
    reportNumber = "UPR-0120T",
    doi = "10.1103/PhysRevLett.43.103",
    journal = "Phys. Rev. Lett.",
    volume = "43",
    pages = "103",
    year = "1979"
}

@article{Shifman:1979if,
    author = "Shifman, Mikhail A. and Vainshtein, A. I. and Zakharov, Valentin I.",
    title = "{Can Confinement Ensure Natural CP Invariance of Strong Interactions?}",
    reportNumber = "ITEP-64-1979",
    doi = "10.1016/0550-3213(80)90209-6",
    journal = "Nucl. Phys. B",
    volume = "166",
    pages = "493--506",
    year = "1980"
}

@article{Marsh:2015xka,
    author = "Marsh, David J. E.",
    title = "{Axion Cosmology}",
    eprint = "1510.07633",
    archivePrefix = "arXiv",
    primaryClass = "astro-ph.CO",
    reportNumber = "KCL-PH-TH-2015-50",
    doi = "10.1016/j.physrep.2016.06.005",
    journal = "Phys. Rept.",
    volume = "643",
    pages = "1--79",
    year = "2016"
}

@article{Sikivie:2020zpn,
    author = "Sikivie, Pierre",
    title = "{Invisible Axion Search Methods}",
    eprint = "2003.02206",
    archivePrefix = "arXiv",
    primaryClass = "hep-ph",
    doi = "10.1103/RevModPhys.93.015004",
    journal = "Rev. Mod. Phys.",
    volume = "93",
    number = "1",
    pages = "015004",
    year = "2021"
}

@article{Irastorza:2021tdu,
    author = "Irastorza, Igor Garcia",
    title = "{An introduction to axions and their detection}",
    eprint = "2109.07376",
    archivePrefix = "arXiv",
    primaryClass = "hep-ph",
    doi = "10.21468/SciPostPhysLectNotes.45",
    journal = "SciPost Phys. Lect. Notes",
    volume = "45",
    pages = "1",
    year = "2022"
}

@article{Co:2019jts,
    author = "Co, Raymond T. and Hall, Lawrence J. and Harigaya, Keisuke",
    title = "{Axion Kinetic Misalignment Mechanism}",
    eprint = "1910.14152",
    archivePrefix = "arXiv",
    primaryClass = "hep-ph",
    reportNumber = "LCTP-19-28",
    doi = "10.1103/PhysRevLett.124.251802",
    journal = "Phys. Rev. Lett.",
    volume = "124",
    number = "25",
    pages = "251802",
    year = "2020"
}

@article{Chang:2019tvx,
    author = "Chang, Chia-Feng and Cui, Yanou",
    title = "{New Perspectives on Axion Misalignment Mechanism}",
    eprint = "1911.11885",
    archivePrefix = "arXiv",
    primaryClass = "hep-ph",
    doi = "10.1103/PhysRevD.102.015003",
    journal = "Phys. Rev. D",
    volume = "102",
    number = "1",
    pages = "015003",
    year = "2020"
}

@article{Co:2019wyp,
    author = "Co, Raymond T. and Harigaya, Keisuke",
    title = "{Axiogenesis}",
    eprint = "1910.02080",
    archivePrefix = "arXiv",
    primaryClass = "hep-ph",
    reportNumber = "LCTP-19-27",
    doi = "10.1103/PhysRevLett.124.111602",
    journal = "Phys. Rev. Lett.",
    volume = "124",
    number = "11",
    pages = "111602",
    year = "2020"
}

@article{Domcke:2020kcp,
    author = "Domcke, Valerie and Ema, Yohei and Mukaida, Kyohei and Yamada, Masaki",
    title = "{Spontaneous Baryogenesis from Axions with Generic Couplings}",
    eprint = "2006.03148",
    archivePrefix = "arXiv",
    primaryClass = "hep-ph",
    reportNumber = "DESY 20-100, DESY-20-100, CERN-TH-2020-088, TU-1102",
    doi = "10.1007/JHEP08(2020)096",
    journal = "JHEP",
    volume = "08",
    pages = "096",
    year = "2020"
}

@article{Co:2020jtv,
    author = "Co, Raymond T. and Fernandez, Nicolas and Ghalsasi, Akshay and Hall, Lawrence J. and Harigaya, Keisuke",
    title = "{Lepto-Axiogenesis}",
    eprint = "2006.05687",
    archivePrefix = "arXiv",
    primaryClass = "hep-ph",
    doi = "10.1007/JHEP03(2021)017",
    journal = "JHEP",
    volume = "03",
    pages = "017",
    year = "2021"
}

@article{Harigaya:2021txz,
    author = "Harigaya, Keisuke and Wang, Isaac R.",
    title = "{Axiogenesis from $SU(2)_R$ phase transition}",
    eprint = "2107.09679",
    archivePrefix = "arXiv",
    primaryClass = "hep-ph",
    doi = "10.1007/JHEP10(2021)022",
    journal = "JHEP",
    volume = "10",
    pages = "022",
    year = "2021",
    note = "[Erratum: JHEP 12, 193 (2021)]"
}

@article{Chakraborty:2021fkp,
    author = "Chakraborty, Sabyasachi and Jung, Tae Hyun and Okui, Takemichi",
    title = "{Composite neutrinos and the QCD axion: Baryogenesis, dark matter, small Dirac neutrino masses, and vanishing neutron electric dipole moment}",
    eprint = "2108.04293",
    archivePrefix = "arXiv",
    primaryClass = "hep-ph",
    reportNumber = "KEK-TH-2341",
    doi = "10.1103/PhysRevD.105.015024",
    journal = "Phys. Rev. D",
    volume = "105",
    number = "1",
    pages = "015024",
    year = "2022"
}

@article{Kawamura:2021xpu,
    author = "Kawamura, Junichiro and Raby, Stuart",
    title = "{Lepto-axiogenesis in minimal SUSY KSVZ model}",
    eprint = "2109.08605",
    archivePrefix = "arXiv",
    primaryClass = "hep-ph",
    reportNumber = "CTPU-PTC-21-33",
    doi = "10.1007/JHEP04(2022)116",
    journal = "JHEP",
    volume = "04",
    pages = "116",
    year = "2022"
}

@article{Co:2021qgl,
    author = "Co, Raymond T. and Harigaya, Keisuke and Johnson, Zachary and Pierce, Aaron",
    title = "{R-parity violation axiogenesis}",
    eprint = "2110.05487",
    archivePrefix = "arXiv",
    primaryClass = "hep-ph",
    reportNumber = "UMN-TH-4104/21, FTPI-MINN-21-21, CERN-TH-2021-147, LCTP-21-25",
    doi = "10.1007/JHEP11(2021)210",
    journal = "JHEP",
    volume = "11",
    pages = "210",
    year = "2021"
}

@article{Co:2021lkc,
    author = "Co, Raymond T. and Dunsky, David and Fernandez, Nicolas and Ghalsasi, Akshay and Hall, Lawrence J. and Harigaya, Keisuke and Shelton, Jessie",
    title = "{Gravitational wave and CMB probes of axion kination}",
    eprint = "2108.09299",
    archivePrefix = "arXiv",
    primaryClass = "hep-ph",
    reportNumber = "UMN-TH-4023/21, FTPI-MINN-21-15, CERN-TH-2021-124",
    doi = "10.1007/JHEP09(2022)116",
    journal = "JHEP",
    volume = "09",
    pages = "116",
    year = "2022"
}

@article{Gouttenoire:2021wzu,
    author = "Gouttenoire, Yann and Servant, G{\'e}raldine and Simakachorn, Peera",
    title = "{Revealing the Primordial Irreducible Inflationary Gravitational-Wave Background with a Spinning Peccei-Quinn Axion}",
    eprint = "2108.10328",
    archivePrefix = "arXiv",
    primaryClass = "hep-ph",
    reportNumber = "DESY 21-126",
    month = "8",
    year = "2021"
}

@article{Gouttenoire:2021jhk,
    author = "Gouttenoire, Yann and Servant, Geraldine and Simakachorn, Peera",
    title = "{Kination cosmology from scalar fields and gravitational-wave signatures}",
    eprint = "2111.01150",
    archivePrefix = "arXiv",
    primaryClass = "hep-ph",
    reportNumber = "DESY 21-134",
    month = "11",
    year = "2021"
}

@article{Fonseca:2019ypl,
    author = "Fonseca, Nayara and Morgante, Enrico and Sato, Ryosuke and Servant, G{\'e}raldine",
    title = "{Axion fragmentation}",
    eprint = "1911.08472",
    archivePrefix = "arXiv",
    primaryClass = "hep-ph",
    reportNumber = "DESY 19-202, DESY-19-202, MITP/19-079",
    doi = "10.1007/JHEP04(2020)010",
    journal = "JHEP",
    volume = "04",
    pages = "010",
    year = "2020"
}

@article{Madge:2021abk,
    author = "Madge, Eric and Ratzinger, Wolfram and Schmitt, Daniel and Schwaller, Pedro",
    title = "{Audible axions with a booster: Stochastic gravitational waves from rotating ALPs}",
    eprint = "2111.12730",
    archivePrefix = "arXiv",
    primaryClass = "hep-ph",
    reportNumber = "MITP-21-063",
    doi = "10.21468/SciPostPhys.12.5.171",
    journal = "SciPost Phys.",
    volume = "12",
    number = "5",
    pages = "171",
    year = "2022"
}

@article{Eroncel:2022vjg,
    author = {Er{\"o}ncel, Cem and Sato, Ryosuke and Servant, Geraldine and S{\o}rensen, Philip},
    title = "{ALP dark matter from kinetic fragmentation: opening up the parameter window}",
    eprint = "2206.14259",
    archivePrefix = "arXiv",
    primaryClass = "hep-ph",
    reportNumber = "DESY 22-106, OU-HET-1148",
    doi = "10.1088/1475-7516/2022/10/053",
    journal = "JCAP",
    volume = "10",
    pages = "053",
    year = "2022"
}

@article{Eroncel:2022efc,
    author = {Er{\"o}ncel, Cem and Servant, G{\'e}raldine},
    title = "{ALP dark matter mini-clusters from kinetic fragmentation}",
    eprint = "2207.10111",
    archivePrefix = "arXiv",
    primaryClass = "hep-ph",
    reportNumber = "DESY 22-115",
    doi = "10.1088/1475-7516/2023/01/009",
    journal = "JCAP",
    volume = "01",
    pages = "009",
    year = "2023"
}

@article{Jeong:2022kdr,
    author = "Jeong, Kwang Sik and Matsukawa, Kohei and Nakagawa, Shota and Takahashi, Fuminobu",
    title = "{Cosmological effects of Peccei-Quinn symmetry breaking on QCD axion dark matter}",
    eprint = "2201.00681",
    archivePrefix = "arXiv",
    primaryClass = "hep-ph",
    reportNumber = "PNUTP-22-A11, TU-1143",
    doi = "10.1088/1475-7516/2022/03/026",
    journal = "JCAP",
    volume = "03",
    number = "03",
    pages = "026",
    year = "2022"
}

@article{DiLuzio:2021gos,
    author = "Di Luzio, Luca and Gavela, Belen and Quilez, Pablo and Ringwald, Andreas",
    title = "{Dark matter from an even lighter QCD axion: trapped misalignment}",
    eprint = "2102.01082",
    archivePrefix = "arXiv",
    primaryClass = "hep-ph",
    reportNumber = "DESY 21-011, DESY-21-011, IFT-UAM/CSIC-20-144, FTUAM-20-21",
    doi = "10.1088/1475-7516/2021/10/001",
    journal = "JCAP",
    volume = "10",
    pages = "001",
    year = "2021"
}

@article{DiLuzio:2024fyt,
    author = "Di Luzio, Luca and S{\o}rensen, Philip",
    title = "{Axion production via trapped misalignment from Peccei-Quinn symmetry breaking}",
    eprint = "2408.04623",
    archivePrefix = "arXiv",
    primaryClass = "hep-ph",
    doi = "10.1007/JHEP10(2024)239",
    journal = "JHEP",
    volume = "10",
    pages = "239",
    year = "2024"
}

@article{Morgante:2021bks,
    author = "Morgante, Enrico and Ratzinger, Wolfram and Sato, Ryosuke and Stefanek, Ben A.",
    title = "{Axion fragmentation on the lattice}",
    eprint = "2109.13823",
    archivePrefix = "arXiv",
    primaryClass = "hep-ph",
    reportNumber = "MITP-21-045",
    doi = "10.1007/JHEP12(2021)037",
    journal = "JHEP",
    volume = "12",
    pages = "037",
    year = "2021"
}

@article{Berges:2019dgr,
    author = {Berges, J{\"u}rgen and Chatrchyan, Aleksandr and Jaeckel, Joerg},
    title = "{Foamy Dark Matter from Monodromies}",
    eprint = "1903.03116",
    archivePrefix = "arXiv",
    primaryClass = "hep-ph",
    doi = "10.1088/1475-7516/2019/08/020",
    journal = "JCAP",
    volume = "08",
    pages = "020",
    year = "2019"
}

@article{Chatrchyan:2020pzh,
    author = "Chatrchyan, Aleksandr and Jaeckel, Joerg",
    title = "{Gravitational waves from the fragmentation of axion-like particle dark matter}",
    eprint = "2004.07844",
    archivePrefix = "arXiv",
    primaryClass = "hep-ph",
    doi = "10.1088/1475-7516/2021/02/003",
    journal = "JCAP",
    volume = "02",
    pages = "003",
    year = "2021"
}

@article{Chatrchyan:2023cmz,
    author = {Chatrchyan, Aleksandr and Er{\"o}ncel, Cem and Koschnitzke, Matthias and Servant, G{\'e}raldine},
    title = "{ALP dark matter with non-periodic potentials: parametric resonance, halo formation and gravitational signatures}",
    eprint = "2305.03756",
    archivePrefix = "arXiv",
    primaryClass = "hep-ph",
    reportNumber = "DESY-23-060",
    doi = "10.1088/1475-7516/2023/10/068",
    journal = "JCAP",
    volume = "10",
    pages = "068",
    year = "2023"
}

@article{Ratzinger:2020oct,
    author = "Ratzinger, Wolfram and Schwaller, Pedro and Stefanek, Ben A.",
    title = "{Gravitational Waves from an Axion-Dark Photon System: A Lattice Study}",
    eprint = "2012.11584",
    archivePrefix = "arXiv",
    primaryClass = "astro-ph.CO",
    reportNumber = "MITP-20-086, ZU-TH-57/20",
    doi = "10.21468/SciPostPhys.11.1.001",
    journal = "SciPost Phys.",
    volume = "11",
    pages = "001",
    year = "2021"
}

@article{Weiner:2020sxn,
    author = "Weiner, Zachary J. and Adshead, Peter and Giblin, John T.",
    title = "{Constraining early dark energy with gravitational waves before recombination}",
    eprint = "2008.01732",
    archivePrefix = "arXiv",
    primaryClass = "astro-ph.CO",
    doi = "10.1103/PhysRevD.103.L021301",
    journal = "Phys. Rev. D",
    volume = "103",
    number = "2",
    pages = "L021301",
    year = "2021"
}

@article{Fasiello:2025ptb,
    author = "Fasiello, Matteo and Lizarraga, Joanes and Papageorgiou, Alexandros and Urio, Ander",
    title = "{Kinetic fragmentation of the QCD axion on the lattice}",
    eprint = "2507.01822",
    archivePrefix = "arXiv",
    primaryClass = "astro-ph.CO",
    doi = "10.1088/1475-7516/2025/09/019",
    journal = "JCAP",
    volume = "09",
    pages = "019",
    year = "2025"
}

@article{Co:2025jnj,
    author = "Co, Raymond T. and Lee, Taegyu and Leonard, Owen P.",
    title = "{Nonperturbative and perturbative dynamics of a light QCD axion: Dark matter and the strong CP problem}",
    eprint = "2508.00979",
    archivePrefix = "arXiv",
    primaryClass = "hep-ph",
    reportNumber = "CETUP2025-00",
    doi = "10.1103/18j2-l2y9",
    journal = "Phys. Rev. D",
    volume = "112",
    number = "11",
    pages = "115007",
    year = "2025"
}

@article{Yamaguchi:1998gx,
    author = "Yamaguchi, Masahide and Kawasaki, M. and Yokoyama, Jun'ichi",
    title = "{Evolution of axionic strings and spectrum of axions radiated from them}",
    eprint = "hep-ph/9811311",
    archivePrefix = "arXiv",
    reportNumber = "UTAP-313, YITP-98-73",
    doi = "10.1103/PhysRevLett.82.4578",
    journal = "Phys. Rev. Lett.",
    volume = "82",
    pages = "4578--4581",
    year = "1999"
}

@inproceedings{Yamaguchi:1999wt,
    author = "Yamaguchi, M. and Kawasaki, M. and Yokoyama, J.",
    title = "{Global string evolution and axion emission}",
    booktitle = "{7th International Symposium on Particles, Strings and Cosmology}",
    pages = "291--294",
    month = "12",
    year = "1999"
}

@inproceedings{Yamaguchi:2000fg,
    author = "Yamaguchi, Masahide and Kawasaki, M. and Yokoyama, J.",
    title = "{Relic axions radiated from axionic strings}",
    booktitle = "{3rd International Workshop on the Identification of Dark Matter}",
    pages = "297--304",
    month = "9",
    year = "2000"
}

@article{Yamaguchi:2002sh,
    author = "Yamaguchi, Masahide and Yokoyama, Jun'ichi",
    title = "{Quantitative evolution of global strings from the Lagrangian view point}",
    eprint = "hep-ph/0210343",
    archivePrefix = "arXiv",
    reportNumber = "BROWN-HET-1332, OU-TAP-188",
    doi = "10.1103/PhysRevD.67.103514",
    journal = "Phys. Rev. D",
    volume = "67",
    pages = "103514",
    year = "2003"
}

@article{Hiramatsu:2010yu,
    author = "Hiramatsu, Takashi and Kawasaki, Masahiro and Sekiguchi, Toyokazu and Yamaguchi, Masahide and Yokoyama, Jun'ichi",
    title = "{Improved estimation of radiated axions from cosmological axionic strings}",
    eprint = "1012.5502",
    archivePrefix = "arXiv",
    primaryClass = "hep-ph",
    reportNumber = "IPMU10-0229, RESCEU-29-10, YITP-10-111",
    doi = "10.1103/PhysRevD.83.123531",
    journal = "Phys. Rev. D",
    volume = "83",
    pages = "123531",
    year = "2011"
}

@article{Hiramatsu:2012gg,
    author = "Hiramatsu, Takashi and Kawasaki, Masahiro and Saikawa, Ken'ichi and Sekiguchi, Toyokazu",
    title = "{Production of dark matter axions from collapse of string-wall systems}",
    eprint = "1202.5851",
    archivePrefix = "arXiv",
    primaryClass = "hep-ph",
    reportNumber = "ICRR-REPORT-608-2011-25, IPMU12-0025, YITP-12-9",
    doi = "10.1103/PhysRevD.85.105020",
    journal = "Phys. Rev. D",
    volume = "85",
    pages = "105020",
    year = "2012",
    note = "[Erratum: Phys.Rev.D 86, 089902 (2012)]"
}

@article{Kawasaki:2018bzv,
    author = "Kawasaki, Masahiro and Sekiguchi, Toyokazu and Yamaguchi, Masahide and Yokoyama, Jun'ichi",
    title = "{Long-term dynamics of cosmological axion strings}",
    eprint = "1806.05566",
    archivePrefix = "arXiv",
    primaryClass = "hep-ph",
    reportNumber = "RESCEU-8/18, RESCEU-8-18",
    doi = "10.1093/ptep/pty098",
    journal = "PTEP",
    volume = "2018",
    number = "9",
    pages = "091E01",
    year = "2018"
}

@article{Buschmann:2019icd,
    author = "Buschmann, Malte and Foster, Joshua W. and Safdi, Benjamin R.",
    title = "{Early-Universe Simulations of the Cosmological Axion}",
    eprint = "1906.00967",
    archivePrefix = "arXiv",
    primaryClass = "astro-ph.CO",
    reportNumber = "LCTP-19-08",
    doi = "10.1103/PhysRevLett.124.161103",
    journal = "Phys. Rev. Lett.",
    volume = "124",
    number = "16",
    pages = "161103",
    year = "2020"
}

@article{Benabou:2023ghl,
    author = "Benabou, Joshua N. and Buschmann, Malte and Kumar, Soubhik and Park, Yujin and Safdi, Benjamin R.",
    title = "{Signatures of primordial energy injection from axion strings}",
    eprint = "2308.01334",
    archivePrefix = "arXiv",
    primaryClass = "hep-ph",
    doi = "10.1103/PhysRevD.109.055005",
    journal = "Phys. Rev. D",
    volume = "109",
    number = "5",
    pages = "055005",
    year = "2024"
}

@article{Kim:2024wku,
    author = "Kim, Heejoo and Park, Junghyeon and Son, Minho",
    title = "{Axion dark matter from cosmic string network}",
    eprint = "2402.00741",
    archivePrefix = "arXiv",
    primaryClass = "hep-ph",
    doi = "10.1007/JHEP07(2024)150",
    journal = "JHEP",
    volume = "07",
    pages = "150",
    year = "2024"
}

@article{Freese:1990rb,
    author = "Freese, Katherine and Frieman, Joshua A. and Olinto, Angela V.",
    title = "{Natural inflation with pseudo - Nambu-Goldstone bosons}",
    reportNumber = "FERMILAB-PUB-90-177-A",
    doi = "10.1103/PhysRevLett.65.3233",
    journal = "Phys. Rev. Lett.",
    volume = "65",
    pages = "3233--3236",
    year = "1990"
}

@article{Adams:1992bn,
    author = "Adams, Fred C. and Bond, J. Richard and Freese, Katherine and Frieman, Joshua A. and Olinto, Angela V.",
    title = "{Natural inflation: Particle physics models, power law spectra for large scale structure, and constraints from COBE}",
    eprint = "hep-ph/9207245",
    archivePrefix = "arXiv",
    reportNumber = "FERMILAB-PUB-92-202-A",
    doi = "10.1103/PhysRevD.47.426",
    journal = "Phys. Rev. D",
    volume = "47",
    pages = "426--455",
    year = "1993"
}

@article{Dimopoulos:2005ac,
    author = "Dimopoulos, S. and Kachru, S. and McGreevy, J. and Wacker, Jay G.",
    title = "{N-flation}",
    eprint = "hep-th/0507205",
    archivePrefix = "arXiv",
    reportNumber = "SLAC-PUB-11016, SU-ITP-05-08",
    doi = "10.1088/1475-7516/2008/08/003",
    journal = "JCAP",
    volume = "08",
    pages = "003",
    year = "2008"
}

@article{Easther:2005zr,
    author = "Easther, Richard and McAllister, Liam",
    title = "{Random matrices and the spectrum of N-flation}",
    eprint = "hep-th/0512102",
    archivePrefix = "arXiv",
    reportNumber = "PUPT-2184",
    doi = "10.1088/1475-7516/2006/05/018",
    journal = "JCAP",
    volume = "05",
    pages = "018",
    year = "2006"
}

@article{Bachlechner:2014hsa,
    author = "Bachlechner, Thomas C. and Dias, Mafalda and Frazer, Jonathan and McAllister, Liam",
    title = "{Chaotic inflation with kinetic alignment of axion fields}",
    eprint = "1404.7496",
    archivePrefix = "arXiv",
    primaryClass = "hep-th",
    doi = "10.1103/PhysRevD.91.023520",
    journal = "Phys. Rev. D",
    volume = "91",
    number = "2",
    pages = "023520",
    year = "2015"
}

@article{Bachlechner:2014gfa,
    author = "Bachlechner, Thomas C. and Long, Cody and McAllister, Liam",
    title = "{Planckian Axions in String Theory}",
    eprint = "1412.1093",
    archivePrefix = "arXiv",
    primaryClass = "hep-th",
    doi = "10.1007/JHEP12(2015)042",
    journal = "JHEP",
    volume = "12",
    pages = "042",
    year = "2015"
}

@article{McAllister:2008hb,
    author = "McAllister, Liam and Silverstein, Eva and Westphal, Alexander",
    title = "{Gravity Waves and Linear Inflation from Axion Monodromy}",
    eprint = "0808.0706",
    archivePrefix = "arXiv",
    primaryClass = "hep-th",
    reportNumber = "SLAC-PUB-13357, SU-ITP-08-15",
    doi = "10.1103/PhysRevD.82.046003",
    journal = "Phys. Rev. D",
    volume = "82",
    pages = "046003",
    year = "2010"
}

@article{Silverstein:2008sg,
    author = "Silverstein, Eva and Westphal, Alexander",
    title = "{Monodromy in the CMB: Gravity Waves and String Inflation}",
    eprint = "0803.3085",
    archivePrefix = "arXiv",
    primaryClass = "hep-th",
    reportNumber = "SU-ITP-08-07, SLAC-PUB-13183",
    doi = "10.1103/PhysRevD.78.106003",
    journal = "Phys. Rev. D",
    volume = "78",
    pages = "106003",
    year = "2008"
}

@article{Marchesano:2014mla,
    author = "Marchesano, Fernando and Shiu, Gary and Uranga, Angel M.",
    title = "{F-term Axion Monodromy Inflation}",
    eprint = "1404.3040",
    archivePrefix = "arXiv",
    primaryClass = "hep-th",
    reportNumber = "IFT-UAM-CSIC-14-032, MAD-TH-04-01",
    doi = "10.1007/JHEP09(2014)184",
    journal = "JHEP",
    volume = "09",
    pages = "184",
    year = "2014"
}

@article{Kappl:2015esy,
    author = "Kappl, Rolf and Nilles, Hans Peter and Winkler, Martin Wolfgang",
    title = "{Modulated Natural Inflation}",
    eprint = "1511.05560",
    archivePrefix = "arXiv",
    primaryClass = "hep-th",
    doi = "10.1016/j.physletb.2015.12.073",
    journal = "Phys. Lett. B",
    volume = "753",
    pages = "653--659",
    year = "2016"
}

@article{Arkani-Hamed:2003xts,
    author = "Arkani-Hamed, Nima and Cheng, Hsin-Chia and Creminelli, Paolo and Randall, Lisa",
    title = "{Extra natural inflation}",
    eprint = "hep-th/0301218",
    archivePrefix = "arXiv",
    reportNumber = "HUTP-03-A006",
    doi = "10.1103/PhysRevLett.90.221302",
    journal = "Phys. Rev. Lett.",
    volume = "90",
    pages = "221302",
    year = "2003"
}

@article{Kim:2004rp,
    author = "Kim, Jihn E. and Nilles, Hans Peter and Peloso, Marco",
    title = "{Completing natural inflation}",
    eprint = "hep-ph/0409138",
    archivePrefix = "arXiv",
    doi = "10.1088/1475-7516/2005/01/005",
    journal = "JCAP",
    volume = "01",
    pages = "005",
    year = "2005"
}

@article{Kaloper:2008fb,
    author = "Kaloper, Nemanja and Sorbo, Lorenzo",
    title = "{A Natural Framework for Chaotic Inflation}",
    eprint = "0811.1989",
    archivePrefix = "arXiv",
    primaryClass = "hep-th",
    doi = "10.1103/PhysRevLett.102.121301",
    journal = "Phys. Rev. Lett.",
    volume = "102",
    pages = "121301",
    year = "2009"
}

@article{Anber:2006xt,
    author = "Anber, Mohamed M. and Sorbo, Lorenzo",
    title = "{N-flationary magnetic fields}",
    eprint = "astro-ph/0606534",
    archivePrefix = "arXiv",
    doi = "10.1088/1475-7516/2006/10/018",
    journal = "JCAP",
    volume = "10",
    pages = "018",
    year = "2006"
}

@article{Anber:2009ua,
    author = "Anber, Mohamed M. and Sorbo, Lorenzo",
    title = "{Naturally inflating on steep potentials through electromagnetic dissipation}",
    eprint = "0908.4089",
    archivePrefix = "arXiv",
    primaryClass = "hep-th",
    doi = "10.1103/PhysRevD.81.043534",
    journal = "Phys. Rev. D",
    volume = "81",
    pages = "043534",
    year = "2010"
}

@article{Barnaby:2010vf,
    author = "Barnaby, Neil and Peloso, Marco",
    title = "{Large Nongaussianity in Axion Inflation}",
    eprint = "1011.1500",
    archivePrefix = "arXiv",
    primaryClass = "hep-ph",
    reportNumber = "UMN-TH-2926-10",
    doi = "10.1103/PhysRevLett.106.181301",
    journal = "Phys. Rev. Lett.",
    volume = "106",
    pages = "181301",
    year = "2011"
}

@article{Adshead:2013qp,
    author = "Adshead, Peter and Martinec, Emil and Wyman, Mark",
    title = "{Gauge fields and inflation: Chiral gravitational waves, fluctuations, and the Lyth bound}",
    eprint = "1301.2598",
    archivePrefix = "arXiv",
    primaryClass = "hep-th",
    doi = "10.1103/PhysRevD.88.021302",
    journal = "Phys. Rev. D",
    volume = "88",
    number = "2",
    pages = "021302",
    year = "2013"
}

@article{Cheng:2015oqa,
    author = "Cheng, Shu-Lin and Lee, Wolung and Ng, Kin-Wang",
    title = "{Numerical study of pseudoscalar inflation with an axion-gauge field coupling}",
    eprint = "1508.00251",
    archivePrefix = "arXiv",
    primaryClass = "astro-ph.CO",
    doi = "10.1103/PhysRevD.93.063510",
    journal = "Phys. Rev. D",
    volume = "93",
    number = "6",
    pages = "063510",
    year = "2016"
}

@article{Barnaby:2011qe,
    author = "Barnaby, Neil and Pajer, Enrico and Peloso, Marco",
    title = "{Gauge Field Production in Axion Inflation: Consequences for Monodromy, non-Gaussianity in the CMB, and Gravitational Waves at Interferometers}",
    eprint = "1110.3327",
    archivePrefix = "arXiv",
    primaryClass = "astro-ph.CO",
    reportNumber = "UMN-TH-3017-11",
    doi = "10.1103/PhysRevD.85.023525",
    journal = "Phys. Rev. D",
    volume = "85",
    pages = "023525",
    year = "2012"
}

@article{Barnaby:2011vw,
    author = "Barnaby, Neil and Namba, Ryo and Peloso, Marco",
    title = "{Phenomenology of a Pseudo-Scalar Inflaton: Naturally Large Nongaussianity}",
    eprint = "1102.4333",
    archivePrefix = "arXiv",
    primaryClass = "astro-ph.CO",
    doi = "10.1088/1475-7516/2011/04/009",
    journal = "JCAP",
    volume = "04",
    pages = "009",
    year = "2011"
}

@article{Cook:2011hg,
    author = "Cook, Jessica L. and Sorbo, Lorenzo",
    title = "{Particle production during inflation and gravitational waves detectable by ground-based interferometers}",
    eprint = "1109.0022",
    archivePrefix = "arXiv",
    primaryClass = "astro-ph.CO",
    doi = "10.1103/PhysRevD.85.023534",
    journal = "Phys. Rev. D",
    volume = "85",
    pages = "023534",
    year = "2012",
    note = "[Erratum: Phys.Rev.D 86, 069901 (2012)]"
}

@article{Pajer:2013fsa,
    author = "Pajer, Enrico and Peloso, Marco",
    title = "{A review of Axion Inflation in the era of Planck}",
    eprint = "1305.3557",
    archivePrefix = "arXiv",
    primaryClass = "hep-th",
    doi = "10.1088/0264-9381/30/21/214002",
    journal = "Class. Quant. Grav.",
    volume = "30",
    pages = "214002",
    year = "2013"
}

@article{Dimastrogiovanni:2018xnn,
    author = "Dimastrogiovanni, Emanuela and Fasiello, Matteo and Hardwick, Robert J. and Assadullahi, Hooshyar and Koyama, Kazuya and Wands, David",
    title = "{Non-Gaussianity from Axion-Gauge Fields Interactions during Inflation}",
    eprint = "1806.05474",
    archivePrefix = "arXiv",
    primaryClass = "astro-ph.CO",
    doi = "10.1088/1475-7516/2018/11/029",
    journal = "JCAP",
    volume = "11",
    pages = "029",
    year = "2018"
}

@article{Maleknejad:2016qjz,
    author = "Maleknejad, Azadeh",
    title = "{Axion Inflation with an SU(2) Gauge Field: Detectable Chiral Gravity Waves}",
    eprint = "1604.03327",
    archivePrefix = "arXiv",
    primaryClass = "hep-ph",
    doi = "10.1007/JHEP07(2016)104",
    journal = "JHEP",
    volume = "07",
    pages = "104",
    year = "2016"
}

@article{Dimastrogiovanni:2016fuu,
    author = "Dimastrogiovanni, Emanuela and Fasiello, Matteo and Fujita, Tomohiro",
    title = "{Primordial Gravitational Waves from Axion-Gauge Fields Dynamics}",
    eprint = "1608.04216",
    archivePrefix = "arXiv",
    primaryClass = "astro-ph.CO",
    doi = "10.1088/1475-7516/2017/01/019",
    journal = "JCAP",
    volume = "01",
    pages = "019",
    year = "2017"
}

@article{Mirzagholi:2020irt,
    author = "Mirzagholi, Leila and Komatsu, Eiichiro and Lozanov, Kaloian D. and Watanabe, Yuki",
    title = "{Effects of Gravitational Chern-Simons during Axion-SU(2) Inflation}",
    eprint = "2003.05931",
    archivePrefix = "arXiv",
    primaryClass = "gr-qc",
    doi = "10.1088/1475-7516/2020/06/024",
    journal = "JCAP",
    volume = "06",
    pages = "024",
    year = "2020"
}

@article{Watanabe:2020ctz,
    author = "Watanabe, Yuki and Komatsu, Eiichiro",
    title = "{Gravitational Wave from Axion-SU(2) Gauge Fields: Effective Field Theory for Kinetically Driven Inflation}",
    eprint = "2004.04350",
    archivePrefix = "arXiv",
    primaryClass = "hep-th",
    month = "4",
    year = "2020"
}

@article{Fujita:2022jkc,
    author = "Fujita, Tomohiro and Imagawa, Kaname and Murai, Kai",
    title = "{Gravitational waves detectable in laser interferometers from axion-SU(2) inflation}",
    eprint = "2203.15273",
    archivePrefix = "arXiv",
    primaryClass = "astro-ph.CO",
    doi = "10.1088/1475-7516/2022/07/046",
    journal = "JCAP",
    volume = "07",
    number = "07",
    pages = "046",
    year = "2022"
}

@article{Badger:2024ekb,
    author = "Badger, Charles and Duval, Hannah and Fujita, Tomohiro and Kuroyanagi, Sachiko and Romero-Rodr{\'\i}guez, Alba and Sakellariadou, Mairi",
    title = "{Detection prospects of gravitational waves from SU(2) axion inflation}",
    eprint = "2406.11742",
    archivePrefix = "arXiv",
    primaryClass = "astro-ph.CO",
    doi = "10.1103/PhysRevD.110.084063",
    journal = "Phys. Rev. D",
    volume = "110",
    number = "8",
    pages = "084063",
    year = "2024"
}

@article{Sorbo:2011rz,
    author = "Sorbo, Lorenzo",
    title = "{Parity violation in the Cosmic Microwave Background from a pseudoscalar inflaton}",
    eprint = "1101.1525",
    archivePrefix = "arXiv",
    primaryClass = "astro-ph.CO",
    doi = "10.1088/1475-7516/2011/06/003",
    journal = "JCAP",
    volume = "06",
    pages = "003",
    year = "2011"
}

@article{Cook:2013xea,
    author = "Cook, Jessica L. and Sorbo, Lorenzo",
    title = "{An inflationary model with small scalar and large tensor nongaussianities}",
    eprint = "1307.7077",
    archivePrefix = "arXiv",
    primaryClass = "astro-ph.CO",
    doi = "10.1088/1475-7516/2013/11/047",
    journal = "JCAP",
    volume = "11",
    pages = "047",
    year = "2013"
}

@article{Bartolo:2016ami,
    author = "Bartolo, Nicola and others",
    title = "{Science with the space-based interferometer LISA. IV: Probing inflation with gravitational waves}",
    eprint = "1610.06481",
    archivePrefix = "arXiv",
    primaryClass = "astro-ph.CO",
    reportNumber = "ACFI-T16-19, UMN-TH-3608-16, CERN-TH-2016-222, KCL-PH-TH-2016-58, IFT-UAM-CSIC-16-104",
    doi = "10.1088/1475-7516/2016/12/026",
    journal = "JCAP",
    volume = "12",
    pages = "026",
    year = "2016"
}

@article{Maggiore:2019uih,
    author = "Maggiore, Michele and others",
    collaboration = "ET",
    title = "{Science Case for the Einstein Telescope}",
    eprint = "1912.02622",
    archivePrefix = "arXiv",
    primaryClass = "astro-ph.CO",
    doi = "10.1088/1475-7516/2020/03/050",
    journal = "JCAP",
    volume = "03",
    pages = "050",
    year = "2020"
}

@article{LISACosmologyWorkingGroup:2022jok,
    author = "Auclair, Pierre and others",
    collaboration = "LISA Cosmology Working Group",
    title = "{Cosmology with the Laser Interferometer Space Antenna}",
    eprint = "2204.05434",
    archivePrefix = "arXiv",
    primaryClass = "astro-ph.CO",
    reportNumber = "LISA CosWG-22-03, FERMILAB-PUB-22-349-SCD",
    doi = "10.1007/s41114-023-00045-2",
    journal = "Living Rev. Rel.",
    volume = "26",
    number = "1",
    pages = "5",
    year = "2023"
}

@article{Bastero-Gil:2022fme,
    author = "Bastero-Gil, Mar and Manso, Ant{\'o}nio Torres",
    title = "{Parity violating gravitational waves at the end of inflation}",
    eprint = "2209.15572",
    archivePrefix = "arXiv",
    primaryClass = "gr-qc",
    doi = "10.1088/1475-7516/2023/08/001",
    journal = "JCAP",
    volume = "08",
    pages = "001",
    year = "2023"
}

@article{Garcia-Bellido:2023ser,
    author = "Garcia-Bellido, Juan and Papageorgiou, Alexandros and Peloso, Marco and Sorbo, Lorenzo",
    title = "{A flashing beacon in axion inflation: recurring bursts of gravitational waves in the strong backreaction regime}",
    eprint = "2303.13425",
    archivePrefix = "arXiv",
    primaryClass = "astro-ph.CO",
    doi = "10.1088/1475-7516/2024/01/034",
    journal = "JCAP",
    volume = "01",
    pages = "034",
    year = "2024"
}

@article{Garretson:1992vt,
    author = "Garretson, W. Daniel and Field, George B. and Carroll, Sean M.",
    title = "{Primordial magnetic fields from pseudoGoldstone bosons}",
    eprint = "hep-ph/9209238",
    archivePrefix = "arXiv",
    reportNumber = "PRINT-92-0448 (CFA,CAMBRIDGE), CFA-3507",
    doi = "10.1103/PhysRevD.46.5346",
    journal = "Phys. Rev. D",
    volume = "46",
    pages = "5346--5351",
    year = "1992"
}

@article{Durrer:2023rhc,
    author = "Durrer, R. and Sobol, O. and Vilchinskii, S.",
    title = "{Backreaction from gauge fields produced during inflation}",
    eprint = "2303.04583",
    archivePrefix = "arXiv",
    primaryClass = "gr-qc",
    reportNumber = "MS-TP-23-06",
    doi = "10.1103/PhysRevD.108.043540",
    journal = "Phys. Rev. D",
    volume = "108",
    number = "4",
    pages = "043540",
    year = "2023"
}

@article{Giovannini:1997eg,
    author = "Giovannini, Massimo and Shaposhnikov, M. E.",
    title = "{Primordial hypermagnetic fields and triangle anomaly}",
    eprint = "hep-ph/9710234",
    archivePrefix = "arXiv",
    reportNumber = "CERN-TH-97-264, DAMTP-97-108",
    doi = "10.1103/PhysRevD.57.2186",
    journal = "Phys. Rev. D",
    volume = "57",
    pages = "2186--2206",
    year = "1998"
}

@article{Anber:2015yca,
    author = "Anber, Mohamed M. and Sabancilar, Eray",
    title = "{Hypermagnetic Fields and Baryon Asymmetry from Pseudoscalar Inflation}",
    eprint = "1507.00744",
    archivePrefix = "arXiv",
    primaryClass = "hep-th",
    doi = "10.1103/PhysRevD.92.101501",
    journal = "Phys. Rev. D",
    volume = "92",
    number = "10",
    pages = "101501",
    year = "2015"
}

@article{Fujita:2016igl,
    author = "Fujita, Tomohiro and Kamada, Kohei",
    title = "{Large-scale magnetic fields can explain the baryon asymmetry of the Universe}",
    eprint = "1602.02109",
    archivePrefix = "arXiv",
    primaryClass = "hep-ph",
    doi = "10.1103/PhysRevD.93.083520",
    journal = "Phys. Rev. D",
    volume = "93",
    number = "8",
    pages = "083520",
    year = "2016"
}

@article{Kamada:2016eeb,
    author = "Kamada, Kohei and Long, Andrew J.",
    title = "{Baryogenesis from decaying magnetic helicity}",
    eprint = "1606.08891",
    archivePrefix = "arXiv",
    primaryClass = "astro-ph.CO",
    doi = "10.1103/PhysRevD.94.063501",
    journal = "Phys. Rev. D",
    volume = "94",
    number = "6",
    pages = "063501",
    year = "2016"
}

@article{Maleknejad:2016dci,
    author = "Maleknejad, Azadeh",
    title = "{Gravitational leptogenesis in axion inflation with SU(2) gauge field}",
    eprint = "1604.06520",
    archivePrefix = "arXiv",
    primaryClass = "hep-ph",
    doi = "10.1088/1475-7516/2016/12/027",
    journal = "JCAP",
    volume = "12",
    pages = "027",
    year = "2016"
}

@article{Jimenez:2017cdr,
    author = "Jim{\'e}nez, Daniel and Kamada, Kohei and Schmitz, Kai and Xu, Xun-Jie",
    title = "{Baryon asymmetry and gravitational waves from pseudoscalar inflation}",
    eprint = "1707.07943",
    archivePrefix = "arXiv",
    primaryClass = "hep-ph",
    doi = "10.1088/1475-7516/2017/12/011",
    journal = "JCAP",
    volume = "12",
    pages = "011",
    year = "2017"
}

@article{Cado:2022evn,
    author = "Cado, Yann and Quir{\'o}s, Mariano",
    title = "{Baryogenesis from combined Higgs{\textendash}scalar field inflation}",
    eprint = "2201.06422",
    archivePrefix = "arXiv",
    primaryClass = "hep-ph",
    doi = "10.1103/PhysRevD.106.055018",
    journal = "Phys. Rev. D",
    volume = "106",
    number = "5",
    pages = "055018",
    year = "2022"
}

@article{Maleknejad:2011sq,
    author = "Maleknejad, A. and Sheikh-Jabbari, M. M.",
    title = "{Non-Abelian Gauge Field Inflation}",
    eprint = "1102.1932",
    archivePrefix = "arXiv",
    primaryClass = "hep-ph",
    reportNumber = "IPM-P-2010-009",
    doi = "10.1103/PhysRevD.84.043515",
    journal = "Phys. Rev. D",
    volume = "84",
    pages = "043515",
    year = "2011"
}

@article{Maleknejad:2011jw,
    author = "Maleknejad, A. and Sheikh-Jabbari, M. M.",
    title = "{Gauge-flation: Inflation From Non-Abelian Gauge Fields}",
    eprint = "1102.1513",
    archivePrefix = "arXiv",
    primaryClass = "hep-ph",
    doi = "10.1016/j.physletb.2013.05.001",
    journal = "Phys. Lett. B",
    volume = "723",
    pages = "224--228",
    year = "2013"
}

@article{Maeda:2013daa,
    author = "Maeda, Kei-ichi and Yamamoto, Kei",
    title = "{Stability analysis of inflation with an SU(2) gauge field}",
    eprint = "1310.6916",
    archivePrefix = "arXiv",
    primaryClass = "gr-qc",
    doi = "10.1088/1475-7516/2013/12/018",
    journal = "JCAP",
    volume = "12",
    pages = "018",
    year = "2013"
}

@article{Adshead:2012kp,
    author = "Adshead, Peter and Wyman, Mark",
    title = "{Chromo-Natural Inflation: Natural inflation on a steep potential with classical non-Abelian gauge fields}",
    eprint = "1202.2366",
    archivePrefix = "arXiv",
    primaryClass = "hep-th",
    doi = "10.1103/PhysRevLett.108.261302",
    journal = "Phys. Rev. Lett.",
    volume = "108",
    pages = "261302",
    year = "2012"
}

@article{Adshead:2013nka,
    author = "Adshead, Peter and Martinec, Emil and Wyman, Mark",
    title = "{Perturbations in Chromo-Natural Inflation}",
    eprint = "1305.2930",
    archivePrefix = "arXiv",
    primaryClass = "hep-th",
    doi = "10.1007/JHEP09(2013)087",
    journal = "JHEP",
    volume = "09",
    pages = "087",
    year = "2013"
}

@article{Adshead:2015kza,
    author = "Adshead, Peter and Sfakianakis, Evangelos I.",
    title = "{Fermion production during and after axion inflation}",
    eprint = "1508.00891",
    archivePrefix = "arXiv",
    primaryClass = "hep-ph",
    doi = "10.1088/1475-7516/2015/11/021",
    journal = "JCAP",
    volume = "11",
    pages = "021",
    year = "2015"
}

@article{Adshead:2018oaa,
    author = "Adshead, Peter and Pearce, Lauren and Peloso, Marco and Roberts, Michael A. and Sorbo, Lorenzo",
    title = "{Phenomenology of fermion production during axion inflation}",
    eprint = "1803.04501",
    archivePrefix = "arXiv",
    primaryClass = "astro-ph.CO",
    doi = "10.1088/1475-7516/2018/06/020",
    journal = "JCAP",
    volume = "06",
    pages = "020",
    year = "2018"
}

@article{Domcke:2018eki,
    author = "Domcke, Valerie and Mukaida, Kyohei",
    title = "{Gauge Field and Fermion Production during Axion Inflation}",
    eprint = "1806.08769",
    archivePrefix = "arXiv",
    primaryClass = "hep-ph",
    reportNumber = "DESY 18-098, DESY-18-098",
    doi = "10.1088/1475-7516/2018/11/020",
    journal = "JCAP",
    volume = "11",
    pages = "020",
    year = "2018"
}

@article{Domcke:2019qmm,
    author = "Domcke, Valerie and Ema, Yohei and Mukaida, Kyohei",
    title = "{Chiral Anomaly, Schwinger Effect, Euler-Heisenberg Lagrangian, and application to axion inflation}",
    eprint = "1910.01205",
    archivePrefix = "arXiv",
    primaryClass = "hep-ph",
    reportNumber = "DESY-19-166, DESY 19-166",
    doi = "10.1007/JHEP02(2020)055",
    journal = "JHEP",
    volume = "02",
    pages = "055",
    year = "2020"
}

@article{Cado:2022pxk,
    author = "Cado, Yann and Quir{\'o}s, Mariano",
    title = "{Numerical study of the Schwinger effect in axion inflation}",
    eprint = "2208.10977",
    archivePrefix = "arXiv",
    primaryClass = "hep-ph",
    doi = "10.1103/PhysRevD.106.123527",
    journal = "Phys. Rev. D",
    volume = "106",
    number = "12",
    pages = "123527",
    year = "2022"
}

@article{Machado:2018nqk,
    author = "Machado, Camila S. and Ratzinger, Wolfram and Schwaller, Pedro and Stefanek, Ben A.",
    title = "{Audible Axions}",
    eprint = "1811.01950",
    archivePrefix = "arXiv",
    primaryClass = "hep-ph",
    reportNumber = "MITP/18-107",
    doi = "10.1007/JHEP01(2019)053",
    journal = "JHEP",
    volume = "01",
    pages = "053",
    year = "2019"
}

@article{Machado:2019xuc,
    author = "Machado, Camila S. and Ratzinger, Wolfram and Schwaller, Pedro and Stefanek, Ben A.",
    title = "{Gravitational wave probes of axionlike particles}",
    eprint = "1912.01007",
    archivePrefix = "arXiv",
    primaryClass = "hep-ph",
    reportNumber = "MITP/19-083",
    doi = "10.1103/PhysRevD.102.075033",
    journal = "Phys. Rev. D",
    volume = "102",
    number = "7",
    pages = "075033",
    year = "2020"
}

@article{Gonzalez:2020fdy,
    author = "Gonzalez, Mark and Hertzberg, Mark P. and Rompineve, Fabrizio",
    title = "{Ultralight Scalar Decay and the Hubble Tension}",
    eprint = "2006.13959",
    archivePrefix = "arXiv",
    primaryClass = "astro-ph.CO",
    doi = "10.1088/1475-7516/2020/10/028",
    journal = "JCAP",
    volume = "10",
    pages = "028",
    year = "2020"
}

@article{Notari:2016npn,
    author = "Notari, Alessio and Tywoniuk, Konrad",
    title = "{Dissipative Axial Inflation}",
    eprint = "1608.06223",
    archivePrefix = "arXiv",
    primaryClass = "hep-th",
    reportNumber = "CERN-TH-2016-189",
    doi = "10.1088/1475-7516/2016/12/038",
    journal = "JCAP",
    volume = "12",
    pages = "038",
    year = "2016"
}

@article{DallAgata:2019yrr,
    author = "Dall'Agata, Gianguido and Gonz{\'a}lez-Mart{\'\i}n, Sergio and Papageorgiou, Alexandros and Peloso, Marco",
    title = "{Warm dark energy}",
    eprint = "1912.09950",
    archivePrefix = "arXiv",
    primaryClass = "hep-th",
    doi = "10.1088/1475-7516/2020/08/032",
    journal = "JCAP",
    volume = "08",
    pages = "032",
    year = "2020"
}

@article{Domcke:2020zez,
    author = "Domcke, Valerie and Guidetti, Veronica and Welling, Yvette and Westphal, Alexander",
    title = "{Resonant backreaction in axion inflation}",
    eprint = "2002.02952",
    archivePrefix = "arXiv",
    primaryClass = "astro-ph.CO",
    reportNumber = "DESY-20-017",
    doi = "10.1088/1475-7516/2020/09/009",
    journal = "JCAP",
    volume = "09",
    pages = "009",
    year = "2020"
}

@article{Gorbar:2021rlt,
    author = "Gorbar, E. V. and Schmitz, K. and Sobol, O. O. and Vilchinskii, S. I.",
    title = "{Gauge-field production during axion inflation in the gradient expansion formalism}",
    eprint = "2109.01651",
    archivePrefix = "arXiv",
    primaryClass = "hep-ph",
    reportNumber = "CERN-TH-2021-128",
    doi = "10.1103/PhysRevD.104.123504",
    journal = "Phys. Rev. D",
    volume = "104",
    number = "12",
    pages = "123504",
    year = "2021"
}

@article{Peloso:2022ovc,
    author = "Peloso, Marco and Sorbo, Lorenzo",
    title = "{Instability in axion inflation with strong backreaction from gauge modes}",
    eprint = "2209.08131",
    archivePrefix = "arXiv",
    primaryClass = "astro-ph.CO",
    doi = "10.1088/1475-7516/2023/01/038",
    journal = "JCAP",
    volume = "01",
    pages = "038",
    year = "2023"
}

@article{Iarygina:2023mtj,
    author = "Iarygina, Oksana and Sfakianakis, Evangelos I. and Sharma, Ramkishor and Brandenburg, Axel",
    title = "{Backreaction of axion-SU(2) dynamics during inflation}",
    eprint = "2311.07557",
    archivePrefix = "arXiv",
    primaryClass = "astro-ph.CO",
    reportNumber = "NORDITA-2023-070",
    doi = "10.1088/1475-7516/2024/04/018",
    journal = "JCAP",
    volume = "04",
    pages = "018",
    year = "2024"
}

@article{Galanti:2024jhw,
    author = "Galanti, Davide Campanella and Conzinu, Pietro and Marozzi, Giovanni and Santos da Costa, Simony",
    title = "{Gauge invariant quantum backreaction in U(1) axion inflation}",
    eprint = "2406.19960",
    archivePrefix = "arXiv",
    primaryClass = "gr-qc",
    doi = "10.1103/PhysRevD.110.123510",
    journal = "Phys. Rev. D",
    volume = "110",
    number = "12",
    pages = "123510",
    year = "2024"
}

@article{Alam:2024fid,
    author = "Alam, Khursid and Dutta, Koushik and Jaman, Nur",
    title = "{CMB constraints on natural inflation with gauge field production}",
    eprint = "2405.10155",
    archivePrefix = "arXiv",
    primaryClass = "astro-ph.CO",
    doi = "10.1088/1475-7516/2024/12/015",
    journal = "JCAP",
    volume = "12",
    pages = "015",
    year = "2024"
}

@article{Dimastrogiovanni:2024lzj,
    author = {Dimastrogiovanni, Ema and Fasiello, Matteo and Michelotti, Martino and {\"O}zsoy, Ogan},
    title = "{A universal constraint on axion non-Abelian dynamics during inflation}",
    eprint = "2405.17411",
    archivePrefix = "arXiv",
    primaryClass = "astro-ph.CO",
    doi = "10.1088/1475-7516/2025/03/007",
    journal = "JCAP",
    volume = "03",
    pages = "007",
    year = "2025"
}

@article{Dimastrogiovanni:2025snj,
    author = "Dimastrogiovanni, Ema and Fasiello, Matteo and Papageorgiou, Alexandros and Gatica, Crist{\'o}bal Zenteno",
    title = "{Pure chromo-natural inflation: signatures of particle production from weak to strong backreaction}",
    eprint = "2504.17750",
    archivePrefix = "arXiv",
    primaryClass = "astro-ph.CO",
    doi = "10.1088/1475-7516/2025/09/042",
    journal = "JCAP",
    volume = "09",
    pages = "042",
    year = "2025"
}

@article{Caravano:2021bfn,
    author = "Caravano, Angelo and Komatsu, Eiichiro and Lozanov, Kaloian D. and Weller, Jochen",
    title = "{Lattice simulations of Abelian gauge fields coupled to axions during inflation}",
    eprint = "2110.10695",
    archivePrefix = "arXiv",
    primaryClass = "astro-ph.CO",
    doi = "10.1103/PhysRevD.105.123530",
    journal = "Phys. Rev. D",
    volume = "105",
    number = "12",
    pages = "123530",
    year = "2022"
}

@article{Moore:1996qs,
    author = "Moore, Guy D.",
    title = "{Motion of Chern-Simons number at high temperatures under a chemical potential}",
    eprint = "hep-ph/9603384",
    archivePrefix = "arXiv",
    reportNumber = "PUPT-1606",
    doi = "10.1016/S0550-3213(96)00445-2",
    journal = "Nucl. Phys. B",
    volume = "480",
    pages = "657--688",
    year = "1996"
}

@article{Moore:1996wn,
    author = "Moore, Guy D.",
    title = "{Improved Hamiltonian for Minkowski Yang-Mills theory}",
    eprint = "hep-lat/9605001",
    archivePrefix = "arXiv",
    reportNumber = "PUPT-1618, PUP-TH-1618",
    doi = "10.1016/S0550-3213(96)00497-X",
    journal = "Nucl. Phys. B",
    volume = "480",
    pages = "689--728",
    year = "1996"
}

@article{Ambjorn:1990pu,
    author = "Ambjorn, Jan and Askgaard, T. and Porter, H. and Shaposhnikov, M. E.",
    title = "{Sphaleron transitions and baryon asymmetry: A Numerical real time analysis}",
    reportNumber = "NBI-HE-90-48",
    doi = "10.1016/0550-3213(91)90341-T",
    journal = "Nucl. Phys. B",
    volume = "353",
    pages = "346--378",
    year = "1991"
}

@article{Sobol:2019xls,
    author = "Sobol, O. O. and Gorbar, E. V. and Vilchinskii, S. I.",
    title = "{Backreaction of electromagnetic fields and the Schwinger effect in pseudoscalar inflation magnetogenesis}",
    eprint = "1907.10443",
    archivePrefix = "arXiv",
    primaryClass = "astro-ph.CO",
    doi = "10.1103/PhysRevD.100.063523",
    journal = "Phys. Rev. D",
    volume = "100",
    number = "6",
    pages = "063523",
    year = "2019"
}

@article{Durrer:2024ibi,
    author = "Durrer, R. and von Eckardstein, R. and Garg, Deepen and Schmitz, K. and Sobol, O. and Vilchinskii, S.",
    title = "{Scalar perturbations from inflation in the presence of gauge fields}",
    eprint = "2404.19694",
    archivePrefix = "arXiv",
    primaryClass = "astro-ph.CO",
    reportNumber = "MS-TP-24-10",
    doi = "10.1103/PhysRevD.110.043533",
    journal = "Phys. Rev. D",
    volume = "110",
    number = "4",
    pages = "043533",
    year = "2024"
}

@article{vonEckardstein:2023gwk,
    author = "von Eckardstein, Richard and Peloso, Marco and Schmitz, Kai and Sobol, Oleksandr and Sorbo, Lorenzo",
    title = "{Axion inflation in the strong-backreaction regime: decay of the Anber-Sorbo solution}",
    eprint = "2309.04254",
    archivePrefix = "arXiv",
    primaryClass = "hep-ph",
    reportNumber = "ACFI-T23-05, MS-TP-23-38",
    doi = "10.1007/JHEP11(2023)183",
    journal = "JHEP",
    volume = "11",
    pages = "183",
    year = "2023"
}

@article{Abe:2014xja,
    author = "Abe, Hiroyuki and Kobayashi, Tatsuo and Otsuka, Hajime",
    title = "{Natural inflation with and without modulations in type IIB string theory}",
    eprint = "1411.4768",
    archivePrefix = "arXiv",
    primaryClass = "hep-th",
    reportNumber = "WU-HEP-14-10, EPHOU-14-019",
    doi = "10.1007/JHEP04(2015)160",
    journal = "JHEP",
    volume = "04",
    pages = "160",
    year = "2015"
}

@article{Garcia-Bellido:2007fiu,
    author = "Garcia-Bellido, Juan and Figueroa, Daniel G. and Sastre, Alfonso",
    title = "{A Gravitational Wave Background from Reheating after Hybrid Inflation}",
    eprint = "0707.0839",
    archivePrefix = "arXiv",
    primaryClass = "hep-ph",
    reportNumber = "IFT-UAM-CSIC-07-38",
    doi = "10.1103/PhysRevD.77.043517",
    journal = "Phys. Rev. D",
    volume = "77",
    pages = "043517",
    year = "2008"
}

@article{Figueroa:2011ye,
    author = "Figueroa, Daniel G. and Garcia-Bellido, Juan and Rajantie, Arttu",
    title = "{On the Transverse-Traceless Projection in Lattice Simulations of Gravitational Wave Production}",
    eprint = "1110.0337",
    archivePrefix = "arXiv",
    primaryClass = "astro-ph.CO",
    reportNumber = "HIP-2011-26, IFT-UAM-CSIC-11-54, IMPERIAL-TP-2011-AR-01",
    doi = "10.1088/1475-7516/2011/11/015",
    journal = "JCAP",
    volume = "11",
    pages = "015",
    year = "2011"
}

@book{Carroll:2004st,
    author = "Carroll, Sean M.",
    title = "{Spacetime and Geometry}: {An Introduction to General Relativity}",
    doi = "10.1017/9781108770385",
    isbn = "978-0-8053-8732-2, 978-1-108-48839-6, 978-1-108-77555-7",
    publisher = "Cambridge University Press",
    month = "7",
    year = "2019"
}

@book{Baumann_2022,
place={Cambridge},
title={Cosmology},
publisher={Cambridge University Press},
author={Baumann, Daniel},
year={2022}}

@Electronic{GWmodule:2022,
  url          = {https://cosmolattice.net},
  author       = "Baeza-Ballesteros, Jorge and Figueroa, Daniel G. and Florio, Adrien and Loayza Romero, Nicolás",
  title        = "{CosmoLattice Technical Note II: Gravitational Waves}",
  year         = {2022},
  
}

@Electronic{GWmodule:2023,
  url          = {https://cosmolattice.net},
  author       = "Baeza-Ballesteros, Jorge and Figueroa, Daniel G. and Loayza Romero, Nicolás",
  title        = "{CosmoLattice Technical Note III: Gravitational Waves from U(1) Gauge Theories}",
  year         = {2023},
  
}

@article{Bassett:1997az,
    author = "Bassett, Bruce A. and Liberati, Stefano",
    title = "{Geometric reheating after inflation}",
    eprint = "hep-ph/9709417",
    archivePrefix = "arXiv",
    reportNumber = "SISSA-120-97-A",
    doi = "10.1103/PhysRevD.60.049902",
    journal = "Phys. Rev. D",
    volume = "58",
    pages = "021302",
    year = "1998",
    note = "[Erratum: Phys.Rev.D 60, 049902 (1999)]"
}

@article{Tsujikawa:1999jh,
    author = "Tsujikawa, Shinji and Maeda, Kei-ichi and Torii, Takashi",
    title = "{Resonant particle production with nonminimally coupled scalar fields in preheating after inflation}",
    eprint = "hep-ph/9901306",
    archivePrefix = "arXiv",
    reportNumber = "WU-AP-75-99",
    doi = "10.1103/PhysRevD.60.063515",
    journal = "Phys. Rev. D",
    volume = "60",
    pages = "063515",
    year = "1999"
}

@article{Tsujikawa:1999iv,
    author = "Tsujikawa, Shinji and Maeda, Kei-ichi and Torii, Takashi",
    title = "{Preheating with nonminimally coupled scalar fields in higher curvature inflation models}",
    eprint = "hep-ph/9906501",
    archivePrefix = "arXiv",
    doi = "10.1103/PhysRevD.60.123505",
    journal = "Phys. Rev. D",
    volume = "60",
    pages = "123505",
    year = "1999"
}

@article{Watanabe:2006ku,
    author = "Watanabe, Yuki and Komatsu, Eiichiro",
    title = "{Reheating of the universe after inflation with f(phi)R gravity}",
    eprint = "gr-qc/0612120",
    archivePrefix = "arXiv",
    doi = "10.1103/PhysRevD.75.061301",
    journal = "Phys. Rev. D",
    volume = "75",
    pages = "061301",
    year = "2007"
}

@article{Garcia-Bellido:2008ycs,
    author = "Garcia-Bellido, Juan and Figueroa, Daniel G. and Rubio, Javier",
    title = "{Preheating in the Standard Model with the Higgs-Inflaton coupled to gravity}",
    eprint = "0812.4624",
    archivePrefix = "arXiv",
    primaryClass = "hep-ph",
    reportNumber = "IFT-UAM-CSIC-08-93",
    doi = "10.1103/PhysRevD.79.063531",
    journal = "Phys. Rev. D",
    volume = "79",
    pages = "063531",
    year = "2009"
}

@article{Figueroa:2016dsc,
    author = "Figueroa, Daniel G. and Byrnes, Christian T.",
    title = "{The Standard Model Higgs as the origin of the hot Big Bang}",
    eprint = "1604.03905",
    archivePrefix = "arXiv",
    primaryClass = "hep-ph",
    reportNumber = "CERN-TH-2016-088",
    doi = "10.1016/j.physletb.2017.01.059",
    journal = "Phys. Lett. B",
    volume = "767",
    pages = "272--277",
    year = "2017"
}

@article{Ema:2016dny,
    author = "Ema, Yohei and Jinno, Ryusuke and Mukaida, Kyohei and Nakayama, Kazunori",
    title = "{Violent Preheating in Inflation with Nonminimal Coupling}",
    eprint = "1609.05209",
    archivePrefix = "arXiv",
    primaryClass = "hep-ph",
    doi = "10.1088/1475-7516/2017/02/045",
    journal = "JCAP",
    volume = "02",
    pages = "045",
    year = "2017"
}

@article{Sfakianakis:2018lzf,
    author = "Sfakianakis, Evangelos I. and van de Vis, Jorinde",
    title = "{Preheating after Higgs Inflation: Self-Resonance and Gauge boson production}",
    eprint = "1810.01304",
    archivePrefix = "arXiv",
    primaryClass = "hep-ph",
    reportNumber = "Nikhef-2018-044",
    doi = "10.1103/PhysRevD.99.083519",
    journal = "Phys. Rev. D",
    volume = "99",
    number = "8",
    pages = "083519",
    year = "2019"
}

@article{Fu:2019qqe,
    author = "Fu, Chengjie and Wu, Puxun and Yu, Hongwei",
    title = "{Nonlinear preheating with nonminimally coupled scalar fields in the Starobinsky model}",
    eprint = "1906.00557",
    archivePrefix = "arXiv",
    primaryClass = "astro-ph.CO",
    doi = "10.1103/PhysRevD.99.123526",
    journal = "Phys. Rev. D",
    volume = "99",
    number = "12",
    pages = "123526",
    year = "2019"
}

@article{Opferkuch:2019zbd,
    author = "Opferkuch, Toby and Schwaller, Pedro and Stefanek, Ben A.",
    title = "{Ricci Reheating}",
    eprint = "1905.06823",
    archivePrefix = "arXiv",
    primaryClass = "gr-qc",
    reportNumber = "CERN-TH-2019-063, MITP/19-032",
    doi = "10.1088/1475-7516/2019/07/016",
    journal = "JCAP",
    volume = "07",
    pages = "016",
    year = "2019"
}

@article{Dimopoulos:2018wfg,
    author = "Dimopoulos, Konstantinos and Markkanen, Tommi",
    title = "{Non-minimal gravitational reheating during kination}",
    eprint = "1803.07399",
    archivePrefix = "arXiv",
    primaryClass = "gr-qc",
    reportNumber = "IMPERIAL-TP-2018-TM-01",
    doi = "10.1088/1475-7516/2018/06/021",
    journal = "JCAP",
    volume = "06",
    pages = "021",
    year = "2018"
}

@article{Bezrukov:2020txg,
    author = "Bezrukov, Fedor and Shepherd, Chris",
    title = "{A heatwave affair: mixed Higgs-$R^2$ preheating on the lattice}",
    eprint = "2007.10978",
    archivePrefix = "arXiv",
    primaryClass = "hep-ph",
    reportNumber = "MAN/HEP/2020/009",
    doi = "10.1088/1475-7516/2020/12/028",
    journal = "JCAP",
    volume = "12",
    pages = "028",
    year = "2020"
}

@article{Bettoni:2021zhq,
    author = "Bettoni, Dario and Lopez-Eiguren, Asier and Rubio, Javier",
    title = "{Hubble-induced phase transitions on the lattice with applications to Ricci reheating}",
    eprint = "2107.09671",
    archivePrefix = "arXiv",
    primaryClass = "hep-ph",
    doi = "10.1088/1475-7516/2022/01/002",
    journal = "JCAP",
    volume = "01",
    number = "01",
    pages = "002",
    year = "2022"
}

@article{Laverda:2023uqv,
    author = "Laverda, Giorgio and Rubio, Javier",
    title = "{Ricci reheating reloaded}",
    eprint = "2307.03774",
    archivePrefix = "arXiv",
    primaryClass = "astro-ph.CO",
    reportNumber = "IPARCOS-UCM-23-042",
    doi = "10.1088/1475-7516/2024/03/033",
    journal = "JCAP",
    volume = "03",
    pages = "033",
    year = "2024",
    note = "[Erratum: JCAP 06, E01 (2024)]"
}

@article{Laverda:2024qjt,
    author = "Laverda, Giorgio and Rubio, Javier",
    title = "{The rise and fall of the Standard-Model Higgs: electroweak vacuum stability during kination}",
    eprint = "2402.06000",
    archivePrefix = "arXiv",
    primaryClass = "hep-ph",
    reportNumber = "IPARCOS-UCM-24-010",
    doi = "10.1007/JHEP05(2024)339",
    journal = "JHEP",
    volume = "05",
    pages = "339",
    year = "2024"
}

@article{Lachapelle:2008sy,
    author = "Lachapelle, Jean and Brandenberger, Robert H.",
    title = "{Preheating with Non-Standard Kinetic Term}",
    eprint = "0808.0936",
    archivePrefix = "arXiv",
    primaryClass = "hep-th",
    doi = "10.1088/1475-7516/2009/04/020",
    journal = "JCAP",
    volume = "04",
    pages = "020",
    year = "2009"
}

@article{Rahmati:2014cwa,
    author = "Rahmati, Shohreh and Seahra, Sanjeev S.",
    title = "{Frustration of resonant preheating by exotic kinetic terms}",
    eprint = "1406.4691",
    archivePrefix = "arXiv",
    primaryClass = "astro-ph.CO",
    doi = "10.1088/1475-7516/2014/10/045",
    journal = "JCAP",
    volume = "10",
    pages = "045",
    year = "2014"
}

@article{Li:2019ncw,
    author = "Li, Xinfei and Fan, Jinbo and Liu, Xin and Deng, Gao-Ming and Huang, Yong-Chang",
    title = "{Preheating in the nonminimal derivative coupling curvaton model with nonstandard kinetic matter term}",
    doi = "10.1142/S0217751X19500799",
    journal = "Int. J. Mod. Phys. A",
    volume = "34",
    number = "15",
    pages = "1950079",
    year = "2019"
}

@article{Adshead:2023nhk,
    author = "Adshead, Peter and Giblin, Jr., John T. and Pfaltzgraff-Carlson, Reid",
    title = "{Kinetic preheating after {\ensuremath{\alpha}}-attractor inflation}",
    eprint = "2311.17237",
    archivePrefix = "arXiv",
    primaryClass = "astro-ph.CO",
    doi = "10.1016/j.physletb.2024.138928",
    journal = "Phys. Lett. B",
    volume = "856",
    pages = "138928",
    year = "2024"
}

@article{Adshead:2024ykw,
    author = "Adshead, Peter and Giblin, Jr., John T. and Tishue, Avery",
    title = "{Gravitational waves from kinetic preheating}",
    eprint = "2402.16152",
    archivePrefix = "arXiv",
    primaryClass = "astro-ph.CO",
    doi = "10.1103/PhysRevD.110.043536",
    journal = "Phys. Rev. D",
    volume = "110",
    number = "4",
    pages = "043536",
    year = "2024"
}

@article{Huang:2024amu,
    author = "Huang, Zhiqi and Ouyang, Xichang and Cui, Yu and Liu, Jianqi and Yao, Yanhong and Qiu, Zehong and Yu, Guangyao and Huang, Lu and Li, Zhuoyang and Wong, Chi-Fong",
    title = "{Curvature perturbations from kinetic preheating after {\ensuremath{\alpha}}-attractor inflation}",
    eprint = "2408.14881",
    archivePrefix = "arXiv",
    primaryClass = "astro-ph.CO",
    reportNumber = "SYSU-SPA-2024",
    doi = "10.1103/djqm-5py3",
    journal = "Phys. Rev. D",
    volume = "112",
    number = "4",
    pages = "043530",
    year = "2025"
}

@article{Brandenberger:2019njw,
    author = "Brandenberger, Robert and Namba, Ryo and Ramos, Rudnei O.",
    title = "{Kinetic Equilibration after Preheating}",
    eprint = "1908.09866",
    archivePrefix = "arXiv",
    primaryClass = "hep-ph",
    month = "8",
    year = "2019"
}

@article{Adshead:2025gka,
    author = "Adshead, Peter and Currens, Eve and Giblin, John T.",
    title = "{Primordial Black Holes from Kinetic Preheating}",
    eprint = "2511.02059",
    archivePrefix = "arXiv",
    primaryClass = "astro-ph.CO",
    month = "11",
    year = "2025"
}

@article{Figueroa:2017slm,
    author = "Figueroa, Daniel G. and Rajantie, Arttu and Torrenti, Francisco",
    title = "{Higgs field-curvature coupling and postinflationary vacuum instability}",
    eprint = "1709.00398",
    archivePrefix = "arXiv",
    primaryClass = "astro-ph.CO",
    reportNumber = "IFT-UAM-CSIC-17-078, IFT-UAM/CSIC-17-078",
    doi = "10.1103/PhysRevD.98.023532",
    journal = "Phys. Rev. D",
    volume = "98",
    number = "2",
    pages = "023532",
    year = "2018"
}

@article{Child:2013ria,
    author = "Child, Hillary L. and Giblin, Jr, John T. and Ribeiro, Raquel H. and Seery, David",
    title = "{Preheating with Non-Minimal Kinetic Terms}",
    eprint = "1305.0561",
    archivePrefix = "arXiv",
    primaryClass = "astro-ph.CO",
    doi = "10.1103/PhysRevLett.111.051301",
    journal = "Phys. Rev. Lett.",
    volume = "111",
    pages = "051301",
    year = "2013"
}

@article{Traschen:1990sw,
      author         = "Traschen, Jennie H. and Brandenberger, Robert H.",
      title          = "{Particle Production During Out-of-equilibrium Phase
                        Transitions}",
      journal        = "Phys. Rev.",
      volume         = "D42",
      year           = "1990",
      pages          = "2491-2504",
      doi            = "10.1103/PhysRevD.42.2491",
      reportNumber   = "BROWN-HET-731",
      SLACcitation   = "%%CITATION = PHRVA,D42,2491;%%"
}

@article{Kofman:1994rk,
      author         = "Kofman, Lev and Linde, Andrei D. and Starobinsky, Alexei
                        A.",
      title          = "{Reheating after inflation}",
      journal        = "Phys. Rev. Lett.",
      volume         = "73",
      year           = "1994",
      pages          = "3195-3198",
      doi            = "10.1103/PhysRevLett.73.3195",
      eprint         = "hep-th/9405187",
      archivePrefix  = "arXiv",
      primaryClass   = "hep-th",
      reportNumber   = "UH-IFA-94-35, SU-ITP-94-13, YITP-U-94-15",
      SLACcitation   = "%%CITATION = HEP-TH/9405187;%%"
}

@article{Shtanov:1994ce,
      author         = "Shtanov, Y. and Traschen, Jennie H. and Brandenberger,
                        Robert H.",
      title          = "{Universe reheating after inflation}",
      journal        = "Phys. Rev.",
      volume         = "D51",
      year           = "1995",
      pages          = "5438-5455",
      doi            = "10.1103/PhysRevD.51.5438",
      eprint         = "hep-ph/9407247",
      archivePrefix  = "arXiv",
      primaryClass   = "hep-ph",
      reportNumber   = "BROWN-HET-957",
      SLACcitation   = "%%CITATION = HEP-PH/9407247;%%"
}

@article{Kaiser:1995fb,
      author         = "Kaiser, David I.",
      title          = "{Post inflation reheating in an expanding universe}",
      journal        = "Phys. Rev.",
      volume         = "D53",
      year           = "1996",
      pages          = "1776-1783",
      doi            = "10.1103/PhysRevD.53.1776",
      eprint         = "astro-ph/9507108",
      archivePrefix  = "arXiv",
      primaryClass   = "astro-ph",
      reportNumber   = "HUTP-95-A027, DART-HEP-95-04",
      SLACcitation   = "%%CITATION = ASTRO-PH/9507108;%%"
}

@article{Kaiser:1997mp,
      author         = "Kaiser, David I.",
      title          = "{Preheating in an expanding universe: Analytic results
                        for the massless case}",
      journal        = "Phys. Rev.",
      volume         = "D56",
      year           = "1997",
      pages          = "706-716",
      doi            = "10.1103/PhysRevD.56.706",
      eprint         = "hep-ph/9702244",
      archivePrefix  = "arXiv",
      primaryClass   = "hep-ph",
      reportNumber   = "HUTP-97-A005",
      SLACcitation   = "%%CITATION = HEP-PH/9702244;%%"
}

@article{Kofman:1997yn,
      author         = "Kofman, Lev and Linde, Andrei D. and Starobinsky, Alexei
                        A.",
      title          = "{Towards the theory of reheating after inflation}",
      journal        = "Phys. Rev.",
      volume         = "D56",
      year           = "1997",
      pages          = "3258-3295",
      doi            = "10.1103/PhysRevD.56.3258",
      eprint         = "hep-ph/9704452",
      archivePrefix  = "arXiv",
      primaryClass   = "hep-ph",
      reportNumber   = "IFA-97-28, SU-ITP-97-18",
      SLACcitation   = "%%CITATION = HEP-PH/9704452;%%"
}

@article{Kaiser:1997hg,
      author         = "Kaiser, David I.",
      title          = "{Resonance structure for preheating with massless
                        fields}",
      journal        = "Phys. Rev.",
      volume         = "D57",
      year           = "1998",
      pages          = "702-711",
      doi            = "10.1103/PhysRevD.57.702",
      eprint         = "hep-ph/9707516",
      archivePrefix  = "arXiv",
      primaryClass   = "hep-ph",
      reportNumber   = "HUTP-97-A041",
      SLACcitation   = "%%CITATION = HEP-PH/9707516;%%"
}

@article{Greene:1997fu,
      author         = "Greene, Patrick B. and Kofman, Lev and Linde, Andrei D.
                        and Starobinsky, Alexei A.",
      title          = "{Structure of resonance in preheating after inflation}",
      journal        = "Phys. Rev.",
      volume         = "D56",
      year           = "1997",
      pages          = "6175-6192",
      doi            = "10.1103/PhysRevD.56.6175",
      eprint         = "hep-ph/9705347",
      archivePrefix  = "arXiv",
      primaryClass   = "hep-ph",
      reportNumber   = "SU-ITP-97-19, IFA-97-29",
      SLACcitation   = "%%CITATION = HEP-PH/9705347;%%"
}

@article{Felder:2000hj,
      author         = "Felder, Gary N. and Garcia-Bellido, Juan and Greene,
                        Patrick B. and Kofman, Lev and Linde, Andrei D. and
                        Tkachev, Igor",
      title          = "{Dynamics of symmetry breaking and tachyonic preheating}",
      journal        = "Phys. Rev. Lett.",
      volume         = "87",
      year           = "2001",
      pages          = "011601",
      doi            = "10.1103/PhysRevLett.87.011601",
      eprint         = "hep-ph/0012142",
      archivePrefix  = "arXiv",
      primaryClass   = "hep-ph",
      reportNumber   = "CITA-2000-60, SU-ITP-00-35, IFT-UAM-CSIC-00-40,
                        FT-UAM-00-26, CERN-TH-2000-365",
      SLACcitation   = "%%CITATION = HEP-PH/0012142;%%"
}

@article{Copeland:2002ku,
      author         = "Copeland, Edmund J. and Pascoli, S. and Rajantie, A.",
      title          = "{Dynamics of tachyonic preheating after hybrid
                        inflation}",
      journal        = "Phys. Rev.",
      volume         = "D65",
      year           = "2002",
      pages          = "103517",
      doi            = "10.1103/PhysRevD.65.103517",
      eprint         = "hep-ph/0202031",
      archivePrefix  = "arXiv",
      primaryClass   = "hep-ph",
      reportNumber   = "DAMTP-2002-1, SUSX-TH-02-002, SISSA-2-2002-EP",
      SLACcitation   = "%%CITATION = HEP-PH/0202031;%%"
}

@article{GarciaBellido:2002aj,
      author         = "Garcia-Bellido, Juan and Garcia Perez, Margarita and
                        Gonzalez-Arroyo, Antonio",
      title          = "{Symmetry breaking and false vacuum decay after hybrid
                        inflation}",
      journal        = "Phys. Rev.",
      volume         = "D67",
      year           = "2003",
      pages          = "103501",
      doi            = "10.1103/PhysRevD.67.103501",
      eprint         = "hep-ph/0208228",
      archivePrefix  = "arXiv",
      primaryClass   = "hep-ph",
      reportNumber   = "CERN-TH-2002-156, IFT-UAM-CSIC-02-30",
      SLACcitation   = "%%CITATION = HEP-PH/0208228;%%"
}

@article{Allahverdi:2010xz,
      author         = "Allahverdi, Rouzbeh and Brandenberger, Robert and
                        Cyr-Racine, Francis-Yan and Mazumdar, Anupam",
      title          = "{Reheating in Inflationary Cosmology: Theory and
                        Applications}",
      journal        = "Ann. Rev. Nucl. Part. Sci.",
      volume         = "60",
      year           = "2010",
      pages          = "27-51",
      doi            = "10.1146/annurev.nucl.012809.104511",
      eprint         = "1001.2600",
      archivePrefix  = "arXiv",
      primaryClass   = "hep-th",
      SLACcitation   = "%%CITATION = ARXIV:1001.2600;%%"
}

@article{Amin:2014eta,
      author         = "Amin, Mustafa A. and Hertzberg, Mark P. and Kaiser, David
                        I. and Karouby, Johanna",
      title          = "{Nonperturbative Dynamics Of Reheating After Inflation: A
                        Review}",
      journal        = "Int. J. Mod. Phys.",
      volume         = "D24",
      year           = "2014",
      pages          = "1530003",
      doi            = "10.1142/S0218271815300037",
      eprint         = "1410.3808",
      archivePrefix  = "arXiv",
      primaryClass   = "hep-ph",
      SLACcitation   = "%%CITATION = ARXIV:1410.3808;%%"
}

@article{Lozanov:2019jxc,
      author         = "Lozanov, Kaloian D.",
      title          = "{Lectures on Reheating after Inflation}",
      year           = "2019",
      eprint         = "1907.04402",
      archivePrefix  = "arXiv",
      primaryClass   = "astro-ph.CO",
      SLACcitation   = "%%CITATION = ARXIV:1907.04402;%%"
}

@article{Khlebnikov:1997di,
      author         = "Khlebnikov, S. Y. and Tkachev, I. I.",
      title          = "{Relic gravitational waves produced after preheating}",
      journal        = "Phys. Rev.",
      volume         = "D56",
      year           = "1997",
      pages          = "653-660",
      doi            = "10.1103/PhysRevD.56.653",
      eprint         = "hep-ph/9701423",
      archivePrefix  = "arXiv",
      primaryClass   = "hep-ph",
      reportNumber   = "PURD-TH-97-02, OSU-TA-01-97",
      SLACcitation   = "%%CITATION = HEP-PH/9701423;%%"
}

@article{Caprini:2018mtu,
      author         = "Caprini, Chiara and Figueroa, Daniel G.",
      title          = "{Cosmological Backgrounds of Gravitational Waves}",
      journal        = "Class. Quant. Grav.",
      volume         = "35",
      year           = "2018",
      number         = "16",
      pages          = "163001",
      doi            = "10.1088/1361-6382/aac608",
      eprint         = "1801.04268",
      archivePrefix  = "arXiv",
      primaryClass   = "astro-ph.CO",
      SLACcitation   = "%%CITATION = ARXIV:1801.04268;%%"
}

@article{Dufaux:2006ee,
      author         = "Dufaux, Jean Francois and Felder, Gary N. and Kofman, L.
                        and Peloso, M. and Podolsky, D.",
      title          = "{Preheating with trilinear interactions: Tachyonic
                        resonance}",
      journal        = "JCAP",
      volume         = "0607",
      year           = "2006",
      pages          = "006",
      doi            = "10.1088/1475-7516/2006/07/006",
      eprint         = "hep-ph/0602144",
      archivePrefix  = "arXiv",
      primaryClass   = "hep-ph",
      reportNumber   = "UMN-TH-2431-06",
      SLACcitation   = "%%CITATION = HEP-PH/0602144;%%"
}

@article{Figueroa:2016wxr,
      author         = "Figueroa, Daniel G. and Torrenti, Francisco",
      title          = "{Parametric resonance in the early Universe—a fitting
                        analysis}",
      journal        = "JCAP",
      volume         = "1702",
      year           = "2017",
      pages          = "001",
      doi            = "10.1088/1475-7516/2017/02/001",
      eprint         = "1609.05197",
      archivePrefix  = "arXiv",
      primaryClass   = "astro-ph.CO",
      reportNumber   = "CERN-TH-2016-202, IFT-UAM-CSIC-16-084",
      SLACcitation   = "%%CITATION = ARXIV:1609.05197;%%"
}

@article{Adshead:2015pva,
      author         = "Adshead, Peter and Giblin, John T. and Scully, Timothy R.
                        and Sfakianakis, Evangelos I.",
      title          = "{Gauge-preheating and the end of axion inflation}",
      journal        = "JCAP",
      volume         = "1512",
      year           = "2015",
      pages          = "034",
      doi            = "10.1088/1475-7516/2015/12/034",
      eprint         = "1502.06506",
      archivePrefix  = "arXiv",
      primaryClass   = "astro-ph.CO",
      SLACcitation   = "%%CITATION = ARXIV:1502.06506;%%"
}

@article{Adshead:2016iae,
      author         = "Adshead, Peter and Giblin, John T. and Scully, Timothy R.
                        and Sfakianakis, Evangelos I.",
      title          = "{Magnetogenesis from axion inflation}",
      journal        = "JCAP",
      volume         = "1610",
      year           = "2016",
      pages          = "039",
      doi            = "10.1088/1475-7516/2016/10/039",
      eprint         = "1606.08474",
      archivePrefix  = "arXiv",
      primaryClass   = "astro-ph.CO",
      SLACcitation   = "%%CITATION = ARXIV:1606.08474;%%"
}

@article{Figueroa:2019jsi,
      author         = "Figueroa, Daniel G. and Florio, Adrien and Shaposhnikov,
                        Mikhail",
      title          = "{Chiral charge dynamics in Abelian gauge theories at
                        finite temperature}",
      journal        = "JHEP",
      volume         = "10",
      year           = "2019",
      pages          = "142",
      doi            = "10.1007/JHEP10(2019)142",
      eprint         = "1904.11892",
      archivePrefix  = "arXiv",
      primaryClass   = "hep-th",
      SLACcitation   = "%%CITATION = ARXIV:1904.11892;%%"
}

@article{Kitajima:2018zco,
      author         = "Kitajima, Naoya and Soda, Jiro and Urakawa, Yuko",
      title          = "{Gravitational wave forest from string axiverse}",
      journal        = "JCAP",
      volume         = "1810",
      year           = "2018",
      pages          = "008",
      doi            = "10.1088/1475-7516/2018/10/008",
      eprint         = "1807.07037",
      archivePrefix  = "arXiv",
      primaryClass   = "astro-ph.CO",
      reportNumber   = "KOBE-COSMO-18-07",
      SLACcitation   = "%%CITATION = ARXIV:1807.07037;%%"
}

@article{Cuissa:2018oiw,
      author         = "Cuissa, Jose Roberto Canivete and Figueroa, Daniel G.",
      title          = "{Lattice formulation of axion inflation. Application to
                        preheating}",
      journal        = "JCAP",
      volume         = "1906",
      year           = "2019",
      pages          = "002",
      doi            = "10.1088/1475-7516/2019/06/002",
      eprint         = "1812.03132",
      archivePrefix  = "arXiv",
      primaryClass   = "astro-ph.CO",
      SLACcitation   = "%%CITATION = ARXIV:1812.03132;%%"
}

@article{Adshead:2019igv,
      author         = "Adshead, Peter and Giblin, John T. and Pieroni, Mauro and
                        Weiner, Zachary J.",
      title          = "{Constraining axion inflation with gravitational waves
                        across 29 decades in frequency}",
      year           = "2019",
      eprint         = "1909.12843",
      archivePrefix  = "arXiv",
      primaryClass   = "astro-ph.CO",
      SLACcitation   = "%%CITATION = ARXIV:1909.12843;%%"
}

@article{Adshead:2019lbr,
      author         = "Adshead, Peter and Giblin, John T. and Pieroni, Mauro and
                        Weiner, Zachary J.",
      title          = "{Constraining axion inflation with gravitational waves
                        from preheating}",
      year           = "2019",
      eprint         = "1909.12842",
      archivePrefix  = "arXiv",
      primaryClass   = "astro-ph.CO",
      SLACcitation   = "%%CITATION = ARXIV:1909.12842;%%"
}

@article{Easther:2006gt,
      author         = "Easther, Richard and Lim, Eugene A.",
      title          = "{Stochastic gravitational wave production after
                        inflation}",
      journal        = "JCAP",
      volume         = "0604",
      year           = "2006",
      pages          = "010",
      doi            = "10.1088/1475-7516/2006/04/010",
      eprint         = "astro-ph/0601617",
      archivePrefix  = "arXiv",
      primaryClass   = "astro-ph",
      SLACcitation   = "%%CITATION = ASTRO-PH/0601617;%%"
}

@article{Easther:2006vd,
      author         = "Easther, Richard and Giblin, Jr., John T. and Lim, Eugene
                        A.",
      title          = "{Gravitational Wave Production At The End Of Inflation}",
      journal        = "Phys. Rev. Lett.",
      volume         = "99",
      year           = "2007",
      pages          = "221301",
      doi            = "10.1103/PhysRevLett.99.221301",
      eprint         = "astro-ph/0612294",
      archivePrefix  = "arXiv",
      primaryClass   = "astro-ph",
      SLACcitation   = "%%CITATION = ASTRO-PH/0612294;%%"
}

@article{GarciaBellido:2007af,
      author         = "Garcia-Bellido, Juan and Figueroa, Daniel G. and Sastre,
                        Alfonso",
      title          = "{A Gravitational Wave Background from Reheating after
                        Hybrid Inflation}",
      journal        = "Phys. Rev.",
      volume         = "D77",
      year           = "2008",
      pages          = "043517",
      doi            = "10.1103/PhysRevD.77.043517",
      eprint         = "0707.0839",
      archivePrefix  = "arXiv",
      primaryClass   = "hep-ph",
      reportNumber   = "IFT-UAM-CSIC-07-38",
      SLACcitation   = "%%CITATION = ARXIV:0707.0839;%%"
}

@article{Dufaux:2007pt,
      author         = "Dufaux, Jean Francois and Bergman, Amanda and Felder,
                        Gary N. and Kofman, Lev and Uzan, Jean-Philippe",
      title          = "{Theory and Numerics of Gravitational Waves from
                        Preheating after Inflation}",
      journal        = "Phys. Rev.",
      volume         = "D76",
      year           = "2007",
      pages          = "123517",
      doi            = "10.1103/PhysRevD.76.123517",
      eprint         = "0707.0875",
      archivePrefix  = "arXiv",
      primaryClass   = "astro-ph",
      SLACcitation   = "%%CITATION = ARXIV:0707.0875;%%"
}

@article{Dufaux:2008dn,
      author         = "Dufaux, Jean-Francois and Felder, Gary and Kofman, Lev
                        and Navros, Olga",
      title          = "{Gravity Waves from Tachyonic Preheating after Hybrid
                        Inflation}",
      journal        = "JCAP",
      volume         = "0903",
      year           = "2009",
      pages          = "001",
      doi            = "10.1088/1475-7516/2009/03/001",
      eprint         = "0812.2917",
      archivePrefix  = "arXiv",
      primaryClass   = "astro-ph",
      reportNumber   = "FTUAM-08-25, IFT-UAM-CSIC-08-90",
      SLACcitation   = "%%CITATION = ARXIV:0812.2917;%%"
}

@article{Bethke:2013aba,
      author         = "Bethke, Laura and Figueroa, Daniel G. and Rajantie,
                        Arttu",
      title          = "{Anisotropies in the Gravitational Wave Background from
                        Preheating}",
      journal        = "Phys. Rev. Lett.",
      volume         = "111",
      year           = "2013",
      number         = "1",
      pages          = "011301",
      doi            = "10.1103/PhysRevLett.111.011301",
      eprint         = "1304.2657",
      archivePrefix  = "arXiv",
      primaryClass   = "astro-ph.CO",
      reportNumber   = "IMPERIAL-TP-2013-LB-1",
      SLACcitation   = "%%CITATION = ARXIV:1304.2657;%%"
}

@article{Bethke:2013vca,
      author         = "Bethke, Laura and Figueroa, Daniel G. and Rajantie,
                        Arttu",
      title          = "{On the Anisotropy of the Gravitational Wave Background
                        from Massless Preheating}",
      journal        = "JCAP",
      volume         = "1406",
      year           = "2014",
      pages          = "047",
      doi            = "10.1088/1475-7516/2014/06/047",
      eprint         = "1309.1148",
      archivePrefix  = "arXiv",
      primaryClass   = "astro-ph.CO",
      SLACcitation   = "%%CITATION = ARXIV:1309.1148;%%"
}

@article{Figueroa:2017vfa,
      author         = "Figueroa, Daniel G. and Torrenti, Francisco",
      title          = "{Gravitational wave production from preheating: parameter
                        dependence}",
      journal        = "JCAP",
      volume         = "1710",
      year           = "2017",
      pages          = "057",
      doi            = "10.1088/1475-7516/2017/10/057",
      eprint         = "1707.04533",
      archivePrefix  = "arXiv",
      primaryClass   = "astro-ph.CO",
      reportNumber   = "CERN-TH-2017-152, IFT-UAM-CSIC-17-069",
      SLACcitation   = "%%CITATION = ARXIV:1707.04533;%%"
}

@article{Dufaux:2010cf,
      author         = "Dufaux, Jean-Francois and Figueroa, Daniel G. and
                        Garcia-Bellido, Juan",
      title          = "{Gravitational Waves from Abelian Gauge Fields and Cosmic
                        Strings at Preheating}",
      journal        = "Phys. Rev.",
      volume         = "D82",
      year           = "2010",
      pages          = "083518",
      doi            = "10.1103/PhysRevD.82.083518",
      eprint         = "1006.0217",
      archivePrefix  = "arXiv",
      primaryClass   = "astro-ph.CO",
      reportNumber   = "IFT-UAM-CSIC-10-38, CERN-PH-TH-2010-121",
      SLACcitation   = "%%CITATION = ARXIV:1006.0217;%%"
}

@article{Figueroa:2016ojl,
      author         = "Figueroa, Daniel G. and García-Bellido, Juan and
                        Torrentí, Francisco",
      title          = "{Gravitational wave production from the decay of the
                        standard model Higgs field after inflation}",
      journal        = "Phys. Rev.",
      volume         = "D93",
      year           = "2016",
      number         = "10",
      pages          = "103521",
      doi            = "10.1103/PhysRevD.93.103521",
      eprint         = "1602.03085",
      archivePrefix  = "arXiv",
      primaryClass   = "astro-ph.CO",
      reportNumber   = "IFT-UAM-CSIC-16-012, CERN-TH-2016-031",
      SLACcitation   = "%%CITATION = ARXIV:1602.03085;%%"
}

@article{Figueroa:2015rqa,
      author         = "Figueroa, Daniel G. and Garcia-Bellido, Juan and
                        Torrenti, Francisco",
      title          = "{Decay of the standard model Higgs field after
                        inflation}",
      journal        = "Phys. Rev.",
      volume         = "D92",
      year           = "2015",
      number         = "8",
      pages          = "083511",
      doi            = "10.1103/PhysRevD.92.083511",
      eprint         = "1504.04600",
      archivePrefix  = "arXiv",
      primaryClass   = "astro-ph.CO",
      reportNumber   = "CERN-PH-TH-2015-090, IFT-UAM-CSIC-15-034",
      SLACcitation   = "%%CITATION = ARXIV:1504.04600;%%"
}

@article{Podolsky:2005bw,
      author         = "Podolsky, Dmitry I. and Felder, Gary N. and Kofman, Lev
                        and Peloso, Marco",
      title          = "{Equation of state and beginning of thermalization after
                        preheating}",
      journal        = "Phys. Rev.",
      volume         = "D73",
      year           = "2006",
      pages          = "023501",
      doi            = "10.1103/PhysRevD.73.023501",
      eprint         = "hep-ph/0507096",
      archivePrefix  = "arXiv",
      primaryClass   = "hep-ph",
      reportNumber   = "UMN-TH-2407-05",
      SLACcitation   = "%%CITATION = HEP-PH/0507096;%%"
}

@article{Lozanov:2016hid,
      author         = "Lozanov, Kaloian D. and Amin, Mustafa A.",
      title          = "{Equation of State and Duration to Radiation Domination
                        after Inflation}",
      journal        = "Phys. Rev. Lett.",
      volume         = "119",
      year           = "2017",
      number         = "6",
      pages          = "061301",
      doi            = "10.1103/PhysRevLett.119.061301",
      eprint         = "1608.01213",
      archivePrefix  = "arXiv",
      primaryClass   = "astro-ph.CO",
      SLACcitation   = "%%CITATION = ARXIV:1608.01213;%%"
}

@article{Krajewski:2018moi,
      author         = "Krajewski, Tomasz and Turzyński, Krzysztof and
                        Wieczorek, Michał",
      title          = "{On preheating in $\alpha$-attractor models of
                        inflation}",
      journal        = "Eur. Phys. J.",
      volume         = "C79",
      year           = "2019",
      number         = "8",
      pages          = "654",
      doi            = "10.1140/epjc/s10052-019-7155-z",
      eprint         = "1801.01786",
      archivePrefix  = "arXiv",
      primaryClass   = "astro-ph.CO",
      SLACcitation   = "%%CITATION = ARXIV:1801.01786;%%"
}

@article{Amin:2011hj,
      author         = "Amin, Mustafa A. and Easther, Richard and Finkel, Hal and
                        Flauger, Raphael and Hertzberg, Mark P.",
      title          = "{Oscillons After Inflation}",
      journal        = "Phys. Rev. Lett.",
      volume         = "108",
      year           = "2012",
      pages          = "241302",
      doi            = "10.1103/PhysRevLett.108.241302",
      eprint         = "1106.3335",
      archivePrefix  = "arXiv",
      primaryClass   = "astro-ph.CO",
      SLACcitation   = "%%CITATION = ARXIV:1106.3335;%%"
}

@article{Zhou:2013tsa,
      author         = "Zhou, Shuang-Yong and Copeland, Edmund J. and Easther,
                        Richard and Finkel, Hal and Mou, Zong-Gang and Saffin,
                        Paul M.",
      title          = "{Gravitational Waves from Oscillon Preheating}",
      journal        = "JHEP",
      volume         = "10",
      year           = "2013",
      pages          = "026",
      doi            = "10.1007/JHEP10(2013)026",
      eprint         = "1304.6094",
      archivePrefix  = "arXiv",
      primaryClass   = "astro-ph.CO",
      SLACcitation   = "%%CITATION = ARXIV:1304.6094;%%"
}

@article{Antusch:2016con,
      author         = "Antusch, Stefan and Cefala, Francesco and Orani, Stefano",
      title          = "{Gravitational waves from oscillons after inflation}",
      journal        = "Phys. Rev. Lett.",
      volume         = "118",
      year           = "2017",
      number         = "1",
      pages          = "011303",
      doi            = "10.1103/PhysRevLett.120.219901,
                        10.1103/PhysRevLett.118.011303",
      note           = "[Erratum: Phys. Rev. Lett.120,no.21,219901(2018)]",
      eprint         = "1607.01314",
      archivePrefix  = "arXiv",
      primaryClass   = "astro-ph.CO",
      SLACcitation   = "%%CITATION = ARXIV:1607.01314;%%"
}

@article{Antusch:2017flz,
      author         = "Antusch, Stefan and Cefala, Francesco and Krippendorf,
                        Sven and Muia, Francesco and Orani, Stefano and Quevedo,
                        Fernando",
      title          = "{Oscillons from String Moduli}",
      journal        = "JHEP",
      volume         = "01",
      year           = "2018",
      pages          = "083",
      doi            = "10.1007/JHEP01(2018)083",
      eprint         = "1708.08922",
      archivePrefix  = "arXiv",
      primaryClass   = "hep-th",
      SLACcitation   = "%%CITATION = ARXIV:1708.08922;%%"
}

@article{Lozanov:2017hjm,
      author         = "Lozanov, Kaloian D. and Amin, Mustafa A.",
      title          = "{Self-resonance after inflation: oscillons, transients
                        and radiation domination}",
      journal        = "Phys. Rev.",
      volume         = "D97",
      year           = "2018",
      number         = "2",
      pages          = "023533",
      doi            = "10.1103/PhysRevD.97.023533",
      eprint         = "1710.06851",
      archivePrefix  = "arXiv",
      primaryClass   = "astro-ph.CO",
      SLACcitation   = "%%CITATION = ARXIV:1710.06851;%%"
}

@article{Amin:2018xfe,
      author         = "Amin, Mustafa A. and Braden, Jonathan and Copeland,
                        Edmund J. and Giblin, John T. and Solorio, Christian and
                        Weiner, Zachary J. and Zhou, Shuang-Yong",
      title          = "{Gravitational waves from asymmetric oscillon dynamics?}",
      journal        = "Phys. Rev.",
      volume         = "D98",
      year           = "2018",
      pages          = "024040",
      doi            = "10.1103/PhysRevD.98.024040",
      eprint         = "1803.08047",
      archivePrefix  = "arXiv",
      primaryClass   = "astro-ph.CO",
      SLACcitation   = "%%CITATION = ARXIV:1803.08047;%%"
}

@article{Liu:2018rrt,
      author         = "Liu, Jing and Guo, Zong-Kuan and Cai, Rong-Gen and Shiu,
                        Gary",
      title          = "{Gravitational wave production after inflation with cuspy
                        potentials}",
      journal        = "Phys. Rev.",
      volume         = "D99",
      year           = "2019",
      number         = "10",
      pages          = "103506",
      doi            = "10.1103/PhysRevD.99.103506",
      eprint         = "1812.09235",
      archivePrefix  = "arXiv",
      primaryClass   = "astro-ph.CO",
      SLACcitation   = "%%CITATION = ARXIV:1812.09235;%%"
}

@article{Lozanov:2019ylm,
      author         = "Lozanov, Kaloian D. and Amin, Mustafa A.",
      title          = "{Gravitational perturbations from oscillons and
                        transients after inflation}",
      journal        = "Phys. Rev.",
      volume         = "D99",
      year           = "2019",
      number         = "12",
      pages          = "123504",
      doi            = "10.1103/PhysRevD.99.123504",
      eprint         = "1902.06736",
      archivePrefix  = "arXiv",
      primaryClass   = "astro-ph.CO",
      SLACcitation   = "%%CITATION = ARXIV:1902.06736;%%"
}

@article{Antusch:2019qrr,
      author         = "Antusch, Stefan and Cefalà, Francesco and Torrentí,
                        Francisco",
      title          = "{Properties of Oscillons in Hilltop Potentials: energies,
                        shapes, and lifetimes}",
      journal        = "JCAP",
      volume         = "1910",
      year           = "2019",
      pages          = "002",
      doi            = "10.1088/1475-7516/2019/10/002",
      eprint         = "1907.00611",
      archivePrefix  = "arXiv",
      primaryClass   = "hep-ph",
      SLACcitation   = "%%CITATION = ARXIV:1907.00611;%%"
}

@article{Fu:2017ero,
      author         = "Fu, Chengjie and Wu, Puxun and Yu, Hongwei",
      title          = "{Production of gravitational waves during preheating with
                        nonminimal coupling}",
      journal        = "Phys. Rev.",
      volume         = "D97",
      year           = "2018",
      number         = "8",
      pages          = "081303",
      doi            = "10.1103/PhysRevD.97.081303",
      eprint         = "1711.10888",
      archivePrefix  = "arXiv",
      primaryClass   = "gr-qc",
      SLACcitation   = "%%CITATION = ARXIV:1711.10888;%%"
}

@article{Giblin:2019nuv,
      author         = "Giblin, John T. and Tishue, Avery J.",
      title          = "{Preheating in Full General Relativity}",
      journal        = "Phys. Rev.",
      volume         = "D100",
      year           = "2019",
      number         = "6",
      pages          = "063543",
      doi            = "10.1103/PhysRevD.100.063543",
      eprint         = "1907.10601",
      archivePrefix  = "arXiv",
      primaryClass   = "gr-qc",
      SLACcitation   = "%%CITATION = ARXIV:1907.10601;%%"
}

@article{Felder:2000hq,
      author         = "Felder, Gary N. and Tkachev, Igor",
      title          = "{LATTICEEASY: A Program for lattice simulations of scalar
                        fields in an expanding universe}",
      journal        = "Comput. Phys. Commun.",
      volume         = "178",
      year           = "2008",
      pages          = "929-932",
      doi            = "10.1016/j.cpc.2008.02.009",
      eprint         = "hep-ph/0011159",
      archivePrefix  = "arXiv",
      primaryClass   = "hep-ph",
      reportNumber   = "SU-ITP-00-29",
      SLACcitation   = "%%CITATION = HEP-PH/0011159;%%"
}

@article{Felder:2007nz,
      author         = "Felder, Gary N.",
      title          = "{CLUSTEREASY: A program for lattice simulations of scalar
                        fields in an expanding universe on parallel computing
                        clusters}",
      journal        = "Comput. Phys. Commun.",
      volume         = "179",
      year           = "2008",
      pages          = "604-606",
      doi            = "10.1016/j.cpc.2008.06.002",
      eprint         = "0712.0813",
      archivePrefix  = "arXiv",
      primaryClass   = "hep-ph",
      SLACcitation   = "%%CITATION = ARXIV:0712.0813;%%"
}

@article{Frolov:2008hy,
      author         = "Frolov, Andrei V.",
      title          = "{DEFROST: A New Code for Simulating Preheating after
                        Inflation}",
      journal        = "JCAP",
      volume         = "0811",
      year           = "2008",
      pages          = "009",
      doi            = "10.1088/1475-7516/2008/11/009",
      eprint         = "0809.4904",
      archivePrefix  = "arXiv",
      primaryClass   = "hep-ph",
      reportNumber   = "SCG-2008-02",
      SLACcitation   = "%%CITATION = ARXIV:0809.4904;%%"
}

@article{Easther:2010qz,
      author         = "Easther, Richard and Finkel, Hal and Roth, Nathaniel",
      title          = "{PSpectRe: A Pseudo-Spectral Code for (P)reheating}",
      journal        = "JCAP",
      volume         = "1010",
      year           = "2010",
      pages          = "025",
      doi            = "10.1088/1475-7516/2010/10/025",
      eprint         = "1005.1921",
      archivePrefix  = "arXiv",
      primaryClass   = "astro-ph.CO",
      SLACcitation   = "%%CITATION = ARXIV:1005.1921;%%"
}

@article{Huang:2011gf,
      author         = "Huang, Zhiqi",
      title          = "{The Art of Lattice and Gravity Waves from Preheating}",
      journal        = "Phys. Rev.",
      volume         = "D83",
      year           = "2011",
      pages          = "123509",
      doi            = "10.1103/PhysRevD.83.123509",
      eprint         = "1102.0227",
      archivePrefix  = "arXiv",
      primaryClass   = "astro-ph.CO",
      SLACcitation   = "%%CITATION = ARXIV:1102.0227;%%"
}

@article{Sainio:2012mw,
      author         = "Sainio, J.",
      title          = "{PyCOOL - a Cosmological Object-Oriented Lattice code
                        written in Python}",
      journal        = "JCAP",
      volume         = "1204",
      year           = "2012",
      pages          = "038",
      doi            = "10.1088/1475-7516/2012/04/038",
      eprint         = "1201.5029",
      archivePrefix  = "arXiv",
      primaryClass   = "astro-ph.IM",
      SLACcitation   = "%%CITATION = ARXIV:1201.5029;%%"
}

@article{Lozanov:2019jff,
      author         = "Lozanov, Kaloian D. and Amin, Mustafa A.",
      title          = "{GFiRe: a Gauge Field integrator for Reheating}",
      year           = "2019",
      eprint         = "1911.06827",
      archivePrefix  = "arXiv",
      primaryClass   = "astro-ph.CO",
      SLACcitation   = "%%CITATION = ARXIV:1911.06827;%%"
}

@article{Aghanim:2018eyx,
      author         = "Aghanim, N. and others",
      title          = "{Planck 2018 results. VI. Cosmological parameters}",
      collaboration  = "Planck",
      year           = "2018",
      eprint         = "1807.06209",
      archivePrefix  = "arXiv",
      primaryClass   = "astro-ph.CO",
      SLACcitation   = "%%CITATION = ARXIV:1807.06209;%%"
}

@article{Akrami:2018odb,
      author         = "Akrami, Y. and others",
      title          = "{Planck 2018 results. X. Constraints on inflation}",
      collaboration  = "Planck",
      year           = "2018",
      eprint         = "1807.06211",
      archivePrefix  = "arXiv",
      primaryClass   = "astro-ph.CO",
      SLACcitation   = "%%CITATION = ARXIV:1807.06211;%%"
}

@article{Antusch:2020iyq,
    author = "Antusch, Stefan and Figueroa, Daniel G. and Marschall, Kenneth and Torrenti, Francisco",
    title = "{Energy distribution and equation of state of the early Universe: matching the end of inflation and the onset of radiation domination}",
    eprint = "2005.07563",
    archivePrefix = "arXiv",
    primaryClass = "astro-ph.CO",
    month = "5",
    year = "2020"
}

@article{Kibble:1976sj,
    author = "Kibble, T. W. B.",
    title = "{Topology of Cosmic Domains and Strings}",
    reportNumber = "ICTP/75/5",
    doi = "10.1088/0305-4470/9/8/029",
    journal = "J. Phys. A",
    volume = "9",
    pages = "1387--1398",
    year = "1976"
}

@article{Kibble:1980mv,
    author = "Kibble, T. W. B.",
    title = "{Some Implications of a Cosmological Phase Transition}",
    reportNumber = "ICTP-79-80-23",
    doi = "10.1016/0370-1573(80)90091-5",
    journal = "Phys. Rept.",
    volume = "67",
    pages = "183",
    year = "1980"
}

@article{Vilenkin:1984ib,
    author = "Vilenkin, Alexander",
    title = "{Cosmic Strings and Domain Walls}",
    reportNumber = "PRINT-84-0840 (TUFTS)",
    doi = "10.1016/0370-1573(85)90033-X",
    journal = "Phys. Rept.",
    volume = "121",
    pages = "263--315",
    year = "1985"
}

@book{Vilenkin:2000jqa,
    author = "Vilenkin, A. and Shellard, E. P. S.",
    title = "{Cosmic Strings and Other Topological Defects}",
    isbn = "978-0-521-65476-0",
    publisher = "Cambridge University Press",
    month = "7",
    year = "2000"
}

@article{Vilenkin:1981kz,
    author = "Vilenkin, A.",
    title = "{Cosmic Strings}",
    doi = "10.1103/PhysRevD.24.2082",
    journal = "Phys. Rev. D",
    volume = "24",
    pages = "2082--2089",
    year = "1981"
}

@article{Vachaspati:1984dz,
    author = "Vachaspati, Tanmay and Vilenkin, Alexander",
    title = "{Formation and Evolution of Cosmic Strings}",
    reportNumber = "TUTP-84-1",
    doi = "10.1103/PhysRevD.30.2036",
    journal = "Phys. Rev. D",
    volume = "30",
    pages = "2036",
    year = "1984"
}

@article{Barriola:1989hx,
    author = "Barriola, Manuel and Vilenkin, Alexander",
    title = "{Gravitational Field of a Global Monopole}",
    reportNumber = "TUTP-89-4",
    doi = "10.1103/PhysRevLett.63.341",
    journal = "Phys. Rev. Lett.",
    volume = "63",
    pages = "341",
    year = "1989"
}

@article{Durrer:2001cg,
    author = "Durrer, R. and Kunz, M. and Melchiorri, A.",
    title = "{Cosmic structure formation with topological defects}",
    eprint = "astro-ph/0110348",
    archivePrefix = "arXiv",
    doi = "10.1016/S0370-1573(02)00014-5",
    journal = "Phys. Rept.",
    volume = "364",
    pages = "1--81",
    year = "2002"
}

@article{Turok:1991qq,
    author = "Turok, Neil and Spergel, David N.",
    title = "{Scaling solution for cosmological sigma models at large N}",
    reportNumber = "PUPT-90-1238",
    doi = "10.1103/PhysRevLett.66.3093",
    journal = "Phys. Rev. Lett.",
    volume = "66",
    pages = "3093--3096",
    year = "1991"
}

@article{Jaffe:1993tt,
    author = "Jaffe, Andrew H.",
    title = "{Quasilinear evolution of compensated cosmological perturbations: The Nonlinear sigma model}",
    eprint = "astro-ph/9311023",
    archivePrefix = "arXiv",
    reportNumber = "FERMILAB-PUB-93-317-A",
    doi = "10.1103/PhysRevD.49.3893",
    journal = "Phys. Rev. D",
    volume = "49",
    pages = "3893--3909",
    year = "1994"
}

@article{Durrer:1998rw,
    author = "Durrer, R. and Kunz, M. and Melchiorri, A.",
    title = "{Cosmic microwave background anisotropies from scaling seeds: Global defect models}",
    eprint = "astro-ph/9811174",
    archivePrefix = "arXiv",
    reportNumber = "UGVA-DPT-98-11-1021",
    doi = "10.1103/PhysRevD.59.123005",
    journal = "Phys. Rev. D",
    volume = "59",
    pages = "123005",
    year = "1999"
}

@article{Fenu:2009qf,
    author = "Fenu, Elisa and Figueroa, Daniel G. and Durrer, Ruth and Garcia-Bellido, Juan",
    title = "{Gravitational waves from self-ordering scalar fields}",
    eprint = "0908.0425",
    archivePrefix = "arXiv",
    primaryClass = "astro-ph.CO",
    reportNumber = "IFT-UAM-CSIC-09-34, CERN-PH-TH-2009-145",
    doi = "10.1088/1475-7516/2009/10/005",
    journal = "JCAP",
    volume = "10",
    pages = "005",
    year = "2009"
}

@article{Garcia-Bellido:2010qjz,
    author = "Garcia-Bellido, Juan and Durrer, Ruth and Fenu, Elisa and Figueroa, Daniel G. and Kunz, Martin",
    title = "{The local B-polarization of the CMB: a very sensitive probe of cosmic defects}",
    eprint = "1003.0299",
    archivePrefix = "arXiv",
    primaryClass = "astro-ph.CO",
    reportNumber = "IFT-UAM-CSIC-10-10, CERN-PH-TH-2010-054",
    doi = "10.1016/j.physletb.2010.11.031",
    journal = "Phys. Lett. B",
    volume = "695",
    pages = "26--29",
    year = "2011"
}

@article{Figueroa:2010zx,
    author = "Figueroa, Daniel G. and Caldwell, Robert R. and Kamionkowski, Marc",
    title = "{Non-Gaussianity from Self-Ordering Scalar Fields}",
    eprint = "1003.0672",
    archivePrefix = "arXiv",
    primaryClass = "astro-ph.CO",
    reportNumber = "IFT-UAM-CSIC-10-09, CERN-PH-TH-2010-039",
    doi = "10.1103/PhysRevD.81.123504",
    journal = "Phys. Rev. D",
    volume = "81",
    pages = "123504",
    year = "2010"
}

@article{Fenu:2013tea,
    author = "Fenu, Elisa and Figueroa, Daniel G. and Durrer, Ruth and Garcia-Bellido, Juan and Kunz, Martin",
    title = "{Cosmic Microwave Background temperature and polarization anisotropies from the large-N limit of global defects}",
    eprint = "1311.3225",
    archivePrefix = "arXiv",
    primaryClass = "astro-ph.CO",
    reportNumber = "IFT-UAM-CSIC-13-121",
    doi = "10.1103/PhysRevD.89.083512",
    journal = "Phys. Rev. D",
    volume = "89",
    number = "8",
    pages = "083512",
    year = "2014"
}

@article{Durrer:2014raa,
    author = "Durrer, Ruth and Figueroa, Daniel G. and Kunz, Martin",
    title = "{Can Self-Ordering Scalar Fields explain the BICEP2 B-mode signal?}",
    eprint = "1404.3855",
    archivePrefix = "arXiv",
    primaryClass = "astro-ph.CO",
    doi = "10.1088/1475-7516/2014/08/029",
    journal = "JCAP",
    volume = "08",
    pages = "029",
    year = "2014"
}

@article{Dvali:2003zj,
    author = "Dvali, Gia and Vilenkin, Alexander",
    title = "{Formation and evolution of cosmic D strings}",
    eprint = "hep-th/0312007",
    archivePrefix = "arXiv",
    doi = "10.1088/1475-7516/2004/03/010",
    journal = "JCAP",
    volume = "03",
    pages = "010",
    year = "2004"
}

@article{Copeland:2003bj,
    author = "Copeland, Edmund J. and Myers, Robert C. and Polchinski, Joseph",
    title = "{Cosmic F and D strings}",
    eprint = "hep-th/0312067",
    archivePrefix = "arXiv",
    doi = "10.1088/1126-6708/2004/06/013",
    journal = "JHEP",
    volume = "06",
    pages = "013",
    year = "2004"
}

@article{Gorghetto:2020qws,
    author = "Gorghetto, Marco and Hardy, Edward and Villadoro, Giovanni",
    title = "{More axions from strings}",
    eprint = "2007.04990",
    archivePrefix = "arXiv",
    primaryClass = "hep-ph",
    doi = "10.21468/SciPostPhys.10.2.050",
    journal = "SciPost Phys.",
    volume = "10",
    number = "2",
    pages = "050",
    year = "2021"
}

@article{Buschmann:2021sdq,
    author = "Buschmann, Malte and Foster, Joshua W. and Hook, Anson and Peterson, Adam and Willcox, Don E. and Zhang, Weiqun and Safdi, Benjamin R.",
    title = "{Dark matter from axion strings with adaptive mesh refinement}",
    eprint = "2108.05368",
    archivePrefix = "arXiv",
    primaryClass = "hep-ph",
    doi = "10.1038/s41467-022-28669-y",
    journal = "Nature Commun.",
    volume = "13",
    number = "1",
    pages = "1049",
    year = "2022"
}

@article{Saikawa:2024bta,
    author = "Saikawa, Ken'ichi and Redondo, Javier and Vaquero, Alejandro and Kaltschmidt, Mathieu",
    title = "{Spectrum of global string networks and the axion dark matter mass}",
    eprint = "2401.17253",
    archivePrefix = "arXiv",
    primaryClass = "hep-ph",
    reportNumber = "KANAZAWA-24-02, MPP-2024-18",
    doi = "10.1088/1475-7516/2024/10/043",
    journal = "JCAP",
    volume = "10",
    pages = "043",
    year = "2024"
}

@article{Benabou:2024msj,
    author = "Benabou, Joshua N. and Buschmann, Malte and Foster, Joshua W. and Safdi, Benjamin R.",
    title = "{Axion Mass Prediction from Adaptive Mesh Refinement Cosmological Lattice Simulations}",
    eprint = "2412.08699",
    archivePrefix = "arXiv",
    primaryClass = "hep-ph",
    reportNumber = "FERMILAB-PUB-24-0912-T",
    doi = "10.1103/6v21-d6sj",
    journal = "Phys. Rev. Lett.",
    volume = "134",
    number = "24",
    pages = "241003",
    year = "2025"
}

@article{Ade:2013xla,
    author = "Ade, P. A. R. and others",
    collaboration = "Planck",
    title = "{Planck 2013 results. XXV. Searches for cosmic strings and other topological defects}",
    eprint = "1303.5085",
    archivePrefix = "arXiv",
    primaryClass = "astro-ph.CO",
    reportNumber = "CERN-PH-TH-2013-138",
    doi = "10.1051/0004-6361/201321621",
    journal = "Astron. Astrophys.",
    volume = "571",
    pages = "A25",
    year = "2014"
}

@article{Lizarraga:2014xza,
    author = "Lizarraga, Joanes and Urrestilla, Jon and Daverio, David and Hindmarsh, Mark and Kunz, Martin and Liddle, Andrew R.",
    title = "{Constraining topological defects with temperature and polarization anisotropies}",
    eprint = "1408.4126",
    archivePrefix = "arXiv",
    primaryClass = "astro-ph.CO",
    doi = "10.1103/PhysRevD.90.103504",
    journal = "Phys. Rev. D",
    volume = "90",
    number = "10",
    pages = "103504",
    year = "2014"
}

@article{Charnock:2016nzm,
    author = "Charnock, Tom and Avgoustidis, Anastasios and Copeland, Edmund J. and Moss, Adam",
    title = "{CMB constraints on cosmic strings and superstrings}",
    eprint = "1603.01275",
    archivePrefix = "arXiv",
    primaryClass = "astro-ph.CO",
    doi = "10.1103/PhysRevD.93.123503",
    journal = "Phys. Rev. D",
    volume = "93",
    number = "12",
    pages = "123503",
    year = "2016"
}

@article{Ringeval:2010ca,
    author = "Ringeval, Christophe",
    title = "{Cosmic strings and their induced non-Gaussianities in the cosmic microwave background}",
    eprint = "1005.4842",
    archivePrefix = "arXiv",
    primaryClass = "astro-ph.CO",
    doi = "10.1155/2010/380507",
    journal = "Adv. Astron.",
    volume = "2010",
    pages = "380507",
    year = "2010"
}

@article{Regan:2014vha,
    author = "Regan, Donough and Hindmarsh, Mark",
    title = "{The bispectrum of matter perturbations from cosmic strings}",
    eprint = "1411.2641",
    archivePrefix = "arXiv",
    primaryClass = "astro-ph.CO",
    doi = "10.1088/1475-7516/2015/03/008",
    journal = "JCAP",
    volume = "03",
    pages = "008",
    year = "2015"
}

@article{Vilenkin:1981bx,
    author = "Vilenkin, A.",
    title = "{Gravitational radiation from cosmic strings}",
    doi = "10.1016/0370-2693(81)91144-8",
    journal = "Phys. Lett. B",
    volume = "107",
    pages = "47--50",
    year = "1981"
}

@article{Vachaspati:1984gt,
    author = "Vachaspati, Tanmay and Vilenkin, Alexander",
    title = "{Gravitational Radiation from Cosmic Strings}",
    reportNumber = "HUTP-84/A065",
    doi = "10.1103/PhysRevD.31.3052",
    journal = "Phys. Rev. D",
    volume = "31",
    pages = "3052",
    year = "1985"
}

@article{Damour:2000wa,
    author = "Damour, Thibault and Vilenkin, Alexander",
    title = "{Gravitational wave bursts from cosmic strings}",
    eprint = "gr-qc/0004075",
    archivePrefix = "arXiv",
    reportNumber = "IHES-P-00-32",
    doi = "10.1103/PhysRevLett.85.3761",
    journal = "Phys. Rev. Lett.",
    volume = "85",
    pages = "3761--3764",
    year = "2000"
}

@article{Damour:2001bk,
    author = "Damour, Thibault and Vilenkin, Alexander",
    title = "{Gravitational wave bursts from cusps and kinks on cosmic strings}",
    eprint = "gr-qc/0104026",
    archivePrefix = "arXiv",
    reportNumber = "IHES-P-01-15",
    doi = "10.1103/PhysRevD.64.064008",
    journal = "Phys. Rev. D",
    volume = "64",
    pages = "064008",
    year = "2001"
}

@article{Damour:2004kw,
    author = "Damour, Thibault and Vilenkin, Alexander",
    title = "{Gravitational radiation from cosmic (super)strings: Bursts, stochastic background, and observational windows}",
    eprint = "hep-th/0410222",
    archivePrefix = "arXiv",
    doi = "10.1103/PhysRevD.71.063510",
    journal = "Phys. Rev. D",
    volume = "71",
    pages = "063510",
    year = "2005"
}

@article{Blanco-Pillado:2017oxo,
    author = "Blanco-Pillado, Jose J. and Olum, Ken D.",
    title = "{Stochastic gravitational wave background from smoothed cosmic string loops}",
    eprint = "1709.02693",
    archivePrefix = "arXiv",
    primaryClass = "astro-ph.CO",
    doi = "10.1103/PhysRevD.96.104046",
    journal = "Phys. Rev. D",
    volume = "96",
    number = "10",
    pages = "104046",
    year = "2017"
}

@article{Auclair:2019wcv,
    author = "Auclair, Pierre and others",
    title = "{Probing the gravitational wave background from cosmic strings with LISA}",
    eprint = "1909.00819",
    archivePrefix = "arXiv",
    primaryClass = "astro-ph.CO",
    doi = "10.1088/1475-7516/2020/04/034",
    journal = "JCAP",
    volume = "04",
    pages = "034",
    year = "2020"
}

@article{Gouttenoire:2019kij,
    author = "Gouttenoire, Yann and Servant, G{\'e}raldine and Simakachorn, Peera",
    title = "{Beyond the Standard Models with Cosmic Strings}",
    eprint = "1912.02569",
    archivePrefix = "arXiv",
    primaryClass = "hep-ph",
    reportNumber = "DESY-19-204",
    doi = "10.1088/1475-7516/2020/07/032",
    journal = "JCAP",
    volume = "07",
    pages = "032",
    year = "2020"
}

@article{Gorghetto:2021fsn,
    author = "Gorghetto, Marco and Hardy, Edward and Nicolaescu, Horia",
    title = "{Observing invisible axions with gravitational waves}",
    eprint = "2101.11007",
    archivePrefix = "arXiv",
    primaryClass = "hep-ph",
    doi = "10.1088/1475-7516/2021/06/034",
    journal = "JCAP",
    volume = "06",
    pages = "034",
    year = "2021"
}

@article{Chang:2021afa,
    author = "Chang, Chia-Feng and Cui, Yanou",
    title = "{Gravitational waves from global cosmic strings and cosmic archaeology}",
    eprint = "2106.09746",
    archivePrefix = "arXiv",
    primaryClass = "hep-ph",
    doi = "10.1007/JHEP03(2022)114",
    journal = "JHEP",
    volume = "03",
    pages = "114",
    year = "2022"
}

@article{Yamada:2022aax,
    author = "Yamada, Masaki and Yonekura, Kazuya",
    title = "{Cosmic F- and D-strings from pure Yang{\textendash}Mills theory}",
    eprint = "2204.13125",
    archivePrefix = "arXiv",
    primaryClass = "hep-th",
    reportNumber = "TU-1153",
    doi = "10.1016/j.physletb.2023.137724",
    journal = "Phys. Lett. B",
    volume = "838",
    pages = "137724",
    year = "2023"
}

@article{Yamada:2022imq,
    author = "Yamada, Masaki and Yonekura, Kazuya",
    title = "{Cosmic strings from pure Yang{\textendash}Mills theory}",
    eprint = "2204.13123",
    archivePrefix = "arXiv",
    primaryClass = "hep-th",
    reportNumber = "TU-1152",
    doi = "10.1103/PhysRevD.106.123515",
    journal = "Phys. Rev. D",
    volume = "106",
    number = "12",
    pages = "123515",
    year = "2022"
}

@article{Servant:2023mwt,
    author = "Servant, G{\'e}raldine and Simakachorn, Peera",
    title = "{Constraining postinflationary axions with pulsar timing arrays}",
    eprint = "2307.03121",
    archivePrefix = "arXiv",
    primaryClass = "hep-ph",
    reportNumber = "DESY-23-094",
    doi = "10.1103/PhysRevD.108.123516",
    journal = "Phys. Rev. D",
    volume = "108",
    number = "12",
    pages = "123516",
    year = "2023"
}

@article{Dimitriou:2025bvq,
    author = "Dimitriou, Androniki and Figueroa, Daniel G. and Simakachorn, Peera and Zaldivar, Bryan",
    title = "{Cosmic string gravitational wave backgrounds at LISA: I. Signal survey, template reconstruction, and model comparison}",
    eprint = "2508.05395",
    archivePrefix = "arXiv",
    primaryClass = "astro-ph.CO",
    month = "8",
    year = "2025"
}

@article{Svrcek:2006yi,
    author = "Svrcek, Peter and Witten, Edward",
    title = "{Axions In String Theory}",
    eprint = "hep-th/0605206",
    archivePrefix = "arXiv",
    reportNumber = "SLAC-PUB-11894",
    doi = "10.1088/1126-6708/2006/06/051",
    journal = "JHEP",
    volume = "06",
    pages = "051",
    year = "2006"
}

@article{Arvanitaki:2009fg,
    author = "Arvanitaki, Asimina and Dimopoulos, Savas and Dubovsky, Sergei and Kaloper, Nemanja and March-Russell, John",
    title = "{String Axiverse}",
    eprint = "0905.4720",
    archivePrefix = "arXiv",
    primaryClass = "hep-th",
    doi = "10.1103/PhysRevD.81.123530",
    journal = "Phys. Rev. D",
    volume = "81",
    pages = "123530",
    year = "2010"
}

@article{Weinberg:1977ma,
    author = "Weinberg, Steven",
    title = "{A New Light Boson?}",
    reportNumber = "HUTP-77/A074",
    doi = "10.1103/PhysRevLett.40.223",
    journal = "Phys. Rev. Lett.",
    volume = "40",
    pages = "223--226",
    year = "1978"
}

@article{Wilczek:1977pj,
    author = "Wilczek, Frank",
    title = "{Problem of Strong  $P$  and  $T$  Invariance in the Presence of Instantons}",
    reportNumber = "Print-77-0939 (COLUMBIA)",
    doi = "10.1103/PhysRevLett.40.279",
    journal = "Phys. Rev. Lett.",
    volume = "40",
    pages = "279--282",
    year = "1978"
}

@article{Peccei:1977hh,
    author = "Peccei, R. D. and Quinn, Helen R.",
    title = "{CP Conservation in the Presence of Instantons}",
    reportNumber = "ITP-568-STANFORD",
    doi = "10.1103/PhysRevLett.38.1440",
    journal = "Phys. Rev. Lett.",
    volume = "38",
    pages = "1440--1443",
    year = "1977"
}

@article{Peccei:1977ur,
    author = "Peccei, R. D. and Quinn, Helen R.",
    title = "{Constraints Imposed by CP Conservation in the Presence of Instantons}",
    reportNumber = "ITP-572-STANFORD",
    doi = "10.1103/PhysRevD.16.1791",
    journal = "Phys. Rev. D",
    volume = "16",
    pages = "1791--1797",
    year = "1977"
}

@article{Hindmarsh:2021vih,
    author = "Hindmarsh, Mark and Lizarraga, Joanes and Lopez-Eiguren, Asier and Urrestilla, Jon",
    title = "{Approach to scaling in axion string networks}",
    eprint = "2102.07723",
    archivePrefix = "arXiv",
    primaryClass = "astro-ph.CO",
    reportNumber = "HIP-2021-7/TH",
    doi = "10.1103/PhysRevD.103.103534",
    journal = "Phys. Rev. D",
    volume = "103",
    number = "10",
    pages = "103534",
    year = "2021"
}

@article{Vaquero:2018tib,
    author = "Vaquero, Alejandro and Redondo, Javier and Stadler, Julia",
    title = "{Early seeds of axion miniclusters}",
    eprint = "1809.09241",
    archivePrefix = "arXiv",
    primaryClass = "astro-ph.CO",
    doi = "10.1088/1475-7516/2019/04/012",
    journal = "JCAP",
    volume = "04",
    pages = "012",
    year = "2019"
}

@article{Klaer:2019fxc,
    author = "Klaer, Vincent B. and Moore, Guy D.",
    title = "{Global cosmic string networks as a function of tension}",
    eprint = "1912.08058",
    archivePrefix = "arXiv",
    primaryClass = "hep-ph",
    doi = "10.1088/1475-7516/2020/06/021",
    journal = "JCAP",
    volume = "06",
    pages = "021",
    year = "2020"
}

@article{Klaer:2017ond,
    author = "Klaer, Vincent B. . and Moore, Guy D.",
    title = "{The dark-matter axion mass}",
    eprint = "1708.07521",
    archivePrefix = "arXiv",
    primaryClass = "hep-ph",
    doi = "10.1088/1475-7516/2017/11/049",
    journal = "JCAP",
    volume = "11",
    pages = "049",
    year = "2017"
}

@article{Klaer:2017qhr,
    author = "Klaer, Vincent B. and Moore, Guy D.",
    title = "{How to simulate global cosmic strings with large string tension}",
    eprint = "1707.05566",
    archivePrefix = "arXiv",
    primaryClass = "hep-ph",
    doi = "10.1088/1475-7516/2017/10/043",
    journal = "JCAP",
    volume = "10",
    pages = "043",
    year = "2017"
}

@article{Copeland:2011dx,
    author = "Copeland, Edmund J. and Pogosian, Levon and Vachaspati, Tanmay",
    title = "{Seeking String Theory in the Cosmos}",
    eprint = "1105.0207",
    archivePrefix = "arXiv",
    primaryClass = "hep-th",
    doi = "10.1088/0264-9381/28/20/204009",
    journal = "Class. Quant. Grav.",
    volume = "28",
    pages = "204009",
    year = "2011"
}

@article{Hindmarsh:2021mnl,
    author = "Hindmarsh, Mark and Lizarraga, Joanes and Urio, Ander and Urrestilla, Jon",
    title = "{Loop decay in Abelian-Higgs string networks}",
    eprint = "2103.16248",
    archivePrefix = "arXiv",
    primaryClass = "astro-ph.CO",
    reportNumber = "HIP-2021-15/TH",
    doi = "10.1103/PhysRevD.104.043519",
    journal = "Phys. Rev. D",
    volume = "104",
    number = "4",
    pages = "043519",
    year = "2021"
}

@article{Hindmarsh:2017qff,
    author = "Hindmarsh, Mark and Lizarraga, Joanes and Urrestilla, Jon and Daverio, David and Kunz, Martin",
    title = "{Scaling from gauge and scalar radiation in Abelian Higgs string networks}",
    eprint = "1703.06696",
    archivePrefix = "arXiv",
    primaryClass = "astro-ph.CO",
    doi = "10.1103/PhysRevD.96.023525",
    journal = "Phys. Rev. D",
    volume = "96",
    number = "2",
    pages = "023525",
    year = "2017"
}

@article{Blanco-Pillado:2023sap,
    author = "Blanco-Pillado, Jose J. and Jim{\'e}nez-Aguilar, Daniel and Lizarraga, Joanes and Lopez-Eiguren, Asier and Olum, Ken D. and Urio, Ander and Urrestilla, Jon",
    title = "{Nambu-Goto dynamics of field theory cosmic string loops}",
    eprint = "2302.03717",
    archivePrefix = "arXiv",
    primaryClass = "hep-th",
    doi = "10.1088/1475-7516/2023/05/035",
    journal = "JCAP",
    volume = "05",
    pages = "035",
    year = "2023"
}

@article{Press:1989yh,
    author = "Press, William H. and Ryden, Barbara S. and Spergel, David N.",
    title = "{Dynamical Evolution of Domain Walls in an Expanding Universe}",
    reportNumber = "NSF-ITP-89-51, CFA-1870",
    doi = "10.1086/168151",
    journal = "Astrophys. J.",
    volume = "347",
    pages = "590--604",
    year = "1989"
}

@article{Garagounis:2002kt,
    author = "Garagounis, Theodore and Hindmarsh, Mark",
    title = "{Scaling in numerical simulations of domain walls}",
    eprint = "hep-ph/0212359",
    archivePrefix = "arXiv",
    reportNumber = "SUSX-TH-02-029",
    doi = "10.1103/PhysRevD.68.103506",
    journal = "Phys. Rev. D",
    volume = "68",
    pages = "103506",
    year = "2003"
}

@article{Hiramatsu:2010yz,
    author = "Hiramatsu, Takashi and Kawasaki, Masahiro and Saikawa, Ken'ichi",
    title = "{Gravitational Waves from Collapsing Domain Walls}",
    eprint = "1002.1555",
    archivePrefix = "arXiv",
    primaryClass = "astro-ph.CO",
    reportNumber = "ICRR-REPORT-559-2009-21, IPMU10-0024",
    doi = "10.1088/1475-7516/2010/05/032",
    journal = "JCAP",
    volume = "05",
    pages = "032",
    year = "2010"
}

@article{Ferreira:2023jbu,
    author = "Ferreira, Ricardo Z. and Gasparotto, Silvia and Hiramatsu, Takashi and Obata, Ippei and Pujolas, Oriol",
    title = "{Axionic defects in the CMB: birefringence and gravitational waves}",
    eprint = "2312.14104",
    archivePrefix = "arXiv",
    primaryClass = "hep-ph",
    reportNumber = "RUP-23-27",
    doi = "10.1088/1475-7516/2024/05/066",
    journal = "JCAP",
    volume = "05",
    pages = "066",
    year = "2024"
}

@article{Heilemann:2025iwv,
    author = "Heilemann, R. and Rosa, M. C. and Correia, J. R. C. C. C. and Martins, C. J. A. P.",
    title = "{Domain wall evolution beyond quartic potentials: The Sine-Gordon potential and Christ-Lee potentials}",
    eprint = "2504.11323",
    archivePrefix = "arXiv",
    primaryClass = "hep-ph",
    doi = "10.1103/rh2q-w1f6",
    journal = "Phys. Rev. D",
    volume = "111",
    number = "12",
    pages = "123550",
    year = "2025"
}

@article{Notari:2025kqq,
    author = "Notari, Alessio and Rompineve, Fabrizio and Torrenti, Francisco",
    title = "{The spectrum of gravitational waves from annihilating domain walls}",
    eprint = "2504.03636",
    archivePrefix = "arXiv",
    primaryClass = "astro-ph.CO",
    doi = "10.1088/1475-7516/2025/07/049",
    journal = "JCAP",
    volume = "07",
    pages = "049",
    year = "2025"
}

@article{Blasi:2025tmn,
    author = {Blasi, Simone and Mariotti, Alberto and Rase, A{\"a}ron and Vanvlasselaer, Miguel},
    title = "{Domain walls in the scaling regime: Equal Time Correlator and Gravitational Waves}",
    eprint = "2511.16649",
    archivePrefix = "arXiv",
    primaryClass = "hep-ph",
    month = "11",
    year = "2025"
}

@article{Zeldovich:1974uw,
    author = "Zeldovich, Ya. B. and Kobzarev, I. Yu. and Okun, L. B.",
    title = "{Cosmological Consequences of the Spontaneous Breakdown of Discrete Symmetry}",
    reportNumber = "SLAC-TRANS-0165, IPM-MOSCOW-15",
    journal = "Zh. Eksp. Teor. Fiz.",
    volume = "67",
    pages = "3--11",
    year = "1974"
}

@article{Lazanu:2015fua,
    author = "Lazanu, A. and Martins, C. J. A. P. and Shellard, E. P. S.",
    title = "{Contribution of domain wall networks to the CMB power spectrum}",
    eprint = "1505.03673",
    archivePrefix = "arXiv",
    primaryClass = "astro-ph.CO",
    doi = "10.1016/j.physletb.2015.06.034",
    journal = "Phys. Lett. B",
    volume = "747",
    pages = "426--432",
    year = "2015"
}

@article{Larsson:1996sp,
    author = "Larsson, Sebastian E. and Sarkar, Subir and White, Peter L.",
    title = "{Evading the cosmological domain wall problem}",
    eprint = "hep-ph/9608319",
    archivePrefix = "arXiv",
    reportNumber = "OUTP-96-11-P",
    doi = "10.1103/PhysRevD.55.5129",
    journal = "Phys. Rev. D",
    volume = "55",
    pages = "5129--5135",
    year = "1997"
}

@article{Correia:2014kqa,
    author = "Correia, J. R. C. C. C. and Leite, I. S. C. R. and Martins, C. J. A. P.",
    title = "{Effects of Biases in Domain Wall Network Evolution}",
    eprint = "1407.3905",
    archivePrefix = "arXiv",
    primaryClass = "hep-ph",
    doi = "10.1103/PhysRevD.90.023521",
    journal = "Phys. Rev. D",
    volume = "90",
    number = "2",
    pages = "023521",
    year = "2014"
}

@article{Correia:2017aqf,
    author = "Correia, J. R. C. C. C. and Martins, C. J. A. P.",
    title = "{General purpose graphics-processing-unit implementation of cosmological domain wall network evolution}",
    eprint = "1710.10420",
    archivePrefix = "arXiv",
    primaryClass = "physics.comp-ph",
    doi = "10.1103/PhysRevE.96.043310",
    journal = "Phys. Rev. E",
    volume = "96",
    number = "4",
    pages = "043310",
    year = "2017"
}

@article{Correia:2018tty,
    author = "Correia, J. R. C. C. C. and Leite, I. S. C. R. and Martins, C. J. A. P.",
    title = "{Effects of biases in domain wall network evolution. II. Quantitative analysis}",
    eprint = "1804.10761",
    archivePrefix = "arXiv",
    primaryClass = "astro-ph.CO",
    doi = "10.1103/PhysRevD.97.083521",
    journal = "Phys. Rev. D",
    volume = "97",
    number = "8",
    pages = "083521",
    year = "2018"
}

@article{Kitajima:2023cek,
    author = "Kitajima, Naoya and Lee, Junseok and Murai, Kai and Takahashi, Fuminobu and Yin, Wen",
    title = "{Gravitational waves from domain wall collapse, and application to nanohertz signals with QCD-coupled axions}",
    eprint = "2306.17146",
    archivePrefix = "arXiv",
    primaryClass = "hep-ph",
    reportNumber = "TU-1198",
    doi = "10.1016/j.physletb.2024.138586",
    journal = "Phys. Lett. B",
    volume = "851",
    pages = "138586",
    year = "2024"
}

@article{Kitajima:2023kzu,
    author = "Kitajima, Naoya and Lee, Junseok and Takahashi, Fuminobu and Yin, Wen",
    title = "{Stability of domain walls with inflationary fluctuations under potential bias, and gravitational wave signatures}",
    eprint = "2311.14590",
    archivePrefix = "arXiv",
    primaryClass = "hep-ph",
    reportNumber = "TU-1215",
    doi = "10.1088/1475-7516/2025/07/053",
    journal = "JCAP",
    volume = "07",
    pages = "053",
    year = "2025"
}

@article{Ferreira:2024eru,
    author = "Ferreira, Ricardo Z. and Notari, Alessio and Pujol{\`a}s, Oriol and Rompineve, Fabrizio",
    title = "{Collapsing domain wall networks: impact on pulsar timing arrays and primordial black holes}",
    eprint = "2401.14331",
    archivePrefix = "arXiv",
    primaryClass = "astro-ph.CO",
    reportNumber = "CERN-TH-2024-020",
    doi = "10.1088/1475-7516/2024/06/020",
    journal = "JCAP",
    volume = "06",
    pages = "020",
    year = "2024"
}

@article{Cyr:2025nzf,
    author = "Cyr, Bryce and Cotterill, Steven and Battye, Richard",
    title = "{Near-Peak Spectrum of Gravitational Waves from Collapsing Domain Walls}",
    eprint = "2504.02076",
    archivePrefix = "arXiv",
    primaryClass = "astro-ph.CO",
    month = "4",
    year = "2025"
}

@article{Babichev:2025stm,
    author = "Babichev, E. and Dankovsky, I. and Gorbunov, D. and Ramazanov, S. and Vikman, A.",
    title = "{Biased domain walls: faster annihilation, weaker gravitational waves}",
    eprint = "2504.07902",
    archivePrefix = "arXiv",
    primaryClass = "hep-ph",
    doi = "10.1088/1475-7516/2025/10/103",
    journal = "JCAP",
    volume = "10",
    pages = "103",
    year = "2025"
}

@article{Dankovsky:2025pjg,
    author = "Dankovsky, I. and Ramazanov, S. and Babichev, E. and Gorbunov, D. and Vikman, A.",
    title = "{Cosmic domain walls on a lattice: Illusive effects of initial conditions}",
    eprint = "2509.25367",
    archivePrefix = "arXiv",
    primaryClass = "hep-ph",
    doi = "10.1103/l9gp-7vp5",
    journal = "Phys. Rev. D",
    volume = "112",
    number = "12",
    pages = "123521",
    year = "2025"
}

@article{Vachaspati:2016abz,
    author = "Vachaspati, Tanmay",
    title = "{Creation of Magnetic Monopoles in Classical Scattering}",
    eprint = "1607.07460",
    archivePrefix = "arXiv",
    primaryClass = "hep-th",
    doi = "10.1103/PhysRevLett.117.181601",
    journal = "Phys. Rev. Lett.",
    volume = "117",
    number = "18",
    pages = "181601",
    year = "2016"
}

@inproceedings{Vachaspati:1997rr,
    author = "Vachaspati, Tanmay",
    title = "{Formation of topological defects}",
    booktitle = "{ICTP Summer School in High-Energy Physics and Cosmology}",
    eprint = "hep-ph/9710292",
    archivePrefix = "arXiv",
    reportNumber = "CWRU-P18-97",
    pages = "446--479",
    month = "10",
    year = "1997"
}

@article{Achucarro:1999it,
    author = "Achucarro, Ana and Vachaspati, Tanmay",
    title = "{Semilocal and electroweak strings}",
    eprint = "hep-ph/9904229",
    archivePrefix = "arXiv",
    reportNumber = "EHU-FT-9808A, CWRU-P34-1998",
    doi = "10.1016/S0370-1573(99)00103-9",
    journal = "Phys. Rept.",
    volume = "327",
    pages = "347--426",
    year = "2000"
}
  
\bibliographystyle{h-physrev4}
\end{multicols}

\end{document}